# What's the (Dark) Matter with Cosmological Bubbles?

by

**Moritz Breitbach**

born in Mainz, Germany

A dissertation submitted for the award of the title

*Doctor of Natural Sciences*

to the Faculty of Physics, Mathematics and Computer Science
of the Johannes Gutenberg University Mainz

April 28, 2022

# Abstract

Despite their tremendous successes, modern-day cosmology and particle physics harbor a variety of unresolved mysteries. Two of the biggest are the origin of the baryon asymmetry of the Universe and the existence and nature of dark matter. In the present thesis, the author addresses these topics in various ways. The first part of the thesis is concerned with cosmological first-order phase transitions that may have occurred shortly after the Big Bang. Such transitions proceed via the nucleation and expansion of true vacuum bubbles and give rise to a rich phenomenology. The author suggests a mechanism to simultaneously explain the baryon asymmetry and dark matter, based on the out-of-equilibrium dynamics at the boundary of a dark phase transition with large order parameter. The same class of phase transitions can, in the parameter regime of small dark matter Yukawa couplings, lead to the production of primordial black holes via the compression of the plasma in shrinking false vacuum regions, as the author demonstrates with a sophisticated numerical simulation. In a third project regarding cosmological phase transitions, the author investigates the possibility of sub-MeV hidden sectors that are decoupled from the remaining plasma and cold enough to be reconciled with cosmological constraints, but at the same time give rise to a detectable gravitational-wave spectrum produced during bubble collisions. In the second part of the thesis, the author assesses the prospects for new physics searches at the *DUNE* near detector, focusing on the *DUNE-PRISM* concept, which suggests consecutive measurements at different on- and off-axis positions. This setup achieves improved signal-to-background ratios and reduces systematic uncertainties.

# List of Publications

[1] M.J. Baker, **M. Breitbach**, J. Kopp, L. Mittnacht, and Y. Soreq, *Filtered Baryogenesis*, *JHEP* **08** (2022) 010 [2112.08987].

Abstract: We propose a new mechanism to simultaneously explain the observed dark matter abundance and the baryon asymmetry of the Universe. The mechanism is based on the Filtered Dark Matter scenario, where dark matter particles acquire a large mass during a first-order phase transition. This implies that only a small fraction of them are energetic enough to enter the advancing true vacuum bubbles and survive until today, while the rest are reflected and annihilate away quickly. We supplement this scenario with a CP-violating interaction, which creates a chiral asymmetry in the population of dark matter particles. In the false vacuum phase, a portal interaction quickly converts the dark sector chiral asymmetry into a Standard Model lepton asymmetry. The lepton asymmetry is then partially converted to a baryon asymmetry by standard electroweak sphaleron processes. We discuss the dependence of the generated asymmetry on the parameters of the model for two different portal interactions and demonstrate successful baryogenesis for both. For one of the portals, it is also possible to simultaneously explain the observed dark matter abundance, over many orders of magnitude in the dark matter mass.

[2] M.J. Baker, **M. Breitbach**, J. Kopp, and L. Mittnacht, *Primordial Black Holes from First-Order Cosmological Phase Transitions*, [2105.07481].

Abstract: We discuss the possibility of forming primordial black holes during a first-order phase transition in the early Universe. As is well known, such a phase transition proceeds through the formation of true-vacuum bubbles in a Universe that is still in a false vacuum. When there is a particle species whose mass increases significantly during the phase transition, transmission of the corresponding particles through the advancing bubble walls is suppressed. Consequently, an overdensity can build up in front of the walls and become sufficiently large to trigger primordial black hole formation. We track this process quantitatively by solving a Boltzmann equation, and we determine the resulting black hole density and mass distribution as a function of the phenomenological parameters of the phase transition.

[3] M.J. Baker, **M. Breitbach**, J. Kopp, and L. Mittnacht, *Detailed Calculation of Primordial Black Hole Formation During First-Order Cosmological Phase Transitions*, [2110.00005].

Abstract: We recently presented a new mechanism for primordial black hole formation during a first-order phase transition in the early Universe, which relies on the build-up of particles which are predominantly reflected from the advancing bubble wall. In this companion paper we provide details of the supporting numerical calculations. After describing the general mechanism, we discuss the criteria that need to be satisfied for a black hole to form. We then set out the Boltzmann equation that describes the evolution of the relevant phase space distribution function, carefully describing our treatment of the Liouville operator and the collision term. Finally, we show that black holes will form for a wide range of parameters.

[4] **M. Breitbach**, J. Kopp, E. Madge, T. Opferkuch, and P. Schwaller, *Dark, Cold, and Noisy: Constraining Secluded Hidden Sectors with Gravitational Waves*, *JCAP* **07** (2019) 007 [1811.11175].

ABSTRACT: We explore gravitational-wave signals arising from first-order phase transitions occurring in a secluded hidden sector, allowing for the possibility that the hidden sector may have a different temperature than the Standard Model sector. We present the sensitivity to such scenarios for both current and future gravitational-wave detectors in a model-independent fashion. Since secluded hidden sectors are of particular interest for dark matter models at the MeV scale or below, we pay special attention to the reach of pulsar timing arrays. Cosmological constraints on light degrees of freedom restrict the number of sub-MeV particles in a hidden sector, as well as the hidden sector temperature. Nevertheless, we find that observable first-order phase transitions can occur. To illustrate our results, we consider two minimal benchmark models: a model with two gauge singlet scalars and a model with a spontaneously broken $U(1)$ gauge symmetry in the hidden sector.

[5] **M. Breitbach**, L. Buonocore, C. Frugiuele, J. Kopp, and L. Mittnacht, *Searching for Physics Beyond the Standard Model in an Off-Axis DUNE Near Detector*, *JHEP* **01** (2022) 048 [2102.03383].

ABSTRACT: Next generation neutrino oscillation experiments like DUNE and T2HK are multi-purpose observatories, with a rich physics program beyond oscillation measurements. A special role is played by their near detector facilities, which are particularly well-suited to search for weakly coupled dark sector particles produced in the primary target. In this paper, we demonstrate this by estimating the sensitivity of the DUNE near detectors to the scattering of sub-GeV dark matter particles and to the decay of sub-GeV sterile neutrinos ("heavy neutral leptons"). We discuss in particular the importance of the DUNE-PRISM design, which allows some of the near detectors to be moved away from the beam axis. At such off-axis locations, the signal-to-background ratio improves for many new physics searches. We find that this leads to a dramatic boost in the sensitivity to boosted dark matter particles interacting mainly with hadrons, while for boosted dark matter interacting with leptons, data taken on-axis leads to marginally stronger exclusion limits. Searches for heavy neutral leptons perform equally well in both configurations.

# Table of Contents

# Prologue

# 1 Introduction

A young physics student in the 21st century is in a truly luxurious position. He or she is presented with a comprehensive theoretical understanding of the fundamental interactions and building blocks of nature down to the subatomic level and with the knowledge of the chronology of our Universe that reaches back to split seconds after it came into existence. The success of modern-day science is based on the pioneering work of 20th century's great minds, who revolutionized our understanding of the world with their novel theories of quantum mechanics and relativity. This paradigm shift was the prerequisite for the great theoretical advances during the second half of the 20th century, where the tremendously successful Standard Model of Particle Physics (SM) and the Lambda-cold-dark-matter Theory ($\Lambda$CDM) were established to describe the microscopic world and the Universe as a whole. The SM is a relativistic quantum field theory that covers three of the four fundamental forces and all known particles. Various particle colliders – the most complex machines ever built – have confirmed the predictions of the SM and are still in operation to determine its parameters at ever-increasing precision. The $\Lambda$CDM model, which emerged in the late 1990s, describes the evolution of the Universe and is based on and supported by the structure of the cosmic microwave background (CMB), the large-scale distribution of galaxies, the existence of dark matter (DM), the observed abundances of light elements, and the accelerating metric expansion.

Despite the unprecedented successes of these theories, there are yet a variety of great mysteries that long to be resolved. While some of them are more of theoretical nature, such as the unknown origin of the hierarchy between the forces, others are more tangible and concern experimental data, like the magnetic moment of the muon which apparently differs from theoretical predictions or the Hubble constant whose measured value varies based on the applied technique. One of the greatest open questions is how the baryon asymmetry of the Universe (BAU) – to which we owe our existence – came about. Another big puzzle is posed by the apparent existence of DM, which constitutes the majority of the Universe's matter content but still remains elusive, theoretically and experimentally. In the present work, we address both of the aforementioned topics.

As fruitful and important as the running of particle colliders such as the *LHC* has been, the technology is being pushed to the limits of what is economically and practically feasible. It is thus questionable if the same approach which has experimentally established the SM will reach the energies that are potentially required to confirm a grand unified theory of fundamental forces or other mechanisms that might be responsible for some of the unexplained phenomena. It is hence exactly at the right time, that a new window to the early Universe was pushed open by the first detection of gravitational waves (GWs) – tiny ripples in spacetime – that originated from the merger of two black holes, by the *LIGO* interferometer. While the mere existence of gravitational radiation confirms Einstein's prediction, tracking the dynamics of massive binaries will allow for precision tests of general relativity. A combination of (future) earth-bound and space-based GW observatories, as well as pulsar timing arrays (PTAs) will map out the GW background present in the Universe across a wide range of frequencies. The resulting power spectrum may carry imprints from mechanisms that were at work during the Big Bang, even long before the epoch of recombination and when the Universe was correspondingly much hotter. This provides a possibility to look into the past even beyond the CMB, to test theories at energies that are far beyond the reach of colliders, or to shed light on dark sectors which might otherwise not interact with the SM at all.

One class of phenomena that may have induced a detectable GW background are cosmological first-order phase transitions (PTs): In the history of the expanding and cooling Universe, the energetically favorable state of the vacuum changes. Under certain circumstances, the transition from the old to the new state occurs abruptly, such that multiple phases temporarily coexist in different patches of the Universe. The energetically favorable phase spreads in the form of bubbles, whose walls eventually collide and thereby source gravitational radiation. The SM predicts two PTs – the breaking of the electroweak symmetry via the Higgs mechanism and the confinement of quarks – which are, however, not first-order. Minimal extensions to the SM can render these transitions first-order or give rise to first-order PTs that occur entirely in dark sectors.

Beyond the possibility to probe such PTs with future GW observatories, they give rise to a rich phenomenology and serve as a basis for new models and theories. As already alluded to, first-order transitions proceed via the nucleation and expansion of bubbles. The boundaries, or walls, of these bubbles may interact with the surrounding plasma and lift particles out of their thermal equilibrium state. This allows for interesting new mechanisms and furthermore provides one of the necessary ingredients to solve the mystery of the BAU. Furthermore, PTs can act as triggers that switch certain interactions on and off and may thereby control and explain how the abundance of DM came about. We apply these techniques in the present work in order to address the BAU and DM puzzles.

This thesis is structured as follows: In Chapter 2, we briefly review the most relevant features and the shortcomings of the SM. The first part of the thesis is devoted to cosmological PTs and applications thereof. We review the theoretical formalism of the effective scalar potential, the mechanics of first-order PTs, and the detection prospects of the associated GW spectrum in Chapter 3. In Chapter 4, we discuss the possibility of generating the BAU and the DM abundance simultaneously at the bubble walls of a PT. In Chapter 5, we propose a novel mechanism for the formation of primordial black holes (which may constitute DM) based on the squeezing of matter between multiple expanding bubbles. In Chapter 6, we investigate how light and cold dark sectors can be reconciled with cosmological observations and if associated dark PTs could be probed by future GW observatories. The second part of the thesis deals with the *DUNE* neutrino experiment, which is currently under construction, with a brief overview of *DUNE* and its science program being given in Chapter 7. In Chapter 8, we evaluate the possibility of probing different DM models at the *DUNE* near detector without interfering with its original purpose. The detector will be mounted on a movable platform, allowing for measurements off the beam axis. We asses the impact of this feature on the sensitivity to new physics.

# 2 The Standard Model and Its Limitations

## 2.1 The Standard Model of Particle Physics

The Standard Model of Particle Physics (SM) is a theory that describes all known elementary particles and their microscopic interactions. It is a relativistic quantum field theory (QFT) in 3+1 dimensions and is built around the symmetry group

$$\underbrace{SU(3)_c}_{\text{strong force}} \times \underbrace{SU(2)_L \times U(1)_Y}_{\text{electroweak force}}, \tag{2.1}$$

which gives rise to the fundamental forces, excluding gravity. The model contains three generations of fermionic fields with intrinsic spin-1/2, a set of spin-1 gauge fields (the symmetry generators), and one spin-0 scalar field (the Higgs). Excitations of these fields are what we call particles, where the fermions constitute regular matter and gauge bosons mediate interactions.

The behavior of the SM fields is described by the Lagrangian density

$$\mathcal{L}_{\text{SM}} = \mathcal{L}_{\text{kin}} + \mathcal{L}_{\text{Yuk}} + \mathcal{L}_{\text{Higgs}}. \tag{2.2}$$

The kinetic term $\mathcal{L}_{\text{kin}}$ contains covariant derivatives of fields, giving rise to their dynamics as well as gauge interactions. The Yukawa term $\mathcal{L}_{\text{Yuk}}$ couples the Higgs field to the fermions. Finally, $\mathcal{L}_{\text{Higgs}}$ contains the Higgs self-couplings and therefore represents the (negative) tree-level Higgs potential.

A lot can be learned about the structure of the SM by just considering the fields and gauge charges of the theory, listed in Table 2.1:

1. Gluons themselves carry $SU(3)_c$ color charges and are thus self-interacting. This gives rise to virtual gluon pairs that dilute bare color charges and results in an anti-screening effect: The strong force is *asymptotically free*, i.e. weaker at short distances or high energies. At low energies, on the other hand, quarks confine to color-neutral objects (baryons and mesons), which is why the nuclear force is short-ranged. In the history of the Universe, the condensation of quarks marks the phase transition (PT) of quantum chromodynamics (QCD) which occurs at a temperature of $\sim 150\,\text{MeV}$. This PT proceeds smoothly, but becomes a first-order transition in some extended versions of the SM [6, 7].

2. Naively, a similar argument could be made for a confining electroweak sector. However, it turns out that the related confinement scale is much smaller than the scale of electroweak symmetry breaking, such that a confining phase never emerges [8].

|  | Name | Field | Charges | | |
|---|---|---|---|---|---|
|  |  |  | $SU(3)_c$ | $SU(2)_L$ | $U(1)_Y$ |
| Spin $\frac{1}{2}$ | quarks | $Q_L^i$ | **3** | **2** | 1/6 |
|  |  | $d_R^i$ | **3** | **1** | −1/3 |
|  |  | $u_R^i$ | **3** | **1** | 2/3 |
|  | leptons | $\ell_L^i$ | **1** | **2** | −1/2 |
|  |  | $e_R^i$ | **1** | **1** | −1 |
| Spin 1 | gluons | $G$ | **8** | **1** | 0 |
|  | electroweak bosons | $W$ | **1** | **3** | 0 |
|  |  | $B$ | **1** | **1** | 0 |
| Spin 0 | Higgs doublet | $H$ | **1** | **2** | 1/2 |

**Table 2.1:** Fields and their gauge charges in the unbroken electroweak phase of the SM. The three fermion generations are represented by the index $i = 1, 2, 3$, the subscripts $L$ and $R$ denote left- and right-handed chiralities, and the **8**, **3**, **2** and **1** stand for octet, triplet, doublet and singlet (i.e. uncharged) representations, respectively. The gauge representations imply a certain multiplicity (there are 8 gluons, for instance).

3. Only particles of left-handed chirality are charged under $SU(2)_L$. This sector of the SM is thus maximally parity-violating [9]. As a consequence, bare mass terms (which couple left- to right-chiral fields) are not gauge invariant and thus forbidden. The gauge bosons are intrinsically massless.

4. The Higgs field has the appropriate structure to couple left- and right-chiral fields to one another in the form of the Yukawa couplings.

**Electroweak symmetry breaking.** The intrinsic absence of fermion and gauge boson masses in the SM is contradictory to what can be observed in nature. The fact that fermions and the weak gauge bosons are indeed massive is attributed to the Higgs mechanism: One degree of freedom (DOF) of the Higgs field has a finite and constant vacuum expectation value (VEV), $v_H$, which *spontaneously breaks* (i.e. "hides") the electroweak symmetry and leaves behind electromagnetism as a residual:

$$\underbrace{SU(2)_L \times U(1)_Y}_{\text{electroweak force}} \rightarrow \underbrace{U(1)_{\text{em}}}_{\text{electromagnetic force}} . \tag{2.3}$$

Expanding the Lagrangian around the broken ground state reveals that only one component of the Higgs doublet, the Higgs boson $h$, is massive. The remaining DOFs are massless Goldstone bosons and take the role of longitudinal gauge boson modes [10]. This gives rise to three massive gauge bosons, $W^{\pm}$ and $Z$, which are mixtures of the gauge bosons of $SU(2)_L \times U(1)_Y$ and mediate the short-range weak force. The remaining superposition of $W$ and $B$ is the massless photon $A$ that corresponds to the residual long-range elec-tromagnetic force. Finally, the Higgs VEV also generates fermion masses via the Yukawa interaction. The fields and charges in the phase of broken electroweak symmetry are listed in Table 2.2.

|  | Name | Field | Charges | |
|---|---|---|---|---|
|  |  |  | $SU(3)_c$ | $U(1)_{\text{em}}$ |
| Spin $\frac{1}{2}$ | quarks | $d_L^i, d_R^i$ | **3** | $-1/3$ |
|  |  | $u_L^i, u_R^i$ | **3** | $2/3$ |
|  | leptons | $e_L^i, e_R^i$ | **1** | $-1$ |
|  |  | $\nu_L^i$ | **1** | $0$ |
| Spin 1 | gluons | $G$ | **8** | $0$ |
|  | weak bosons | $W^\pm$ | **1** | $\pm 1$ |
|  |  | $Z$ | **1** | $0$ |
|  | photon | $A$ | **1** | $0$ |
| Spin 0 | Higgs boson | $h$ | **1** | $0$ |

**Table 2.2:** Fields and their gauge charges after the breaking of the electroweak symmetry.

When the Universe was younger than $\sim 10\,\text{ps}$, thermal corrections made the scalar potential parabolic such that the electroweak symmetry was intact.[1] When the particle plasma in the expanding Universe cooled below the temperature of $\sim 100\,\text{GeV}$, the tree-level Higgs potential – which favors a finite VEV – started to dominate and the electroweak symmetry has been broken. Lattice calculations have shown that the transition between these two phases occurred smoothly and simultaneously everywhere in the Universe. An abrupt electroweak phase transition (EWPT), on the other hand, is predicted in various extensions to the SM [11–16]. We discuss the theory and phenomenology of such first-order PTs in Chapter 3.

**Fermion mass mixing.** The three generations of fermion fields appear both in the kinetic terms as well as the Yukawa part of the SM Lagrangian. It is fundamentally allowed that the mass eigenstates (which propagate freely) are linear combinations of the flavor eigenstates (which is the interacting basis). This mismatch between mass and flavor states can be eliminated *individually* in the upper and lower components of the fermion $SU(2)$ doublets. However, the charged weak bosons, $W^\pm$, give rise to flavor-changing currents that connect the components of the doublets and thereby make it impossible to eliminate the mismatch entirely [17]. As a result, flavor-changing weak interactions allow for a mixing among the three quark generations, an effect observed at particle colliders. The mass mixing is described by the $3 \times 3$ unitary *CKM matrix*, which can be parameterized by three mixing angles and one complex phase that violates charge-parity symmetry (CP) [18]. The observation of neutrino oscillations [19, 20] – implying neutrino masses that are not part of the SM – suggests a similar mechanism in the lepton sector, analogously described by the *PMNS matrix* [21]. Measurements show that the CKM matrix is almost diagonal,

---

[1] This assumes that the reheating temperature after inflation was sufficiently large.

whereas the PMNS matrix encompasses a much more sizable mixing [22, 23]. While the CP-violating phase of the CKM matrix is known to a precision of $\sim 5°$, it is almost undetermined in case of the PMNS matrix but will be a subject of future neutrino experiments such as *DUNE*, which we discuss in Chapter 7.

## 2.2 Unsolved Puzzles

The SM is tremendously successful in describing the fundamental interactions of nature. Famously, its prediction of the electron magnetic dipole moment perfectly agrees with experimental data at a precision of 1 per $10^{12}$ [24]. The SM (and its preceding theories) presaged the existence of the weak bosons, the gluons, the heavy quarks, as well as the Higgs boson, which have all been confirmed experimentally later on [25, 26]. However, there is a variety of open questions and experimental anomalies which strongly suggest the existence of physics beyond the SM and new fundamental theories in general, some of which are:

- ◆ How are the neutrino masses generated [19]?
- ◆ What protects the cosmological constant and the Higgs mass from Planck-scale corrections [27, 28]?
- ◆ Is the current vacuum state of the Universe stable [29]?
- ◆ Is there a theory that unifies all fundamental forces and reduces the number of fundamental constants [30]?
- ◆ How can gravity be reconciled with QFT [31]?
- ◆ Why does the QCD sector conserve CP (aka "strong CP problem") [32]?
- ◆ What causes the observed accelerated expansion of the Universe [33]?
- ◆ Why is there a discrepancy between theory prediction and measurement of the muon magnetic moment [34]?

Two further puzzles that are of particular importance in the context of the present thesis are presented in the sections below.

### 2.2.1 The Baryon Asymmetry of the Universe

"Why is there something rather than nothing?" is a famous philosophical question, which was declared "the fundamental question of metaphysics" by Martin Heidegger. In the realm of *actual physics*, there exists a quite analogous unanswered question: "Why is there matter rather than just radiation?" The fact that we are surrounded by and made of matter, while antimatter appears to be entirely absent, raises the question how and when this asymmetry emerged. During the inflationary phase of the Big Bang, any preexisting asymmetry has been diluted, so naively one would expect equal amounts of matter and antimatter which would ultimately annihilate to leave only radiation behind. Reality is different: Baryonic matter is all around us and gives rise to the impressive complexity of our Universe. As can be inferred from the observed light element abundances produced during Big Bang nucleosynthesis (BBN) and the analyses of anisotropies in the cosmic microwave background (CMB), there are [35]

$$\frac{n_B - n_{\bar{B}}}{n_\gamma} \approx 6.2 \times 10^{-10} \tag{2.4}$$

baryons over antibaryons per photon in today's Universe, where $n_B$, $n_{\bar{B}}$, and $n_\gamma$ denote the number densities of baryons, antibaryons, and photons, respectively. A more useful quantification of the baryon asymmetry of the Universe (BAU) is based on a normalization w.r.t. the comoving entropy density $s$, which dilutes in the same way as matter in course of the expansion of the Universe. The quantity [36]

$$Y_B \equiv \frac{n_B - n_{\bar{B}}}{s} \approx 8.7 \times 10^{-11} \tag{2.5}$$

is constant and applies today as well as during the hot Big Bang. One solution to the puzzle would be the existence of entire galaxies made of antimatter, such that the asymmetry exists only locally but averages out over larger distances. However, no gamma radiation is observed that would indicate annihilation occurring at the matter–antimatter interfaces, which rules out this scenario [37]. The seeming absence of antimatter in today's Universe implies $n_{\bar{B}} \ll n_B$.

Sakharov famously derived three necessary conditions for the emergence of a baryon–antibaryon asymmetry [38]:

1. Violation of baryon number.

2. Violation of C and CP.

3. Departure from thermal equilibrium.

The first condition is obvious but necessary to demand, as the baryon number (and lepton number) is accidentally[2] conserved in the SM at the classical level. However, the number of baryons-plus-leptons is anomalous, i.e. not conserved in the full quantum theory [40]: In the electroweak sector of the SM, an infinite number of degenerate but topologically distinct vacuum states coexist. The different vacua are connected by non-perturbative, unstable solutions to the electroweak field equations – *electroweak sphalerons* – which give rise to processes that violate baryon-plus-lepton number [41]. The first of Sakharov's conditions is hence provided by the SM, at least prior to electroweak symmetry breaking, when the sphaleron transitions were efficient [42].

The second condition is also easy to comprehend: If charge symmetry (C) would be conserved, every process that produces a baryon would be countered by one that produces an antibaryon, while the conservation of CP would imply equal numbers of left-chiral baryons and right-chiral antibaryons. While the non-conservation of C is a characteristic feature of the electroweak sector – as can be seen by the lack of C-conjugated weak decays [9] – the violation of CP is less obvious. The latter has been observed experimentally in decays of various types of mesons [43–48] and originates from a complex phase in the mass mixing of quarks via flavor-changing weak interactions (described by the CKM matrix within the SM). Recent neutrino oscillation measurements hint at a violation in the lepton sector (a complex phase in the PMNS matrix) as well [49], but the precise amount is yet to be determined. Overall, however, the amount of CP violation in the SM is far from sufficient to account for the entire observed baryon asymmetry [50].

---
[2]A global symmetry is called accidental if every allowed renormalizable operator respects it [39].

Finally, the third condition is required because in a thermal equilibrium the charge-parity-time symmetry (CPT) – which follows from Lorentz invariance – ensures a balance between interactions and their inverses on average, preventing any net baryon number from emerging. In the history of the expanding Universe, particle species drop out of equilibrium when their interaction rates drop below the Hubble rate, which allows for baryogenesis in the context of grand unified theorys (GUTs) [51]. Another phenomenon involving out-of-equilibrium dynamics are cosmological first-order PTs, which are discussed in detail in Chapter 3. In such a PT, the phase boundaries advance through space and interact with the surrounding plasma, which removes some of the particles from thermal equilibrium. While the SM is found to feature no first-order transitions [6, 7, 11–16], they can be achieved via minimal theoretical extensions. If the EWPT was of first order and additional CP violation was present, the BAU could be explained via electroweak baryogenesis (EWBG) [52–54].

In Chapter 4, we present a novel mechanism that produces the observed BAU at the bubble walls of a dark sector first-order PT – partly analogous to conventional EWBG – and sets the observed dark matter (DM) abundance at the same time.

### 2.2.2 Dark Matter

DM is a phenomenon that might sound a lot like science fiction to a non-specialist. While the detailed nature of DM is indeed still "fictional", there is compelling evidence for the existence of an unidentified energy content in the Universe that behaves like matter, but interacts only very weakly, if at all, through the forces described by the SM. First hints for DM were found over a century ago, when it was observed that the velocity dispersion of stars in the Milky Way deviates from what the visible mass alone would imply [55]. Since then, this phenomenon has been observed in the dynamics of numerous galaxies and galaxy clusters and suggests the existence of large spherical DM halos in which galaxies are embedded [56–58]. The presence of DM in galaxy clusters and in the large-scale structure (LSS) of the Universe was furthermore confirmed via strong [59,60] and weak [61] gravitational lensing, respectively. In the history of the expanding Universe, DM played an important role in structure formation: Without DM, any density perturbation would have been washed out during the era of radiation domination. It requires the gravitational wells provided by DM – which does not interact with radiation – to form galaxies and clusters fast enough. The distinct evolution of regular vs. dark matter is well described by the Lambda-cold-dark-matter Theory (ΛCDM) and leaves visible imprints in the observed baryon acoustic oscillations (BAO) and in the power spectrum of the CMB, confirming the existence of DM [62]. Finally, the light element abundances generated during BBN are sensitive to the amount of visible matter and constrain it to be less than the total amount of observed matter [63].

Despite these numerous unambiguous but indirect hints for the existence of DM, its detailed nature and properties are yet undiscovered. What we know about DM can be summarized as follows:

- ◆ Approximately 27% of the total energy and 85% of matter in the Universe is constituted by DM [36].

- ◆ In our proximity, DM has a density and velocity (due to the motion of our solar system) of roughly $0.3\,\text{GeV}/\text{cm}^3$ and $200\,\text{km/s}$, respectively [64]. Since we live in a DM halo, the local density is $\sim 10^5$ times larger than the global average.

- ◆ DM must be electrically neutral, except for a possible millicharge [65, 66].

- ◆ DM must be non-relativistic ("cold") and have negligible pressure to not spoil bottom-up structure formation where galaxies form *before* galaxy clusters, as is observed in our Universe. This rules out the light SM neutrinos as DM candidates.

- ◆ DM must have a lifetime of at least the age of the Universe.

- ◆ DM must interact only very weakly with matter and with itself, as can be inferred, for instance, from observations of the Bullet Cluster [67].

While an exhaustive review of the topic is beyond the scope of this thesis, we list some of the most prominent DM candidates and theories in the following.

**Weakly interacting massive particles (WIMPs).** This theorized form of DM, which is a prediction of supersymmetry [68], was the longstanding favorite candidate because of its natural properties: A particle with a mass and annihilation cross section typical for the SM electroweak sector – roughly $100\,\text{GeV}$ and $3\times 10^{-26}\,\text{cm}^3/\text{s}$ – would yield the observed relic density via thermal freeze-out, i.e. a departure from thermal equilibrium which fixes the DM density and typically occurs when the temperature of the Universe drops below the DM mass by a factor of 20 [69]. In recent years, however, the simplest forms of supersymmetry and the vanilla WIMP have been severely constrained by collider searches [70] and direct detection experiments [71], respectively. For this reason, a much wider mass range and non-standard production mechanisms – e.g. via freeze-in [72], co-annihilation [73], or first-order PTs [74–77] – are nowadays being discussed.

**Massive astrophysical compact halo objects (MACHOs).** DM could be constituted of non-luminous macroscopic objects such as black holes (BHs), neutron stars, brown dwarfs, white dwarfs, or faint red dwarfs. Searches for this form of DM are in general based on gravitational lensing [78]. Baryonic MACHOs are mostly ruled out as an explanation for DM by observations of the BAO, the CMB, and the products of BBN [79–81]. This leaves primordial black holes (PBHs) formed of non-baryonic matter as viable candidates, which could account for the entirety of DM if they weigh around $10^{-13}$ solar masses [82].

- **Sterile neutrinos.** While SM neutrinos have proven to be too light, there is the possibility that heavy, right-handed SM singlets – "sterile neutrinos" or "heavy neutral leptons" – constitute DM. Their existence is motivated by the seesaw mechanism, which provides an explanation for the SM neutrinos masses [83]. These are not part of the SM but necessary to explain their experimentally observed oscillating behavior [19].

- **Axions and axion-like particles (ALPs).** The pseudoscalar axion particle, which could have masses up to $\sim$ eV or be much lighter, is highly motivated by the strong CP problem of QCD to which it provides an elegant dynamic solution [64, 84].

- **Modified Newtonian dynamics (MOND).** An entirely different explanation for the observed phenomena could be adjustments to the known laws of gravity [85–87]. These theorized modifications would affect the long-range behavior (galaxy scale) of gravity but leave the dynamics on shorter distances (solar system) unchanged. However, this approach has trouble explaining the CMB anisotropies as well as the behavior of galaxy clusters [88]. Furthermore, the fact that gravitational waves (GWs) travel at light speed – as multi-messenger observation of a neutron star merger suggest – rules out the majority of MOND theories [89].

The DM mass, $m_{\text{DM}}$, can be narrowed down based on theoretical considerations:

- ◆ $m_{\text{DM}} \gtrsim 10^{-21}$ eV for scalar DM, which would otherwise spoil the formation of the smallest observed structures – dwarf galaxies – due to the uncertainty principle [90].

- ◆ $m_{\text{DM}} \gtrsim 100$ eV for fermionic DM, which is also set by the observation of dwarf galaxies, but under the consideration of the Pauli exclusion principle [91].

- ◆ $m_{\text{DM}} \gtrsim 10$ keV for thermally produced DM, which would otherwise be "warm" and wash out small-scale structures [92].

- ◆ $m_{\text{DM}} \gtrsim 10$ MeV for thermally produced DM, to be compatible with the measured anisotropies of the CMB and the observed light element abundances produced during BBN [93].

- ◆ $m_{\text{DM}} \gtrsim 1$ GeV for thermally produced fermionic DM, to avoid an overclosure of the Universe via thermal freeze-out [94]. This bound can be evaded if the DM couples via additional mediators [95].

- ◆ $m_{\text{DM}} \lesssim 100$ TeV for thermally produced DM. Above this *Griest–Kaminkowski bound*, obtaining the observed relic density via thermal freeze-out would require unitarity-violating DM cross sections [96].

- ◆ $m_{\text{DM}}$ must not exceed ten solar masses to prevent a disruption of dwarf galaxies [97].

Tremendous experimental efforts have been made in order to establish a direct proof of DM. The various approaches – ranging from the attempt of a direct detection via the recoil of atomic nuclei or electrons [98–101], over the search for gamma and cosmic rays as DM annihilation or decay products [98, 102–104], to missing energy searches at particle colliders [105–108] – predominantly aim at the detection of WIMP or WIMP-like DM. Despite some isolated and inconclusive hints [109–111], none of these attempts has led to a clear discovery up to this day [112–114]. Instead, increasingly strong constraints are being placed, especially in the MeV to TeV WIMP landscape.

The DM problem is approached in various ways in the present thesis. In Chapter 4, we develop a baryogenesis mechanisms that is based on the Filtered DM scenario, where the relic abundance settles non-thermally at the boundary of a cosmological first-order PT. The mechanism can explain DM over a wide mass range that extends far beyond the discussed Griest–Kaminkowski bound. Also based on cosmological PTs, in Chapter 5 we propose a novel mechanism for the formation of PBHs, which may constitute parts or even the entirety of DM in a certain mass regime. In Chapter 6, we investigate the generic possibility of decoupled dark sectors with sub-MeV particle masses in the light of cosmological constraints (from BBN and the CMB) and survey the detection prospects of GWs produced in a related dark PT. Finally, in Chapter 8, we compute the expected sensitivity of the near detector of *DUNE* – which is currently under constructions – with regards to light DM mediated by a dark photon as well as leptophobic DM.

# Part I

# Cosmological Phase Transitions

# 3 Theoretical Background

A phase transition (PT) is a thermodynamical phenomenon that describes the change of characteristic properties of a medium, e.g. the state of matter or the magnetization, as a result of changing external conditions, such as temperature or pressure. Typical examples are the boiling of water, the demagnetization of a ferromagnet, the emergence of superconductivity, or the crystallization of a supercooled liquid. Formally, a PT can be characterized by the non-smoothness of a system's free energy w.r.t. an external variable and it is often accompanied by the restoration or breaking of a symmetry. For instance, the continuous translational symmetry in a liquid is reduced to a set of discrete symmetries in a solid crystal, giving rise to Goldstone bosons in the form of collective excitations ("phonons"). As its name suggests, the *order parameter* of a system characterizes its structure, and is, for instance, given by the density or magnetization. The (non-)smoothness of the order parameter classifies the PT: A first-order PT is the most abrupt form of a transition and defined by a discontinuous change of the order parameter. Discontinuities in derivatives of the order parameter indicate second- and higher-order transitions.

The physics of PTs – usually a topic of solid-state physics and thermodynamics – is of great relevance also in the context of the evolution of the early Universe. If the inflaton field reheated the Universe to sufficiently high temperatures, it is expected that the vacuum underwent the electroweak phase transition (EWPT) at a temperature $T \sim 100\,\text{GeV}$, breaking the electroweak symmetry. Another PT is the condensation of quarks when the Universe entered the confining phase of quantum chromodynamics (QCD) at $T \sim 150\,\text{MeV}$, breaking chiral symmetry (which was already slightly violated since the EWPT). The corresponding order parameters are the vacuum expectation values (VEVs) of the Higgs field, $\langle H \rangle$ (or $\langle H^\dagger H \rangle$ for a gauge-invariant formulation), and of the quark condensate, $\langle \overline{Q}Q \rangle$, respectively. Theoretical and numerical investigations have shown that both of these transitions are *smooth crossovers* – PTs of order infinity – in the Standard Model of Particle Physics (SM) [6, 7, 11–16], but can be rendered first-order by imposing simple theoretical extensions. We are specifically interested in first-order transitions, which give rise to interesting out-of-equilibrium dynamics at the phase boundaries and encompass the emission of gravitational radiation during the collisions of vacuum bubbles.

The physics of cosmological PTs has recently gained a lot of attention thanks to the advent of gravitational-wave (GW) astronomy [115] which gifts us with new prospect of probing first-order PTs in the foreseeable future [116–122]. Significant advances have been made in the understanding of PT dynamics [123–127] and in the related computational techniques [128–134]. Furthermore, the phenomenology of first-order PTs provides a useful tool for the construction of theories concerning baryogenesis [52–54, 135–150] and dark matter (DM) [74–77], which are also subject of the present thesis.

For a detailed discussion of cosmological first-order PTs we refer the reader to the author's master thesis [151]. In the following, we summarize the main results and formulas required for our analyses: We will address the effective scalar potential and its thermal evolution in Section 3.1 and discuss the GW spectrum sourced during vacuum bubble collisions and the detection prospects in Section 3.2.

## 3.1 The Effective Scalar Potential

The free energy density assigned to a scalar field $\phi$ is represented by the *effective scalar potential* $V_{\text{eff}}(\phi)$. The mean field value of $\phi$ – also called VEV – is denoted by $\langle\phi\rangle$ and will always occupy a (local or global) minimum of the potential. Diagrammatically, the effective potential can be understood as the sum of all one-particle-irreducible diagrams with any number of external legs:

$$V_{\text{eff}}(\phi) = -\sum_{n=0}^{\infty} \frac{\phi^n}{n!} \Gamma^{(n)}(p=0) \tag{3.1}$$

$$= \underset{\Gamma^{(0)}}{\bigcirc} + \underset{\Gamma^{(1)}}{\bigcirc}\text{---} + \text{---}\underset{\Gamma^{(2)}}{\bigcirc}\text{---} + \underset{\Gamma^{(3)}}{\bigcirc} + \ldots$$

Here, $\Gamma^{(n)}(p=0)$ is the effective $n$-point vertex that contains all possible interactions of any loop order, for vanishing external momenta [152, 153]. In position space, this corresponds to vanishing spatial derivatives, reflecting the mean-field description that disregards excitations. As a consequence, the effective potential is technically a function of $\langle\phi\rangle$ (but we will write $V_{\text{eff}}(\phi)$ for brevity), and the external legs in Eq. (3.1) are to be understood as vacuum insertions. We will dissect the effective potential and discuss the most important contributions in the loop expansion in the following.

### 3.1.1 Loop Expansion

**Tree-level potential.** The tree-level contribution (zeroth loop order) to the effective potential is part of the Lagrangian density that defines the theory. The most generic renormalizable tree-level potential for a scalar field charged under a $\mathbb{Z}_2$ symmetry is given by[1]

$$V_{\text{tree}}(\phi) = \mu^2 \phi^2 + \lambda \phi^4 \tag{3.2}$$

$$= \text{-----} + \times .$$

This potential gives rise to a spontaneous breaking of the $\mathbb{Z}_2$ symmetry if $\mu^2 < 0$.

---

[1] Without loss of generality, we work with a *real* scalar field at this point. The formalism works similarly for *complex* fields after replacements like $\phi^2 \to \phi^\dagger \phi$. Note that the breaking of continuous gauge symmetries requires complex scalar fields.

**Coleman–Weinberg one-loop potential.** The *Coleman–Weinberg potential* includes a sum of all *n*-point scalar interactions at one-loop level [154]. The particles running in the loops can be scalars, fermions, and vector bosons, depending on the underlying theory:

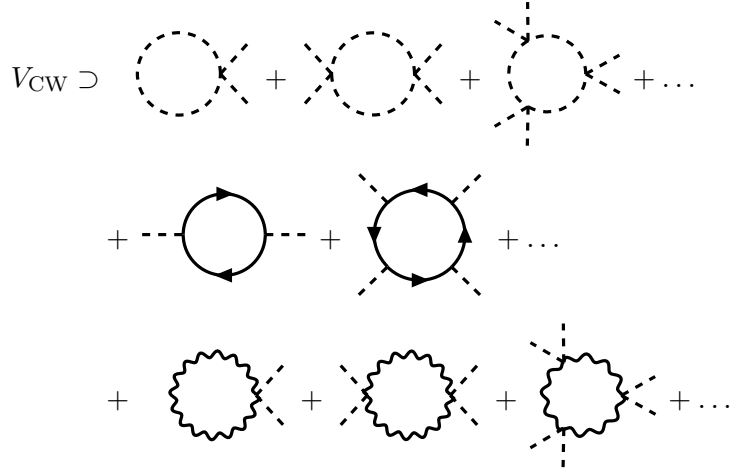

Note that ghost propagators as well as mixed scalar–gauge loop diagrams vanish in Landau gauge. The infinite series of diagrams can be expressed as a sum of logarithms, and the Coleman–Weinberg potential can be written as

$$V_{\mathrm{CW}}(\phi) = \sum_a \pm \frac{m_a^4(\phi)}{64\pi^2}\left[\log\left(\frac{m_a^2(\phi)}{\Lambda^2}\right) - C_a\right] + V_{\mathrm{ct}}(\phi)\,, \tag{3.3}$$

where the sum runs over all degrees of freedom (DOFs) that interact with $\phi$, with bosons (fermions) contributing positively (negatively), and longitudinal gauge modes and Goldstones must both be included separately [155]. Note that the 2- and 4-point diagrams in $V_{\mathrm{CW}}$ are UV-divergent. To arrive at the finite expression in Eq. (3.3), the singularities have been separated via dimensional regularization and then removed by adding appropriate counter terms [156]. This procedure introduces a renormalization scale $\Lambda$ and induces the constant terms $C_a = 3/2$ (5/6) for scalars and fermions (gauge bosons). From a model-building perspective, it is useful if the tree-level structure of the potential (e.g. minima and curvature) is preserved by the one-loop contribution. This motivates the use of *finite counterterms*, $V_{\mathrm{ct}}$, in Eq. (3.3). These terms exhibit the same structure as the tree-level potential, but the coupling constants are determined based on a set of renormalization conditions which, for instance, impose vanishing derivatives of $V_{\mathrm{CW}}$ at certain field values. The field-dependent masses $m_a(\phi)$ appearing in the Coleman–Weinberg potential are determined by the second derivatives of the Lagrangian w.r.t. the field operator $a$. In our example based on Eq. (3.2), the mass of $\phi$ amounts to $m_\phi^2(\phi) = 2\mu^2 + 12\lambda\phi^2$. Remember that we work in the mean-field description, so $m_a(\phi)$ is the short form of $m_a(\langle\phi\rangle)$.

**Thermal one-loop potential.** Our considerations up to this point concern interactions in the realm of conventional quantum field theory (QFT), which are present also in the absence of a thermal bath. In the hot expanding Universe, however, the effective scalar potential receives temperature dependent corrections [157]. The thermal QFT formalism can be derived from conventional QFT by performing a Wick rotation and replacing the imaginary time variable with the inverse of the plasma temperature $T$. This is based on the (anti-)periodicity of thermal field operators in the imaginary time direction. As a consequence, energy loop integrals are replaced by discrete sums over multiples of $\pi T$ – the *Matsubara frequencies* – which in turn can be rewritten as closed-form expressions by virtue of the residue theorem [158].

The one-loop potential in the thermal formalism contains a zero-temperature part, which is identical to $V_{\rm CW}$, and a finite-temperature part

$$V_T(\phi) = \sum_a \pm \frac{T^4}{2\pi^2} \int_0^\infty \mathrm{d}x \, x^2 \log\left[1 \mp \exp\left(-\sqrt{x^2 + \frac{m_a^2(\phi)}{T^2}}\right)\right], \quad (3.4)$$

which again sums over all DOFs interacting with $\phi$ and the sign depend on the spin of the respective particle. The high-temperature expansion is given by

$$V_T(\phi) \approx T^2 \left[ \sum_{\rm b} \left( \frac{m_{\rm b}^2(\phi)}{24} - \frac{m_{\rm b}^3(\phi)}{12\pi T} \right) + \sum_{\rm f} \frac{m_{\rm f}^2(\phi)}{48} \right], \quad (3.5)$$

where the individual sums run over all bosonic and fermionic DOFs, respectively. With masses of the form $m^2 \sim \phi^2$, we see that the thermal potential becomes parabolic in hot settings. This results in vanishing VEVs and restored symmetries in the early Universe.

**Thermal masses and daisy resummation.** In a thermal environment, particle propagators receive loop corrections: Hard thermal loops induce so-called Debye masses and give rise to another significant contribution to the effective potential. As it turns out, the thermal masses of fermions [159] are irrelevant in this context, and we will focus on the bosonic propagators

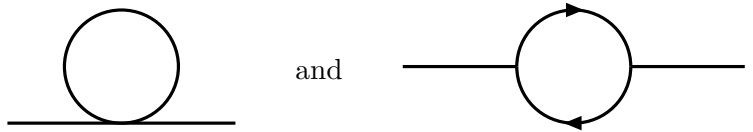

where solid lines (without arrowheads) represent scalars or vector bosons. These diagrams can be evaluated using the thermal QFT formalism to obtain the corresponding mass corrections, which are in general proportional to $T$ and depend on the number of DOFs that run in the thermal loops as well as the involved coupling constants. As can be shown by explicit calculation, the Debye mass of the scalar $\phi$ is identical to the second derivative of the thermal one-loop potential in the $T \to \infty$ limit,

$$\Pi_\phi(T) = V_T''(\phi) \quad (3.6)$$

and amounts to $2\lambda T^2$ in the considered example setup where only two real scalar DOFs run in the hard thermal loop. The mass corrections of longitudinal Abelian and non-Abelian gauge boson modes amount to [160]

$$\Pi^{\mathrm{L}}_{U(1)} = g_1^2 T^2 \left[\frac{1}{6}\sum_{\mathrm{s}} Y_{\mathrm{s}}^2 + \frac{1}{12}\sum_{\mathrm{f}} Y_{\mathrm{f}}^2\right],$$
$$\Pi^{\mathrm{L}}_{SU(N)} = g_N^2 T^2 \left[\frac{N}{3} + \frac{1}{6}\sum_{\mathrm{s}} C(r_{\mathrm{s}}) + \frac{1}{12}\sum_{\mathrm{f}} C(r_{\mathrm{f}})\right], \quad (3.7)$$

while the transverse polarization modes remain massless. The sums run over all scalar and fermionic DOFs that are charged under the respective gauge symmetry, and $g_1$ and $g_N$ are the gauge couplings. The gauge charge or representation of species $a$ is denoted by $Y_a$ or $r_a$, respectively, and $C(r)$ is the characteristic constant of representation $r$.

The presence of thermal propagator corrections has important implications for the effective potential. Any bosonic loop in the thermal one-loop potential can be garnished with an arbitrary number of thermal corrections, which results in a *daisy diagram*:

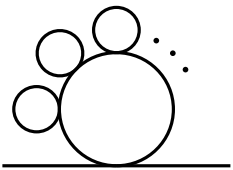

Each of these diagrams individually belongs to a higher order of the loop expansion. However, the resummation of all diagrams – from zeroth to infinite loop order – makes for a contribution to the effective potential that scales with the 3/2-th power of the coupling constants [161]. The propagator resummation is a geometric series and can hence be rewritten in a compact form. For a boson $a$, we have

$$\frac{1}{p^2 - m_a^2} + \frac{\Pi_a}{(p^2 - m_a^2)^2} + \frac{\Pi_a^2}{(p^2 - m_a^2)^3} + \ldots = \frac{1}{p^2 - m_a^2 - \Pi_a} \quad (3.8)$$

or, diagrammatically,

$$\underline{\phantom{xxx}} + \underline{\bigcirc} + \underline{\bigcirc\bigcirc} + \ldots = \underline{\phantom{x}\bullet\phantom{x}},$$

with a "dressed propagator" as result. The resummation can thus be incorporated in the effective potential by performing the simple replacement[2]

$$m_a^2(\phi) \to m_a^2(\phi) + \Pi_a(T)$$

for all scalar and longitudinal gauge boson DOFs in Eqs. (3.3) and (3.4). Note that daisy diagrams with two or more loops are IR-divergent [155]. The smallest loop energies

---
[2]See Ref. [162] for a more sophisticated procedure with theoretical uncertainties that are smaller and understood better.

– corresponding to the lowest Matsubara frequencies – hence make the most significant contribution. It is therefore a good approximation to restrict the mass replacement to the zeroth Matsubara frequency, which has two technical advantages: Firstly, the zero mode is temperature independent and allows for a proper UV regularization of the Coleman–Weinberg potential. Secondly, the fermionic Debye masses are irrelevant due to the absence of corresponding zero modes and must not be determined.

By performing the thermal mass replacement in the one-loop effective potential, each bosonic propagator is substituted by its "dressed" version:

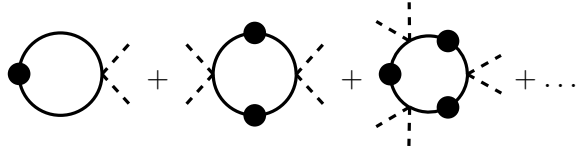

Algebraically, the replacement reads

$$\left[V_{\text{CW}}(\phi) + V_T(\phi)\right]_{m^2 \to m^2 + \Pi} = V_{\text{CW}}(\phi) + V_T(\phi) + V_{\text{daisy}}(\phi) \qquad (3.9)$$

and introduces a new correction to the effective potential – the *daisy contribution* – which can be written as a separate term [163] if the involved gauge couplings are not too large:

$$V_{\text{daisy}}(\phi) = -\frac{T}{12\pi} \sum_a \left\{ [m_a^2(\phi) + \Pi_a(T)]^{-\frac{3}{2}} - [m_a^2(\phi)]^{-\frac{3}{2}} \right\}. \qquad (3.10)$$

∗ ∗ ∗

We remind the reader that an in-depth derivation and discussion of the different contributions is provided in the author's master thesis [151]. In summary, the effective potential up to the first loop order plus the daisy resummation is given by

$$V_{\text{eff}}(\phi) \approx V_{\text{tree}}(\phi) + V_{\text{CW}}(\phi) + V_T(\phi) + V_{\text{daisy}}(\phi). \qquad (3.11)$$

This is the state-of-the-art treatment of the effective scalar potential in the literature on PTs and GW phenomenology. Efforts have been made to establish a gauge-invariant formulation of the effective potential,[3] to improve the understanding of the loop expansion and its validity, and to derive higher-order corrections [165–170]. While these refinements are crucial to make predictions about the stability of today's vacuum state [171], they are beyond the scope of this thesis and not necessary for our analyses.

### 3.1.2 Thermal Evolution

Based on the different contributions to the effective potential presented above, we can inspect the temperature-dependent behavior. In early times of the Universe, when the plasma temperature is much larger than any dimensionful quantity in the scalar sector, the effective potential is dominated by the thermal one-loop potential, $V_T$. As can be seen by Eq. (3.4), this contribution is proportional to $\phi^2$ in the high-temperature limit. At this stage, the background field $\langle\phi\rangle$ exhibits the value zero in the entire Universe. As the latter expands and cools down, the other contributions to the effective potential become more and more dominant. In today's cold Universe, the tree-level potential, $V_{\text{tree}}$, entirely dictates the structure. (Remember that we imposed finite counterterms that cancel any impact of $V_{\text{CW}}$ on the main features of the potential.) If the tree-level potential of a given theory has a minimum at non-zero field values – like Eq. (3.2) with $\mu^2 < 0$ – the vacuum will undergo a PT at some point in the thermal evolution. The temperature at which the non-zero minimum becomes energetically favorable defines the *critical temperature* $T_c$.

---

[3] In the usual description, the effective potential is only gauge invariant at its minima [164].

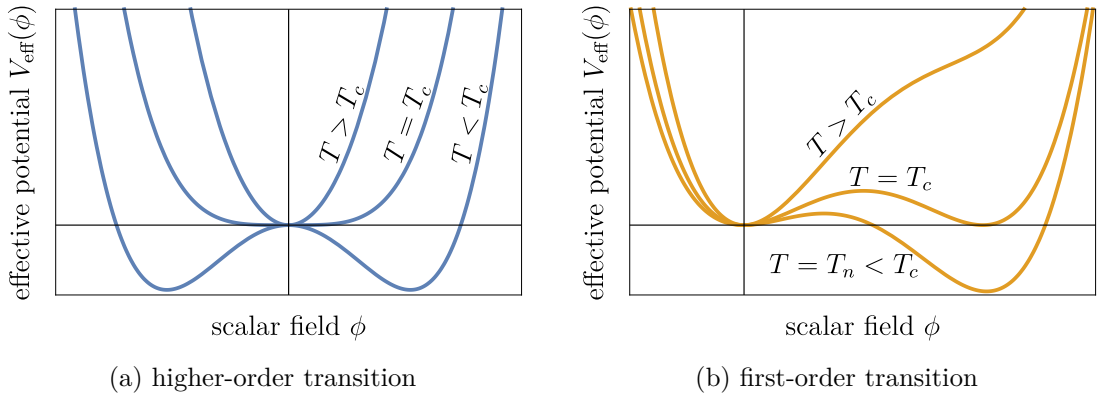

(a) higher-order transition

(b) first-order transition

**Figure 3.1:** Exemplary thermal evolution of the effective scalar potential with decreasing temperature, assuming a tree-level potential with negative quadratic term. Details of the theory decide about the order of the transition.

Interestingly the Coleman–Weinberg potential *alone* can induce a PT, even with $\mu = 0$ in the Lagrangian, which is then called *conformal transition* [154]. In this scenario, the one-loop potential introduces a mass in the otherwise scale-free theory. These kinds of transitions tend to be very energetic and can induce large order parameters [172–191], a feature we make use of in Chapters 4 and 5.

**Higher-order transitions.** Figure 3.1a illustrates the behavior of an effective potential that gives rise to a higher-order PT. In this scenario, the minimum at $\phi = 0$ starts to shift towards finite field values as soon as the temperature drops below $T_c$. As there is no barrier present, the vacuum continuously evolves from $\langle\phi\rangle = 0$ to $\langle\phi\rangle \neq 0$, which marks a transition of second or higher order that occurs in the entire Universe at once.[4] This behavior is expected in case of the $\mathbb{Z}_2$ symmetric tree-level potential defined in Eq. (3.2).

**First-order transitions.** A richer phenomenology comes with a behavior as illustrated in Fig. 3.1b. As the Universe cools down, in this scenario, the vacuum is temporarily trapped in a local minimum at $\phi = 0$ by a potential barrier. The two minima separated by the barrier become degenerate at the critical temperature, $T_c$. Only after a certain degree of *supercooling*, when the *nucleation temperature* $T_n$ is reached, a tunneling to the global minimum occurs. This abrupt change of the background field marks a first-order transition. The tunneling is a stochastic event and hence occurs – by chance – randomly distributed in the entire Universe. The released energy triggers the transition in the surrounding vacuum, resulting in expanding bubbles of *true vacuum* (where $\langle\phi\rangle \neq 0$) which eventually replace the surrounding *false vacuum* ($\langle\phi\rangle = 0$). An illustration of a first-order PT is provided in Fig. 3.2.

---

[4]In our analyses in Chapter 6, we track only the VEV (and not its derivatives) and can therefore not distinguish second-order from higher-order transitions. This is sufficient, as we are interested in the GW spectrum related to first-order transitions.

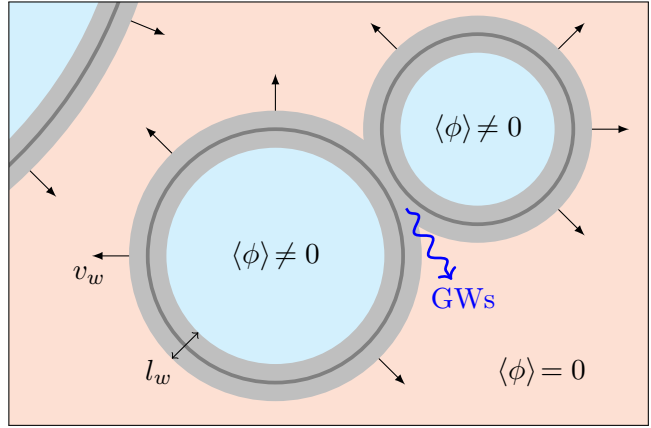

**Figure 3.2:** An illustration of a cosmological first-order PT: Bubbles of true vacuum (cyan) nucleate, expand, and eventually replace the false vacuum (red). The thickness and velocity of the phase boundaries are indicated as $l_w$ and $v_w$, respectively. Collisions of the bubble walls cause anisotropies and thereby generate a stochastic GW background that may be detectable today.

Based on our considerations, a PT first-order can be induced in two ways:

1. The scalar tree-level potential may contain a cubic term ($\propto \phi^3$) which explicitly breaks the $\mathbb{Z}_2$. In conjunction with an appropriate choice of model parameters, the new term can give rise to a potential barrier.

2. The potential barrier may instead arise at loop level: The expansion of the thermal one-loop potential, given in Eq. (3.5), contains cubic terms ($\propto m_b^3 \propto \phi^3$) for each boson that interacts with $\phi$. An expansion of the daisy contribution in Eq. (3.10), on the other hand, contains the same cubic terms but with opposite sign. This cancellation of cubic terms is, however, incomplete because transverse gauge polarization modes do not contribute to the daisy correction by virtue of gauge invariance (i.e. the vanishing of transverse Debye masses). As a consequence, each gauged scalar theory naturally allows for a first-order PT.

In Chapter 6, we discuss and analyze first-order PTs – in the context of different toy models – that are based on both of these mechanisms.

### 3.1.3 Vacuum Decay

From now on, we will focus on the first-order scenario and discuss the concomitant false-vacuum decay in more detail. The behavior of the background scalar field in presence of an effective potential is described by the $d$-dimensional Euclidean action [192]

$$S_d[\phi] \propto \int \mathrm{d}r\, r^{d-1} \left[\frac{1}{2}\left(\frac{\mathrm{d}\phi}{\mathrm{d}r}\right)^2 + V_{\mathrm{eff}}(\phi)\right], \tag{3.12}$$

where $d=4$ ($d=3$) applies to a vacuum decay via zero-temperature quantum tunneling (via thermal fluctuations). In this expression, we combined all $d$ dimensions into a single radial coordinate $r$ by making use of the $O(d)$ symmetry of a single true-vacuum bubble.[5] Applying the principal of stationary action yields the so-called *bounce equation*

$$\frac{\mathrm{d}^2\langle\phi\rangle}{\mathrm{d}r^2} + \frac{d-1}{r}\frac{\mathrm{d}\langle\phi\rangle}{\mathrm{d}r} = V'_{\mathrm{eff}}(\langle\phi\rangle), \tag{3.13}$$

where we made the background-field interpretation explicit by writing $\langle\phi\rangle$ instead of $\phi$. A solution to this differential equation with boundary conditions[6]

$$\langle\phi\rangle(r\to\infty) = 0, \qquad \langle\phi\rangle'(r=0) = 0,$$

represents the field configuration during the PT – i.e. the bubble wall profile – and may be obtained analytically or numerically depending on the underlying theory [193]. Note that multiple solutions may coexist in case of an effective potential with more than two minima. In such cases, bubbles that correspond to the solution with the smallest action ($S_4$ or $S_3/T$) are the most likely ones to nucleate.

We are particularly interested in PTs in the early Universe and will hence focus on the bubble nucleation via thermal fluctuations. As a function of the action that corresponds to the solution of the 3-dimensional bounce equation, the *rate of bubble nucleation* amounts to [194, 195]

$$\Gamma_n \sim T^4 \exp\left(-\frac{S_3}{T}\right). \tag{3.14}$$

For a PT to progress, bubbles must be nucleated faster than the Universe expands. The *nucleation temperature* $T_n$ is accordingly defined by $\Gamma_n(T_n) \sim H(T_n)$, where the Hubble rate in a radiation dominated Universe is given by the first Friedmann equation, $H^2 = \rho_{\mathrm{rad}}/(3M_{\mathrm{Pl}}^2)$, with the reduced Planck mass $M_{\mathrm{Pl}}$. The radiation energy density is [196]

$$\rho_{\mathrm{rad}} = \frac{\pi^2}{30}g_\star(T)T^4, \tag{3.15}$$

where $g_\star$ counts the effective number of relativistic DOFs. The nucleation condition can

---

[5] An Euclidean $O(4)$ symmetry corresponds to a spacetime parabola, describing an $O(3)$ symmetric bubble with an expansion velocity that approaches the speed of light.

[6] Note that these boundary conditions do not fix the field value immediately after tunneling. In fact, this value is determined in the course of solving of the bounce equation and is (slightly) smaller than the final true-vacuum VEV.

be rearranged and posed as a requirement on the thermal tunneling action:

$$\frac{S_3(T_n)}{T_n} \sim 146 - 4\log\left(\frac{T_n}{100\,\text{GeV}}\right) - 2\log\left(\frac{g_\star(T_n)}{100}\right). \tag{3.16}$$

## 3.2 Bubble Collisions and Gravitational Waves

The previous section reviewed the theoretical formalism describing cosmological first-order PTs which encompass the nucleation of bubbles. If the transition proceeds successfully, multiple expanding bubbles will eventually collide, which represents an anisotropic process that sources gravitational radiation. The collisions occur randomly distributed throughout the Universe and generate a stochastic, isotropic, unpolarized, and Gaussian GW background – similar to the cosmic microwave background (CMB) – which can potentially be observed today. Note that the temperature at which the bubble collisions occur is smaller than the nucleation temperature due to the intermediate expansion of the Universe, especially in case of highly supercooled transitions [120]. For simplicity, we neglect this distinction and assume that the GWs are produced immediately at $T_n$. In this section, we discuss the characterizing parameters of cosmological PTs as well as the GW spectrum and its detectability.

### 3.2.1 Characteristic Parameters

**Transition strength** $\alpha$. The amplitude of the GW spectrum produced by bubble collisions depends on the energy that is liberated during the PT. The *latent heat* is defined as [197]

$$\varepsilon \equiv \Delta V_\text{eff} - T_n \left[\frac{\partial}{\partial T}\Delta V_\text{eff}\right]_{T_n}, \tag{3.17}$$

i.e. the potential difference between the two vacua, $\Delta V_\text{eff}$, minus the energy consumed by the entropy increase in course of the transition. The latent heat represents the energy that is available for the acceleration of the bubble walls and the heating of the surrounding particle plasma. As we will see further below, the GW amplitude is determined by the latent heat normalized w.r.t. the total energy density of the (radiation-dominated) Universe at the time of the transition. The *transition strength* parameter is thus defined as

$$\alpha \equiv \frac{\varepsilon}{\rho_\text{rad}(T_n)}. \tag{3.18}$$

**Inverse time scale** $\beta$. Another characteristic of the PT that impacts the GW spectrum is the duration of the PT. Bubbles start to nucleate as soon as the nucleation condition – Eq. (3.16) – is fulfilled. If the tunneling action decreases quickly, the bubble nucleation rate – Eq. (3.14) – may increase beyond the Hubble rate before the PT concludes. (Remember that we defined the equality of these rates as the onset of nucleation.) In such a scenario, many small bubbles are generated per Hubble volume and the transition proceeds much

*3.2 Bubble Collisions and Gravitational Waves*faster. This behavior is quantified by the *inverse time scale* of the PT [198]

$$\beta \equiv -\left[\frac{\mathrm{d}}{\mathrm{d}t}\frac{S_3}{T}\right]_{T_n} = T_n\left[\frac{\mathrm{d}}{\mathrm{d}T}\frac{S_3}{T}\right]_{T_n} \times H\,, \tag{3.19}$$

which we express in terms of the Hubble rate using $H = -\dot{T}/T$ for an adiabatically expanding Universe.

**Bubble wall velocity** $v_w$. The walls of the expanding bubbles are driven by the latent heat release of the PT. In empty space, the bubble walls will quickly reach luminal velocities, $v_w \sim 1$, with continuously increasing Lorentz factor $\gamma_w$. The situation can be different if the scalar field that undergoes the PT couples to thermally abundant particles, either itself or other species, which induces a decelerating friction acting on the wall. Plasma particles that gain mass across the wall (due to the increasing scalar VEV) are accelerated in the direction of wall movement, drawing on the the energy budget that drives the bubble expansion. This effect can hinder the walls from reaching large $\gamma_w$ or luminal velocities at all. At the same time, this generates a bulk motion in the particle plasma which turns into density waves. If the wall remains slower than sound speed, $v_w < c_\mathrm{s} = 1/\sqrt{3}$, a plasma shock front builds up ahead of the wall ("deflagration"), while in the case of faster walls, the density wave travels partly or entirely behind the wall ("hybrid" or "detonation") [197, 199–202].

Sufficiently strong PTs usually result in luminal wall velocities [124]. In the context of GW production, it is useful to distinguish between walls with limited Lorentz factor, which can however still move at close-to-luminal speeds, and walls that accelerate indefinitely. The two regimes can be distinguished based on the friction estimator

$$\alpha_\mathrm{run} \equiv \frac{T_n^2}{\rho_\mathrm{rad}}\left[\sum_\mathrm{b}\frac{\Delta m_\mathrm{b}^2}{24} + \sum_\mathrm{f}\frac{\Delta m_\mathrm{f}^2}{48}\right]\,, \tag{3.20}$$

which includes the sum of all bosonic and fermionic mass differences between the false and true vacua (excluding Goldstones [197]). If $\alpha > \alpha_\mathrm{run}$, the latent heat release is large enough to overcome friction and $\gamma_w$ grows indefinitely ("runaway walls"), while otherwise, $\gamma_w$ is limited. Note that this distinction only applies in the absence of vector bosons. The latter induce transition radiation which limits $\gamma_w$ and hence rules out the runaway regime in this case [123, 124, 203].

The exact determination of the wall velocity is highly model-dependent, non-trivial, and beyond the scope of this thesis [204]. In our analyses in Chapters 4 and 6, we assume ad-hoc values for $v_w$ and apply the runaway distinction via $\alpha_\mathrm{run}$ in Chapter 6. In Chapter 5, we test our results for a variety of different $v_w$ and explicitly calculate the wall friction to ensure that the imposed velocities are consistent with other assumptions.

From a phenomenological standpoint, slow walls are beneficial for models of baryogenesis, where out-of-equilibrium processes require time to occur on both sides of the wall. Fast walls, on the other hand, correspond to particularly violent bubble collisions, resulting in a strong GW spectrum that is more likely to be noticed by today's (or future) observatories.

*37*

### 3.2.2 The Gravitational-Wave Spectrum

According to general relativity, gravitational radiation is sourced by anisotropic (i.e. non-spherical) motion of mass or energy. While each individual true-vacuum bubble is spherically symmetric and thus incapable of producing GWs, the symmetry is broken when the bubbles collide. Both, the collision of the domain walls and of the plasma density waves, if present, act as GW sources. In his master thesis [151], the author reviewed the mathematical description of GW emission based on the Einstein equations and linearized gravity. In the following, we will estimate the amplitude and frequency scaling of the produced GW spectrum before we present quantitative results from the literature.

**Scaling estimate and redshift.** First, note that the approximate bubble size at the time of collision is given by $v_w/\beta$, with wall velocity $v_w$ and inverse transition time scale $\beta$. It can be shown that the energy density of the produced GW background scales as [198]

$$\rho_{\text{GW}} \sim G \frac{(\varepsilon\, v_w^3 \beta^{-2})^2 \beta^{-1}}{(v_w^3 \beta^{-3})}\,, \tag{3.21}$$

with the total GW energy in the numerator and the characteristic volume of the source in the denominator. $\varepsilon$ is the latent heat defined in Eq. (3.17). Trading the gravitational constant $G$ for the Hubble rate $H$ via the Friedmann equation and normalizing w.r.t. the critical energy density (i.e. the radiation density plus the latent heat) yields the *dimensionless amplitude* of the GW spectrum

$$\Omega_{\text{GW}} \equiv \frac{\rho_{\text{GW}}}{\rho_{\text{crit}}} \sim \left(\frac{\alpha}{1+\alpha}\right)^2 \left(\frac{H}{\beta}\right)^2 v_w^3\,, \tag{3.22}$$

where we used the definition of the transition strength $\alpha$ in Eq. (3.18). Conveniently, when redshifting the dimensionless amplitude to obtain its value in today's Universe, the explicit dependence on $T_n$ cancels out. Based on energy and entropy conservation during the expansion of the Universe, the *redshifted amplitude* becomes [119]

$$\begin{aligned} h^2 \Omega_{\text{GW}}^0 &= \left(\frac{T_0}{T_n}\right)^4 \left(\frac{g_{\star s}^0}{g_{\star s}(T_n)}\right)^{\frac{4}{3}} \frac{\rho_{\text{rad}}(T_n)}{3 M_{\text{Pl}}^2 H_0^2} \times h^2 \Omega_{\text{GW}} \\ &\approx 1.67 \times 10^{-5} \left(\frac{100}{g_\star(T_n)}\right)^{\frac{1}{3}} \times \Omega_{\text{GW}}\,, \end{aligned} \tag{3.23}$$

where "0" marks today's quantities and $h$ is the Hubble constant divided by $100\,\text{km/s/Mpc}$. Note that we used $g_{\star s} = g_\star$, which is valid prior to the decoupling of the SM neutrinos, i.e. for $T_n \gtrsim 1\,\text{MeV}$. The temperature and relativistic entropy DOFs today amount to $T_0 \approx 2.4 \times 10^{-13}\,\text{GeV}$ [205] and $g_{\star s}^0 \approx 3.9$ [206], respectively.

The *peak frequency* of the GW spectrum depends on the size of the source. A naive estimate is thus given by the inverse bubble size at the time of collision:

$$f_{\text{p}} \sim \frac{\beta}{v_w}\,. \tag{3.24}$$

Considering the conservation of entropy again, today's *redshifted peak frequency* is found to be

$$f_{\rm p}^0 = \frac{T_0}{T_n}\left(\frac{g_{\star s}^0}{g_{\star s}(T_n)}\right)^{\frac{1}{3}} \times f_{\rm p}$$

$$\approx 16.5\,\mu{\rm Hz}\left(\frac{T_n}{100\,{\rm GeV}}\right)\left(\frac{g_\star(T_n)}{100}\right)^{\frac{1}{6}} \times \frac{f_{\rm p}}{H}\,. \tag{3.25}$$

From the derived scaling behavior of amplitude and redshift, we can conclude that a sizable GW signal is expected from strong transitions (large $\alpha$, corresponding to sizable latent heat) that proceed slowly ($\beta \sim H$, allowing the first nucleated bubbles to grow unimpeded). The peak frequency is set by the inverse time scale of the transition in terms of Hubble, $\beta/H$. For fixed $\beta/H$, the redshifted peak frequency today is proportional to the temperature at the time of collision (which we set equal to the nucleation temperature, $T_n$).

**Contributions to the spectrum.** With the expected scaling and redshift behavior at hand, we can attend to a more detailed and quantitative discussion of the GW spectrum. It is expected that three mechanisms contribute to the generation of gravitational radiation during the collision of vacuum bubbles:

1. The *scalar field contribution* ($\Omega_\phi$) from the collisions of the actual bubble walls, i.e. the moving scalar VEV gradients. The corresponding spectral properties are usually derived in the envelope approximation which assumes thin walls, a quick dispersion after the collision, and focuses on the bubble intersections [207].

2. The *sound-wave contribution* ($\Omega_{\rm sw}$) is caused by the collisions of density waves that are evoked by the bubble walls, if they couple to the plasma. This effect lasts longer than the scalar contribution and is thus enhanced by a factor of $\beta/H$.

3. A small fraction of the energy injected into the plasma generates *magnetohydrodynamic turbulence*, which last for several Hubble times and also acts as a source of gravitational radiation ($\Omega_{\rm turb}$).

With these contributions, the total frequency-dependent GW power spectrum is given by

$$h^2\Omega_{\rm GW}(f) = h^2\Omega_\phi(f) + h^2\Omega_{\rm sw}(f) + h^2\Omega_{\rm turb}(f)\,. \tag{3.26}$$

In recent years, a lot of effort has been put into the prediction of the GW spectrum from the different contributions based on analytical and numerical methods as well as simulations. The redshifted GW power spectrum of a single contribution $i \in \{\phi, {\rm sw}, {\rm turb}\}$ can be parameterized as

$$h^2\Omega_i(f) = \Omega\left(\frac{100}{g_\star(T_n)}\right)^{\frac{1}{3}}\left(\frac{\kappa\,\alpha}{1+\alpha}\right)^a\left(\frac{H}{\beta}\right)^b \mathcal{V}(v_w)\,\mathcal{S}(f)\,, \tag{3.27}$$

| Contribution | $\Omega_\phi$ | $\Omega_{\rm sw}$ | $\Omega_{\rm turb}$ |
|---|---|---|---|
| $\Omega$ | $1.67 \times 10^{-5}$ | $2.65 \times 10^{-6}$ | $3.35 \times 10^{-4}$ |
| $f_{\rm p}$ | $\frac{0.62\beta}{1.8-0.1v_w+v_w^2}$ | $\frac{2\beta}{\sqrt{3}v_w}$ | $\frac{3.5\beta}{2v_w}$ |
| $\kappa$ ($\alpha > \alpha_{\rm run}$) | $1 - \frac{\alpha_{\rm run}}{\alpha}$ | $\frac{\alpha_{\rm run}}{0.73+0.083\sqrt{\alpha_{\rm run}}+\alpha_{\rm run}}$ | $\sim 10\%\,\kappa_{\rm sw}$ |
| $\kappa$ ($\alpha < \alpha_{\rm run}$) | $0$ | $\frac{\alpha}{0.73+0.083\sqrt{\alpha}+\alpha}$ | $\sim 10\%\,\kappa_{\rm sw}$ |
| $a$ | $2$ | $2$ | $\frac{3}{2}$ |
| $b$ | $2$ | $1$ | $1$ |
| $\mathcal{V}(v_w)$ | $\frac{0.11v_w^3}{0.42+v_w^2}$ | $v_w$ | $v_w$ |
| $\mathcal{S}(f)$ | $\frac{3.8(f/f_{\rm p}^0)^{2.8}}{1+2.8(f/f_{\rm p}^0)^{3.8}}$ | $\left(\frac{f}{f_{\rm p}^0}\right)^3\left(\frac{7}{4+3(f/f_{\rm p}^0)^2}\right)^{\frac{7}{2}}$ | $\frac{(f/f_{\rm p}^0)^3}{(1+f/f_{\rm p}^0)^{11/3}[1+8\pi(f/f_{\rm p}^0)(f_{\rm p}/H)]}$ |
| Reference | [208] | [209] | [210] |

**Table 3.1:** Spectral parameters to be used in Eq. (3.27) for the three GW production mechanisms during bubble collisions, as stated by the given references. Note that the conversion efficiency, $\kappa$, differs depending on the bubble wall behavior (runaway vs. non-runaway). The peak frequency today, $f_{\rm p}^0$, is obtained based on the original peak frequency, $f_{\rm p}$, according to the relation in Eq. (3.25). It currently remains unclear whether the sound-wave simulation is valid for strong transitions with $\alpha \gtrsim 0.1$. Furthermore, the conversion efficiency to turbulent motion during very fast transitions with $\beta/H \gtrsim 100$ is unknown and may be smaller than the indicated 10% [181, 211].

where we write $\Omega_{\rm GW}$ and $f$ instead of $\Omega_{\rm GW}^0$ and $f^0$, for brevity. The amplitude $\Omega$, efficiency factors $\kappa$, exponents $a$ and $b$, wall velocity factor $\mathcal{V}(v_w)$, and the spectral shape $\mathcal{S}(f)$ have been determined in different analytical and numerical studies and are listed in Table 3.1.

The relevance of the different contributions depends on the bubble wall behavior, as can be seen by the listed expressions for the efficiency factors $\kappa$: In the case of runaway walls with negligible plasma interactions ($\alpha \gg \alpha_{\rm run}$), we observe $\kappa = 1$ for the scalar field contribution and $\kappa = 0$ for the others. Bubble walls in the runaway regime but with sizable plasma interactions ($\alpha > \alpha_{\rm run}$) result in a shared energy budget among all three contributions. In the non-runaway regime ($\alpha < \alpha_{\rm run}$), the scalar field contribution is negligible as the walls are non-relativistic and most of the latent heat is transferred to the plasma.

The low-frequency end of the spectrum from any of the three contributions scales roughly with the third power of $f$, which is due to the causal disconnectedness of super-Hubblean distances. Furthermore, note the additional suppression of the turbulence compared to the sound-wave contribution in case of fast transitions, represented by the factor $f_{\rm p}/H \sim \beta/H$ in the denominator of the turbulent spectral shape $\mathcal{S}(f)$. The interplay of the different contributions can, depending on the PT parameters, result in a characteristic double-bump signature. Two typical GW spectra, corresponding to runaway and non-runaway bubble walls, are shown in the right panel of Fig. 3.3.

In the following section and in Chapter 6, we employ the discussed expressions for the GW power spectrum to assess the detectability of first-order PT – in general and in the context of specific toy models – via future GW observatories. Note that, at the time our analyses were conducted, the presented GW spectra were state of the art. In the meantime, further progress has been made in the simulation of the different contributions and the understanding of the PT energy budget [181, 186, 212–215]. This, however, does not invalidate our general findings and conclusions.

### 3.2.3 Experimental Sensitivity

The first observation of a black hole (BH) merger via the gravitational radiation it generates was a groundbreaking discovery that started a new era of experimental astrophysics and cosmology [115]. Several highly sensitive next-generation GW observatories of different kinds will become operational in the upcoming decades and aim to observe GWs from astrophysical sources such as supernovae [216], rotating neutron stars [217], and binaries with different masses and mass ratios [218–226]. Furthermore, it will be possible to establish a CMB-like map of the gravitational background radiation. Considering that the theorized graviton would decouple at Planck-scale temperatures – as opposed to the photon which decouples around $1\,\text{eV}$ to produce the CMB – a GW background will provide insight into eras that date back much prior to recombination. Cosmological phenomena in the early Universe such as cosmic inflation, cosmic strings, or first-order PTs may have left their imprints in the GW power spectrum [227, 228]. However, also astrophysical sources can contribute to a stochastic GW background, for instance the overlapping signals of numerous, individually indistinguishable weak binaries.

Various experimental approaches cover a wide range of GW frequencies: The high-frequency end of the spectrum ($10^1 \sim 10^5\,\text{Hz}$) is relevant for the observation of supernovae as well as compact inspirals and is covered by existing *Earth-bound interferometers* with arm lengths of $\mathcal{O}(\text{km})$ such as *LIGO* [229], *Virgo* [230] and *KAGRA* [231], as well as planned experiments like *ET* [232]. Massive and extreme-mass-ratio binaries source GWs in the mid-frequency range ($10^{-5} \sim 10^1\,\text{Hz}$) and will be probed by future *space-based interferometers* such as *LISA* [233], *BBO* [234], *B-DECIGO* [235], and *DECIGO* [236, 237] with arm lengths of $10^5 \sim 10^6\,\text{km}$. Another possibility for the detection of gravitational radiation is given by the observation and timing of millisecond pulsars. A pulsar is a magnetized rotating neutron star that emits beams of electromagnetic radiation along its rotational axis. If the light cone periodically hits Earth, the pulsar acts as a precise and stable astrophysical clock which can be read off by radio telescopes on Earth. Tiny variations in the arrival time of the pulses are hints for metric perturbations, indicating the presence of gravitational radiation between the pulsar and Earth. Many pulsars in multiple directions of the sky are tracked in *pulsar timing arrays (PTAs)* such as *EPTA* [238, 239], *NANOGrav* [240], or *SKA* [241, 242] (under construction), probing stochastic backgrounds in the low-frequency regime ($10^{-9} \sim 10^{-6}\,\text{Hz}$), where the upper and lower ends of the sensitive band are set by the timing interval and the total observation time, respectively. In this frequency band, a stochastic background of SMBHBs is expected, with first hints already being reported by *NANOGrav* [243]. Very recent proposals such as an $\mathcal{O}(10^9\,\text{km})$ space-based interferometer [244] or the possibility of using asteroids as test masses [245] aim to bridge the experimental µHz gap. Due to their novelty, we were unable to incorporate these promising approaches into our plots and analyses.

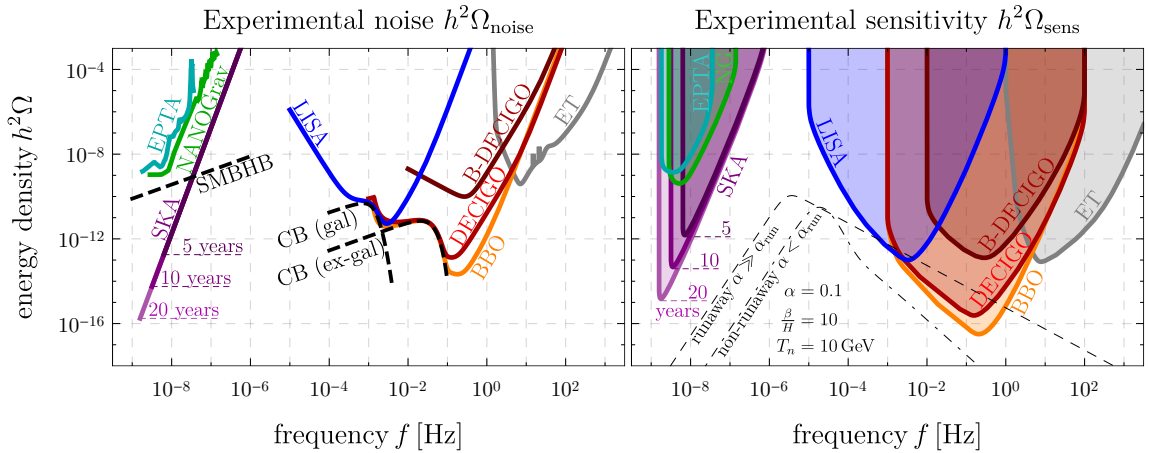

**Figure 3.3:** Experimental sky-averaged noise (left) and sensitivity (right) in terms of the dimensionless GW energy density as a function of frequency for different current and future GW observatories. The noise curves of space-based interferometers contain visible bumps caused by galactic and extragalactic compact binaries (CBs). The noise peak at $f \approx 1/\text{year} \approx 3 \times 10^{-8}$ Hz, visible for *EPTA*, is caused by the Earth's rotation around the Sun. For *SKA*, we include projections for three different observation periods and assume that the predicted supermassive black hole binary (SMBHB) background will be resolved. The right plot also displays exemplary GW spectra that would be expected from first-order PT with bubbles deep in the runaway regime and with non-runaway bubbles, respectively.

In the left panel of Fig. 3.3, we plot the frequency-dependent noise curves of several GW experiments, focusing on the most relevant frequencies for first-order PTs, covered by PTAs and space-based observatories. In order to make actual predictions about the detectability of a continuous, stochastic GW signal, we must consider the *integrated signal-to-noise ratio (SNR)*, defined as

$$\varrho \equiv \sqrt{t_{\text{obs}} \int_{-\infty}^{\infty} \mathrm{d}f \left( \frac{\Omega_{\text{GW}}(f)}{\Omega_{\text{noise}}(f)} \right)^2}, \tag{3.28}$$

with $t_{\text{obs}}$ denoting the duration of the GW observation. A signal is visible if $\varrho$ exceeds a certain threshold value $\varrho_{\text{thr}}$, which we give in Table 3.2 for the different GW experiments.

A pure power-law signal of the form $\Omega_\gamma(f) = \Omega_\gamma \times (f/\text{Hz})^\gamma$ will be detected if its amplitude $\Omega_\gamma$ exceeds [248]

$$\Omega_\gamma^{\text{thr}} = \varrho_{\text{thr}} \left[ t_{\text{obs}} \int_{-\infty}^{\infty} \mathrm{d}f \left( \frac{(f/\text{Hz})^\gamma}{\Omega_{\text{noise}}(f)} \right)^2 \right]^{-\frac{1}{2}}, \tag{3.29}$$

| Experiment | $t_{\text{obs}}$ in years | $\varrho_{\text{thr}}$ | Reference |
|---|---|---|---|
| *EPTA* | 18 | 1.19 | [239] |
| *NANOGrav* | 11 | 0.697 | [246] |
| *SKA* | 5, 10, 20 | 4 | [241] |
| *LISA* | 4 | 10 | [120] |
| *B-DECIGO* | 4 | 8 | [235] |
| *DECIGO* | 4 | 25 | [247] |
| *BBO* | 4 | 25 | [247] |
| *ET* | 5 | 5 | [232] |

**Table 3.2:** GW observation periods and SNR thresholds for the considered observatories, used to assess the sensitivity to (broken) power-law GW signals.

as can be easily inferred from Eq. (3.28). The envelope curve $\Omega_{\text{sens}}(f) \equiv \max_{\gamma \in \mathbb{R}}[\Omega_\gamma^{\text{thr}}(f)]$ is the *power-law integrated sensitivity* – plotted in the right panel of Fig. 3.3 – and represents the sensitivity to an arbitrary (broken) power-law signal.

Finally, with the expected GW spectra and the experimental sensitivities at hand, we can determine numerically which PT temperatures, strengths, and time scales can be probed. We present the results in Fig. 3.4 by plotting the regions where $\varrho > \varrho_{\text{thr}}$ in the $T_n$–$\alpha$- and $T_n$–$\beta$-plane for bubbles deep in the runaway regime ($\alpha \gg \alpha_{\text{run}}$) and for non-runaway bubbles ($\alpha < \alpha_{\text{run}}$). As can be gathered from the figure, PTAs are most sensitive to PTs occurring in the temperature range keV $\sim$ GeV, while the sensitive region extends to much lower temperatures in the case of runaway bubbles. Space-based observatories, on the other hand, cover the transition temperatures $10\,\text{GeV} \sim 10^9\,\text{GeV}$, again with an increased sensitivity for runaway bubbles. As expected from our earlier considerations, strong and slow transitions (with large $\alpha$ and small $\beta/H$) have the best detection prospects. Furthermore, the high-temperature boundary of the sensitivities shifts towards lower $T_n$ if $\beta/H$ is increased, which is because faster transitions and smaller temperatures have opposite and canceling effects on the signal's peak frequency. The kink of the sensitive regions in the $T_n$–$\beta$-plane for non-runaway bubbles is due to the different dependence of the two signal components – sound waves and turbulence – on $\beta/H$.

In Chapter 6, we apply the presented procedures to assess the possibility of probing first-order PTs in generic decoupled hidden sectors and in the context of specific toy models featuring sub-MeV masses and nucleation temperatures.

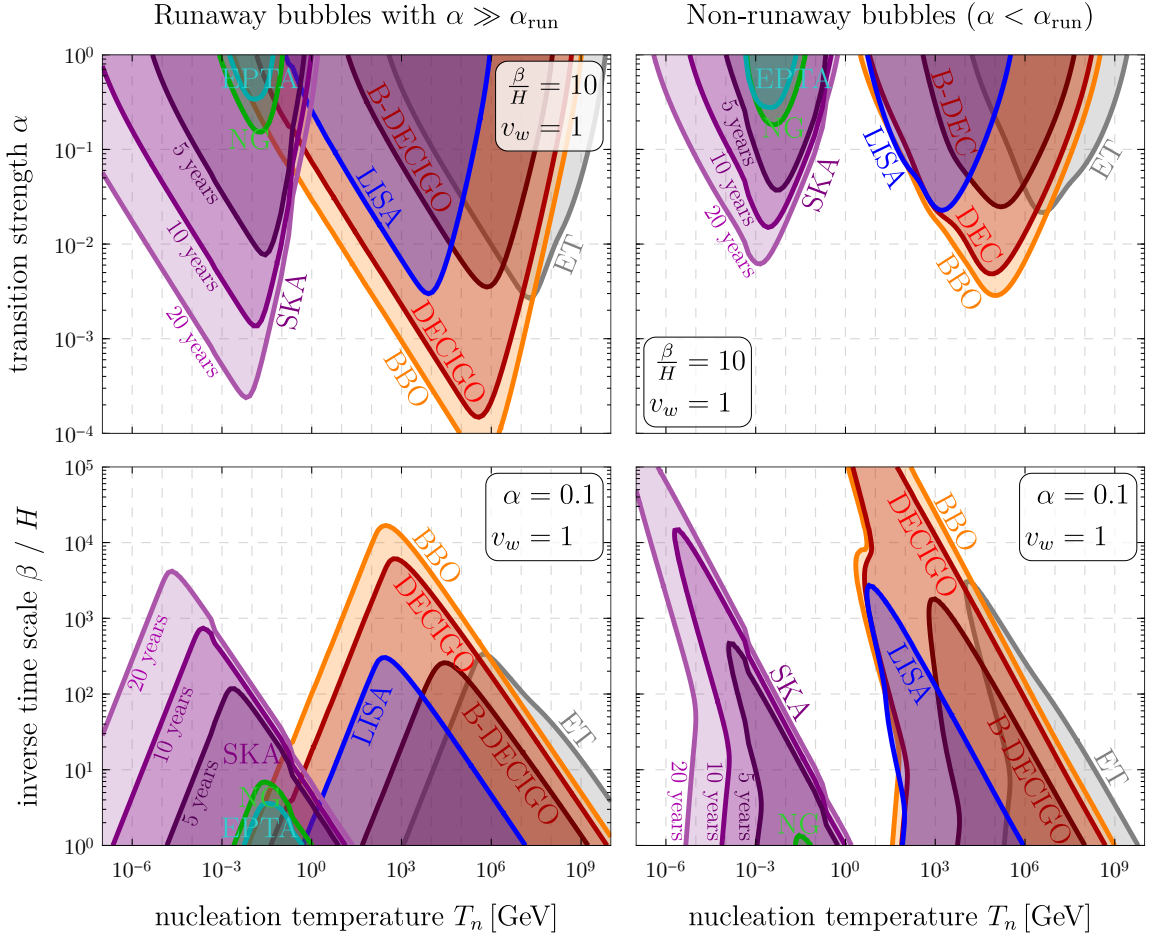

**Figure 3.4:** Anticipated sensitivity of various existing and future GW observatories. In the left panel we assume bubble walls that are deep in the runaway regime, while the right panel shows the non-runaway case. We plot the sensitivities as functions of the PT nucleation temperature, $T_n$, the transition strength $\alpha$, and the inverse time scale $\beta$ (in terms of the Hubble rate $H$).

# 4 Dark Matter and Baryogenesis at the Phase Boundary

*This chapter is based on the publication [1] of the author and his collaborators. In this project, the author played a central role in the development of the physical formalism and the writing of the publication. He implemented and performed all of the required numerical calculations, which where partly crosschecked with results obtained by his colleague LM. The author produced all figures that appear in the publication and in the following chapter, except Fig. 4.6, which was done by MJB.*

## 4.1 Introduction

In the previous section, we established the theoretical backgrounds of first-order cosmological phase transitions (PTs). In the following, we develop a mechanism that operates at the boundary of such a PT and is capable of generating the observed dark matter (DM) abundance and the baryon asymmetry of the Universe (BAU) at the same time.

Our mechanism builds on the recently studied Filtered DM scenario [77, 249], which requires a first-order PT during which a DM candidate obtains a mass that is multiple times larger than the temperature of the Universe. This hierarchy ensures that most of the DM particles – which we assume to be in thermal equilibrium with the particles of the Standard Model of Particle Physics (SM) prior to the PT – are reflected by the wall due to energy conservation. These particles are "filtered out" and stay in the false vacuum where they eventually annihilate. Only few particles – from the high-energy tail of the thermal distribution – reach the true vacuum, where they freeze out immediately due to the large gained mass and constitute the DM abundance observed today. This mechanism is different from the standard weakly interacting massive particle (WIMP) paradigm (see Section 2.2.2) – as the DM freeze-out occurs not in course of the expansion and cooling of the Universe but is instead triggered by a cosmological PT. This allows for DM masses and couplings that are beyond the typical WIMP regime.

A first-order PT is an out-of-equilibrium process and thereby already fulfills one of the three Sakharov conditions for baryogenesis (see Section 2.2.1). We augment the Filtered DM scenario with a coupling – between the scalar field that undergoes the PT and the DM field – that violates charge-parity symmetry (CP) at dimension five in an effective field theory (EFT) approach. As a consequence, DM particles of a certain chirality are predominantly reflected by the bubble wall. In the true vacuum, any emerging chiral asymmetry is quickly washed out by the DM mass term which mixes chiralities. In the false vacuum, where the DM is massless, the asymmetry is converted into a lepton–antilepton asymmetry by either a dimension-6 or a dimension-8 portal operator. In the true vacuum, the rate of these portal interactions is exponentially suppressed by the large DM mass. Therefore, a net lepton number persists after the PT completes. Finally, electroweak sphalerons convert part of the lepton number into a baryon number, which then constitutes the BAU observed today.

The proposed mechanism works similar to electroweak baryogenesis (EWBG), see Section 2.2.1, which also operates at the boundary of a first-order PT. In our case, however, an asymmetry is first generated entirely in the dark sector and the mechanism occurs prior to the electroweak phase transition (EWPT).

Recently, there have been other interesting approaches to baryogenesis. Refs. [145–147] propose a mechanism that is almost identical to EWBG but occurs entirely in a dark sector, including sphalerons which are introduced by non-Abelian dark forces. Ref. [148] extends the SM by a scalar field to render the EWPT first-order [250–252]. In this approach, a chiral asymmetry is first generated in the dark sector and then transferred to the SM via a DM–$\tau$ coupling. The DM abundance in this model is set through conventional thermal freeze-out, as opposed to our mechanism, where the freeze-out is triggered by the PT. Ref. [150] investigates the possibility of baryogenesis at the boundary of a dark first-order PT with large mass jump (similar to our approach, but without explaining the DM abundance) and compares the relevance of annihilation vs. decay in the false vacuum regions. Successful baryogenesis via a strong dark PT with relativistic bubble walls has been demonstrated in Refs. [253, 254]. Other approaches that employ dark sector dynamics in order to explain the BAU can be found in Refs. [255–281].

This chapter is structured as follows: We start by introducing a minimal toy model, review the Filtered DM mechanism, and add the ingredients for baryogenesis in Section 4.2. In Section 4.3, we develop a formalism to test our ideas quantitatively and explain all of its components. We illustrate the workings of our mechanism in detail for individual benchmark points, explore entire slices of the model parameter space, and evaluate the DM direct detection prospects in Section 4.4. Our findings are summarized in Section 4.5.

## 4.2 The Mechanism

To demonstrate our ideas, we propose a minimal dark sector containing a real scalar field $\phi$ and a Dirac fermion $\chi$ – the DM candidate – both singlets under the SM gauge groups. When the expanding and cooling Universe reaches the nucleation temperature, $T_n$, we assume that the scalar field $\phi$ undergoes a first-order PT: Bubbles of true vacuum – where the vacuum expectation value (VEV) of $\phi$ has the finite value $\langle\phi\rangle = \langle\phi\rangle^\infty$ – nucleate and expand. For simplicity, we illustrate our mechanism at a bubble wall that is flat in the $x$–$y$-plane and moves in the direction orthogonal to it. The physical field configuration is given by the solution to the bounce equation – see Eq. (3.13) – and depends on details of the scalar sector which we leave unspecified. For the bubble wall profile we hence use the generic *kink solution* [282]

$$\langle\phi\rangle(z) = \frac{\langle\phi\rangle^\infty}{2}\left[1 + \tanh\left(\frac{3z}{l_w}\right)\right] = \begin{cases} 0 & \text{at } z \to -\infty \\ \langle\phi\rangle^\infty & \text{at } z \to +\infty \end{cases}, \qquad (4.1)$$

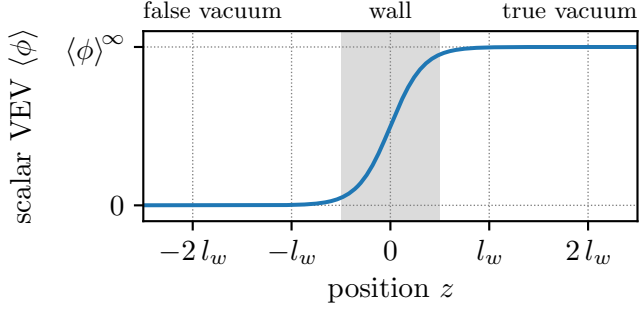

**Figure 4.1:** Interpolating function used to model the scalar VEV across the bubble wall. The width parameter $l_w$ is indicated by a vertical gray band.

which is plotted in Fig. 4.1. The coordinate $z$ represents the non-trivial direction in the wall's rest frame and $l_w$ parameterizes the wall width.

While we assume $\chi$ to be initially massless (or very light), a Yukawa coupling,

$$\mathcal{L}_\chi = -y_\chi \phi \overline{\chi} \chi \,, \tag{4.2}$$

generates a mass, $m_\chi \sim y_\chi \langle \phi \rangle$, that is non-zero in the true vacuum.[1] The $U(1)$ symmetry that $\chi$ enjoys in the Lagrangian ensures its stability. Note that we use the superscript "$\infty$" to denote the values that are assumed after the PT (i.e. deep in the true vacuum), while e.g. $m_\chi(z)$ or $m_\chi$ (without the "$\infty$") refer to quantities that vary from zero to their final values across the bubble wall.

### 4.2.1 Filtered Dark Matter

We employ the Filtered DM mechanism suggested in Refs. [77,249] to generate the observed DM abundance. The main requirement for this scenario is a first-order PT during which the DM particles gains a large mass

$$m_\chi^\infty \equiv y_\chi \langle \phi \rangle^\infty \gg T_n \,. \tag{4.3}$$

This hierarchy requires a transition with large order parameter, $\langle \phi \rangle^\infty \gg T_n/y_\chi$, a typical feature of quasi-conformal or dilaton-like setups [172–191]. We do not specify a scalar potential but work with Eq. (4.3) as an assumption, which is sufficient for our purposes. A particle can only traverse the bubble wall if its kinetic energy is large enough to compensate for the mass increase from (approximately) zero to $m_\chi^\infty$. Consequentially, most of the thermally distributed $\chi$'s at temperature $T_n$ that approach the bubble wall are reflected. These particles are trapped in the false vacuum, where their overabundance depletes via annihilation.

---

[1] We here use "$\sim$" instead of "$=$" to leave room for the mass corrections from higher-dimensional operators that we will later add.

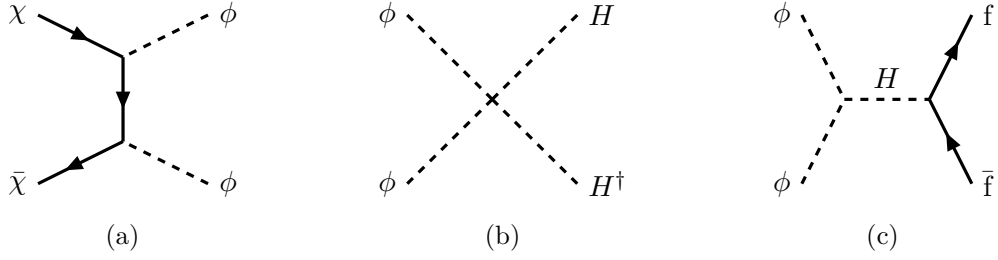

**Figure 4.2:** The relevant processes for Filtered DM: In the false vacuum, the massless $\chi$ is thermalized via (a), but freezes out in the true vacuum due to the large mass obtained during the PT. (Note that $\chi\bar\chi \leftrightarrow \phi$ is typically forbidden by the thermal mass of $\chi$.) The processes (b) and (c) keep the scalar $\phi$ at a common temperature with the SM, which is one of our simplifying assumptions. $H$ and f represent the Higgs field and an arbitrary SM fermion, respectively.

**Thermalization in the false vacuum.** To ensure that $\chi$ is initially in thermal equilibrium and that any arising overabundance annihilates away via the process shown in Fig. 4.2a, we require that its interaction with $\phi$ is faster than the completion of the PT, i.e. [77]

$$\begin{aligned}\Gamma^{\text{false}}_{\chi\bar\chi\leftrightarrow\phi\phi} &= \langle\sigma v\rangle_{\chi\bar\chi\leftrightarrow\phi\phi}\, n^{\text{eq}}_\chi \\ &\approx \frac{y_\chi^4}{256\pi T_n^2}\left[8\log\left(6\frac{T_n}{m_\phi}\right)-3\right]\times 2\frac{3}{4}\frac{\zeta(3)}{\pi^2}T_n^3 \\ &\gtrsim \beta\end{aligned} \tag{4.4}$$

in the false vacuum.[2] Here, $\langle\sigma v\rangle$ is a thermally averaged cross section [283] and $n^{\text{eq}}_\chi$ is the equilibrium number density of $\chi$. The parameter $\beta$ is the inverse time scale of the PT, defined in Eq. (3.19), and highly depends on the (unspecified) scalar sector. We conservatively assume $\beta = 1000\,H$ for our numerical analyses, corresponding to a rather fast transition [284]. To evaluate Eq. (4.4) we use $m_\phi \sim T_n$ and $H \propto T_n^2$ (the first Friedmann equation) for a radiation-dominated Universe.

**Freeze-out in the true vacuum.** A tiny fraction of $\chi$ particles in the tail of the thermal distribution is energetic enough to permeate the bubble wall and reach the true vacuum, where they drop out of equilibrium and instantaneously freeze out. The DM abundance is hence set by the PT, a distinctive feature of the Filtered DM mechanism which sets it apart from the conventional WIMP scenario. To make sure that the freeze-out actually

---

[2]Note that a violation of this condition would imply an inefficient annihilation and lead to the accumulation of $\chi$ overdensities in front of the bubble walls. This could in turn slow down the walls and even result in the formation of black holes. Chapter 5 is devoted to this interesting scenario.

occurs, we demand that the process in Fig. 4.2a becomes inefficient, i.e. [77]

$$\Gamma^{\text{true}}_{\chi\bar{\chi}\leftrightarrow\phi\phi} \approx \frac{9y_\chi^4 T_n}{64\pi(m_\chi^\infty)^3} \times 2\left(\frac{m_\chi^\infty T_n}{2\pi}\right)^{\frac{3}{2}} \exp\left(-\frac{m_\chi^\infty}{T_n}\right)$$
$$\lesssim H \tag{4.5}$$

in the true vacuum. Notice the large exponential suppression of the number density thanks to $m_\chi^\infty \gg T_n$. The resulting DM abundance normalized to the critical density has been estimated analytically in Ref. [77] and amounts to

$$\Omega_{\text{DM}}h^2 \approx 0.17\left(\frac{T_n}{\text{TeV}}\right)\left(\frac{m_\chi^\infty}{30T_n}\right)^{-\frac{5}{2}} \exp\left(-\frac{m_\chi^\infty}{30T_n}\right), \tag{4.6}$$

where $h$ is the Hubble constant divided by $100\,\text{km/s/Mpc}$. This expression can be matched to the observed abundance, $\Omega_{\text{DM}}h^2 \approx 0.12$ [36], to obtain a relation between $m_\chi^\infty/T_n$ and $T_n$. At $T_n = 1\,\text{TeV}$, for instance, a final mass of $m_\chi^\infty \sim 30\,T_n$ yields the observed relic abundance. Thanks to the exponential in Eq. (4.6), the Filtered DM mechanism is viable across wide ranges of temperatures and DM masses, exceeding even the Griest–Kamionkowski unitarity bound of WIMP DM at $m_\chi \sim 100\,\text{TeV}$ (see Section 2.2.2).

**Further assumptions and constraints.** The connection between the dark sector and the SM is established via a Higgs portal [285–287],

$$V(\phi, H) \supset \lambda_{\phi H}\phi^2 H^\dagger H, \tag{4.7}$$

which is part of the scalar potential. One of the simplifying assumptions that went into the calculation of Eq. (4.6) is that $\phi$ has a common temperature with the SM plasma and that any overabundance of $\phi$ generated by $\chi$ annihilation is immediately depleted.[3] This implies the requirement

$$\Gamma_{\phi\phi\leftrightarrow\text{SM}} > \max(H, \Gamma_{\chi\bar{\chi}\leftrightarrow\phi\phi}), \tag{4.8}$$

where the interaction rate $\Gamma_{\phi\phi\leftrightarrow\text{SM}}$ encompasses the processes in Figs. 4.2b and 4.2c and is proportional to $\lambda_{\phi H}^2$ (see Ref. [77] for details).[4] This constraint places a lower bound on $\lambda_{\phi H}$ and thereby also on the DM–nucleon cross section relevant for direct detection. In our main region of interest, $m_\phi \gg m_h$, the bound is given by $\lambda_{\phi H}^2 \gtrsim 10^5\, T_n/M_{\text{Pl}}$. Collider searches for invisible Higgs decays set an upper bound of $\lambda_{\phi H} \lesssim 0.007(1 - 4m_\phi^2/m_h^2)^{-1/4}$ in the regime $m_\phi < m_h/2$ [288], where $m_h \approx 125\,\text{GeV}$ is the Higgs mass.

Finally, we require that all couplings are perturbative, i.e. $y_\chi, \lambda_{\phi H} < \sqrt{4\pi}$, and that the scalar mass matrix contains the physical Higgs mass as an eigenvalue, which translates

---

[3] The heating of the SM bath caused by the depletion of $\phi$ is negligible due to the large number of SM degrees of freedom (DOFs) in the early Universe, $g_\star \sim 100$, compared to the $\mathcal{O}(1)$ number of dark DOFs.

[4] The process in Fig. 4.2c is only relevant below the electroweak scale, where our baryogenesis mechanism does not function anyhow.

to [289, 290]

$$\lambda_{\phi H} < \frac{|m_\phi^2 - m_h^2|}{\langle\phi\rangle^\infty \langle H\rangle^\infty}, \tag{4.9}$$

where $\langle H\rangle^\infty \approx 246\,\text{GeV}$ is the Higgs VEV in the broken electroweak phase.

### 4.2.2 Filtered Baryogenesis

In the following, we augment the Filtered DM scenario with the missing ingredients for baryogenesis by imposing a set of EFT operators.

**Generation of a dark chiral asymmetry.** According to Sakharov (see Section 2.2.1), one crucial requirement for baryogenesis is the violation of CP. As the amount of CP violation in the SM is too small to explain the observed BAU, we introduce a violation in the dark sector via the extended dimension-5 Yukawa operator

$$\mathcal{L}_\chi = -y_\chi \left(\phi + \frac{T_R^\chi + i T_I^\chi}{\langle\phi\rangle^\infty}\phi^2\right)\overline{\chi_R}\chi_L + h.c., \tag{4.10}$$

which induces a complex mass of $\chi$,

$$\overline{m}_\chi(z) = y_\chi \langle\phi\rangle(z)\left(1 + (T_R^\chi + i T_I^\chi)\frac{\langle\phi\rangle(z)}{\langle\phi\rangle^\infty}\right). \tag{4.11}$$

Throughout this work, we use the definitions $m_\chi \equiv \text{abs}(\overline{m}_\chi)$ and $\theta_\chi \equiv \arg(\overline{m}_\chi)$. As a consequence of the complex coupling, the interactions of $\chi$ with $\phi$ – i.e. with the boundary of the PT – occur in an asymmetric way: $\chi$ particles of a certain chirality, depending on the sign of $T_I^\chi$, are preferentially reflected or transmitted by the wall. Together with the violation of parity symmetry (P) due to the existence of a bubble wall, this results in an overabundance of $\chi_{\text{RH}} + \bar\chi_{\text{RH}}$ vs. $\chi_{\text{LH}} + \bar\chi_{\text{LH}}$ on one side of the wall and vice versa. The subscripts "RH" and "LH" denote the particles' chiralities.[5]

In order to reduce the number of parameters, we assume a maximal phase by setting $T_R^\chi = 0$, so that the amount of CP violation is parameterized by $T_I^\chi$ alone. Note that an EFT approach requires that the operator suppression scale, in this case $\langle\phi\rangle^\infty/|T_I^\chi|$, is larger than all other energy scales in the effective theory, with the largest being $m_\chi^\infty$ in our setup. Together with the fact that our mechanism operates around $y_\chi \sim 1$, as we will find, the EFT is valid if $|T_I^\chi| \lesssim 1$.

**Transfer to the visible sector.** In the context of baryogenesis, an asymmetry is of little use if it is stuck in the dark sector. One of the simplest EFT operators connecting $\chi$ to the SM is $(H^\dagger H)(\overline{\chi}\chi)$ at dimension 5. This operator can, for instance, turn the pair $\chi_{\text{LH}}\bar\chi_{\text{LH}}$ into $H^\dagger H$. However, the latter does not carry any type of asymmetry. Therefore, this operator can only deplete the chiral asymmetry but not convert or transfer it.

---

[5]In our notation, $\chi_L \equiv P_L\chi$ is the field operator that annihilates $\chi_{\text{LH}}$ and creates $\bar\chi_{\text{RH}}$, while $\overline{\chi_R} \equiv (P_R\chi)^\dagger\gamma^0$ annihilates $\bar\chi_{\text{LH}}$ and creates $\chi_{\text{RH}}$, and $P_{L/R} \equiv (1 \mp \gamma^5)/2$ are the chiral projectors. Note that the charge conjugation (parity conjugation) of $\chi_L$ annihilates and creates quanta of $\bar\chi_{\text{LH}}$ and $\chi_{\text{RH}}$ ($\chi_{\text{RH}}$ and $\bar\chi_{\text{LH}}$) instead.

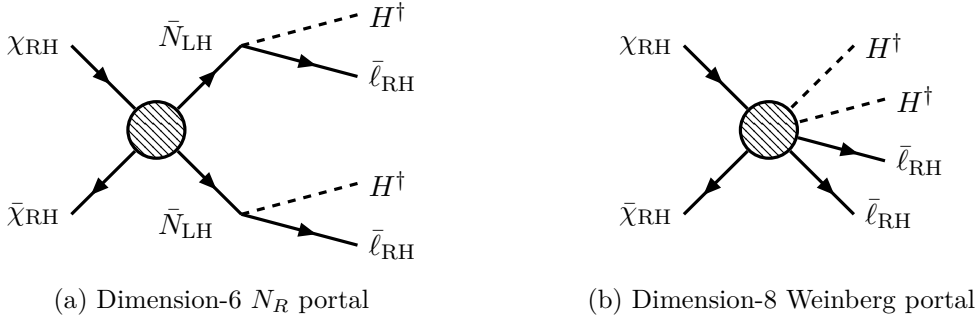

(a) Dimension-6 $N_R$ portal  (b) Dimension-8 Weinberg portal

**Figure 4.3:** Portals that convert the dark chiral asymmetry into a SM lepton asymmetry. The blobs represent unspecified UV physics captured as EFT operators.

We thus introduce the "dimension-6 $N_R$ portal",

$$\mathcal{L}_\mathrm{p} = \frac{1}{\Lambda_\mathrm{p}^2} \sum_{j=1,2,3} (\overline{N_R^j} N_R^{jc})(\overline{\chi_R}\chi_L) + \sum_{j,k=1,2,3} y_\nu^{jk} \overline{\ell^j} \tilde{H} N_R^k + h.c.,  \quad (4.12)$$

which involves three generations of heavy right-handed Majorana neutrino fields $N_R^j$ as well as left-handed SM lepton doublets $\ell^j$, both carrying lepton number. We assume the same suppression scale $\Lambda_\mathrm{p}$ for all generations and furthermore set $y_\nu^{jk} \equiv y_\nu \delta^{jk}$. With these simplifying assumptions, the system is flavor universal and we will drop the generation indices from now on. Thanks to the structure of the first term, which contains the charge conjugated operator $N_R^c \equiv i\gamma^2 N_R^*$, a chiral asymmetry in $\chi$ is turned into an overabundance of $\bar{N}_\mathrm{LH}$ vs. $N_\mathrm{RH}$ or vice versa. The second term in Eq. (4.12) – a neutrino Yukawa term – then lets the right-handed neutrinos decay into Higgs bosons and leptons or antileptons. The decay occurs at the rate [42, 291, 292]

$$\Gamma_Y^N \approx 7.9 \times 10^{-3} \, y_\nu^2 \, T_n \quad (4.13)$$

and requires a sizable Yukawa coupling $y_\nu$ in order to proceed efficiently. Taken together, the dimension-6 $N_R$ portal conveys the chiral asymmetry from the dark to the visible sector and thereby generates a net lepton number, which is accidentally conserved in the SM. The portal itself does also approximately conserve the produced lepton number after the PT concludes. This is due to the large mass of $\chi$ in the true vacuum, which suppresses its number density and thereby also the portal interaction rate (similar to the triggered freeze-out in Filtered DM).

As an alternative to Eq. (4.12) we introduce the "dimension-8 Weinberg portal",

$$\mathcal{L}_\mathrm{p} = \frac{1}{\Lambda_\mathrm{p}^4} \sum_{j=1,2,3} (\overline{\ell^{jc}} \tilde{H}^*)(\tilde{H}^\dagger \ell^j)(\overline{\chi_R}\chi_L) + h.c. \quad (4.14)$$

Again, we assume a universal suppression scale and from now on hide the generation indices. This portal is based on the standard dimension-5 Weinberg operator [293] – typically considered as a generator of neutrino masses – coupled to the $\chi$ current. Note that it is basically identical to the dimension-6 $N_R$ portal, but with the heavy right-handed neutrinos integrated out. The dimension-8 Weinberg portal directly transfers the dark chiral asymmetry to the SM and thereby produces a lepton–antilepton asymmetry. Figure 4.3 shows diagrams of the suggested portal processes.

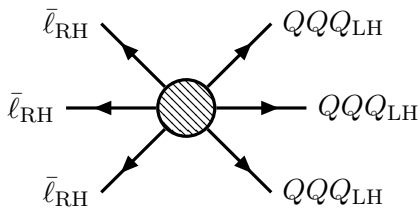

**Figure 4.4:** The Filtered Baryogenesis mechanism relies on electroweak sphaleron transitions which convert right-handed antileptons into left-handed quarks, in a way that conserves $B - L$ but depletes $B + L$. The process turns half of the present lepton asymmetry ($L < 0$) into a baryon asymmetry ($B > 0$).

What both portals have in common is the violation of charge symmetry (C) – one of Sakharov's criteria – and of lepton number. Furthermore, both portal interactions are exponentially suppressed in the true vacuum, where the produced lepton asymmetry thus freezes out. On dimensional grounds we estimate the rate for the $d$-dimensional portal interaction to be

$$\Gamma_\text{p} = T_n \left(\frac{T_n}{\Lambda_\text{p}}\right)^{2(d-4)} \exp\left(-\frac{m_\chi(z)}{T_n}\right), \tag{4.15}$$

which contains the discussed exponential suppression in the true vacuum.

**From lepton number to baryon number.** While baryon and lepton number ($B$ and $L$) are classically conserved in the SM at any finite order of perturbation theory, they are anomalous at the quantum level [40]. The SM hence conveniently provides the necessary ingredient to convert the generated $L$ into $B$: Electroweak sphalerons – non-perturbative, unstable solutions to the electroweak field equations [41] – give rise to a process that interconnects all generations of left-handed quarks and leptons simultaneously, as illustrated in Fig. 4.4. This results in the violation of $B + L$, while $B - L$ is conserved. The rate of electroweak sphaleron transitions is exponentially suppressed in today's Universe – wherefore they are yet undetected and hypothetical – but is expected to be efficient above the electroweak scale with the interaction rate [42]

$$\Gamma_\text{ws} \approx 5.2 \times 10^{-6}\, T_n\,. \tag{4.16}$$

Our mechanism thus has to operate prior to the EWPT, i.e. $T_n \gtrsim 100\,\text{GeV}$, so that the electroweak sphalerons can efficiently relax $B + L$ by turning half of the overabundance of antileptons vs. leptons into an overabundance of baryons vs. antibaryons. The residual lepton asymmetry is unproblematic, as it is only loosely constrained experimentally [294–

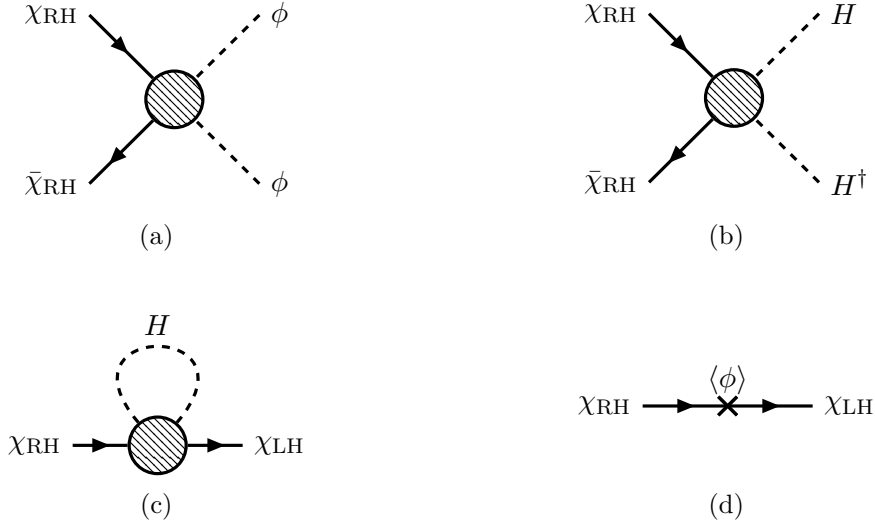

**Figure 4.5:** Washout processes in the Filtered Baryogenesis mechanism. The chiral $\chi$ asymmetry is partly depleted via the dimension-5 part of the Yukawa interaction, (a), as well as the dimension-5 Higgs–DM coupling, (b) and (c). Also the VEV-induced mass of $\chi$ leads to a washout of the chiral asymmetry (d).

296]. Note that electroweak sphalerons couple to left-handed fermions or right-handed antifermions exclusively, which is perfectly compatible with our portals that produce an overabundance of $\bar{\ell}_\text{RH}$ vs. $\ell_\text{LH}$.

**Chiral washout.** Finally, we have to consider the depletion of the generated chiral asymmetry before it is converted. One source of washout is the CP-violating dimension-5 part of the Yukawa interaction in Eq. (4.10). It gives rise to the process shown in Fig. 4.5a and we estimate a rate of

$$\Gamma_5^{(a)} = T_n \left( y_\chi^2 T_I^\chi \frac{T_n}{m_\chi^\infty} \right)^2 \exp\left(-\frac{m_\chi(z)}{T_n}\right) \tag{4.17}$$

for this washout contribution on dimensional grounds. Again, as for the rates of our portal processes, an exponential factor arises due to the Boltzmann suppression for massive $\chi$.

For a consistent EFT approach, all operators up to the considered dimension (6 or 8 in this case) must be included. Most operators, such as the usual Weinberg operator, do however not affect any of the asymmetries we are interested in. Nevertheless, the dimension-5 operator

$$\mathcal{L}_5 = \frac{\lambda_5}{\Lambda_\text{p}} (H^\dagger H)(\bar{\chi}\chi) \tag{4.18}$$

contributes to chiral washout via the processes shown in Figs. 4.5b and 4.5c and must be included. We parameterize the corresponding EFT suppression scale as $\Lambda_\text{p}/\lambda_5$ and

estimate the rates

$$\Gamma_5^{(b)} = T_n \left(\lambda_5 \frac{T_n}{\Lambda_p}\right)^2 \exp\left(-\frac{m_\chi(z)}{T_n}\right), \tag{4.19}$$

$$\Gamma_5^{(c)} = T_n \left(\lambda_5 \frac{T_n}{\Lambda_p}\right)^2. \tag{4.20}$$

Finally, we have to consider that the mass term of $\chi$ allows for chirality flips, as shown in Fig. 4.5d. In the false vacuum (where $\chi$ is massless), this process is inactive, but in the true vacuum (where $\chi$ has a large mass) it leads to a strong chiral washout.[6] We find that the rate of this contribution amounts to

$$\Gamma_M^\chi \approx 2.1 \times 10^{-1} \frac{m_\chi^2(z)}{T_n}, \tag{4.21}$$

based on the formulas given in Refs. [42, 144, 292, 297].

Chiral washout is relevant *during* the occurrence of the mechanism – in vicinity of the bubble wall and in the false vacuum – but also *after* the PT: The portal interaction rate, Eq. (4.15), is exponentially suppressed but not entirely zero in the true vacuum. At a slow rate, the generated lepton and baryon asymmetries are partly converted back to a dark chiral asymmetry, which is then immediately erased by the strong chiral washout. This puts temperature-dependent constraints on the model parameter space, which we will discuss as part of our results. Another contribution to the washout of the produced baryon asymmetry can be the depletion of lepton number due to a possible Majorana mass term for $N_R$. Our mechanism thus requires a tiny Majorana mass term, which, together with the requirement of sizable $y_\nu$, can be satisfied by inverse seesaw models [298, 299]. In any case, the electroweak sphaleron transitions are switched off by the EWPT after which the generated baryon asymmetry is conserved until today.

∗ ∗ ∗

The proposed Filtered Baryogenesis mechanism is illustrated in Fig. 4.6. The combined Lagrangian consists of the CP-violating Yukawa coupling, one of the two proposed portals, the dimension-5 coupling contributing to washout, and the effective scalar potential containing the Higgs portal:

$$\mathcal{L} \subset \mathcal{L}_\chi + \mathcal{L}_p + \mathcal{L}_5 - V(\phi, H). \tag{4.22}$$

In summary, we start with the generation of a dark chiral asymmetry via a CP-violating coupling at the P-violating phase boundary of a first-order PT. The asymmetry is then converted into a SM lepton asymmetry via one of two possible lepton number and C-violating portals. Finally, the baryon and lepton number violating electroweak sphalerons turn the lepton asymmetry into a baryon asymmetry which persists until today. While our explanations suggest a chronological order of the different steps, they are in reality intertwined and occur partly simultaneously. One objection to the functionality of the mechanism might be that the generated asymmetries are equal and opposite on both sides of the wall and should therefore cancel out. However, the PT's out-of-equilibrium dynamics together with diffusion and interaction rates that differ on both sides of the bubble wall allow a net asymmetry to survive, as we will see in our detailed numerical analyses. Note that all three of Sakharov's baryogenesis conditions are fulfilled in the described setup.

---

[6] The thermal mass of $\chi$ is given by the self-energy diagram in which $H$ and $\chi$ both run in a thermal loop and is present in both phases. However, it involves *two* chiral flips and thus conserves chirality overall. We hence neglect the thermal mass of $\chi$ in our analyses.

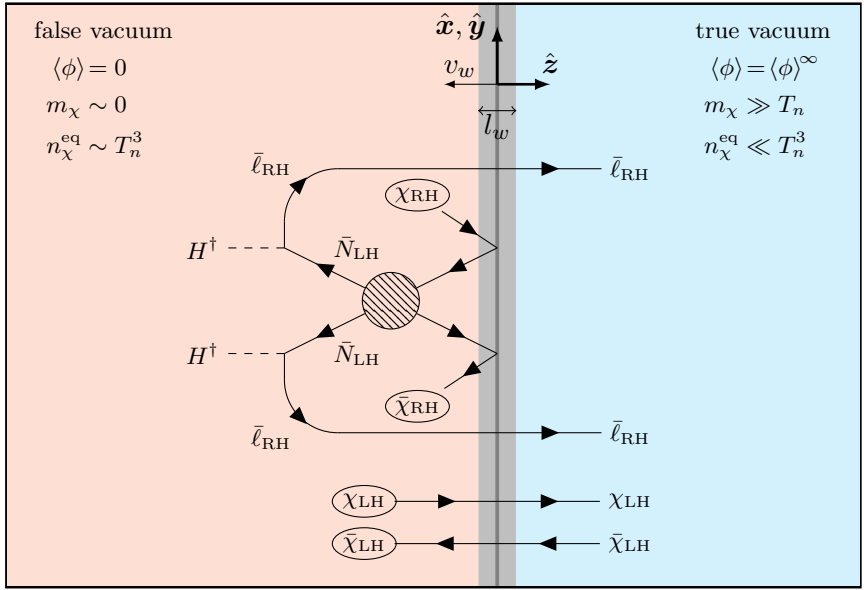

**Figure 4.6:** Illustration of Filtered Baryogenesis. We consider a planar bubble wall (gray band) which moves to the left (negative *z*-direction) when viewed from the rest frame of the plasma. Thanks to the CP-violating Yukawa interaction, right-handed $\chi$ have a higher probability of being reflected than left-handed ones. The chiral asymmetry is transferred to the SM via one of the proposed portal operators (the dimension-6 $N_R$ portal in shown the figure). The generated lepton asymmetry diffuses into all directions and is slowly converted into baryons via electroweak sphalerons (not shown).

## 4.3 Tracking Asymmetries via Transport Equations

In order to verify that the proposed mechanism works as intended, a quantitative analysis is required. The formula that most generally describes the evolution of a phase space density $f$ is the Boltzmann equation,

$$\mathbf{L}[f] = \mathbf{C}[f] \,. \tag{4.23}$$

The Liouville operator on the l.h.s. is the total time derivative of $f$ and describes the dynamics of the free theory, while the collision term on the r.h.s. captures interactions and source terms. Note that Eq. (4.23) may represent a whole system of coupled equations for different species and their individual $f$'s. In general, the phase space density is a function of momentum, space, and time. However, we apply the following simplifications:

1. We assume that the bubble walls are thin compared to the bubble radii. This is justified because $l_w \sim 1/T_n$, whereas the bubble radii quickly reach $1/H \sim M_{\text{Pl}}/T_n^2$ with Planck mass $M_{\text{Pl}}$. We can thus describe the wall as flat in the $x$–$y$-plane, leaving the $z$-direction as the only non-trivial spatial coordinate.

2. Quickly after the true-vacuum bubbles have formed, the particle densities in the vicinity of the wall reach a steady state. This fact eliminates any time dependence if the system is viewed from the rest frame of the bubble wall, in which the plasma moves towards positive $z$. In this description, the wall profile and the mass of $\chi$ are functions of the wall-frame coordinate $z$ only.

3. Finally, we make use of the diffusion ansatz which assumes that the equilibration among the different momentum modes via self-scattering is fast compared to other interaction rates. This is equivalent to demanding that the diffusion coefficients (i.e. the mean free paths) of the involved particles are the largest length scales of the problem [300]. With the diffusion approximation, the momentum dependence of the Boltzmann equation can be integrated out. This yields a set of transport equations, which are partial differential equations that apply to number densities. (We discuss the implications of going beyond the diffusion approximation in Appendix 4.C.)

We define the asymmetries

$$
\begin{aligned}
\delta\chi &\equiv n_{\chi_{\text{RH}}} - n_{\bar\chi_{\text{LH}}}, \\
\delta\bar\chi &\equiv n_{\chi_{\text{LH}}} - n_{\bar\chi_{\text{RH}}}, \\
\delta N &\equiv n_{N_{\text{RH}}} - n_{\bar\ell_{\text{LH}}}, \\
\delta\ell &\equiv n_{\ell_{\text{LH}}} - n_{\bar\ell_{\text{RH}}}, \\
\delta Q &\equiv n_{Q_{\text{LH}}} - n_{\bar Q_{\text{RH}}}, \\
\delta H &\equiv n_H - n_{H^\dagger},
\end{aligned}
\tag{4.24}
$$

based on the $z$-dependent number densities $n_a$ of particle species $a$ (where $a$ is a placeholder). Here, $N$, $\ell$, and $Q$ are the right-handed neutrinos, and the left-handed lepton and quark doublets, respectively. The defined asymmetries apply to any of the three fermion generations (remember the assumed flavor universality). $H$ denotes the Higgs doublet. Note that $\delta\chi = -\delta\bar\chi$ at all times, as there are no $\chi$ number violating interactions and we assume a symmetric initial state. Furthermore, we neglect the SM lepton Yukawa couplings, allowing us to disregard right-handed SM leptons entirely. Right-handed quarks, on the other hand, are expected to develop an asymmetry due to the fast strong sphaleron transitions and the more sizable Yukawas, two effects which both mix chiralities. The asymmetry in any of the six $SU(2)_L$ singlet quarks can thus be approximated as $\delta Q/2$. (See Appendix 4.A for an extended version of the transport equations, including SM lepton Yukawas and considering a finite strong sphaleron rate.) Based on these definitions and relations, the net baryon and lepton number densities are given by[7]

$$n_B = \left(3 + \frac{6}{2}\right)\frac{\delta Q}{3} = 2\,\delta Q\,,$$
$$n_L = 3\,\delta\ell\,. \tag{4.25}$$

Finally, the transport equations that apply to the asymmetries are given by

$$\begin{aligned}
\partial_\mu j^\mu_{\delta\chi} &= 3\Gamma_\mathrm{p}\mu_\mathrm{p} + (\Gamma_5^{(\mathrm{a})} + \Gamma_5^{(\mathrm{b})} + \Gamma_5^{(\mathrm{c})} + \Gamma_M^\chi)\mu_M^\chi + S_\chi\,, \\
\partial_\mu j^\mu_{\delta N} &= 2\Gamma_\mathrm{p}\mu_\mathrm{p} - \Gamma_Y^N \mu_Y^N\,, \\
\partial_\mu j^\mu_{\delta\ell} &= \Gamma_Y^N \mu_Y^N - \Gamma_\mathrm{ws}\mu_\mathrm{ws}\,, \\
\partial_\mu j^\mu_{\delta H} &= 3\Gamma_Y^N \mu_Y^N\,, \\
\partial_\mu j^\mu_{\delta Q} &= -\frac{3}{2}\Gamma_\mathrm{ws}\mu_\mathrm{ws}\,,
\end{aligned} \tag{4.26}$$

in case of the dimension-6 $N_R$ portal and

$$\begin{aligned}
\partial_\mu j^\mu_{\delta\chi} &= 3\Gamma_\mathrm{p}\mu_\mathrm{p} + (\Gamma_5^{(\mathrm{a})} + \Gamma_5^{(\mathrm{b})} + \Gamma_5^{(\mathrm{c})} + \Gamma_M^\chi)\mu_M^\chi + S_\chi\,, \\
\partial_\mu j^\mu_{\delta\ell} &= 2\Gamma_\mathrm{p}\mu_\mathrm{p} - \Gamma_\mathrm{ws}\mu_\mathrm{ws}\,, \\
\partial_\mu j^\mu_{\delta H} &= 6\Gamma_\mathrm{p}\mu_\mathrm{p}\,, \\
\partial_\mu j^\mu_{\delta Q} &= -\frac{3}{2}\Gamma_\mathrm{ws}\mu_\mathrm{ws}\,,
\end{aligned} \tag{4.27}$$

for the dimension-8 Weinberg portal, which we derived in analogy to Refs. [42,142,144,301]. We will dissect these equations in the following.

### 4.3.1 Particle Dynamics

The l.h.s. of the transport equations – the remnant of the Liouville operator – is a set of divergences of the asymmetry 4-currents $j^\mu_{\delta a}$ of particle species $a$, and can be written as

$$\partial_\mu j^\mu_{\delta a} \equiv \frac{\partial}{\partial t}\delta a + \boldsymbol{\nabla}\cdot\boldsymbol{j}_{\delta a} = [v_w - D'_a(z)]\delta a'(z) - D_a(z)\delta a''(z) - D''_a(z)\delta a(z)\,, \tag{4.28}$$

where the primes denote derivatives w.r.t. $z$. According to our steady state description, we rewrote the time derivative of the asymmetry $\delta a$ as the bubble wall velocity, $v_w$, times the spatial derivative of $\delta a$. Furthermore, we used a generalized version of Fick's law to write the asymmetry 3-current as $\boldsymbol{j}_{\delta a} = -\boldsymbol{\nabla}(D_a \delta a)$. The latter introduces the diffusion coefficient $D_a$, which represents the mean free path of species $a$. While in our setup this quantity is constant for all SM particles, it depends on $m_\chi$ and thus on $z$ for $\chi$. In

---

[7] In the introductory part of this thesis, in Eq. (2.5), we denoted the baryon and antibaryon number densities separately. Here, $n_B$ is meant to be the difference of the two.

Appendix 4.B we list the used values and conduct the calculation of $D_\chi(z)$. The terms proportional to $D'_a(z)$ and $D''_a(z)$ usually do not appear in the literature on electroweak baryogenesis, where all $D_a$ are typically taken to be independent of $z$.[8] Setting Eq. (4.28) to zero yields Fokker–Planck equations in one spatial dimension for a stationary system. Neglecting the $z$-dependence of $D_\chi$ would reduce these to a the more commonly used form of equations that follow Fick's diffusion law (see Ref. [302] for a comparison of the two approaches).

### 4.3.2 Particle Interactions

The r.h.s. of the transport equations, Eqs. (4.26) and (4.27), originates from the collision terms of the Boltzmann equation and describes interactions between the particle asymmetries. The $\Gamma$'s are interaction rates and have been discussed in Section 4.2.2. The $\mu$'s consist of linear combinations of the effective chemical potentials of the particle asymmetries involved in the respective interactions, i.e.

$$\begin{aligned}
\mu_{\rm p} &\equiv -2\mu_{\delta\chi} - 2\mu_{\delta N}\,, \\
\mu^\chi_M &\equiv -2\mu_{\delta\chi}\,, \\
\mu^N_Y &\equiv \mu_{\delta N} - \mu_{\delta\ell} - \mu_{\delta H}\,, \\
\mu_{\rm ws} &\equiv 3\mu_{\delta\ell} + 9\mu_{\delta Q}\,,
\end{aligned} \quad (4.29)$$

for the dimension-6 $N_R$ portal and

$$\begin{aligned}
\mu_{\rm p} &\equiv -2\mu_{\delta\chi} - 2\mu_{\delta\ell} - 2\mu_{\delta H}\,, \\
\mu^\chi_M &\equiv -2\mu_{\delta\chi}\,, \\
\mu_{\rm ws} &\equiv 3\mu_{\delta\ell} + 9\mu_{\delta Q}\,,
\end{aligned} \quad (4.30)$$

for the dimension-8 Weinberg portal. The integer factors in the above expressions reflect how the interactions depend on the different particle asymmetries. For instance, the factors in $\mu_{\rm p}$ represent the multiplicity of particles in the portal processes, see Fig. 4.3, using that $\mu_{\delta\bar{\chi}} = -\mu_{\delta\chi}$. The effective chemical potentials, $\mu_{\delta a}$, are related to the number density asymmetries, $\delta a$, via [144]

$$\delta a = \mu_{\delta a}\, k_a(\tfrac{m_a}{T_n}) + \mathcal{O}(\mu^3_{\delta a})\,, \quad (4.31)$$

which we express as an expansion around small $\mu_{\delta a}$. Here, we absorbed a factor $T_n^2/6$ in the definition of $\mu_{\delta a}$ (hence the denotation "effective"), and the factors $k_a(m_a/T_n)$ are momentum integrals of the respective phase space distribution functions with the chemical potentials set to zero, see for instance Refs. [42,144]. For massless particles, $k_a(0)$ is simply the number of Weyl fermion or real scalar DOFs: $k_\chi(0) = k_N(0) = 1$, $k_\ell(0) = 2$, $k_H(0) = 4$, and $k_Q(0) = 6$. Considering the thermal masses of the SM species (which are their only masses prior to the EWPT) gives rise to slight corrections compared to these integer numbers. The factor $k_\chi$ is $z$-dependent and changes from 1 in the false vacuum (where $\chi$ is massless) to $\sim 0$ in the true vacuum (where $m_\chi/T_n$ is large).

---

[8]Neglecting the $z$-dependence of the diffusion coefficient is justified, if the mass change compared to the temperature is not too large. This is (more or less) fulfilled by all SM species at the EWPT, for instance. In our PT, however, $\chi$ gains a very large mass, $m^\infty_\chi \gg T_n$, so that the variation of the diffusion coefficient should be taken into account.

The integer factors in front the $\Gamma\mu$ terms in the transport equations describe how the processes change the asymmetries. For instance, the coefficients in front of the $\Gamma_\mathrm{p}\mu_\mathrm{p}$ terms in Eq. (4.27) can be understood as follows: The portal process (Fig. 4.3b) occurs for each of the three lepton generations, so it increases $\delta\chi$ by three, decreases $\delta\bar\chi$ by three (which we do not track separately due to $\delta\bar\chi = -\delta\chi$), increases $\delta\ell$ by two (note that our $\delta\ell$ counts the asymmetry *per generation*), and increases $\delta H$ by six.

### 4.3.3 The Source Term

The final and most crucial ingredient on the r.h.s. of the transport equations, Eqs. (4.26) and (4.27), is the source term $S_\chi$, which quantifies the generation of a chiral asymmetry $\delta\chi$ originating from the CP-violating interaction of $\chi$ with the bubble wall. After applying the semi-classical WKB approximation and in the limit of small wall velocities, $v_w \ll 1$, the source takes the form [143, 303–310]

$$S_\chi(z) = v_w \frac{D_\chi}{K_4} \left[ K_8 \left(m_\chi^2 \theta'_\chi\right)' - K_9\, m_\chi^2 \left(m_\chi^2\right)' \theta'_\chi \right]'. \qquad (4.32)$$

Note that all quantities except $v_w$ on the r.h.s. depend on $z$. The functions $K_n$ represent thermal averages of different combinations of $\chi$'s momentum, energy, and phase space density (and its derivatives), and are listed in Ref. [311]. The complex nature of the CP-violating Yukawa interaction, Eq. (4.10), enters the source term via $\theta_\chi$, the phase of $\chi$'s complex mass. The WKB source as a function of $z$ exhibits multiple sign flips due to the interplay of its different terms. Note that the WKB approximation is valid as long as the involved mean free paths are smaller than the width of the bubble wall [312]. We will thus constrain the model parameter space according to the condition $D_\chi < l_w$ at $z = 0$ (roughly at the position where the source is active) in our numerical results.

∗ ∗ ∗

Now that we have discussed all components and quantities, the transport equation in Eqs. (4.26) and (4.27) can be solved numerically. To do so, we first rewrite the second-order differential equations in terms of twice as many first-order differential equations. We then solve these equations numerically using the boundary-value-problem solver provided by the PYTHON library SCIPY [313], with boundary conditions $\delta a(z \to -\infty) = 0$ and $\delta a'(z \to \infty) = 0$.[9] The first condition ensures vanishing asymmetries deep in the false vacuum, i.e. prior to the PT. The second condition, on the other hand, demands that all asymmetries freeze out deep in the true vacuum, i.e. after the PT. The latter neglects possible washout effects, which we will consider separately in our results in the form of parameter space constraints. Note that our system involves vastly different length scales: the large mean free path of $\chi$ and the slow weak sphaleron rate on the one hand, and the comparably small $1/\Gamma_M^\chi$ and $l_w$ on the other hand. To improve numerical stability in light of this hierarchy, we solve the transport equations in terms of the variable $\tilde{z} \equiv \mathrm{arsinh}(3z/l_w)$, which depends linearly on $z$ near bubble wall (of width $l_w$) and logarithmically on $z$ far away from it.

---

[9]In simplified situations, where all $z$-dependent quantities are modeled as step functions, the system is analytically solvable [144]. We have verified that our numeric results coincide with analytically obtained ones in applicable cases.

## 4.4 Results

With the appropriate transport equations at hand, the Filtered Baryogenesis mechanism can be tested quantitatively. We first present and discuss the solutions of the transport equations for exemplary parameter points. Next, we explore slices of the model parameter space and point out the region in which both the Filtered DM and Filtered Baryogenesis can work simultaneously. We finally assess the prospects of direct detection.

### 4.4.1 Benchmark Points

In Fig. 4.7 we present the $z$-dependent solutions to the transport equations for two exemplary parameter points, one for the dimension-6 $N_R$ portal and one for the dimension-8 Weinberg portal. The asymmetries are plotted normalized w.r.t. the entropy density, $s = 2\pi^2 g_\star T_n^3/45$ with $g_\star = 106.75$ effective relativistic DOFs prior to the EWPT, i.e.

$$\begin{aligned} Y_{\delta a} &\equiv \frac{\delta a}{s}\,, \\ Y_B &\equiv \frac{n_B}{s} = 2\frac{\delta Q}{s}\,. \end{aligned} \tag{4.33}$$

The chosen normalization is particularly convenient, because both $n_B$ and $s$ redshift equally and are otherwise constant after the generation of the baryon asymmetry. While the presented benchmark points are picked arbitrarily, we choose the amount of CP violation (parameterized by $T_I^\chi$) such that deep in the true vacuum, $Y_B$ reaches the experimentally observed value [36]

$$Y_B^{\mathrm{obs}} = (8.65 \pm 0.04) \times 10^{-11}\,. \tag{4.34}$$

At both benchmark points presented in Fig. 4.7, $\chi$ develops a chiral asymmetry (blue curve) in the vicinity of the bubble wall (at $z \sim 0$). The spikes in the asymmetry are caused by the source term, $S_\chi$, and its interplay with the term proportional to $D_\chi''$ in the transport equations, which acts similar to a source. (The quantities $S_\chi$ and $D_\chi''$ are shown in the lower plots of both panels of the figure.) The sizable diffusion coefficient $D_\chi$ spreads the generated chiral asymmetry over distance scales much larger than the wall width, $l_w$. In case of the dimension-6 benchmark point, this spread reaches even further due to the larger mean free path of $\chi$, which is in turn caused by the choice of a smaller Yukawa coupling (compared to the dimension-8 benchmark point). In the true vacuum, the large mass of $\chi$ leads to an immediate washout of the chiral asymmetry via the $\Gamma_M^\chi$ rate. In the false vacuum, where the chiral asymmetry persists (up to some minor washout via the $\Gamma_5$ rates), our portal operators transfer part of the asymmetry to the SM (green and red curves), taking a detour via right-handed neutrinos (orange curve) in case of the dimension-6 $N_R$ portal. The bubble wall has enough time to move past the generated lepton asymmetry, so that the latter spreads far into the true vacuum. Finally, the electroweak sphalerons turn part of the negative lepton number into the desired baryon asymmetry (black curve). Because the sphaleron rate, given in Eq. (4.16), is slow compared to all other rates in the mechanism, it takes a comparably long time, $1/\Gamma_{\text{ws}} \sim 10^5/T_n$, until the sphalerons show a sizable effect. This is why in our parametrization of the spatial coordinate, which works logarithmically far from the origin, the increase of $B$ becomes visible only at distances of around $10^5/T_n = 10^4\, l_w$ away from the bubble wall.

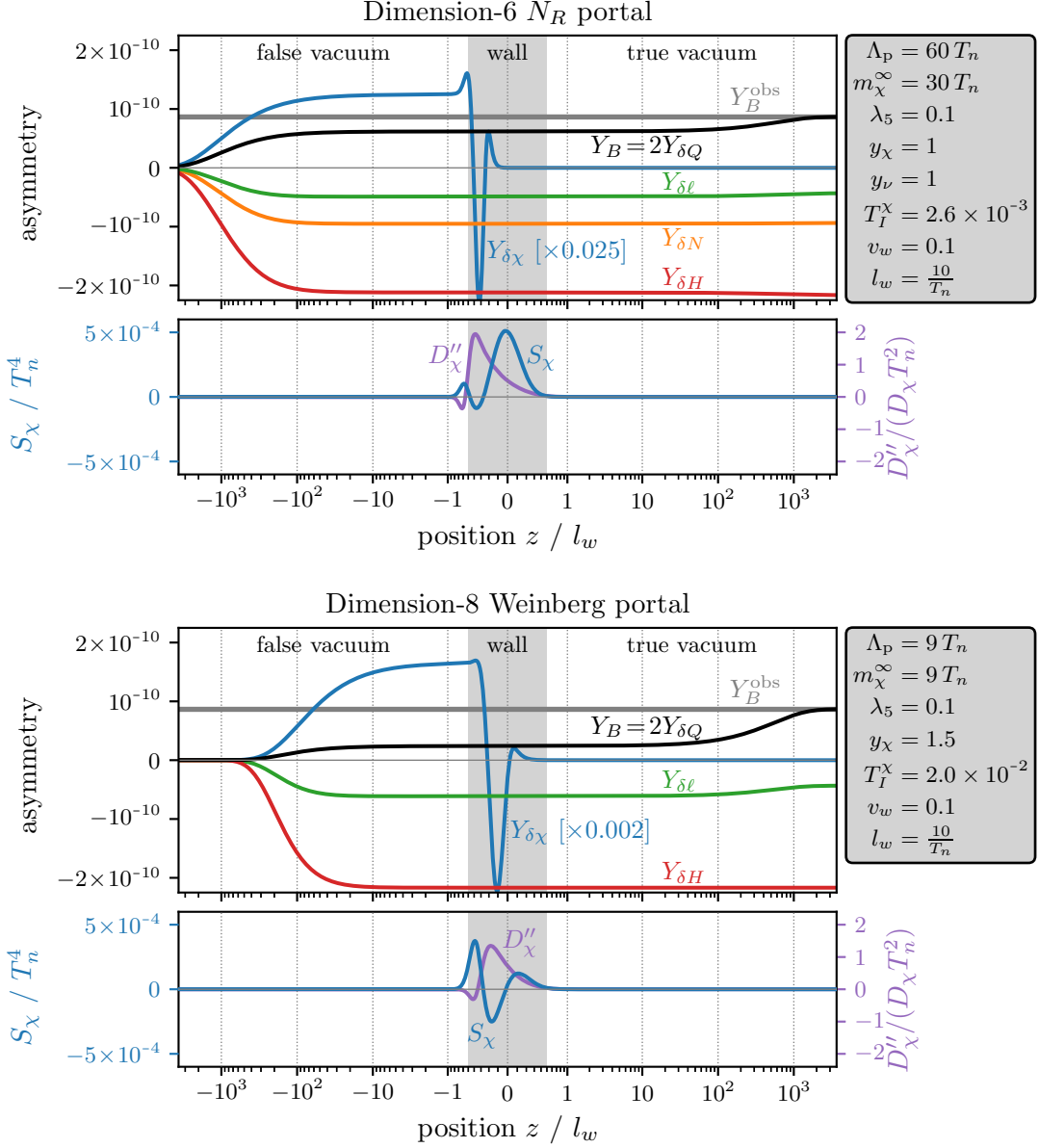

**Figure 4.7:** Solution to the steady state transport equations for two exemplary parameter points of Filtered Baryogenesis, one for the dimension-6 $N_R$ portal (top), and one for the dimension-8 Weinberg portal (bottom). The resulting dark chiral asymmetry and the SM asymmetries are shown in the main plots of both panels, together with the observed value of the baryon asymmetry, which is successfully generated by our mechanism in both cases. The two small plots show the source term $S_\chi$ and the second derivative of the diffusion coefficient $D''_\chi$, which in the transport equations are both responsible for the generation of $\chi$ asymmetry. The wall width is highlighted as a vertical gray band. The horizontal axis is linear in $\tilde{z} = \mathrm{arsinh}(3z/l_w)$, which in turn scales linearly with $z$ close to the wall and logarithmically with $z$ far away from it.

Note that we plotted the $\chi$ asymmetries scaled down (by factors stated in the figure) compared to the SM asymmetries. This reflects the fact that the portal rates, given in Eq. (4.15), are suppressed by the large EFT scale, $\Lambda_{\text{p}} \gg T_n$. As a consequence, in order to obtain the desired baryon asymmetry, a much larger chiral asymmetry must be generated in the dark sector, i.e. $Y_{\delta\chi} \gg Y_B^{\text{obs}} \sim 10^{-10}$.

### 4.4.2 Model Parameter Space

We will now broaden our scope by exploring entire slices of the model parameter space. In Fig. 4.8 we indicate by a color gradient (blue and green shading) the sign and magnitude of CP violation (parameterized by $T_I^\chi$) required to obtain $Y_B = Y_B^{\text{obs}}$ via the Filtered Baryogenesis mechanism, for both of the suggested portals. In the white regions, no value satisfying $|T_I^\chi| < 1$ (to ensure EFT validity) yields the experimentally observed baryon asymmetry.

In the left panels of Fig. 4.8, the Yukawa coupling $y_\chi$ and the DM mass $m_\chi^\infty$ are varied along the axes. As the plots reveal, the mechanism works in wide regions of the parameter slice. Increasing $m_\chi^\infty$ has two effects: Firstly, since we fix $\Lambda_{\text{p}}$ to a multiple of $m_\chi^\infty$, the portal rate becomes more suppressed. Secondly, $\chi$ particles that develop the chiral asymmetry – the ones with momenta close to the reflection threshold $p_z \sim m_\chi^\infty \gg T_n$ – are thermally less abundant. Above certain values of $m_\chi^\infty$, the mechanism is incapable of generating the observed baryon asymmetry. Varying $y_\chi$, on the other hand, leads to multiple sign flips in the required value for $T_I^\chi$. This is due to the strong dependence of the diffusion coefficient $D_\chi$ on $y_\chi$ together with the subtle interplay between the source term, $S_\chi$, and the term proportional to $D_\chi''$. As indicated by the black dotted contours, the validity of our results is constrained to values $y_\chi \gtrsim 1$. For smaller values, the mean free path of $\chi$ around the wall becomes larger than the wall width, $D_\chi(z=0) > l_w$, implying that the WKB approximation used in the derivation of the source term is inappropriate. The proposed baryogenesis mechanism potentially still works in this regime, but our numerical results may become inaccurate.

The right panels of Fig. 4.8 show the effects of varying the portal suppression scale, $\Lambda_{\text{p}}$, and the dimension-5 coupling, $\lambda_5$. We observe that for larger $\Lambda_{\text{p}}$ – i.e. slower asymmetry transfer to the SM – more CP violation is needed. Above a certain threshold, the required $|T_I^\chi|$ would again be too large and spoil the EFT validity. Also very small $\Lambda_{\text{p}}$ can render the EFT invalid, namely in the regime $\Lambda_{\text{p}} < m_\chi^\infty$ (orange hatched area). However, one could argue that it suffices to have a consistent EFT description in the true vacuum, where $\chi$ is massless and the portal is at work. Regardless, $\Lambda_\mathrm{p} \sim T_n$ constitutes a hard cutoff below which the EFT is undeniably invalid (orange area). The dimension-5 coupling, $\lambda_5$, controls the rate of chiral washout via the processes in Figs. 4.5b and 4.5c, an effect that becomes mostly negligible for $\lambda_5 \ll 1$.

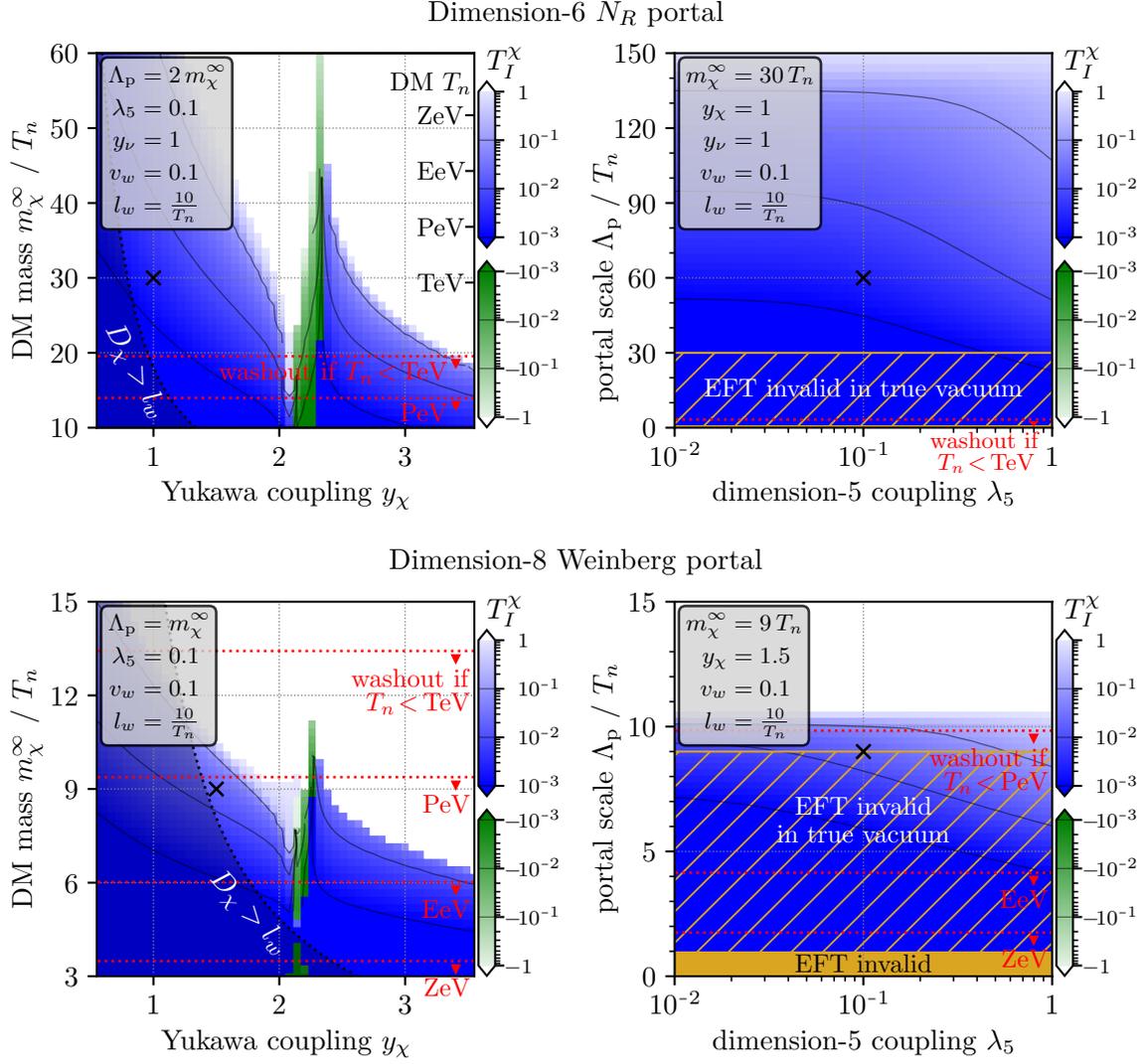

**Figure 4.8:** Slices of the Filtered Baryogenesis model parameter space with the dimension-6 $N_R$ portal (top) and the dimension-8 Weinberg portal (bottom). The color gradient (blue and green shading) indicates the sign and magnitude of the CP violation (in terms of the parameter $T_I^\chi$) required to obtain the observed baryon asymmetry. In the white regions, no value $|T_I^\chi| < 1$ (required for EFT validity) yields the target asymmetry. The orange hatched and orange filled regions indicates where the EFT approach breaks down in the true vacuum or entirely because $\Lambda_p > m_\chi^\infty$ or $\Lambda_p > T_n$, respectively. The WKB source term may be inaccurate to the left of the black dotted contour, where $D_\chi(z=0) > l_w$. Red dotted contours indicate the washout constraints at different temperatures. Black crosses show the exemplary parameter points presented in Fig. 4.7. The second vertical axis in the upper left plot indicates the temperatures $T_n$ at which the correct DM density is obtained via Filtered DM.

As already alluded to, the produced baryon asymmetry may be washed out by an effect that we neglect in the transport equations: While the portal rate $\Gamma_\mathrm{p}$, given in Eq. (4.15), is exponentially suppressed after the PT, it is not entirely zero. The lepton asymmetry and via the electroweak sphalerons also the baryon asymmetry can slowly "leak" back into the dark sector, where any chiral asymmetry is immediately erased by the large mass of $\chi$. We thus demand that the suppressed portal is slower than the Hubble rate, $\Gamma_\mathrm{p}(z \to \infty) < H$, which yields temperature dependent washout bounds (red dotted contours in Fig. 4.8). We show these limits only for temperatures above the electroweak scale, which constitutes a general lower limit for our mechanism.

Finally, the connection to the Filtered DM mechanism can be made. The requirement to produce the observed DM density, according to Eq. (4.6), can be translated into a relation between $m_\chi^\infty/T_n$ and $T_n$. The second vertical axis in the top left panel of Fig. 4.8 indicates the values of $T_n$ at which the Filtered DM mechanism would be successful. As we can see, there is plenty of viable parameter space in which the DM and baryogenesis mechanisms can work simultaneously in case of the dimension-6 $N_R$ portal. The dimension-8 Weinberg portal suffers from a larger EFT suppression and is hence limited to smaller $m_\chi^\infty/T_n$. In this regime, the Filtered DM would require temperatures much below the electroweak scale to not overclose the Universe. This is incompatible with Filtered Baryogenesis, which relies on efficient electroweak sphalerons.

Note that we keep the value of $y_\nu$ fixed, since our results are basically independent of it as long as $y_\chi \gtrsim 0.1$ (to ensure that right-handed neutrinos decay into leptons fast enough). Furthermore, we also fixed the wall velocity and width, $v_w$ and $l_w$, because we observed only a slight dependence on these parameters as well. While there is no reason for our mechanism to stop working at luminal velocities, $v_w \sim 1$, our results would probably be inaccurate in this regime due to the $v_w \ll 1$ approximations that entered the used version of the source term.

An inaccuracy may arise from the fact that the transport equations are based on the diffusion ansatz, i.e. the assumption of local thermal equilibrium. Especially in case of the large $D_\chi$ that we observe near the wall and in the false vacuum, the different momentum modes equilibrate slower than other rates, the largest being $\Gamma_M^\chi$. A more careful treatment would consider the dynamics of the entire phase space (including the momentum modes), similar to the approach that is pursued in Chapter 5. In Appendix 4.C, we estimate how such a treatment would impact our results, finding that the mechanism probably remains viable.

### 4.4.3 Direct Detection Cross Section

The previous section demonstrated that Filtered Baryogenesis is a viable mechanism over substantial regions of the model parameter space and even exhibits an overlap with Filtered

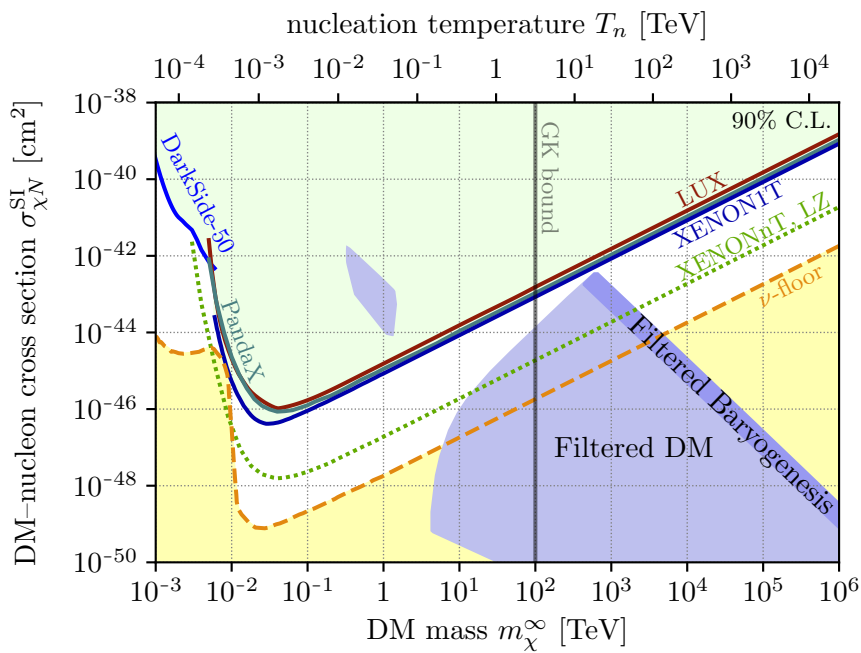

**Figure 4.9:** Spin-independent DM–nucleon cross section as a function of the DM mass for the Filtered DM mechanism alone (light blue regions) and combined with Filtered Baryogenesis (blue region). A second axis on top shows the nucleation temperature at which Filtered DM yields the correct relic density. Also shown in the plot are current experimental limits (solid), future sensitivities (dotted), and the neutrino floor (yellow region) [314–316]. The vertical line indicates the upper limit for the mass of conventional WIMP DM, i.e. the Griest–Kamionkowski bound.

DM. It now remains to assess the direct detection prospects of our DM candidate $\chi$. Possible interactions with the SM occur via the exchange of $\phi$ and are thus controlled by the Higgs portal coupling, $\lambda_{\phi H}$. (Remember that the Filtered DM requirement, that $\phi$ must be in thermal equilibrium, placed a lower limit on $\lambda_{\phi H}$.) The spin-independent DM–nucleon cross section is given by [317, 318]

$$\sigma_{\chi N}^{\text{SI}} = \frac{\lambda_{\phi H}^2 f_N^2 m_N^4 (m_\chi^\infty)^4}{4\pi\, m_h^4 m_\phi^4 (m_N + m_\chi^\infty)^2} \tag{4.35}$$

with the form factor $f_N \approx 0.326$ [319], the nucleon mass $m_N \approx 940\,\text{GeV}$, the Higgs mass $m_h \approx 125\,\text{GeV}$, and the scalar mass $m_\phi \sim T_n$.

In Fig. 4.9 we plot in light blue the viable regions of Filtered DM in the plane of cross section vs. true vacuum DM mass. As discussed, the requirement to obtain the observed DM density imposes a direct correspondence between $m_\chi^\infty$ and $T_n$, which we reflect by adding a second axis on top of the plot. The viable region is determined by scanning the model parameters $m_\chi^\infty/T_n$, $y_\chi$, and $\lambda_{\phi H}$ over the ranges of interest and plotting only the points for which all Filtered DM conditions, discussed in Section 4.2.1, are fulfilled. Well below the electroweak scale, $T_n \sim m_\phi \ll 100\,\text{GeV}$, collider searches for invisible Higgs decays strongly constrain the Higgs portal coupling, $\lambda_{\phi H}$. This prohibits thermal contact between $\phi$ and the SM, invalidating the Filtered DM mechanism. At temperatures $T_n \gtrsim 10\,\text{GeV}$, heavier SM fermions become kinematically accessible, which facilitates the required $\phi$–SM equilibration and the mechanism starts to be successful. Around $T_n \sim m_\phi \sim 125\,\text{GeV}$, the $\phi$–$H$ mass mixing highly constrains $\lambda_{\phi H}$ again, which causes the gap that splits the viable region into two patches. The upper right edge of the larger patch is dictated by perturbativity and the lower left edge reflects the condition that $\chi$ should be thermalized prior to the PT. Note that Filtered DM remains viable even beyond the Griest–Kamionkowski bound at $m_\chi^\infty \sim 100\,\text{TeV}$ which represents a hard upper limit for conventional thermal DM (see Section 2.2.2).

We now turn to the possibility that both Filtered DM and Filtered Baryogenesis are realized within the same model. As can be seen from our parameter scans in Fig. 4.8, such an overlap exists in case of the dimension-6 $N_R$ portal. The blue diagonal strip in Fig. 4.9, extending from $m_\chi^\infty \sim 500\,\text{TeV}$ to the right edge, indicates the combined viable region. It is constrained from the lower left by our demand $y_\chi \gtrsim 1$ to ensure the validity of the WKB approximation. As discussed, this is not a hard cutoff, but our results may become inaccurate for smaller $y_\chi$. Note that the narrowness of the combined region is not a sign of fine tuning but just a result of projecting the higher dimensional parameter space onto the $\sigma_{\chi N}^{\text{SI}}$–$m_\chi^\infty$-plane.

## 4.5 Conclusions

The objective of this chapter was nothing less than the development of a mechanism that simultaneously explains the DM abundance and the BAU. We started by reviewing the Filtered DM mechanism from Ref. [77], which operates at the boundary of a first-order PT during which the DM gains a mass that is large compared to the temperature. As a consequence, the bulk of the initial thermal DM abundance is "filtered out" and annihilates away. The tiny fraction that reaches the true vacuum immediately freezes out – due to the number density suppression corresponding to the large obtained mass – and constitutes the DM abundance observed today.

On top of Filtered DM, we added the missing ingredients for successful baryogenesis using an EFT approach. We started by imposing a CP-violating Yukawa interaction that generates a chiral DM asymmetry in the vicinity of the bubble wall. We discussed two candidates for higher-dimensional portal operators that could turn the chiral asymmetry into a SM lepton asymmetry. Electroweak sphalerons turn the latter into the desired baryon asymmetry. A key feature of the imposed portals is that they are efficient in the false vacuum, but suppressed after the PT has concluded. This is guaranteed by the large ratio between DM mass and temperature, which goes hand in hand with Filtered DM.

We tested the proposed mechanism for individual benchmark points as well as entire slices of the model parameter space. For the dimension-6 $N_R$ portal, a substantial overlap of the DM and baryogenesis mechanisms could be identified for DM masses $30 \sim 60$ times larger than the nucleation temperature of the PT. In case of the dimension-8 Weinberg portal, on the other hand, the mass–temperature ratios that lead to successful baryogenesis turned out to be too small for Filtered DM to occur prior to the EWPT (where the required electroweak sphalerons are active).

The combined mechanism can explain DM masses from $500\,\text{TeV}$ upwards. The corresponding DM–nucleon cross section exceeds the neutrino floor up to masses of $\sim 4\,\text{PeV}$ and could be detected by the future *XENONnT* or *LZ* experiments if the DM mass is below $\sim 2\,\text{PeV}$.

The validity of our analyses might be limited in the case of small DM Yukawa couplings. This is due to the resulting large mean free path which in turn invalidates the WKB approach and the diffusion approximation. We expect that the mechanism can still work in this regime, however, a more sophisticated treatment would be required to make quantitative claims. In the next chapter such an approach will be pursued in a different context.

Finally, we note that we left the details of the first-order PT mostly unspecified and instead imposed only minimal assumptions, in particular a large order parameter. A future work could direct its attention to the scalar sector and derive properties such as strength and time scale of the PT as well as the width and velocity of the bubble walls.

# Appendix of Chapter 4

## 4.A Extended Version of the Transport Equations

The transport equations used in our main analysis, Eqs. (4.26) and (4.27), neglect all Yukawa couplings and assume instantaneous strong sphaleron transitions. To test the validity of these assumptions, we set up an extended set of transport equation for the dimension-8 Weinberg portal, in which the third generation Yukawa couplings and the finite strong sphaleron rate are included:

$$\begin{aligned}
\partial_\mu j^\mu_{\delta\chi} &= \Gamma_{\rm p}(2\mu_{\rm p}^{1,2} + \mu_{\rm p}^3) + (\Gamma_5^{\rm (a)} + \Gamma_5^{\rm (b)} + \Gamma_5^{\rm (c)} + \Gamma_M^\chi)\mu_M^\chi + S_\chi\,, \\
\partial_\mu j^\mu_{\delta\tau} &= -\Gamma_Y^\tau \mu_Y^\tau\,, \\
\partial_\mu j^\mu_{\delta\ell_{1,2}} &= 2\Gamma_{\rm p}\mu_{\rm p}^{1,2} - \Gamma_{\rm ws}\mu_{\rm ws}\,, \\
\partial_\mu j^\mu_{\delta\ell_3} &= 2\Gamma_{\rm p}\mu_{\rm p}^3 - \Gamma_{\rm ws}\mu_{\rm ws} + \Gamma_Y^\tau \mu_Y^\tau\,, \\
\partial_\mu j^\mu_{\delta H} &= 2\Gamma_{\rm p}(2\mu_{\rm p}^{1,2} + \mu_{\rm p}^3) - \Gamma_Y^\tau \mu_Y^\tau + \Gamma_Y^t \mu_Y^t - \Gamma_Y^b \mu_Y^b\,, \\
\partial_\mu j^\mu_{\delta u} &= \Gamma_{\rm ss}\mu_{\rm ss}\,, \\
\partial_\mu j^\mu_{\delta t} &= \Gamma_{\rm ss}\mu_{\rm ss} - \Gamma_Y^t \mu_Y^t\,, \\
\partial_\mu j^\mu_{\delta b} &= \Gamma_{\rm ss}\mu_{\rm ss} - \Gamma_Y^b \mu_Y^b\,, \\
\partial_\mu j^\mu_{\delta Q_{1,2}} &= -3\Gamma_{\rm ws}\mu_{\rm ws} - 2\Gamma_{\rm ss}\mu_{\rm ss}\,, \\
\partial_\mu j^\mu_{\delta Q_3} &= -3\Gamma_{\rm ws}\mu_{\rm ws} - 2\Gamma_{\rm ss}\mu_{\rm ss} + \Gamma_Y^t \mu_Y^t + \Gamma_Y^b \mu_Y^b\,,
\end{aligned} \quad (4.36)$$

with

$$\begin{aligned}
\mu_{\rm p}^i &= -2\mu_{\delta\chi} - 2\mu_{\delta\ell_i} - 2\mu_{\delta H}\,, \qquad \mu_M^\chi = -2\mu_{\delta\chi}\,, \\
\mu_Y^\tau &= \mu_{\delta\tau} - \mu_{\delta\ell_3} + \mu_{\delta H}\,, \qquad \mu_Y^t = \mu_{\delta t} - \mu_{\delta Q_3} - \mu_{\delta H}\,, \qquad \mu_Y^b = \mu_{\delta b} - \mu_{\delta Q_3} + \mu_{\delta H}\,, \\
\mu_{\rm ws} &= 2\mu_{\delta\ell_{1,2}} + \mu_{\delta\ell_3} + 6\mu_{\delta Q_{1,2}} + 3\mu_{\delta Q_3}\,, \qquad \mu_{\rm ss} = -4\mu_{\delta u} - \mu_{\delta t} - \mu_{\delta b} + 4\mu_{\delta Q_{1,2}} + 2\mu_{\delta Q_3}\,.
\end{aligned} \quad (4.37)$$

We introduced the asymmetries $\delta\tau$ (for the $\tau$ lepton), $\delta u$ (as a proxy for the right-handed $u$, $d$, $c$, and $s$ quarks), $\delta t$ and $\delta b$ (for the right-handed $t$ and $b$ quarks), and we split up the left-handed lepton and quark doublets into individual generations. All asymmetries are defined in analogy to Eq. (4.24). The value of the newly introduced diffusion coefficient $D_\tau$ is given in Eq. (4.38), while all quarks share the same diffusion coefficient $D_Q$. The new

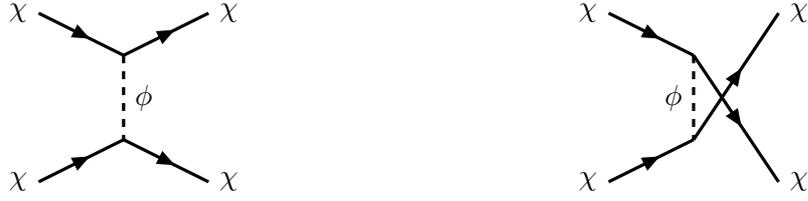

**Figure 4.10:** Scattering processes relevant for the calculation of the diffusion coefficient.

Yukawa rates amount to $\Gamma_Y^\tau \approx 1.4 \times 10^{-6} T_n$, $\Gamma_Y^t \approx 2.9 \times 10^{-2} T_n$, $\Gamma_Y^b \approx 1.9 \times 10^{-5} T_n$ [42, 291, 292], and the strong sphaleron rate is given by $\Gamma_{ss} \approx 5.2 \times 10^{-6} T_n$ [42, 320]. The quark asymmetries now combine to the baryon number density according to $n_B = \frac{1}{3}(4\delta u + \delta t + \delta b + 2\delta Q_{1,2} + \delta Q_3)$.

As we have tested for several benchmark points, the final $Y_B$ obtained from the extended version of the transport equation shows only per-cent level deviations from the results obtained by the minimal version. We conclude that the more extensive treatment presented in this chapter is not necessary for our purposes.

## 4.B Diffusion Coefficients

In order to describe the asymmetries near the PT boundary, we use transport equations based on the diffusion approximation. Rewriting the dynamical part of these equations via Eq. (4.28) introduces diffusion coefficients, $D_a$. This quantity measures the mean free path of the respective species $a$ in the thermal bath and thus inversely depends on its total scattering rate with the surrounding plasma. In our mechanism, all SM particles and the right-handed neutrinos have constant and zero masses. In these cases, we use the values found in the literature on electroweak baryogenesis [291, 306, 307, 321]

$$D_N \approx \frac{16\pi^2}{T_n}\frac{1}{y_\nu^4}, \qquad D_\tau \approx \frac{380}{T_n}, \qquad D_\ell \approx D_H \approx \frac{100}{T_n}, \qquad D_Q \approx \frac{6}{T_n}. \qquad (4.38)$$

Our DM candidate $\chi$, however, requires a more careful treatment. An essential feature of the Filtered Baryogenesis mechanism is the large DM mass gain at the bubble wall, with a final mass $m_\chi^\infty \gg T_n$ in the true vacuum. We thus consider a $z$-dependent diffusion coefficient, which gives rise to additional terms proportional to its first and second derivatives in the transport equations. For the computation of $D_\chi(z)$, we modify the calculation in Ref. [291] to include a finite mass of the interacting fermion. The starting point is the spin-averaged squared matrix element for the $t$- and $u$-channel scattering diagrams shown in Fig. 4.10,

$$\overline{|\mathcal{M}|}^2 = y_\chi^4 \left( 2\frac{(\frac{t}{4} - m_\chi^2)^2}{(t - m_\phi^2)^2} + 2\frac{(\frac{u}{4} - m_\chi^2)^2}{(u - m_\phi^2)^2} + \frac{(\frac{s}{4} - m_\chi^2)^2 - (\frac{t}{4} - m_\chi^2)^2 - (\frac{u}{4} - m_\chi^2)^2}{(t - m_\phi^2)(u - m_\phi^2)} \right). \quad (4.39)$$

Note that a $z$-dependence enters this expression via $m_\chi$. For the scalar mediator we assume a constant mass of $m_\phi = T_n$. Based on the matrix element, the total scattering rate for a

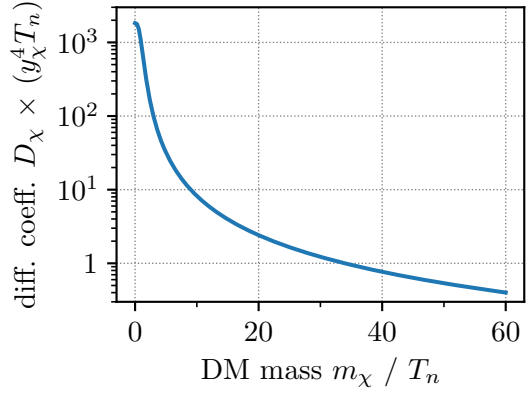

**Figure 4.11:** Diffusion coefficient of $\chi$ as a function of its mass.

$\chi$ particle with 4-momentum $p$ and energy $E_p$ is given by

$$\Gamma_\chi^{\text{tot}}(p) = \frac{1}{2E_p} \int d\Pi_k d\Pi_{p'} d\Pi_{k'} (2\pi)^4 \delta^{(4)}(p+k-p'-k') \overline{|\mathcal{M}|}^2 \left(-f'_\chi(E_k)\right)(1-\cos\theta), \quad (4.40)$$

where $d\Pi_k \equiv d^3k/(16\pi^3 E_k)$, $f'_\chi \equiv -T_n \frac{\partial f_\chi}{\partial E}$, $f_\chi$ is a Fermi–Dirac distribution, and $\theta$ is the scattering angle. The integral can most easily be evaluated in the center-of-mass frame. Finally, the diffusion coefficient is given by

$$D_\chi(z) = D_\chi(m_\chi) = \frac{12}{T_n^3} \int \frac{d^3p}{(2\pi)^3} \frac{-f'_\chi(E_p)}{\Gamma_\chi^{\text{tot}}} \left(\frac{p_z}{E_p}\right)^2 \propto \frac{1}{y_\chi^4}. \quad (4.41)$$

A plot of the diffusion coefficient as a function of $m_\chi$ is displayed in Fig. 4.11.

## 4.C Validity of the Diffusion Approximation

The diffusion approximation, which our transport equations are based on, assumes local thermal equilibrium. The rate at which the different momentum modes equilibrate is given by the inverse of the mean free path and it should be the fastest rate in the system [322]. However, we find that $1/D_\chi$ is smaller than the chiral relaxation rate, $\Gamma_M^\chi$. In other words, the change in the asymmetry over one diffusion length is large, which invalidates the diffusion ansatz.

In order to estimate how a full kinematic treatment – which does not rely on the diffusion approximation – would impact our results, we proceed as follows: First, note that the calculation of $\Gamma_M^\chi$ (see Ref. [42]) involves a momentum integral over a combination of equilibrium distribution functions. This, however, is based on the assumption that momentum modes equilibrate efficiently. In our scenario, the chiral asymmetry is generated mostly around momenta close to the reflection threshold, $p_z \sim m_\chi^\infty \gg T_n$, and the equilibration across momentum modes is not efficient because $1/D_\chi \lesssim \Gamma_M^\chi$, as discussed. To account

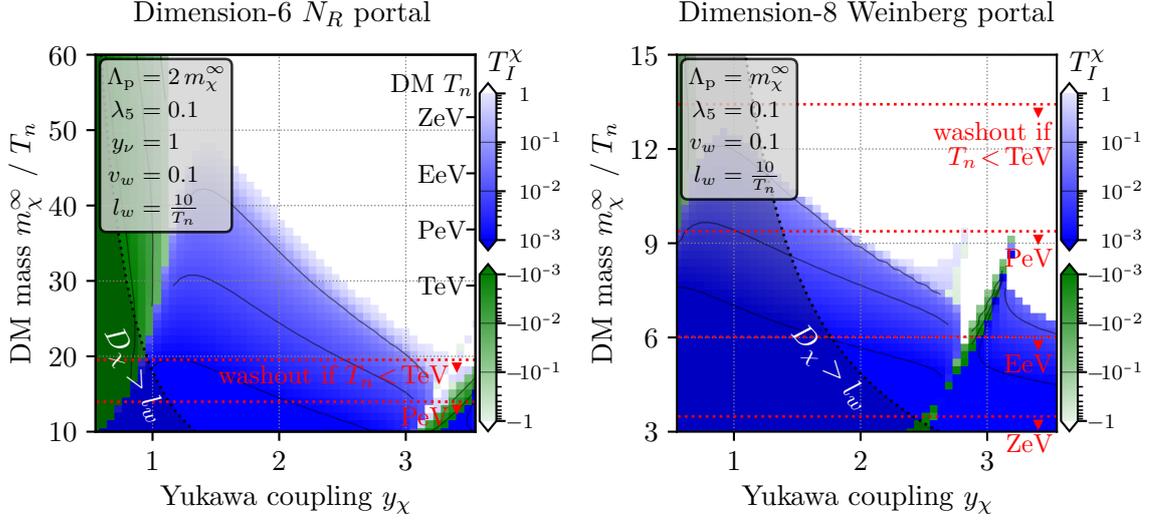

**Figure 4.12:** Slices of Filtered Baryogenesis model parameter space in analogy to Fig. 4.8. In this version, however, we use a modified $\Gamma_M^\chi$ to estimate the outcome of an analysis that goes beyond the diffusion approximation.

for this, we define a modified chiral relaxation rate,

$$\tilde{\Gamma}_M^\chi(z) \equiv \Gamma_M^\chi(z) \frac{\mathcal{I}\left(\sqrt{(m_\chi^\infty)^2 - m_\chi^2(z)}\right)}{\mathcal{I}(T_n)} \,, \tag{4.42}$$

where $\mathcal{I}(k)$ is the integrand in the definition of $\Gamma_M^\chi$ in Ref. [42], without the $k^2$ Jacobian, and divided by a Fermi–Dirac distribution (in order to obtain an expression that is not exponentially dependent on the integral momentum $k$). This modification shifts the peak of the momentum distribution from $\sim T_n$ (i.e. equilibrium) to $\sim m_\chi$ (where the chiral asymmetry is generated in our scenario). Because the modified relaxation rate is in general smaller than the original one, the diffusion approximation is now justified. Figure 4.12 shows slices of the model parameter space, in direct analogy to the ones presented in Fig. 4.8, but using $\tilde{\Gamma}_M^\chi$ instead of $\Gamma_M^\chi$. As can be seen in the figure, the modification deforms the regions of successful baryogenesis but leaves the Filtered Baryogenesis mechanism intact in general.

# 5 Primordial Black Holes from Squeezed Matter

*This chapter is based on the publications [2, 3] of the author and his collaborators. The author played a major role in the derivation of the formulas required for the numerical analysis. He implemented the entire mechanism in the form of a comprehensive numerical simulation. The enormous scope of the problem made it necessary to have the results crosschecked against ones obtained from an entirely independent implementation, done by the collaborator LM. The author produced all figures that appear in the publication and in the following sections, except Fig. 5.9, which was created by MJB.*

## 5.1 Introduction

The idea of primordial black holes (PBHs) – black holes (BHs) that emerged during or soon after the Big Bang – dates back to the 1960s [323]. Several phenomena provide theoretical motivation for the existence of PBHs in different regimes of mass and abundance: PBHs could account for parts or even the entirety of dark matter (DM) in the Universe either directly [324–328] or as a product of their evaporation into Hawking radiation [329–340]. They can alter the expansion history of the Universe [335, 341], destroy monopoles or domain walls [342, 343], evolve into the supermassive BHs found in the center of most galaxies [344], or seed the formation of the Universe's large-scale structure (LSS) [82, 345–348]. By today, the abundance of PBHs is constrained mostly by the non-observation of Hawking radiation, gravitational lensing, distortions in the cosmic microwave background (CMB), and LSS [82]. In the future, gravitational-wave observatories with increasing sensitivity will be able to probe individual or stochastic BH merger events and will thus also play an important role in the research related to PBH [349].

Theorized PBH formation mechanisms involve the collapse of density perturbations arising during inflation [350–356], of topological defects [357–363], or of scalar condensates [364, 365]. This chapter presents a new mechanism for the formation of PBHs: We imagine a situation in which a particle species $\chi$ is thermally abundant during the Big Bang and obtains a mass in a cosmological phase transition (PT). If the latter is a first-order transition, it proceeds via the nucleation and expansion of bubbles filled with the energetically favorable true vacuum. Similar to the Filtered Baryogenesis mechanism described in the previous chapter, this mechanism relies on a setup where $\chi$ experiences a sizable mass jump at the phase boundary, so that most $\chi$ particles are reflected off the advancing bubble walls by virtue of energy conservation. However, in this scenario we focus on the regions of model parameter space where $\chi$ cannot annihilate efficiently. As a consequence, overdensities accumulate in front of the walls and are ultimately squeezed together between several expanding bubbles, i.e. inside shrinking false-vacuum pockets. The resulting accumulations of $\chi$ particles may become sufficiently dense to form BHs.

We show the plausibility of such a scenario using analytic arguments first, but then develop a much more detailed description of the problem. In particular, we go beyond simplifications such as the diffusion approximation and the steady state ansatz used in the previous chapter. Quantitative results obtained from a sophisticated numerical simulation will underpin our proposal.

Note that PBH formation due to bubble collisions during first-order PTs has already been discussed [366–372]. These considerations, however, focus on the energy density in the scalar field itself. In our scenario, the particle plasma compressed by the bubble walls is responsible for the BH formation. Recently, a similar mechanism has been proposed for the formation of "DM nuggets" [373–375].

In section Section 5.2, we lay out the groundwork for our mechanism by setting up a minimal toy model, stating our assumptions, and discussing the criteria for successful BH formation. We elaborate further in Section 5.3 by giving a full kinetic description of the problem in terms of a Boltzmann equation and describe the method used to solve it numerically. Section 5.4 gives insights into our complex numerical simulation by providing exploratory visualizations of our results. In Section 5.5, we finally show that our mechanism can indeed generate PBHs, study the dependence on different parameters, and discuss the PBH mass landscape and its constraints. A summary of our approach, its possible shortcomings, and of our results and findings is given in Section 5.6.

## 5.2 The Black Hole Formation Mechanism

The proposed PBH formation mechanism may in principle be realized in a variety of different ways. In the following, we introduce a minimal model to make our idea work: We impose a dark sector that is comprised of an initially massless Dirac fermion $\chi$ and a real scalar $\phi$, both gauge singlets. The relevant parts of the Lagrangian are

$$\mathcal{L} \supset -y_\chi \phi \bar{\chi} \chi - V(\phi) \,, \tag{5.1}$$

where $y_\chi$ is a Yukawa coupling and $V(\phi)$ denotes the effective scalar potential. While the explicit form of the potential is irrelevant to us, we require that it gives rise to a first-order PT. In this scenario, the scalar potential initially favors a vacuum expectation value (VEV) of zero, $\langle\phi\rangle = 0$, but with the expansion and cooling of the Universe develops a separate minimum which becomes energetically favorable. When the nucleation temperature $T_n$ is reached, bubbles the true vacuum, $\langle\phi\rangle \neq 0$, nucleate and expand to fill the Universe. (See Chapter 3 for a detailed review of first-order PTs.) As the bubble walls finally coalesce, the remaining regions of false vacuum become approximately spherical due to surface tension. Note that in other contexts, first-order PTs are usually described as expanding true-vacuum bubbles. We instead consider a shrinking false-vacuum bubble. Figure 5.1 illustrates the imagined setup.

In what follows, we make use of this spherical symmetry and define a radius $r$ with respect to the center of a shrinking false-vacuum bubble. In this coordinate system, we

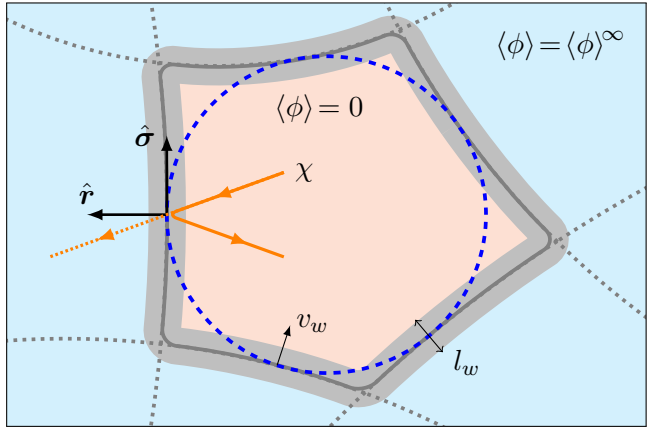

**Figure 5.1:** An illustration of our PBH formation mechanism: In a first-order PT, regions of true vacuum (cyan) expand and finally coalesce, resulting in shrinking regions of false vacuum (red), which we approximate as spherical. While some $\chi$ particles with high momentum pass through the wall (dotted orange), most are reflected and trapped in the false vacuum (solid orange). An overdensity builds up and may eventually form a BH. The bubble walls have a thickness $l_w$ and move inwards with velocity $v_w$. We also indicate the local $r$–$\sigma$ coordinate system used in our formal description.

model the time dependent position of the wall as

$$r_w(t) = r_w^0 - v_w t\,, \tag{5.2}$$

where $r_w^0$ is the initial bubble radius, $v_w$ is the wall velocity, and $r_w = 0$ is reached after the time $r_w^0/v_w$. To describe the VEV at the phase boundary, we use an analytic function as a generic proxy for a solution to the bounce equation – Eq. (3.13) – that interpolates smoothly between 0 and $\langle\phi\rangle^\infty$:

$$\langle\phi\rangle(r,t) = \frac{\langle\phi\rangle^\infty}{2}\left[1 + \tanh\left(\frac{3\gamma_w(r - r_w(t))}{l_w}\right)\right]. \tag{5.3}$$

The wall has a width of $l_w \sim 1/T_n$ but appears Lorentz contracted to an observer in the plasma frame, hence the gamma factor $\gamma_w$. Thanks to the Yukawa coupling, $\chi$ becomes massive during the PT,

$$m_\chi(r,t) = y_\chi \langle\phi\rangle(r,t)\,. \tag{5.4}$$

In the following, we will often shorten $m_\chi(r,t)$ to $m_\chi$ and use $m_\chi^\infty \equiv y_\chi \langle\phi\rangle^\infty$ to denote $\chi$'s final mass in the true vacuum.

A crucial ingredient of our mechanism is the hierarchy $m_\chi^\infty \gg T_n$ which ensures that only very energetic $\chi$ particles – in the tail of the thermal distribution – can pass through the wall and exit the shrinking bubble. Most of the $\chi$ particles in the plasma have energies $\sim T_n$ and thus bounce off the wall. They are trapped inside the shrinking bubble and are squeezed together until a BH may form. The large required order parameter of the PT, $\langle\phi\rangle^\infty \gg T_n/y_\chi$, can be realized in quasi-conformal or dilaton-like setups [172–191]. We focus on the regime of small $y_\chi$ to prevent a depletion of the overdensity via annihilation. This feature sets our PBH formation apart from the Filtered Baryogenesis mechanism discussed in Chapter 4.

### 5.2.1 Simplifying Assumptions

In addition to the general features described in the previous section, we make the following simplifying assumptions:

1. The PT occurs during the radiation-dominated era of the Universe, $T_n \gtrsim 1\,\text{eV}$. This allows us to work with the same relation between Hubble rate and temperature, $H \propto T_n^2$ (the first Friedmann equation), throughout our entire analysis.

2. Prior to the transition, $\chi$ follows a thermal distribution with temperature $T_n$ but is disconnected from the Standard Model of Particle Physics (SM). This scenario could arise if $\chi$ was in contact with the SM at some high temperature, but decoupled later on (similar to the SM neutrino decoupling). From that point on, the $\chi$ abundance would share a common temperature with the SM plasma (up to some possible corrections from the freeze-out of SM species) because both sectors are relativistic and redshift like radiation.

3. Throughout the whole transition, $\phi$ is in thermal and chemical equilibrium with the SM. Making this assumption, we can focus on the $\chi$ overdensities alone. At the same time, we require a $\phi$–Higgs portal coupling that is small enough to allow for an independent treatment of the dark and the electroweak PT.[1]

4. The temperature of the SM thermal bath stays constant throughout the PT, i.e. $T = T_n$. In reality we would expect $T \propto 1/\sqrt{t}$. However, we find that if BHs form in our mechanism, they do so after not much more than one Hubble time.

5. The latent heat release of the PT is large enough to compress the $\chi$ abundance to the point of BH formation, but not too large, to avoid a period of inflation. We quantify this requirement in Section 5.3.3.

6. The wall velocity $v_w$ and thickness $l_w$ stay constant throughout the PT. While this might sound hard to justify, we argue in Section 5.5.2 based on our findings that changes in these quantities do not invalidate the PBH formation mechanism.

We expect that none of these assumptions is strictly necessary for our mechanism to work, but relaxing them would make our analysis unnecessarily complex and merely constrain the viable model parameter space.

---

[1]Portal couplings as small as $10^{-3}$ are sufficient to maintain thermal contact for $T_n \gtrsim 1\,\text{GeV}$ [75]. However, if $T_n \gg 1\,\text{TeV}$, these portal couplings would give contributions to the Higgs mass and thus had to be smaller. In that case, additional new particles are required to keep $\phi$ in equilibrium.

### 5.2.2 Black Hole Formation Criteria

The formation of a black hole can occur in two ways: Firstly, a region of space can become so dense that it forms a Jeans instability and starts to collapse gravitationally, or, at even higher densities, a Schwarzschild horizon may form immediately. In the following we derive the criteria for both possibilities.

**Collapse of a Jeans instability.** A Jeans instability arises when the free-fall time of a spherically symmetric volume [376] falls below the sound-crossing time,

$$t_{\text{ff}} \equiv \sqrt{\frac{3\pi}{32G\Delta\rho}} \quad < \quad t_{\text{s}} \equiv \frac{r_w(t)}{c_{\text{s}}}, \tag{5.5}$$

where $G$ is the gravitational constant, $c_{\text{s}}$ is the sound speed for ultra-relativistic particles, and $\Delta\rho$ is a local energy overdensity w.r.t. the exterior. For our purposes it is appropriate to express physical distances in terms of the Hubble radius. Using the Friedmann equation for radiation domination, we can write the Hubble radius (at some initial time $t_0$) as

$$r_H^0 \equiv \frac{1}{H(t_0)} = \sqrt{\frac{3}{8\pi G} \frac{30}{\pi^2 g_\star} \frac{1}{T_n^2}}, \tag{5.6}$$

where $g_\star$ is the total effective number of relativistic degrees of freedom (DOFs) in the Universe at temperature $T_n$, including $\chi$ and $\phi$. The Jeans instability criterion can now be rewritten as

$$\Delta\rho \approx \rho_\chi(t) \quad > \quad \frac{\pi^4}{360} g_\star T_n^4 \left(\frac{r_H^0}{r_w(t)}\right)^2. \tag{5.7}$$

Note that we approximate $\Delta\rho$ as $\rho_\chi$, neglecting the energy carried by the wall itself as well as the exponentially suppressed density of $\chi$ particles outside the bubble (due to the large mass of $\chi$). Interestingly, the derived criterion is independent of the physical scale (when disregarding the slight temperature dependence of $g_\star$ for $T_n \lesssim 100\,\text{GeV}$).

After a Jeans instability has formed, it collapses into a BH over a time $\sim t_{\text{ff}}$, given that no internal pressure prevents it from doing so.[2] As $\chi$ is a fermion, we indeed have to worry about degeneracy pressure. In addition to the Jeans instability criterion, we thus demand that the gravitational pull supersedes degeneracy pressure. This can be formalized by computing the total energy of the system,

$$E_{\text{tot}} = E_{\text{grav}} + E_{\text{kin}}, \tag{5.8}$$

where

$$E_{\text{grav}} = -G \int_0^{r_w(t)} \left(\frac{E_{\text{kin}}}{\frac{4}{3}\pi r_w^3(t)}\right)^2 \frac{(\frac{4}{3}\pi r^3)(4\pi r^2 \text{d}r)}{r} = -\frac{3}{5}\frac{GE_{\text{kin}}^2}{r_w(t)} \tag{5.9}$$

---

[2] Another requirement for a successful collapse is the dissipation of kinetic energy and angular momentum, e.g. via scattering. We do not investigate this aspect further, since we find that degeneracy pressure alone prevents the collapse in our case.

is the gravitational energy and

$$E_{\text{kin}} = \rho \cdot V = \frac{\rho}{n^{\frac{4}{3}}} \frac{N^{\frac{4}{3}}}{V^{\frac{1}{3}}} = \frac{3}{4}\left(\frac{9\pi}{2g_\chi}\right)^{\frac{1}{3}} \frac{N^{\frac{4}{3}}}{r_w(t)} \tag{5.10}$$

is the kinetic energy of a degenerate Fermi gas with $g_\chi = 4$ relativistic DOFs in $\chi + \bar{\chi}$. In the above expression, $V$ is the volume of the bubble, and the energy and number densities $\rho$ and $n$ are both evaluated in the relativistic degeneracy limit. Assuming for now that no particles escape the shrinking bubble, the total number of $\chi$ is constant and given by

$$N = \frac{3}{4} g_\chi \frac{\zeta(3)}{\pi^2} T_n^3 \cdot \frac{4}{3}\pi (r_w^0)^3 \,. \tag{5.11}$$

We now require that, despite a possibly arising degeneracy pressure, further collapse is still energetically favorable, i.e.

$$\frac{\mathrm{d}E_{\text{tot}}}{\mathrm{d}t} < 0$$
$$\Leftrightarrow \quad \frac{9}{5}\frac{GE_{\text{kin}}^2}{r_w^2(t)} - \frac{E_{\text{kin}}}{r_w(t)} > 0$$
$$\Leftrightarrow \quad \frac{r_w^0}{r_w(t)} \gtrsim \sqrt{\frac{330}{g_\chi}} \frac{r_H^0}{r_w^0} \,. \tag{5.12}$$

In the last step we used the Friedmann equation, Eq. (5.6), together with $g_\star \sim 110$. Note that this derivation assumes that the $\chi$ particles are degenerate right from the start of the collapse, when they are still spread over the whole bubble of radius $r_w^0$. This is a conservative assumption, since one could imagine that the degenerate gas forms only at a later stage of the collapse which would make Eq. (5.12) easier to satisfy.

**Formation of a Schwarzschild horizon.** In the case that degeneracy pressure halts the collapse of a Jeans instability, we can still consider the possibility that the $\chi$ abundance is compressed by the shrinking bubble even further, until an event horizon forms. Formally, this requires the compressed region to become smaller than its Schwarzschild radius, i.e.

$$r_w(t) \quad < \quad r_s \equiv 2G\,\Delta\rho\,\frac{4}{3}\pi r_w^3(t) \,, \tag{5.13}$$

where $\Delta\rho$ is again the local energy overdensity. Using the Friedmann equation again, this can be rewritten as

$$\Delta\rho \approx \rho_\chi(t) \quad > \quad \frac{\pi^2}{30}g_\star T_n^4 \left(\frac{r_H^0}{r_w(t)}\right)^2 \,. \tag{5.14}$$

Note that the Schwarzschild criterion is more stringent than the Jeans instability criterion in Eq. (5.7). This is expected, because when compressing the $\chi$ abundance, a Jeans instability is created *before* a Schwarzschild horizon may form.

| Initial radius $r_w^0$ | Required compression $r_w^0/r_w(t)$ to fulfill criterion | | |
|---|---|---|---|
| | Jeans instability | Degeneracy pressure | Schwarzschild horizon |
| $0.1\,r_H^0$ | 51 | 87 | 56 |
| $1.0\,r_H^0$ | 5.1 | 8.7 | 5.6 |
| $1.5\,r_H^0$ | 3.4 | 5.8 | 3.8 |
| $2.0\,r_H^0$ | 2.6 | 4.3 | 2.8 |

**Table 5.1:** Analytic estimates for the compression ratios $r_w^0/r_w(t)$ required to fulfill the BH formation criteria for different initial radii $r_w^0$ and assuming $T_n \gtrsim 100\,\text{GeV}$ (so that $g_\star \sim 110$). Note that a Jeans instability can only collapse if the gravitational pull overcomes degeneracy pressure.

### 5.2.3 Analytic Estimates

In this section we perform rough estimates to check if and how our mechanism could work in light of the criteria derived in the previous sections. For this purpose we do, for now,

1. neglect annihilation and scattering,
2. neglect highly energetic particles that pass through the wall and escape the bubble,
3. assume non-relativistic wall velocities, $v_w \ll 1$.

Given these assumptions, the number density of $\chi$ particles inside the shrinking bubble of radius $r_w(t)$ increases as $[r_w^0/r_w(t)]^3$. Furthermore, in each reflection off the wall, a $\chi$ particle gains an energy of $dE = 2v_w E$ and the time between two reflections is $dt = 2r_w(t)$ (for a radial trajectory). Solving $dE/E = dt\,v_w/r_w(t)$ shows that the energy scales as $r_w^0/r_w(t)$. Taking both effects together, we expect that the energy density is initially in thermal equilibrium,

$$\rho_\chi(t_0) = \rho_\chi^{\text{eq}} = \frac{7\pi^2}{240} g_\chi T_n^4\,, \tag{5.15}$$

and behaves as

$$\rho_\chi(t) \sim \left(\frac{r_w^0}{r_w(t)}\right)^4 \rho_\chi(t_0) \tag{5.16}$$

when squeezed together by the shrinking bubble.

We can compare this estimate to the criteria for the formation of a Jeans instability, Eq. (5.7), or a Schwarzschild horizon, Eq. (5.14). In Table 5.1 we present the compression ratios $r_w^0/r_w(t)$ required to fulfill the different criteria for various exemplary initial radii $r_w^0$. The obtained numbers reveal the following:

1. As expected, a Jeans instability always forms before a Schwarzschild horizon. However, a Jeans instability can only collapse if degeneracy pressure does not prevent it. According to our roughly estimated degeneracy criterion for the case of relativistic particles, Eq. (5.12), this would only happen for compression ratios larger than those needed to from a Schwarzschild horizon. Consequentially, our PBH formation mechanism has to rely on the Schwarzschild criterion.[3]

2. Small initial radii, $r_w^0 \ll r_H^0$, require large compression ratios to fulfill any of the criteria. This would involve huge $\chi$ overdensities which in turn exert a strong pressure on the bubble wall. As we will quantify in Section 5.3.3, the required latent heat to keep the wall moving despite such large amounts of friction would lead to an era of inflation, making BH formation less likely. In our analysis we will therefore focus on large initial radii, $r_w^0 \sim r_H^0$.

## 5.3 A Kinetic Description via the Boltzmann Equation

Our analytic estimates in the previous section show that our proposed PBH formation mechanism is plausible. In the following we will develop a detailed kinematic description of the problem which allows us to evolve the phase space of $\chi$ in presence of a shrinking false-vacuum bubble. This enables us to include the effects of annihilation, particles escaping the bubble, and relativistic wall velocities. Note that, in contrast to the approach taken for Filtered Baryogenesis in Chapter 4, we will neither rely on the diffusion approximation nor make a steady state ansatz. The accumulating overdensities crudely violate the assumption of local thermal equilibrium and a shrinking bubble as a whole is no stationary system.

The phase space distribution function $f_\chi$ of $\chi$ – a function of momentum, space, and time – evolves according to the Boltzmann equation,

$$\mathbf{L}[f_\chi] = \mathbf{C}[f_\chi]. \tag{5.17}$$

The Liouville operator $\mathbf{L}[f_\chi]$ captures the dynamical evolution while the collision term $\mathbf{C}[f_\chi]$ accounts for scattering and annihilation. (Note that decay is not an issue since $\chi$ is stabilized by a global $U(1)$ symmetry.) Although $\chi$ is assumed to be at a temperature $T_n$ at the beginning of the PT, it will quickly depart from a thermal distribution when being compressed by the shrinking bubble. We can still parameterize $f_\chi$ in terms of the equilibrium distribution by introducing the *phase space enhancement factor* $\mathcal{A}$ and writing

$$f_\chi(\boldsymbol{p}, r, t) \equiv \mathcal{A}(\boldsymbol{p}, r, t) \, f_\chi^{\mathrm{eq}}(\boldsymbol{p}, r, t), \tag{5.18}$$

with the Fermi–Dirac equilibrium distribution

$$f_\chi^{\mathrm{eq}}(\boldsymbol{p}, r, t) = \left\{ 1 + \exp\left[ \frac{(|\boldsymbol{p}|^2 + m_\chi^2(r, t))^{\frac{1}{2}}}{T_n} \right] \right\}^{-1}. \tag{5.19}$$

In this ansatz, we already made the spherical symmetry explicit by using $r$, the radial position w.r.t. the bubble's center. The dependence on time $t$ enters through the mass of $\chi$, given in Eq. (5.4), which changes across the moving bubble wall. Finding the $\mathcal{A}(\boldsymbol{p}, r, t)$ that solves the Boltzmann equation will be the main subject of our efforts. The system starts with $\mathcal{A} = 1$ everywhere prior to the PT, by assumption. As soon as the walls form and start to move, $\chi$ particles are reflected and accumulate, resulting in $\mathcal{A} \gg 1$ inside the shrinking bubble.

---

[3]Our mechanism could very well be realized by the compression of scalar or vector particles instead of fermions. In that case, degeneracy pressure would not be an issue and forming a Jeans instability could be enough.

Note that our definition of $f_\chi$ represents the phase space of a single DOF. For symmetry reasons, the distribution is identical for all $g_\chi = 4$ DOFs of $\chi + \bar{\chi}$.

### 5.3.1 Particle Dynamics: The Liouville Operator

The Liouville operator is the kinetic part of the Boltzmann equation and given by the total time derivative of the phase space distribution function,

$$\mathbf{L}[f_\chi] \equiv \frac{\mathrm{d} f_\chi}{\mathrm{d} t} \,. \tag{5.20}$$

Before we proceed, recall an important observation from our analytic estimates: We expect that our PBH formation mechanism requires Hubble-sized bubbles, $r_w \sim r_H \sim M_{\mathrm{Pl}}/T_n^2$, where $M_{\mathrm{Pl}}$ is the Planck mass. On the other hand, the typical width of a bubble wall is naturally $l_w \sim 1/T_n$. As a consequence, if our mechanism occurs at a temperature not too close to the Planck scale, $T_n \ll M_{\mathrm{Pl}}$, then the required bubbles have radii much larger than the wall width, $r_w \gg l_w$. To deal with this multi-scale problem, we split our system into a *near-wall regime*, which describes the vicinity of the wall, and a *bulk regime*, which refers to the interior of the bubble.

**The near-wall regime.** A key aspect of our mechanism is the significant variation of $\chi$'s mass across the bubble wall in the near-wall regime, an effect that is captured by the Liouville operator. Gravitational effects as well as annihilation and scattering, on the other hand, are negligible here. This is because the wall is thin compared to the interior of the bubble, so that particles spend only very little time in this regime.

We thus start with the relativistic Lagrangian of a non-interacting massive particle,

$$\mathcal{L}_{\mathrm{free}} = -\frac{m_\chi}{\gamma} \,, \tag{5.21}$$

with the Lorentz factor (in spherical coordinates)

$$\gamma = \left(1 - \dot{r}^2 - r^2 \dot{\theta}^2 - r^2 \sin^2\theta \, \dot{\varphi}^2\right)^{\frac{1}{2}} \,. \tag{5.22}$$

By applying the Euler–Lagrange equations and using the relations[4]

$$p_r = E \dot{r} \,, \qquad p_\theta = E \, r \, \dot{\theta} \,, \qquad p_\varphi = E \, r \sin\theta \, \dot{\varphi} \,, \qquad E = \gamma \, m_\chi \,, \tag{5.23}$$

---

[4] Note that we work with physical momenta (also in case of the angular components $p_\theta$ and $p_\phi$) and not with the canonical ones of the usual Hamiltonian formalism.

we obtain the forces

$$\begin{aligned}
\frac{\mathrm{d}p_r}{\mathrm{d}t} &= -\frac{m_\chi}{E}\frac{\partial m_\chi}{\partial r} + \frac{1}{r}\frac{p_\theta^2 + p_\phi^2}{E}, \\
\frac{\mathrm{d}p_\theta}{\mathrm{d}t} &= -\frac{1}{r}\frac{p_\theta p_r - p_\varphi^2 \cot\theta}{E}, \\
\frac{\mathrm{d}p_\varphi}{\mathrm{d}t} &= -\frac{1}{r}\frac{p_\varphi p_r + p_\varphi p_\theta \cot\theta}{E}.
\end{aligned} \qquad (5.24)$$

These expressions hint at another symmetry we can make use of: The trajectory of any non-interacting particle will always be contained in a plane which passes through the center of the bubble. Due to the spherical symmetry, the phase space distribution is identical in all possible planes. As a consequence, one trivial dimension can be removed from the system by defining a tangential momentum[5]

$$p_\sigma \equiv \sqrt{p_\theta^2 + p_\varphi^2}. \qquad (5.25)$$

The forces in this coordinate system are given by

$$\begin{aligned}
\frac{\mathrm{d}p_r}{\mathrm{d}t} &= -\frac{m_\chi}{E}\frac{\partial m_\chi}{\partial r} + \frac{1}{r}\frac{p_\sigma^2}{E}, \\
\frac{\mathrm{d}p_\sigma}{\mathrm{d}t} &= \frac{p_\theta}{p_\sigma}\frac{\mathrm{d}p_\theta}{\mathrm{d}t} + \frac{p_\varphi}{p_\sigma}\frac{\mathrm{d}p_\varphi}{\mathrm{d}t} = -\frac{1}{r}\frac{p_\sigma p_r}{E}.
\end{aligned} \qquad (5.26)$$

The term $\propto \partial m_\chi/\partial r$ is crucial for our mechanism: It changes a particle's radial momentum, $p_r$, as it interacts with the wall. The terms $\propto p_\sigma$ represent a Coriolis-type force: They constantly shuffle the momentum of a non-radially moving particle between $p_r$ and $p_\sigma$, while keeping the total momentum constant. In the near-wall regime, these Coriolis-type terms $\propto 1/r \sim 1/r_w$ are negligible compared to the term $\propto \partial m_\chi/\partial r \propto 1/l_w$ and can thus be dropped, again due to the hierarchy $r_w \gg l_w$.

Equation (5.26) can be used to express the Liouville operator as

$$\begin{aligned}
\mathbf{L}[f_\chi] \equiv \frac{\mathrm{d}f_\chi}{\mathrm{d}t} &= \frac{\mathrm{d}p_r}{\mathrm{d}t}\frac{\partial f_\chi}{\partial p_r} + \frac{\mathrm{d}p_\sigma}{\mathrm{d}t}\frac{\partial f_\chi}{\partial p_\sigma} + \frac{\mathrm{d}r}{\mathrm{d}t}\frac{\partial f_\chi}{\partial r} + \frac{\partial f_\chi}{\partial t} \\
&= -\frac{m_\chi}{E}\frac{\partial m_\chi}{\partial r}\frac{\partial f_\chi}{\partial p_r} + \frac{p_r}{E}\frac{\partial f_\chi}{\partial r} + \frac{\partial f_\chi}{\partial t}.
\end{aligned} \qquad (5.27)$$

where $f_\chi = f_\chi(p_r, p_\sigma, r, t)$. Note that, since we neglect the Coriolis-type force, $p_\sigma$ is constant near the wall. However, we still have to keep track of this dimension because it behaves non-trivially in the bulk of the bubble.

---

[5]Technically, the momentum is now expressed in a cylindrical coordinate system, with axial component $p_r$, radial component $p_\sigma$, and an angle (the orientation of the trajectory's plane) under which the system is invariant. In terms of spatial coordinates, $p_r$ is a projection of the momentum on the radial direction and $p_\sigma$ is a projection on the direction orthogonal to it. The coordinate system is indicated in Fig. 5.1.

Finally, we insert our factorization ansatz for $f_\chi$, Eq. (5.18), to arrive at

$$\mathbf{L}[f_\chi] = f_\chi^{\text{eq}} \left[ -\frac{m_\chi}{E}\frac{\partial m_\chi}{\partial r}\frac{\partial}{\partial p_r} + \frac{p_r}{E}\frac{\partial}{\partial r} + \frac{\partial}{\partial t} - (1 - f_\chi^{\text{eq}})\frac{m_\chi}{T_n E}v_w\frac{\partial m_\chi}{\partial r} \right] \mathcal{A}(p_r, p_\sigma, r, t), \quad (5.28)$$

where we have used $\partial m_\chi/\partial t = v_w \partial m_\chi/\partial r$.

**The bulk regime.** Deep inside the shrinking bubble, in the false vacuum, we assume $m_\chi \approx 0$. The energy of a non-interacting particle in this region is almost conserved, with only minor changes caused by gravity. This motivates a rewriting of the momentum components as

$$E = \sqrt{p_r^2 + p_\sigma^2} \qquad \text{and} \qquad \xi \equiv \arctan\left(\frac{p_r}{p_\sigma}\right). \quad (5.29)$$

To derive the forces in the bulk regime, we consider the dynamics of a Schwarzschild geodesic. The energy of a massless particle in a spherically symmetric gravitational potential blue-/redshifts according to

$$\frac{E'}{E} = \sqrt{\frac{1 - \frac{r_s}{r}}{1 - \frac{r_s}{r'}}}, \quad (5.30)$$

where the Schwarzschild radius $r_s$ is given in Eq. (5.13). Considering infinitesimal shifts, $r' = r + \mathrm{d}r$ and $E' = E + \mathrm{d}E$, yields

$$\frac{\mathrm{d}E}{\mathrm{d}t} = -\frac{r_s E \sin\xi}{2r(r - r_s)}. \quad (5.31)$$

The effect of gravitational deflection of a massless particle is described by

$$\frac{\mathrm{d}r}{\mathrm{d}\varphi} = \pm r^2 \sqrt{\frac{1}{b^2} - \frac{1}{r^2}\left(1 - \frac{r_s}{r}\right)}, \quad (5.32)$$

where $\varphi$ is the polar angle w.r.t. the bubble center (setting the azimuth angle $\theta$ to $\pi/2$ w.l.o.g.) and the impact parameter is [377]

$$b = \frac{r\cos\xi}{\sqrt{1 - \frac{r_s}{r}}}. \quad (5.33)$$

By using the relations in Eq. (5.23) and applying implicit differentiation, we can also write $\mathrm{d}r/\mathrm{d}\varphi = r\tan\xi$. Combining this with Eq. (5.32) and taking the time derivative on both sides yields

$$\frac{\mathrm{d}\xi}{\mathrm{d}t} = \frac{|\sin\xi|}{2r^2}\frac{2r^3 - r_s b^2}{r^3 + r_s b^2}\left(\frac{1}{b^2} - \frac{r - r_s}{r^3}\right)^{-\frac{1}{2}}. \quad (5.34)$$

Interestingly, Eq. (5.26) in the limit $m_\chi \to 0$ (and transformed to the coordinates $E$ and $\xi$) is identical to Eqs. (5.31) and (5.34) in the limit $r_s \to 0$, which can be viewed as a validation of our derivations.

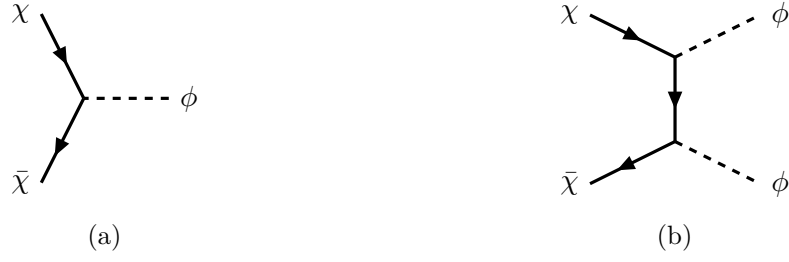

**Figure 5.2:** Annihilation processes contributing to the depletion of the accumulated $\chi$ overdensity. Note that in the Filtered DM mechanism presented in Chapter 4, process (a) was forbidden by $\chi$'s thermal mass. However, the latter is negligible in the PBH scenario, which relies on small Yukawa couplings.

Putting everything together, the Liouville operator in the bulk of the bubble can be written as[6]

$$\mathbf{L}[f_\chi] = f_\chi^{\text{eq}} \left[ \sin\xi \frac{\partial}{\partial r} - \frac{r_s E \sin\xi}{2r(r-r_s)} \frac{\partial}{\partial E} + \frac{|\sin\xi|}{2r^2} \frac{2r^3 - r_s b^2}{r^3 + r_s b^2} \left( \frac{1}{b^2} - \frac{r-r_s}{r^3} \right)^{-\frac{1}{2}} \frac{\partial}{\partial \xi} \right.$$
$$\left. + \frac{\partial}{\partial t} + \frac{1 - f_\chi^{\text{eq}}}{T_n} \frac{r_s E \sin\xi}{2r(r-r_s)} \right] \mathcal{A}(E, \xi, r, t) \,. \quad (5.35)$$

Here, we already inserted our factorization ansatz for $f_\chi$.

### 5.3.2 Particle Interactions: The Collision Term

For non-interacting particles, the Liouville operator is sufficient to describe the phase space dynamics. However, in the case at hand, $\chi$ interacts via the imposed Yukawa coupling and a collision term $\mathbf{C}[f_\chi]$ must be considered. The accumulated $\chi$ overdensity may annihilate into $\phi$ particles or scatter off each other. As we already argued, we expect thin walls, $l_w \ll r_w$, so that the overall effect is negligible in the near-wall regime. We will therefore include the collision term only in the bulk regime, where $m_\chi \approx 0$. The different contributions are discussed in the following.

**Annihilation via $\chi\bar\chi \to \phi$.** The process $\chi(p) + \bar\chi(q) \leftrightarrow \phi(k)$ shown in Fig. 5.2a scales as $y_\chi^2$ and is thus expected to be the dominant annihilation channel. $p$, $q$, and $k$ represent the 4-momenta of the interacting particles. The general form of the collision term for this type of interaction reads [283]

$$\mathbf{C}[f_\chi] \supset -\frac{1}{2} \frac{1}{2E_p} \int d\Pi_q d\Pi_k \, (2\pi)^4 \delta^{(4)}(p+q-k) \, \overline{|\mathcal{M}|}^2$$
$$\times \left[ f_{\chi,p} f_{\bar\chi,q}(1 + f_{\phi,k}) - f_{\phi,k}(1 - f_{\chi,p})(1 - f_{\bar\chi,q}) \right], \quad (5.36)$$

---

[6]In our actual implementation, we work with logarithmic reparametrizations of $r$ and $t$. This is required to maintain numerical precision in late stages of the simulation, where $r_w$ may shrink to values $\ll r_w^0$.

where $\mathcal{M}$ is the matrix element, $E_p$ is the energy of $\chi$, and $\mathrm{d}\Pi_q \equiv \mathrm{d}^3q/(16\pi^3 E_q)$ is the phase space integration measure belonging to $\bar{\chi}$. Analogous definitions apply to all interaction partners. Note that $\chi$ and $\bar{\chi}$ are massless in the bulk, but $m_\phi \sim T_n$. In the above expression we introduced short-hand notations for the phase space distribution functions: $f_{\chi,p} \equiv f_\chi(E_p, \xi_p, r, t)$, $f_{\bar{\chi},q} \equiv f_\chi(E_q, \xi_q, r, t)$ and $f_{\phi,k} \equiv f_\phi^{\mathrm{eq}}(\boldsymbol{k})$. The prefactor $1/2$ is required because our $f_\chi$ represents only a single DOF.

Despite the large overdensities generated by our mechanism, we find that neglecting Fermi blocking and Bose enhancement by setting $1 \pm f = 1$ is a good approximation.[7] After inserting the spin-averaged squared matrix element,

$$\overline{|\mathcal{M}|}^2 = 4 y_\chi^2 (p \cdot q), \tag{5.37}$$

the collision term becomes

$$\begin{aligned}
\mathbf{C}[f_\chi] &\supset -\frac{y_\chi^2}{E_p} \int \mathrm{d}\Pi_q \mathrm{d}\Pi_k \, (2\pi)^4 \delta^{(4)}(p+q-k)\,(p\cdot q)\left[f_{\chi,p} f_{\bar{\chi},q} - f_{\phi,k}\right] \\
&= -\frac{y_\chi^2 m_\phi^2}{32\pi^2} \frac{f_{\chi,p}^{\mathrm{eq}}}{E_p} \int \frac{\mathrm{d}\alpha_{pq} \mathrm{d}E_q \mathrm{d}\xi_q\, E_q \cos\xi_q}{(|\boldsymbol{p}+\boldsymbol{q}|^2 + m_\phi^2)^{\frac{1}{2}}} \, \delta(E_p + E_q - (|\boldsymbol{p}+\boldsymbol{q}|^2 + m_\phi^2)^{\frac{1}{2}}) \\
&\qquad\qquad\qquad\qquad \times \left[\mathcal{A}(E_p, \xi_p, r, t)\mathcal{A}(E_q, \xi_q, r, t) - 1\right] f_{\bar{\chi},q}^{\mathrm{eq}} \\
&= -\frac{y_\chi^2 m_\phi^2}{32\pi^2} \frac{f_{\chi,p}^{\mathrm{eq}}}{E_p^2 \cos\xi_p} \int \mathrm{d}E_q \mathrm{d}\xi_q \left[1 - \left(\frac{1 - \sin\xi_p \sin\xi_q - m_\phi^2/(2E_p E_q)}{\cos\xi_p \cos\xi_q}\right)^2\right]^{-\frac{1}{2}} \\
&\qquad\qquad\qquad\qquad \times \left[\mathcal{A}(E_p, \xi_p, r, t)\mathcal{A}(E_q, \xi_q, r, t) - 1\right] f_{\bar{\chi},q}^{\mathrm{eq}}.
\end{aligned} \tag{5.38}$$

In the above calculation we made use of detailed balance, $f_{\phi,k} = f_{\chi,p}^{\mathrm{eq}} f_{\bar{\chi},q}^{\mathrm{eq}}$, and introduced $\alpha_{pq}$, the angle between the two planes that contain the trajectories of $\chi(p)$ and $\bar{\chi}(q)$. The integration over $\alpha_{pq}$ reflects the fact that all possible relative orientations of these two planes contribute to the interaction. The final form of the collision term implies the kinematic requirement $|1 - \sin\xi_p \sin\xi_q - m_\phi^2/(2E_p E_q)| < \cos\xi_p \cos\xi_q$.

Note that our mechanism involves *overdensities* of $\chi$, which are represented by a phase space enhancement factor $\mathcal{A} > 1$. The derived collision term is thus negative, corresponding to annihilation (rather than production) of $\chi\bar{\chi}$. This effect reduces the accumulated overdensities and drives the $\chi$ abundance partly back towards equilibrium, possibly hindering BH formation.

**Annihilation via $\chi\bar{\chi} \to \phi\phi$.** The process $\chi(p) + \bar{\chi}(q) \leftrightarrow \phi(k) + \phi(l)$ shown in Fig. 5.2b scales as $y_\chi^4$ and is expected to be less dominant than the one previously discussed (espe-

---

[7] The overdensities that build up due to reflections off the wall are mostly localized around large momenta $\sim m_\chi^\infty$. For this reason, we observe $f_\chi \gg f_\chi^{\mathrm{eq}}$ but at the same time $f_\chi \ll 1$.

*5 Primordial Black Holes from Squeezed Matter*

cially for $y_\chi \ll 1$). This contribution to the collision term is given by [283]

$$\mathbf{C}[f_\chi] \supset -\frac{1}{2}\frac{1}{2E_p}\int d\Pi_q d\Pi_k d\Pi_l \, (2\pi)^4 \delta^{(4)}(p+q-k-l)\, \overline{|\mathcal{M}|}^2 \\ \times \left[f_{\chi,p}f_{\bar\chi,q}(1+f_{\phi,k})(1+f_{\phi,l}) - f_{\phi,k}f_{\phi,l}(1-f_{\chi,p})(1-f_{\bar\chi,q})\right]. \quad (5.39)$$

We again start by setting $1\pm f = 1$ and making use of detailed balance, $f_{\phi,k}f_{\phi,l} = f^{\rm eq}_{\chi,p}f^{\rm eq}_{\bar\chi,q}$. By carrying out the $\phi$ phase space integrals, the matrix element can be rewritten in terms of the kinematic factor $E_p E_q |v_\chi - v_{\bar\chi}|$ times the cross section[8]

$$\sigma(\chi\bar\chi \to \phi\phi) = \frac{y_\chi^4}{32\pi s}\left[2\log\!\left(\frac{s}{m_\phi^2}\right) - 3\right] + \mathcal{O}\!\left(\frac{m_\phi^2}{s}\right), \quad (5.40)$$

with the squared center-of-mass energy

$$s = 2E_p E_q(1 - \sin\xi_p \sin\xi_q - \cos\xi_p \cos\xi_q \cos\alpha_{pq}). \quad (5.41)$$

The contribution to the collision term then becomes

$$\begin{aligned}
\mathbf{C}[f_\chi] &\supset -2\frac{f^{\rm eq}_{\chi,p}}{E_p}\int d\Pi_q \sqrt{(s-2m_\chi^2)^2 - 4m_\chi^4}\, \sigma(\chi\bar\chi \to \phi\phi) \\
&\qquad\times \left[\mathcal{A}(E_p,\xi_p,r,t)\mathcal{A}(E_q,\xi_q,r,t) - 1\right]f^{\rm eq}_{\bar\chi,q} \\
&\approx -\frac{y_\chi^4}{128\pi^3}\frac{f^{\rm eq}_{\chi,p}}{E_p}\int dE_q d\xi_q\, E_q \cos\xi_q \left[\mathcal{A}(E_p,\xi_p,r,t)\mathcal{A}(E_q,\xi_q,r,t) - 1\right]f^{\rm eq}_{\bar\chi,q} \\
&\qquad\times \left[2\log\!\left(\frac{E_p E_q \cos\xi_p \cos\xi_q}{m_\phi^2}\right) + 4\,\mathrm{arsinh}\!\left(\sqrt{\frac{1-\cos(\xi_p-\xi_q)}{2\cos\xi_p \cos\xi_q}}\right) - 3\right].
\end{aligned} \quad (5.42)$$

Note that integrating over the angle $\alpha_{pq}$ is impossible in this case, as it enters the the kinematic requirement $s > 4m_\phi^2$. We carried out the integral nevertheless, using $2E_p E_q[1 - \cos(\xi_p - \xi_q)] > 4m_\phi^2$ as an approximate integration bound. The sign of Eq. (5.42) together with the fact that our mechanism involves $\mathcal{A} > 1$ implies that this term contributes to $\chi\bar\chi$ annihilation (rather than production) as well.

**Scattering.** The collision term does in principle also include the effect of the scattering with $\chi$, $\bar\chi$, or $\phi$, as shown in Fig. 5.3. However, we do not include these contributions in our calculations for the following reasons:

1. The scattering rates scale as $y_\chi^4$ and are thus suppressed compared to the $\chi\bar\chi \to \phi$ channel. On top of that, scattering conserves the total number of $\chi + \bar\chi$. Therefore, the effect only plays a secondary role.

2. The only effect of scattering is a redistribution of momentum, moving overdensities

---

[8]The terms of order $m_\phi^2/s$ can be neglected because the overdensities in our mechanism are generated around large momenta $\sqrt{s} \sim m_\chi^\infty \gg T_n \sim m_\phi$.

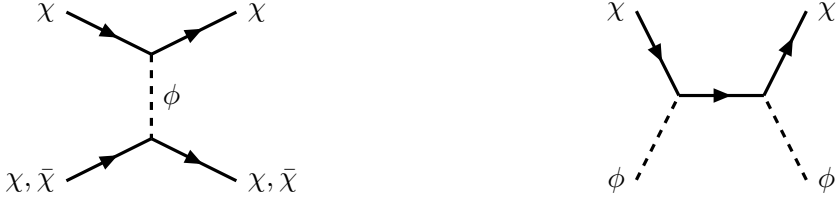

**Figure 5.3:** Scattering processes, which we do *not* consider.

from regions of large momenta $\sim m_\chi^\infty$ (where they are produced) to smaller momenta $\sim T_n$ (thermal equilibrium). This increases the effect of annihilation, which is most efficient at small momenta, but at the same time decreases the amount of particles that become energetic enough to escape the bubble.

3. Contrary to the annihilation terms, the scattering terms cannot be simplified as much analytically. A high-dimensional phase space integral would have to be computed which would drastically impact the performance of our simulation.

### 5.3.3 Wall Friction

With the accumulation of $\chi$ overdensities inside the bubble, more and more particles interact with the wall and draw energy from it. In other words, the moving wall feels an increasing friction force which might eventually slow it down or even stop it. The pressure the overdensity exerts on the wall amounts to [204, 378]

$$P_\chi(t) = g_\chi \int \mathrm{d}r \int \frac{\mathrm{d}^3 p}{(2\pi)^3} \frac{\partial E}{\partial m_\chi} \frac{\partial m_\chi}{\partial r} \delta f_\chi \,, \tag{5.43}$$

where $\delta f_\chi \equiv f_\chi - f_\chi^{\mathrm{eq}}$ is the deviation from the equilibrium distribution (increasing with time). The pressure that drives the wall forward, on the other hand, is given by the latent heat $\Delta V$ of the PT.[9] As we do not specify any details of the effective potential, the latent heat is treated as a free parameter. For our mechanism to work, we demand that

$$P_\chi(t) < \Delta V \tag{5.44}$$

at all times, so that the wall never stops.

If $\Delta V$ is too large, on the other hand, the vacuum energy inside the bubble may dominate and cause an intermediate period of inflation.[10] This would redshift and dilute the $\chi$ abundance, hampering the BH formation mechanism. We thus demand that the vacuum energy inside the bubble is smaller than the radiation density in the Universe, i.e.

---

[9] In case of the strongly supercooled PTs we are interested in, the entropy change in the plasma is negligible and the latent heat coincides with $\Delta V$. Furthermore, $\Delta V$ is meant to be the difference of the effective potential, *including* thermal corrections, between the true and false vacua. Using this standard definition of the latent heat, $\Delta V$ already accounts for the pressure due to the difference in $f_\chi^{\mathrm{eq}}$ itself [379]. For this reason the definition of the pressure $P_\chi$ involves $\delta f_\chi$ instead of $f_\chi$.

[10] We assume that the vacuum energy after the PT, i.e. outside the shrinking bubble, is negligible. Hence, the vacuum energy inside the bubble can be approximated as $\Delta V$.

$$\Delta V < \rho_{\rm rad} = \frac{\pi^2}{30} g_\star T_n^4 \,. \tag{5.45}$$

Taking both requirements together, we have to make sure that $P_\chi(t)$ – which is computed based on the solution of the Boltzmann equation – stays always smaller than $\rho_{\rm rad}$. Ideally, we want $P_\chi(t) \ll \rho_{\rm rad}$ at all times to leave an adequate window for $\Delta V$ and avoid fine-tuning.

### 5.3.4 Numerical Method

All ingredients of the Boltzmann equation have been discussed: the Liouville operator in both regimes and the different contributions to the collision term. We will now set forth the method employed to determine the solution. In the near-wall regime, for instance, the Boltzmann equation can be written in the form

$$\Omega_{p_r}(p_r, p_\sigma, r, t)\frac{\partial \mathcal{A}}{\partial p_r} + \Omega_r(p_r, p_\sigma, r, t)\frac{\partial \mathcal{A}}{\partial r} + \frac{\partial \mathcal{A}}{\partial t} = \Omega_\mathcal{A}(p_r, p_\sigma, r, t, \mathcal{A}) \,, \tag{5.46}$$

where the $\Omega$'s are defined to match Eq. (5.28). The above expression is a partial differential equation (PDE), which we ultimately need to solve in order to obtain $\mathcal{A}(p_r, p_\sigma, r, t)$.

The specific situation in Eq. (5.46), where the coefficients on the l.h.s. are independent of $\mathcal{A}$ and the ones on the r.h.s. independent of derivatives of $\mathcal{A}$, is eligible for the application of the *method of characteristics*. In this approach, the PDE is reduced to a system of ordinary differential equations (ODEs). Applied to the present scenario, we start by solving the parametric equations

$$\frac{{\rm d}p_r(t)}{{\rm d}t} = \Omega_{p_r}\big(p_r(t), p_\sigma(t), r(t), t\big)\,, \qquad \frac{{\rm d}p_\sigma(t)}{{\rm d}t} = 0\,, \qquad \frac{{\rm d}r(t)}{{\rm d}t} = \Omega_r\big(p_r(t), p_\sigma(t), r(t), t\big)\,. \tag{5.47}$$

Each set of initial conditions yields a *characteristic curve* $(r(t), p_r(t), p_\sigma(t))$, which can be interpreted as the physical trajectory of a single particle that is never annihilated. The value of $\mathcal{A}$ along the characteristic curve is then determined by solving

$$\frac{{\rm d}\mathcal{A}(t)}{{\rm d}t} = \Omega_\mathcal{A}\big(p_r(t), p_\sigma(t), r(t), t, \mathcal{A}(t)\big)\,, \tag{5.48}$$

where we start in equilibrium, $\mathcal{A}(t_0) = 1$. As soon as the trajectory reaches a certain distance from the wall (we make the cut at $|r - r_w(t)| > 3 l_w$), the effect of the wall – caused by the $\partial m_\chi / \partial r$ term – becomes negligible. From this point on, we consider the trajectory to evolve in the bulk regime, where similar ODEs – now based on Eqs. (5.35), (5.38), and (5.42) – apply. If the trajectory does now traverse the bubble. When it reaches $r = r_w(t)$ on the opposite side, its evolution continues in the near-wall regime, and so forth. At the interface between the two regimes, the last value of $\mathcal{A}$ from the previous regime is carried over to the new regime as an initial condition. As soon as a trajectory penetrates the wall, it escapes to $r \to \infty$ and can be discarded from that point on, because it cannot contribute to the further increasing overdensity inside the bubble.

This procedure is repeated for a large set of trajectories with different starting points in order to densely populate the entire phase space with characteristic curves. The solutions $\mathcal{A}(p_r, p_\sigma, r, t)$ and $\mathcal{A}(E, \xi, r, t)$ are then obtained by discretizing $p_r$–$p_\sigma$–$r$–$t$- and $E$–$\xi$–$r$–$t$-space and averaging over all curve segments that pass through the individual grid cells. In practice, it is unavoidable that some cells are never visited by a trajectory. These gaps will conservatively be interpreted as if they were in equilibrium ($\mathcal{A} = 1$).

A subtlety arises in the bulk regime, where the collision term is included: Equations (5.38) and (5.42) contain momentum integrals over the interaction partner's phase space. The interaction partner, however, is $\bar\chi$ and thus follows the same phase space distribution as $\chi$. At a given $r$ and $t$, this effectively couples together all trajectories across the entire momentum space. As an approximation, we evolve all trajectories independently for one time step $\Delta t$ and evaluate the phase space integrals of $\bar\chi$ using the result from the previous timestep, i.e. using $\mathcal{A}(E_q, \xi_q, r, t - \Delta t)$.

After $\mathcal{A}$ is determined, the following quantities can be computed[11]

$$
\begin{aligned}
n_\chi(r,t) &= g_\chi \int \frac{\mathrm{d}^3 p}{(2\pi)^3} f_\chi(\boldsymbol{p}, r, t)\,, \\
\rho_\chi(r,t) &= g_\chi \int \frac{\mathrm{d}^3 p}{(2\pi)^3} \left(|\boldsymbol{p}|^2 + m_\chi^2(r,t)\right)^{\frac{1}{2}} f_\chi(\boldsymbol{p}, r, t)\,, \\
\langle\rho_\chi\rangle(t) &= \frac{4\pi}{\frac{4}{3}\pi r_w^3(t)} \int_0^{r_w(t)} \mathrm{d}r\, r^2 \rho_\chi(r, t)\,,
\end{aligned}
\qquad (5.49)
$$

where, again, $f_\chi = \mathcal{A} f_\chi^{\mathrm{eq}}$ and $g_\chi = 4$. Tracking the number density, $n_\chi$, enables us to test the conservation of particle number in our simulation, as will be discussed in the subsequent section. The energy density, $\rho_\chi$, is relevant for the evaluation of the Schwarzschild criterion, Eq. (5.14). For simplicity, we will instead use the average energy density in the bulk of the bubble, $\langle\rho_\chi\rangle$, as our indicator for BH formation, which is conservative.

We solve the Boltzmann equation using the error-controlled explicit solver algorithm Runge–Kutta–Dormand–Prince-5 [380] provided by the BOOST library for C++. Our implementation makes intensive use of multithreading and is capable of obtaining fine grained solutions based on millions of characteristic curves in a reasonable amount of time (a few hours per simulation on a desktop machine).

### 5.3.5 Consistency Check

Our numerical implementation of the Boltzmann equation requires the discretization of phase space. To ensure that the chosen grids spacing and the curve density are appropriate, we monitor the conservation of particle number. To do so, we first calculate the total

---

[11] We remain general by writing $f_\chi$ as a function of $\boldsymbol{p}$ at this point. In practice, it is a function of $(p_r, p_\sigma)$ or $(E, \xi)$, depending on the considered regime.

number of $\chi$ particles inside the shrinking bubble,

$$N_\chi(t) = 4\pi \int_0^{r_w(t)} \mathrm{d}r\, r^2\, n_\chi(r,t)\,. \tag{5.50}$$

Now we demand that any change in the total particle number is accounted for by either annihilation or particles escaping the bubble. In other words, we test if the integrated continuity equation,

$$N_\chi(t) - \int_{t_0}^t \mathrm{d}t' \left([\dot N_\chi(t')]_{\text{escaping}} + [\dot N_\chi(t')]_{\text{annihilation}}\right) = N_\chi(t_0)\,, \tag{5.51}$$

with

$$[\dot N_\chi(t)]_{\text{escaping}} = -g_\chi \left[4\pi r^2 \int \frac{\mathrm{d}^3 p}{(2\pi)^3} \frac{p_r + v_w E}{E} f_\chi\right]_{r=r_w(t)+3l_w} \tag{5.52}$$

and

$$[\dot N_\chi(t)]_{\text{annihilation}} = g_\chi\, 4\pi \int_0^{r_w(t)} \mathrm{d}r\, r^2 \int \frac{\mathrm{d}^3 p}{(2\pi)^3} \mathbf{C}[f_\chi] \tag{5.53}$$

is fulfilled. Note that, to obtain the rate at which particles escape, we evaluate the phase space at a distance of $3l_w$ from the wall center in the true vacuum, where $\partial m_\chi/\partial r \approx 0$ and reflections occurring beyond this point are negligible. Equation (5.52) can be derived by first expressing the rate in the wall's rest frame and then transforming it to the plasma frame. We will demonstrate this consistency check as part of our results in Section 5.5.1.

## 5.4 Visualizing Phase Space

Before discussing the outcome of entire simulations of shrinking bubbles, we want to check the plausibility of our solutions to the Boltzmann equation by inspecting individual particle trajectories as well as slices of the phase space.

### 5.4.1 Particle Trajectories

Using the method of characteristics for solving the Boltzmann equation has a useful side effect: Individual particle trajectories can be visualized, which helps our understanding of the microphysics that underlie our mechanism. A characteristic curve can be interpreted as the physical trajectory of a hypothetical particle that never annihilates. The phase space density along the characteristic curves is encoded in the enhancement factor $\mathcal{A}$. This quantity is affected by the wall (increasing $\mathcal{A}$), by annihilation (decreasing $\mathcal{A}$), and by gravity (blue-/redshifting $\mathcal{A}$). Two exemplary particle trajectories are visualized in Fig. 5.4, each in positional $x$–$y$-space[12] (left panels) and in $r$–$p_r$-space (right panels):

---

[12] We determine the characteristic curves in the discussed radial coordinate system, but here we plot $x$ and $y$ for a more intuitive illustration.

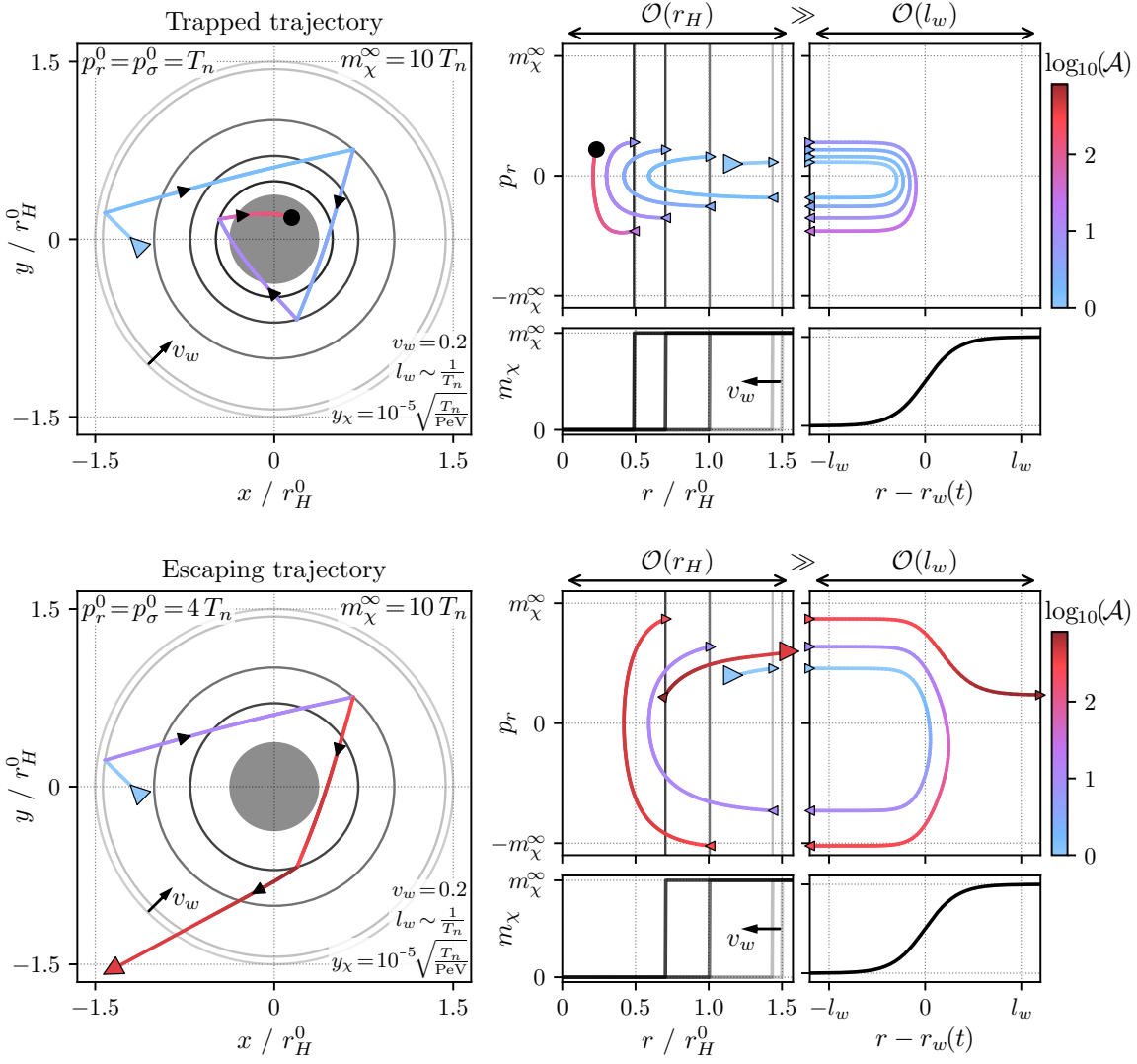

**Figure 5.4:** Visualization of two exemplary characteristic curves, which can be interpreted as particle trajectories. The color indicates the value of the corresponding phase space enhancement factor $\mathcal{A}$. On the left, we show the trajectories in positional $x$–$y$-space in the bulk regime. The gray circles indicate the shrinking bubble and the gray disk represents the BH that ultimately forms. On the right, we show the same trajectories in $r$–$p_r$-space, in the bulk regime and zoomed in on the near-wall regime. The switching between the two regimes is indicated by small colored arrowheads. Also on the right, we illustrate how the mass $m_\chi$ changes across the bubble wall. Note that we plot the near-wall regime as a function of $r - r_w(t)$, i.e. centered around the moving wall (but the momenta are still plasma-frame quantities). The upper panels show a trajectory with small initial momentum, which is trapped inside the bubble and ultimately ends when the BH forms (black dot). The lower panels show a trajectory from the same simulation but with large initial momentum. After it gained enough energy from a few reflections, it passes through the wall and escapes the bubble (large red arrowhead), not contributing to the BH formation.

1. The upper panels show the path of a $\chi$ particle starting inside the bubble and with initial momentum $p_r^0 = p_\sigma^0 = T_n$, i.e. a typical representative of the initial thermal distribution. Due to its small radial momentum, $p_r \ll m_\chi^\infty$, it bounces off the shrinking bubble's wall several times. Each reflection increases the energy and the phase space enhancement factor $\mathcal{A}$ along the trajectory. As the particle traverses the bulk of the bubble, $\mathcal{A}$ slightly decreases due to annihilation. After the bubble radius has shrunk by about a factor of three, we find that the overdensity inside the bubble (determined from *all* trajectories, not just the shown one) is large enough to form a BH. The bending of the trajectory prior to BH formation is due to gravity.

2. The lower panels show a $\chi$ particle with a larger initial momentum $p_r^0 = p_\sigma^0 = 4\,T_n$, i.e. one from the tail of the thermal distribution. After only two reflections it gains enough momentum, $p_r \gtrsim m_\chi^\infty$, to pass through the wall and to escape the bubble. As a consequence, this particle will not become part of the BH that eventually forms.

In our simulation we neglect trajectories that start outside the bubble, as those can never become part of the squeezed $\chi$ abundance. (By entering the bubble from outside, particles gain a momentum $p_r \gtrsim m_\chi^\infty$ and will thus escape the bubble on the opposite side.).

The model parameters used for the visualization of the trajectories are indicated in the figure. Note that we express values of the Yukawa coupling in terms of $\sqrt{T_n/\text{PeV}}$. This makes our results scale independent: For small Yukawa couplings, the only relevant annihilation process is $\chi\bar\chi \to \phi$. Its rate scales as $y_\chi^2 T_n$, whereas $r_H^0$ scales as $1/T_n$ during radiation domination. Indicating the Yukawa coupling in terms of $\sqrt{T_n/\text{PeV}}$ thus ensures that the amount of annihilation per Hubble length $r_H^0$ is independent of the temperature. However, this scale invariance cannot be maintained for large $T_n$, where the corresponding $y_\chi$ would be of $\mathcal{O}(1)$ and annihilation via $\chi\bar\chi \to \phi\phi$ becomes relevant. This would be the case if $T_n \gtrsim 10^{10}$ TeV for the value of $y_\chi$ chosen in Fig. 5.4.

### 5.4.2 Slices of Phase Space

Now that we gained an intuition for the behavior of individual trajectories, we broaden our view by investigating the evolution of phase space as a whole, which in turn is constructed from a large number of characteristic curves. Figure 5.5 visualizes slices of the phase space enhancement factor $\mathcal{A}(p_r, p_\sigma, r, t)$ in the near-wall regime at different fixed $p_\sigma$ and $t$:

1. The left panels visualize an early time $t_1$, shortly after the start of the simulation, where the bubble has still almost its initial size, $r_w(t_1) \approx r_w^0$. The part of the $\chi$ abundance that approaches the wall ($p_r > 0$, $r < r_w(t_1)$) starts in equilibrium ($\mathcal{A} = 1$) but develops overdensities ($\mathcal{A} > 1$) when interacting with the wall. The largest $\mathcal{A}$ are generated in front of the wall ($p_r \sim -m_\chi^\infty$, $r < r_w(t_1)$) and correspond to $\chi$ particles that are reflected. We also observe an overdensity behind the wall, consisting of particles which were energetic enough to escape the bubble. However, recall that $\mathcal{A}$ measures the *relative* overdensity w.r.t. the equilibrium abundance, which is highly suppressed outside the bubble ($r > r_w(t_1)$) where $m_\chi \gg T_n$. This means that the absolute number of particles escaping the bubble is tiny. This suppression is also visible in the small energy density, $\rho_\chi \ll T_n^4$, outside of the bubble. The average energy density in the bulk of the bubble, $\langle \rho_\chi \rangle$, our indicator for BH formation, is also shown in the plot. As the PT just started, most of the bubble's interior is still in equilibrium and we observe $\langle \rho_\chi \rangle \sim T_n^4$ accordingly.

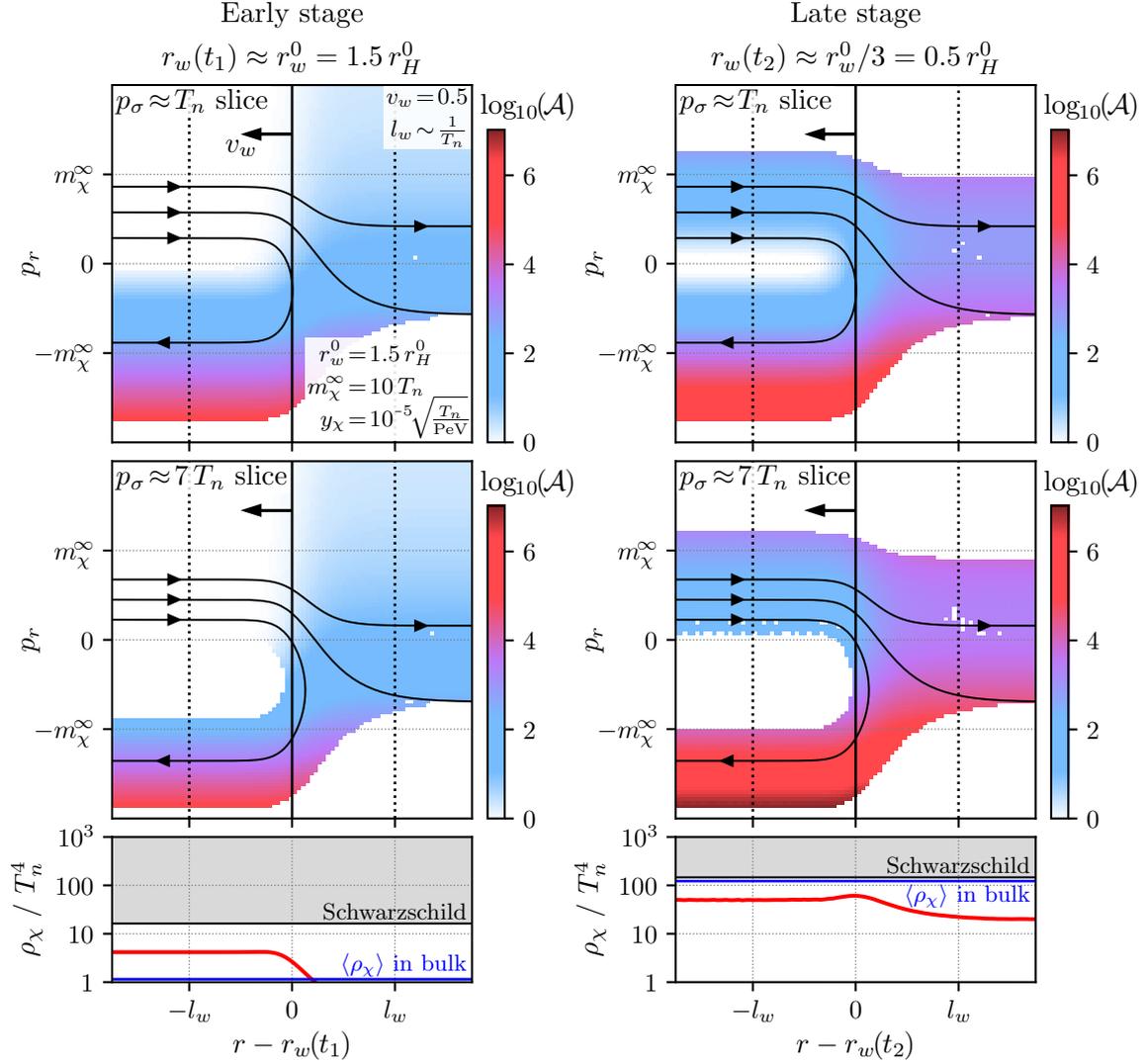

**Figure 5.5:** Visualization of the phase space in the near-wall regime: The upper and middle panels show slices of the phase space enhancement factor $\mathcal{A}$ at different fixed $p_\sigma$ and $t$. The bubble wall moves to the left and its center is indicated by a black vertical line. Also shown are three exemplary particle trajectories. The left panels show the situation at a time $t_1$, shortly after the start of the simulation. As can be seen by the color gradient, an overdensity accumulates in front of the wall. The right panels show a later time $t_2$, when the bubble radius has shrunk by a factor of three. At this point, part of the once reflected $\chi$ abundance has traversed the bubble and interacts with the wall again, increasing the overdensity even further. The lower panels show the corresponding energy density $\rho_\chi$ (red) together with the average energy density in the bulk of the bubble $\langle\rho_\chi\rangle$ (blue). The Schwarzschild criterion (gray region) is almost fulfilled by $\langle\rho_\chi\rangle$, indicating that a BH may form shortly after $t_2$.

2. The right panels show the same simulation at a later time $t_2$, at which the bubble has shrunk to a third of its initial size, $r_w(t_2) \approx r_w^0/3$. The $\chi$ abundance approaching the wall ($p_r > 0$, $r < r_w(t_2)$) is now out of equilibrium ($\mathcal{A} \gg 1$). This represents particles that have already been reflected by the wall, then traversed the bulk of the bubble and finally reached the opposite side. Subsequent reflections then lead to a further increase of $\mathcal{A}$. The energy density, $\rho_\chi$, is now much larger than in the beginning of the simulation.[13] Note that $\langle \rho_\chi \rangle$ is now almost large enough to fulfill the Schwarzschild criterion, indicating that a BH may form shortly after $t_2$.

Each slice in Fig. 5.5 also contains three exemplary trajectories. The middle one represents a "threshold trajectory", which has just enough radial momentum to pass through the wall. In the rest frame of the wall, this threshold would be exactly given by $p_r \sim m_\chi^\infty$. Boosted to the plasma frame, the threshold is smaller by a factor $\gamma_w(1 + v_w)$ for $p_\sigma = 0$, for instance. (A similar but lengthier expression applies to the case $p_\sigma > 0$.) Another interesting observation is that trajectories approaching the wall with $p_r \approx 0$ but sizable $p_\sigma$ will gain a large negative $p_r$ upon reflection. This is also a relativistic effect and results in an unpopulated region around small negative $p_r$, as can be seen by the white areas in the $p_\sigma \approx 7\, T_n$ slices.

The main takeaway of Fig. 5.5 is that the major part of the initial thermal $\chi$ abundance is reflected by the wall, leading to a growing overdensity inside the shrinking false-vacuum bubble. This behavior leads to BH formation in certain regions of the model parameter space, as we will see in the following section.

## 5.5 Results

In the previous section we established a detailed understanding of the phase space evolution in a typical solution of the Boltzmann equation. We will now shift our focus to the behavior of the average energy density in the bulk of the bubble, $\langle \rho_\chi \rangle$, for different model parameters. By comparing this quantity to the Schwarzschild criterion, Eq. (5.14), we can test if PBH formation occurs. We vary $r_w^0$, $m_\chi^\infty$, and $y_\chi$ in Section 5.5.1 and discuss the dependence on $v_w$ and $l_w$ in Section 5.5.2.

### 5.5.1 Model Parameter Dependence

Figure 5.6 depicts the evolution of $\langle \rho_\chi \rangle$ as a function of the shrinking bubble radius, $r_w(t)$, for different initial radii $r_w^0$, true-vacuum masses $m_\chi^\infty$, and Yukawa couplings $y_\chi$. Here, we keep the wall velocity and thickness constant at $v_w = 0.5$ and $l_w \sim 1/T_n$, as we discuss the impact of these parameters in the next section. The gray area indicates where the Schwarzschild criterion is fulfilled. In the left plot, we express $y_\chi$ in terms of $\sqrt{T_n/\mathrm{PeV}}$ to make our results scale invariant in the regime $T_n \lesssim 10^{10}\,\mathrm{TeV}$ (see the discussion at the end of Section 5.4.1).[14] The right plot shows parameter points for the fixed nucleation temperature $T_n = 10^{12}\,\mathrm{TeV}$.

---

[13]The average energy density in the bubble, $\langle \rho_\chi \rangle$, is now even larger than the density close to the wall. This is because the nearly radial modes have not yet crossed the entire bubble after their first reflection. Part of the $\chi$ abundance impinging on the wall is thus still in equilibrium, while most of the bubble is already filled with particles that were reflected at least once.

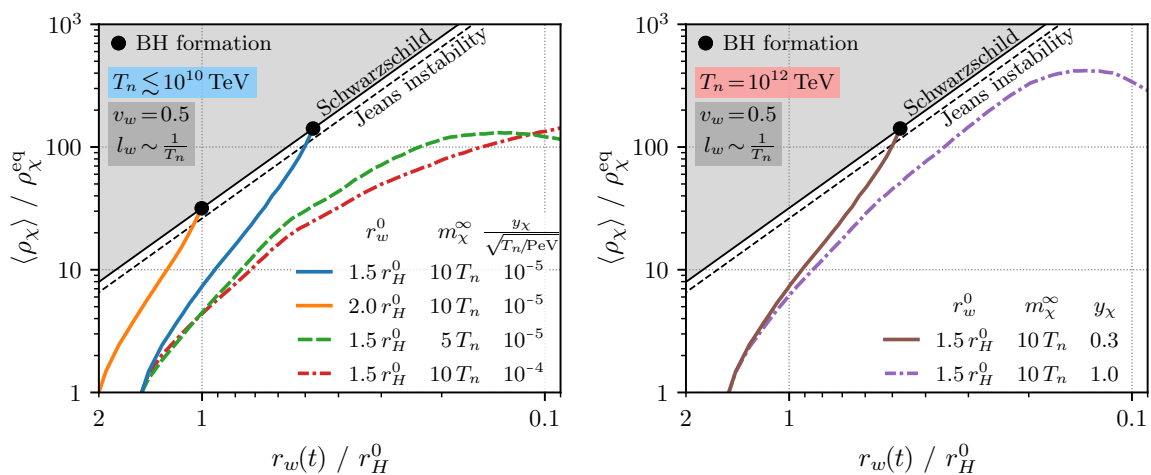

**Figure 5.6:** The $\chi$ energy density averaged over the bubble $\langle\rho_\chi\rangle$ as a function of the shrinking bubble radius, $r_w(t)$, for different model parameter points. When the density becomes large enough to fulfill the Schwarzschild criterion (gray region), a BH forms (black dot). Small true-vacuum masses, $m_\chi^\infty$, or large Yukawa couplings, $y_\chi$, can prevent BH formation. We also show where the systems develop Jeans instabilities (dashed black line). However, as discussed in Section 5.2.2, this criterion is not sufficient to conclude BH formation because degeneracy pressure halts the collapse.

The solid blue curve in the left plot represents our main benchmark point, with $r_w^0 = 1.5\,r_H^0$, $m_\chi^\infty = 10\,T_n$, and $y_\chi = 10^{-5}\sqrt{T_n/\mathrm{PeV}}$, which we already examined more closely in the previous section. In the beginning of the simulation, $\langle\rho_\chi\rangle$ increases faster than the $[r_w^0/r_w(t)]^4$ scaling that we naively anticipated in Section 5.2.3. This is due to the relativistic wall velocity leading to an additional energy gain on each reflection, an effect that we did not consider in our analytic estimate. Another relativistic effect is the large increase of radial momentum when a mostly tangential modes is reflected (see discussion at the end of Section 5.4.2). These modes thereby become more radial and require more time to reach the opposite side of the bubble, where they interact with the wall a second time. This leads to a phase of slightly slower density growth. Finally, $\langle\rho_\chi\rangle$ increases faster again because the second wave of reflections commences, but also because the effects of gravity become noticeable. Shortly after the bubble radius reaches a third of its initial value, i.e. at $r_w \approx 0.5\,r_H^0$, a Schwarzschild horizon forms. Our non-relativistic estimate in Table 5.1 predicted a required compression of 3.8.

The benchmark point represented by the solid orange curve is similar to the first one,

---

[14] Note that our results assume $g_\star \sim 110$. When SM species become heavy, this number decreases and the Schwarzschild criterion becomes easier to satisfy. Our results are thus conservative below $T_n \sim 100\,\mathrm{GeV}$.

except for the larger initial radius, $r_w^0 = 2\,r_H^0$. In this case, the Schwarzschild criterion is fulfilled even earlier, after the radius has shrunk by a factor of only two, i.e. at $r_w \approx r_H^0$.[15]

We also show parameter points with a smaller true-vacuum mass, $m_\chi^\infty = 5\,T_n$ (dashed green), and with a larger Yukawa coupling, $y_\chi = 10^{-4}\sqrt{T_n/\text{PeV}}$ (dot-dashed red). In these scenarios, the growth of the energy density eventually stagnates and a BH does not form.

In the $T_n = 10^{12}\,\text{TeV}$ plot on the right of Fig. 5.6 we show one successful parameter point with $y_\chi = 0.3$ (solid brown) and one where a larger Yukawa coupling, $y_\chi = 1$, again prevents BH formation (dot-dashed purple).

To better understand why no BH is formed at some of the parameter points, it is helpful to examine the evolution of the particle budget. In Fig. 5.7 we plot the number of particles, $N_\chi$, that are trapped inside the bubble (solid), that have escaped through the wall (dashed), and that have annihilated (dotted & dot-dashed). These curves are obtained by evaluating Eq. (5.50) and by integrating Eqs. (5.52) and (5.53) over time, respectively. Each panel in the figure corresponds to one parameter point in Fig. 5.6. In case of successful BH formation (top left, top right & bottom left panels) the amount of escaping and annihilating particles is negligible. The situation is different in the scenario with a smaller true-vacuum mass (middle left panel). At this parameter point, particles can pass through the wall more easily and a significant fraction of the $\chi$ abundance escapes the bubble. The two steps in the "escaping" curve correspond to the first and second reflection waves, each of which is accompanied by a fraction of particles that passes through the wall. In case of the large Yukawa couplings (middle right & bottom right panels) a sizable number of particles annihilate. For $T_n \lesssim 10^{10}\,\text{TeV}$, where our results correspond to $y_\chi \ll 1$, the $\chi\bar{\chi} \to \phi$ contribution dominates annihilation. At $T_n = 10^{12}\,\text{TeV}$ and $y_\chi = 1$, the $\chi\bar{\chi} \to \phi\phi$ annihilation channel becomes relevant, too.

The thin solid line in each panel represents the sum of the trapped, escaped and annihilated particles, i.e. the l.h.s. of Eq. (5.51), and serves as a consistency check for our simulation. The curve is almost constant and thus indicates that the overall particle budget is consistent. Only shortly before BH formation, when gravity takes over, the simulation starts to become numerically unstable and we observe a slight unphysical increase.

Finally, we want to verify if there is a window for the parameter $\Delta V$ – the latent heat of the PT – such that friction cannot stop the wall ($P_\chi(t) < \Delta V$ at all times) and that the interior of the bubble does not inflate ($\Delta V < \rho_{\text{rad}}$), as discussed in Section 5.3.3. For our benchmark points starting with initial radii $r_w^0 = \{1.5\,r_H^0, 2\,r_H^0\}$ we observe $P_\chi/\rho_{\text{rad}} \lesssim \{0.6, 0.03\}$. We conclude that it is possible to satisfy both requirements without fine-tuning. In contrast, scenarios involving smaller initial radii would violate at least one of the two assumptions, i.e. the wall would be stopped or there would be a period of inflation.

Summarizing the findings presented in this section, we saw that the proposed PBH for-mation mechanism is viable for large enough initial radii ($r_w^0 \gtrsim 1.5\, r_H^0$), large enough true-vacuum masses ($m_\chi^\infty \gtrsim 10\, T_n$), and small enough Yukawa couplings ($y_\chi \lesssim 10^{-5}\sqrt{T_n/\text{PeV}}$ for $T_n \lesssim 10^{10}$ TeV and $y_\chi \lesssim 0.3$ at $T_n = 10^{12}$ TeV).

---

[15] The reader may be worried about the causal connectedness of the region forming the BH. There is no problem, however, as the Hubble radius grows linearly with time, $r_H(t) = r_H^0(1 + t - t_0)$. The time it takes until the Schwarzschild criterion is fulfilled, on the other hand, is $t - t_0 \approx (2r_w^0 - r_w^0)/v_w$ for the discussed parameter point. When the BH forms, we thus have $r_H = 3r_H^0$, which is larger than the full diameter of the bubble ($\approx 2r_H^0$) at that time.

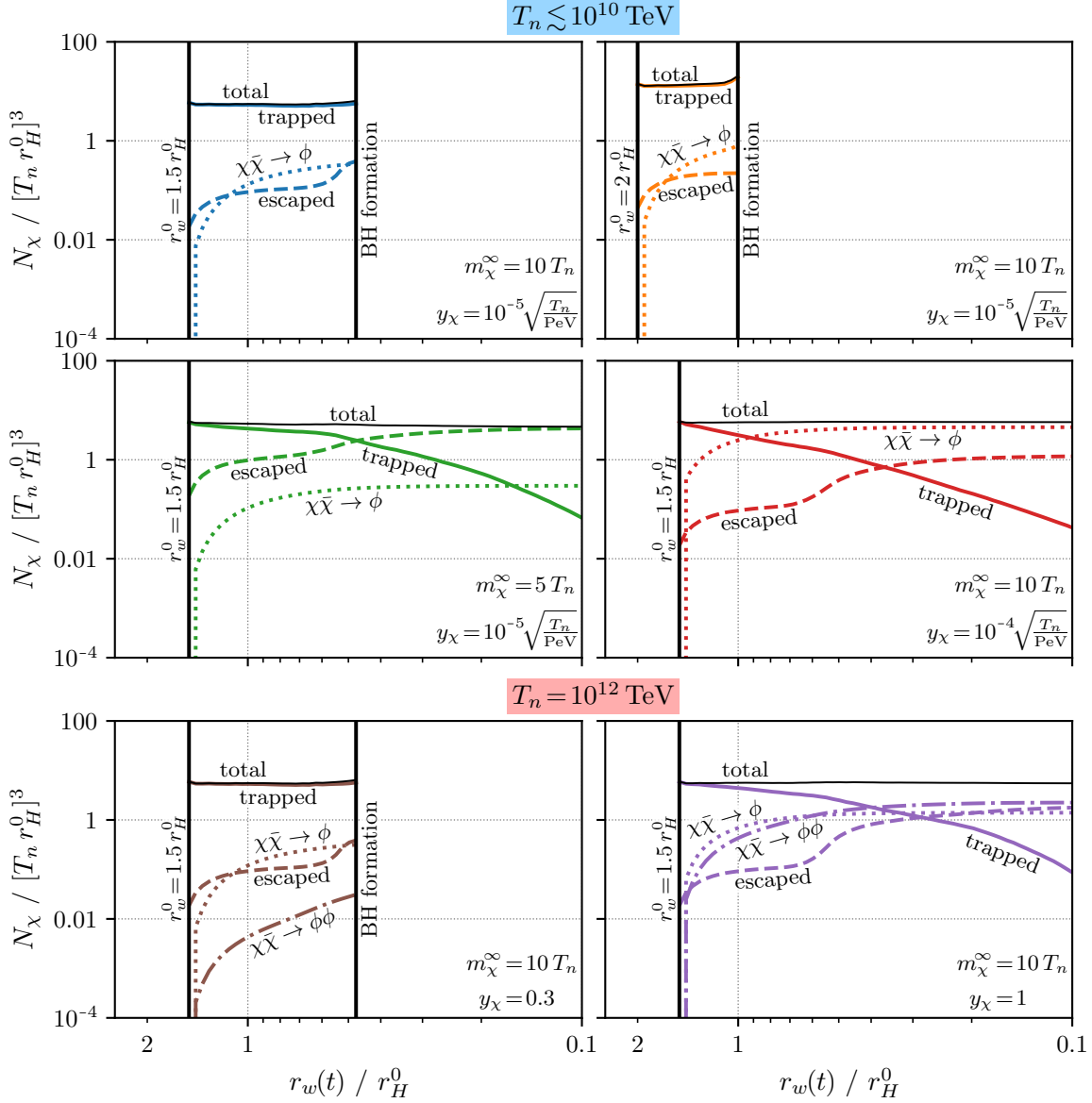

**Figure 5.7:** Number of particles $N_\chi$ that are trapped inside the shrinking bubble (solid), that have escaped through the wall (dashed), and that have annihilated (dotted & dot-dashed) for all six parameter points presented in Fig. 5.6. The total of all colored curves (thin black line) is almost constant, which shows that the overall particle budget in our simulation is consistent.

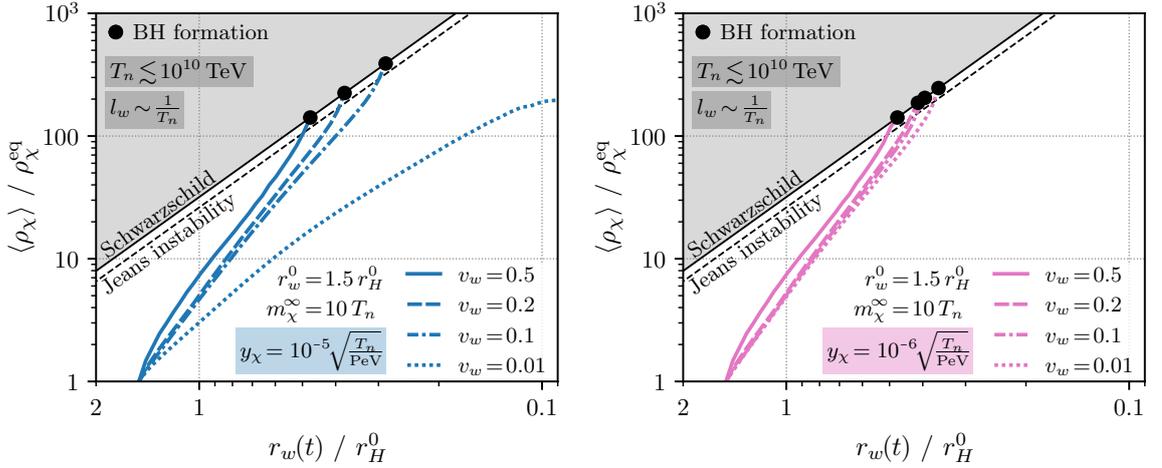

**Figure 5.8:** The $\chi$ energy density averaged over the bubble, $\langle\rho_\chi\rangle$, as a function of the shrinking bubble radius, $r_w(t)$, for different wall velocities. The parameter points in the left panel contain our default value of the Yukawa coupling. The right panel, on the other hand, shows scenarios with reduced annihilation, in which our mechanism is viable even in the case of very slow walls.

### 5.5.2 Wall Velocity and Wall Thickness

So far, we assumed specific values for the wall velocity and the wall thickness and kept them constant during our simulations. In reality, these quantities may change due to the large overdensities building up inside the bubble. For instance, the wall will be slowed down in response to the friction exerted by the $\chi$ particles, as discussed in Section 5.3.3. In what follows, we will discuss different regimes of the wall velocity and furthermore argue that our results are independent of the wall thickness.

**Varying wall velocity.** As the wall velocity, $v_w$, is highly model dependent and will most likely not stay constant throughout the PT, we have to make sure that the mechanism works over a range of possible values. In the left panel of Fig. 5.8 we show the evolution of the energy density $\langle\rho_\chi\rangle$ at our main benchmark point, but for different wall velocities. For smaller velocities we observe that the density grows slower in the beginning. In fact, the initial slopes of the $v_w = \{0.2, 0.1\}$ curves almost follow the $[r_w^0/r_w(t)]^4$ scaling expected by our non-relativistic estimate. The further evolution of the density is qualitatively similar for the different velocities. Slower walls lead to a later BH formation, as the energy input by the wall is reduced.

**Slow walls ($v_w \ll 0.1$).** To understand the behavior of the $v_w = 0.01$ curve, note that the time it takes the PTs to proceed is $\sim r_H^0/v_w$. In the $v_w = 0.01$ scenario there is, despite the small Yukawa coupling, enough time to allow for significant annihilation, resulting in a flattening of the density curve and no BH formation. Imposing a smaller $y_\chi$, as done in the right panel of Fig. 5.8, again leads to successful BH formation even with such slow walls. The slow wall scenario, however, is highly unlikely anyways: It would require a delicate balance between the forward driving and the retaining pressure, i.e. $\Delta V \sim P_\chi$. It is more likely that the wall entirely stops (and the PT does not complete) or that it reaches relativistic velocities.

**Relativistic walls ($v_w \sim 1$).** Given a large enough latent heat, $\Delta V$, the wall does likely reach luminal velocities. This is in principal beneficial for our mechanism, as we can see in Fig. 5.8. We limited our analysis to values $v_w \leq 0.5$, because our numerical implementation is especially well-suited for this regime.[16] Nevertheless, we expect that larger velocities lead to even faster BH formation. Note that, if the wall reaches large velocities, it does so very quickly. This is, because on dimensional grounds, the distance scale over which the wall accelerates is $\sim T_n^3/\Delta V$, which is much smaller than the distance the wall has to travel until a BH forms, $\sim r_H \sim M_{\rm Pl}/T_n^2$, because $T_n \ll M_{\rm Pl}$ and $\Delta V \sim T_n^4$.

While $v_w \sim 1$ is in general compatible with our mechanism, a problem arises if the bubble wall becomes too energetic. Viewed from the wall's rest frame, the typical energy of a thermally abundant $\chi$ particle is $\sim \gamma_w T_n$. If this energy is larger than $m_\chi^\infty$, most particles will be able to pass through the wall. In other words, the bubble wall becomes mostly transparent if the Lorentz factor reaches $\gamma_w \sim m_\chi^\infty/T_n$, spoiling our mechanism. Fortunately, the friction induced by $\chi$ can prevent the wall from reaching large $\gamma_w$. As we argued above, large wall velocities are reached quickly after the onset of the PT, implying that until BH formation, each $\chi$ particle in the shrinking bubble is reflected only once. This in turn means, that the $\chi$-induced friction is approximately constant until the BH forms, making a scenario plausible where $\chi$ prevents $\gamma_w$ from growing indefinitely. Nevertheless, if required, additional friction can be introduced by adding vector bosons which induce transition radiation and naturally limit $\gamma_w$ [124, 203].

**Varying wall thickness.** Our results are entirely independent of the wall width, $l_w$, if the walls are thin compared to the bubble radius, $l_w \ll r_w$. This is a valid assumption because typically $l_w \sim 1/T_n$ as well as $r_w \sim r_H \sim M_{\rm Pl}/T_n^2$, and we consider the regime $T_n \ll M_{\rm Pl}$. Thanks to the resulting scale separation, $l_w$ is the only length scale in the near-wall regime and can thus be arbitrarily rescaled. Under the given assumptions, our mechanism is thus entirely independent of the wall width. In fact, in our numerical implementation we treat the near-wall regime as a black box of width zero when viewed from the bulk regime's perspective.

---

[16]At velocities $v_w \sim 1$, reflected particles travel alongside the advancing bubble wall over long distances. This would spoil a key feature of our implementation: the segmentation into near-wall and bulk regime.

### 5.5.3 The Black Hole Mass Landscape

In the previous sections, we have examined under which assumptions and in which regions of parameter space the proposed PBH mechanism is viable. We found that the mechanism may occur at any temperature and that our results apply universally, up to the discussed limitations. It now remains to determine the properties of the resulting BH population and to consider existing experimental constraints.

As we have learned from our analyses, Hubble-sized bubbles are required to accumulate enough energy for BH formation while avoiding intermediate inflation at the same time. By rearranging the criterion for the formation of a Schwarzschild horizon, Eq. (5.14), and using $r_w \sim r_H$, we obtain

$$m_{\mathrm{BH}} \sim r_H^3 \frac{1}{G\,r_H^2} = \frac{r_H}{G} \tag{5.54}$$

for the typical mass of a produced BH. The BH density at the time of production, i.e. at temperature $T_n$, is given by

$$n_{\mathrm{BH}}(T_n) = \frac{\mathfrak{p}}{\frac{4}{3}\pi r_H^3}\,. \tag{5.55}$$

Here, $\mathfrak{p}$ is the probability that a BH forms per Hubble volume, which in turn depends on the details of the PT: If new, smaller bubbles of true vacuum nucleate inside the shrinking false-vacuum bubble, the accumulating $\chi$ abundance is split into separate populations which are too small individually to form BHs. In Eq. (3.19) we defined the parameter $\beta$ as the inverse time scale of the PT, or equivalently, the rate at which the bounce action changes. In a fast transition with $\beta \gg H$, the bubble nucleation rate increases drastically before the PT has time to complete. This leads to the nucleation of many tiny bubbles and results in a small $\mathfrak{p}$. In case of a slow transition with $\beta \sim H$, on the other hand, it is more likely that a Hubble-sized region will be compressed unimpeded, corresponding to a larger $\mathfrak{p}$. The latter situation is typically realized in supercooled PTs [173, 190]. In any case, we expect the absolute value of $\mathfrak{p}$ to be small. But even if BH formation is rare, the resulting PBH density today can still be sizable, as we will see further below.

Based on the mass and density, the expected PBH abundance relative to the observed DM abundance is given by

$$\begin{aligned} f_{\mathrm{PBH}} \equiv \frac{\Omega_{\mathrm{PBH}}}{\Omega_{\mathrm{DM}}} &= \frac{1}{\Omega_{\mathrm{DM}}^0}\frac{m_{\mathrm{BH}}\,n_{\mathrm{BH}}^0}{\rho_{\mathrm{crit}}^0} \\ &= \frac{1}{\Omega_{\mathrm{DM}}^0}\frac{m_{\mathrm{BH}}\,n_{\mathrm{BH}}(T_n)}{\rho_{\mathrm{crit}}^0}\frac{g_{\star s}^0 T_0^3}{g_{\star s}(T_n)T_n^3} \\ &\approx 3.7 \times 10^9 \left(\frac{T_n}{\mathrm{GeV}}\right)\mathfrak{p}\,, \end{aligned} \tag{5.56}$$

where "0" labels today's values. Note that, when neglecting the effects of accretion and evaporation, both the PBH and DM abundances redshift equally, implying that $f_{\mathrm{PBH}}$ stays constant. We can thus evaluate this observable at any scale, for instance at the Universe's temperature today, $T_0$, as we did in the first line of Eq. (5.56). In the second line, we rewrote $n_{\mathrm{BH}}^0$ in terms of the known quantity $n_{\mathrm{BH}}(T_n)$, and to obtain the numerical value in the last line we used $h^2 \Omega_{\text{DM}}^0 \approx 0.12$ [36], $\rho_{\text{crit}}(T_0) = 3h^2 (100\,\text{km/s/Mpc})^2/(8\pi G)$, $T_0 \approx 2.4 \times 10^{-13}\,\text{GeV}$ [205], $g_{\star s}^0 \approx 3.9$ [206], and $g_{\star s}(T_n) \sim 110$.

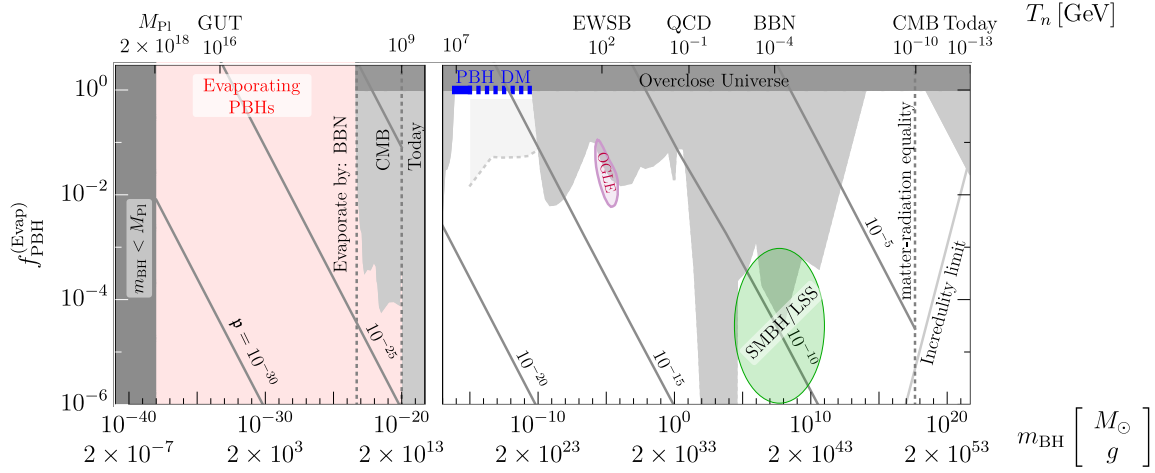

**Figure 5.9:** PBH abundance relative to the DM abundance, $f_{\text{PBH}}$, as a function of the PT temperature (top axis) and the corresponding BH mass (bottom axis) for different formation probabilities $\mathfrak{p}$ as gray lines. The red region corresponds to PBHs that have evaporated until today, in which case we show the abundance prior to evaporation. Experimentally excluded regions are grayed out and regions of particular interest are highlighted in color.

In Fig. 5.9 we show the relative PBH abundance as a function of the PT temperature $T_n$ (top axis) for different values of $\mathfrak{p}$ as gray lines. The bottom axis shows the corresponding BH mass in solar masses and grams. Note that PBHs lighter than $\sim 10^{-19} M_\odot$ would already have evaporated via Hawking radiation and we show the abundance prior to evaporation in this regime. Some regions of particular interest are highlighted:

1. The gray regions are experimentally excluded based on evaporation constraints, gravitational lensing, gravitational-wave measurements, CMB distortions, LSS, and other methods [82].

2. The blue band around $m_{\text{BH}} \sim 10^{-13} M_\odot$ indicates where the PBH abundance could account for all DM in the Universe. This region might be ruled out by observations of white dwarfs, neutron stars, and supernovae [381–383]. These constraints, drawn in light gray, are however currently being disputed [82].

3. Recent findings of the microlensing experiment *OGLE* could be explained by PBHs in the purple region around $m_{\text{BH}} \sim 10^{-5} M_\odot$ (rougly the Earth's mass) [384, 385].

4. PBHs in the green region around $m_{\text{BH}} \sim 10^8 M_\odot$ could be the seeds for supermassive black holes (SMBHs) and/or the LSS of the Universe [82, 344–348].

Other regions that are excluded in general are those that imply an overclosed Universe ($f_{\text{PBH}} > 1$), trans-Planckian masses ($m_{\text{BH}} < M_{\text{Pl}}$), or less than one BH in today's observ-able Universe ("incredulity limit"). It is noteworthy that even with tiny BH formation probabilities ($\mathfrak{p} \ll 1$), large abundances ($f_\text{PBH} \sim 1$) of almost any mass can be generated.

## 5.6 Conclusions

In the present chapter we suggested a new mechanism for the formation of PBHs which may occur in course of a first-order cosmological PT driven by either a scalar or a confining sector. The basic conditions are similar to those required for the Filtered Baryogenesis mechanism presented in Chapter 4: A large order parameter gives rise to a sizable mass jump of a dark sector particle at the PT boundary. As a consequence, the bulk of its thermal abundance is reflected by the bubble walls and thus trapped and squeezed together in the remaining false-vacuum regions, eventually reaching energy densities large enough to form BHs. By analytic estimates we found that the BH formation in our mechanism relies on the generation of a Schwarzschild horizon. A Jeans instability alone – which always forms prior to a Schwarzschild horizon – would *not* collapse, due to degeneracy pressure and the lacking possibility to shed angular momentum.

Our approach to obtain numerical results is based on tracking the evolution of phase space by solving the Boltzmann equation for a dark sector fermion (however, a scalar particle could equally work) in presence of moving bubble walls. We modeled a single, spherically symmetric shrinking bubble of false vacuum, which we expect to be a realistic assumptions in the presence of surface tension. This symmetry allowed us to describe the system using only two momentum, one spatial, and one temporal dimension. Motivated by the thinness of the bubble walls, we divided the system into a region near the wall – where the wall is described as locally flat – and the interior of the bubble. For both regimes, we derived the relevant terms describing the particle dynamics using different appropriate coordinate systems. We also took the annihilation of the accumulating overdensities into account, but we neglected scattering effects which we argued to be subdominant. To finally solve the Boltzmann equation, we employed the method of characteristics, which constructs the phase space from individual solution curves.

We have inspected the evolution of single trajectories and entire slices of phase space to convince ourselves of the physical plausibility of the obtained numerical results. By testing our mechanism at different model parameter points, we found that (i) a broken phase mass of at least ten times the Universe's temperature is required to trap sufficient amounts of particles inside the bubble, that (ii) a large initial bubble size of a few Hubble radii is required to avoid the large overdensities from stopping the bubble walls, and that (iii) a large range of Yukawa couplings is viable as long as annihilation in the false vacuum is suppressed. The latter requirement is the main distinctive feature compared to the Filtered Baryogenesis mechanism, which relies on sizable Yukawa couplings in order to keep the DM candidate in thermal equilibrium in the false vacuum. For all presented benchmark points, we tracked the overall particle number budget as a crosscheck of the numerical stability of our simulations.

Leaving the detailed nature of the PT and the underlying scalar potential unspecified (except the assumption of a large order parameter), we also treated the wall width and velocity as free parameters. We found that our results are entirely independent of the width, as long as thin walls are a good approximation. Furthermore, we demonstrated that the mechanism is viable across a wide range of wall velocities in the range of $0.01 \sim 1$ times the speed of light, with the limitation that the relativistic gamma factor must be limited to prevent the walls from becoming transparent. Disallowing a mini-inflation inside the bubble, on the one hand, and requiring that the bubble walls are not stopped by the friction caused by the overdensities, on the other hand, imposes lower and upper limits on the latent heat of the PT. Given a large enough initial bubble size, we find that no fine-tuning is required.

Our mechanism is not bound to any particular temperature scale and could correspondingly form PBHs of any mass. The requirement of a large order parameter is fulfilled by supercooled PTs, which usually proceed at a rate not much faster than Hubble. Slow transitions result in an increased probability of BH formation per Hubble volume, which in turn allows for the production of sizable PBH densities. We discussed experimental and theoretical constraints that apply to different BH mass regimes and highlighted the regions where the PBH population could account for all DM or contribute to the emergence of the Universe's LSS.

For a future project we could envision to enhance our numerical simulation by additional physical effects. One could, for instance, allow for a varying wall velocity in response to the friction induced by the overdense plasma. Moreover, particle scattering could be included to check if the dissipation of angular momentum would be sufficient to already let the arising Jeans instabilities collapse (in case of a realization of the mechanism with a scalar particle, to also avoid degeneracy pressure). Also some of the other simplifying assumptions could be dropped, for instance by evaluating the mechanism during matter domination or by incorporating the Hubble expansion during the PT. It would also be interesting to relate the properties of the PT to the actual model parameters and to study the underlying dark scalar (or confining) sector in more detail.

# 6 Gravitational Waves from Hidden Sectors

*This chapter is based on the publication [4] of the author and his collaborators. In this project, the author derived all details of the toy models, implemented and conducted the numerical analysis of the scalar potential and phase transition dynamics, and determined the experimental gravitational-wave sensitivity. All results were crosschecked by an independent implementation of the collaborator EM. The author produced all figures that appear in the following sections.*

## 6.1 Introduction

Among all dark matter (DM) candidates, the hypothetical weakly interacting massive particle (WIMP) with masses in the region around 100 GeV is probably the most prominent one. However, so far, extensive searches of various kinds have not lead to a discovery. As a consequence, alternative explanations are increasingly gaining attention. In Chapter 4 we reviewed a mechanism that sets the DM abundance via a phase transition (PT), and in Chapter 5 we discussed the possibility that primordial black holes could (partly) constitute DM. Both of these considerations are not bound to the WIMP scale at all. The present chapter will focus on the possibility of non-standard thermal histories of dark sectors accompanied by dark PTs that leave gravitational-wave (GW) imprints. More specifically, we will consider generic hidden[1] sectors that

1. feature sub-MeV PT dynamics and particle masses and

2. are thermally decoupled from the visible sector.

We take into account the complications of such scenarios with regard to cosmological observations and assess the detection prospects of GWs that are emitted during bubble collisions caused by a dark first-order PT. In the worst case scenario, DM couples to the visible part of the Universe exclusively via gravity. In that case, the observation of stochastic GWs might be the only remaining possibility to directly probe a hidden sector.

In contrary to the PTs we focused on in Chapters 4 and 5 – with large order parameters – we will now consider a more ordinary type of first-order PTs where the hidden sector masses are naturally of the same order as the nucleation temperature. We will not expound any specific DM mechanism, but in a more general sense investigate different thermal scenarios that may occur at the time of sub-MeV photon temperatures. Note that thermal freeze-out scenarios with DM masses below $\sim$ GeV require additional mediators to not overclose the Universe [94, 95].

---

[1] We use "dark" and "hidden" interchangeably.

The reasons for our focus on sub-MeV scales are manifold: Firstly, current direct detection experiments, depending on the technique, reach DM masses as low as $\sim$ GeV or even $\sim$ MeV [386], but are unable to constrain DM lighter than that [387]. Methods for the exploration of the sub-MeV regime via direct detection have been developed only quite recently [388, 389]. Secondly, the GW spectra produced during first-order PTs are more likely to be detectable if they occur at a later stage during the expansion of the Universe, i.e. at lower temperatures. This is, because the energy released in the form of gravitational radiation relative to the total energy density becomes larger, the fewer particles of the Standard Model of Particle Physics (SM) are relativistic. GW signals produced at the sub-MeV scale, redshifted until today, exhibit frequencies below $10^{-7}$ Hz and can thus be probed by pulsar timing arrays (PTAs).

If the DM abundance is set thermally – i.e. not via a mechanism such as the earlier discussed Filtered DM – then the freeze-out occurs naturally shortly after the temperature of the thermal bath has dropped below the DM mass [69]. However, any relativistic particle content additional to the SM is highly constrained at sub-MeV temperatures by the observed abundances of light elements produced during Big Bang nucleosynthesis (BBN) and measurements of the cosmic microwave background (CMB). Therefore, this regime is usually not given much attention. To reconcile additional relativistic degrees of freedom (DOFs) with the stringent cosmological bounds, we consider a hidden sector that is *thermally decoupled* from the SM particle content. This is a well motivated possibility, considering the lack of direct detection evidence, which hints at a very small or even no coupling between DM and the SM at all. In addition to the hidden sector being decoupled, we assume that it is *colder* than the visible sector by a certain degree. This corresponds to a reduced amount of energy in the hidden sector and possibly evades the BBN and CMB constraints. At the same time, any GW signal emitted in a cold hidden sector is less energetic compared to the total energy in the Universe and thus harder to detect.

This chapter is structured as follows: In Section 6.2, we derive the implications of cold and light hidden sectors for cosmological observations and for the expected GW spectrum. We apply these general findings to specific toy models in Section 6.3, where we point out the regions of parameter space that are reconcilable with the cosmological constraints at sub-MeV scales and, at the same time, can be probed by future GW observatories. We summarize our findings in Section 6.4.

## 6.2 Cosmology of Cold and Light Hidden Sectors

Relativistic DOFs in sub-MeV hidden sectors are constrained by cosmological observations. Two phenomena are highly sensitive to the ratio between the total relativistic energy density $\rho_{\rm rad}$, driving the Hubble expansion, and the energy density of the relativistic thermal bath of SM particles $\rho_\gamma$, controlling the timing of freeze-out processes and recombination:

1. **Production of light element abundances during BBN** starting at temperatures around $T_\gamma \sim 1\,\text{MeV}$. A modified expansion rate leads to an earlier or delayed freeze-out of neutron–proton interactions. This affects the final helium abundance relative to the total nucleon abundance, which is constrained by observations [390–394].

2. **Anisotropies and polarization imprinted in the CMB at the time of recombination** at $T_\gamma \sim 0.3\,\text{eV}$. An alteration of the expansion rate impacts the photon diffusion scale and changes the timings of matter–radiation-equality and photon decoupling. These effects would be visible in the acoustic peaks and in the tail of the CMB power spectrum [196, 395].

A hidden sector may alter the ratio $\rho_\text{rad}/\rho_\gamma$ in two ways. Firstly, if the temperature of the Universe drops below the mass of a hidden species *after* $\nu$-decoupling at $T_\gamma \sim 1\,\text{MeV}$, the annihilation of dark sector particles could heat up either the photons or the neutrinos exclusively. Secondly, a light hidden species that stays relativistic ("dark radiation") could directly contribute to $\rho_\text{rad}$.

The ratio $\rho_\text{rad}/\rho_\gamma$ is usually parameterized in terms of the *effective number of neutrino species*, defined as

$$N_\text{eff} \equiv \frac{8}{7}\left(\frac{\rho_\text{rad}}{\rho_\gamma} - 1\right)\left(\frac{11}{4}\right)^\frac{4}{3}, \tag{6.1}$$

which is a meaningful quantity after both $\nu$-decoupling and $e^\pm$-annihilation have occurred, i.e. for $T_\gamma \lesssim 511\,\text{keV}$. In a scenario with the SM alone, the relativistic energy density is given by

$$\rho_\text{rad}^\text{SM} = \rho_\nu + \rho_\gamma = \left(\frac{7}{8}N_\nu(\xi_\nu^\text{SM})^4 + 1\right)\rho_\gamma, \tag{6.2}$$

where $N_\nu = 3$ is the number of SM neutrino species and $\xi_\nu \equiv T_\nu/T_\gamma$ is the temperature ratio between neutrinos and photons. The $e^\pm$-annihilation heats up the photons and results in the ratio $\xi_\nu^\text{SM} \approx (4/11)^{1/3}$, as dictated by the conservation of comoving entropy. The exact value is slightly larger due to fact that $\nu$-decoupling is not entirely complete at the time of $e^\pm$-annihilation.[2] Taking this effect into account, the theoretical effective number of neutrino species in the SM amounts to [396]

$$N_\text{eff}^\text{SM} = 3.046. \tag{6.3}$$

The observed light element abundances generated during BBN suggest the value [36]

$$N_\text{eff}^\text{BBN} = 2.95^{+0.56}_{-0.52}, \tag{6.4}$$

with 95% confidence level (CL) uncertainties and assuming a constant $N_\text{eff}$ during BBN [392, 393, 397, 398] (see Ref. [399] for the impact of relaxing this assumption). Complementary constraints are obtained from 2018 *Planck* data. Combining CMB anisotropies and polarization, CMB lensing effects and baryon acoustic oscillations (BAO) determined from galaxy surveys yields the value [36]

$$N_\text{eff}^\text{CMB} = 2.99^{+0.34}_{-0.33}, \tag{6.5}$$

which applies at the time of recombination at $T_\gamma \sim 0.3\,\text{eV}$. The above value is obtained from a Lambda-cold-dark-matter Theory (ΛCDM) fit which, among many other parameters, also yields the value of today's Hubble constant, $H_0 = (67.7 \pm 0.4)\,\text{km/s/Mpc}$ at 68% CL. This value is currently in tension with low redshift measurements reporting $H_0 = (73.5 \pm 1.7)\,\text{km/s/Mpc}$ [400]. Therefore, Eq. (6.5) might represent a too stringent upper bound. Incorporating the low-redshift value of $H_0$ in the ΛCDM fit yields [36]

$$N_{\text{eff}}^{\text{CMB}+H_0} = 3.27 \pm 0.30\,. \tag{6.6}$$

---

[2]Note that the partial reheating of the neutrinos leads to a slight deviation from a thermal distribution [396]. Therefore, strictly speaking, $T_\nu$ and $\xi_\nu$ are not well defined. We will tolerate this inaccuracy and nevertheless use these quantities in our calculations.

In light of these observational constraints, the accommodation of additional light DOFs is only possible if they are less energetic, i.e. colder than the SM thermal bath. We quantify this by defining

$$\xi_h \equiv \frac{T_h}{T_\gamma}\,, \tag{6.7}$$

where $T_h$ and $T_\gamma$ are the hidden and visible sector (photon) temperatures, respectively. One possible explanation for a temperature difference could be that the two sectors were never in thermal equilibrium and during the epoch of reheating, the inflaton decayed preferentially into one of the two sectors. Another possibility is that the two sectors were initially in thermal contact but decoupled early on. Whenever a thermalized species drops out of equilibrium, it heats up the remaining plasma of its sector. Consequentially, if one sector contains more heavy species than the other, $\xi_h$ will deviate from unity. We are particularly interested in the case $\xi_h < 1$, which possibly evades the aforementioned cosmological constraints.

### 6.2.1 Hidden Sector Scenarios

We will now study the cosmological implications of different thermal scenarios for generic hidden sectors with $g_h$ effective DOFs.[3] We begin by showing that light hidden sectors with $\xi_h = 1$ are experimentally excluded and then move on to scenarios where $\xi_h < 1$ is assumed.

**Thermal contact with the visible sector.** If the hidden sector is in thermal contact with the SM particle content, then $\xi_h = T_\gamma/T_h = 1$. We consider two possible scenarios with two stages each, illustrated in Fig. 6.1:

(A1) hidden sector in thermal contact with photons,

(A2) hidden sector has reheated photons (after $\nu$-decoupling),

(B1) hidden sector in thermal contact with neutrinos,

(B2) hidden sector has reheated neutrinos (after $\nu$-decoupling).

---

[3]"Effective" here means that each bosonic (fermionic) DOF contributes 1 (7/8) [196].

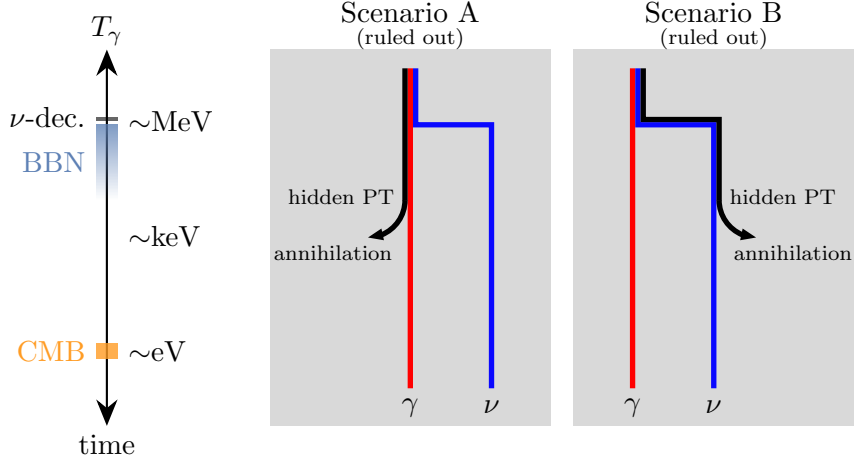

**Figure 6.1:** Thermal evolution of a hidden sector (black path) initially in thermal contact with (A) the photons (red path) or (B) the neutrinos (blue path). As the temperature drops below the mass of hidden sector particles, the latter annihilate or decay and reheat either the photons or the neutrinos. Assuming that the hidden sector annihilates after $\nu$-decoupling, both scenarios are ruled out by experimental constraints on $N_{\text{eff}}$.

To determine $N_{\text{eff}}$ as a function of $g_h$, we start with the relativistic energy density

$$\text{(A1)} \quad \rho_{\text{rad}} = \rho_\nu + \rho_{\gamma+h} = \left(\frac{7}{8} N_\nu \xi_\nu^4 + 1 + \frac{g_h}{g_\gamma}\right)\rho_\gamma \,,$$

$$\text{(B1)} \quad \rho_{\text{rad}} = \rho_{\nu+h} + \rho_\gamma = \left(\left(\frac{7}{8} N_\nu + \frac{g_h}{g_\gamma}\right)(\xi_{\nu+h})^4 + 1\right)\rho_\gamma \,,$$

$$\text{(A2), (B2)} \quad \rho_{\text{rad}} = \rho_\nu + \rho_\gamma = \left(\frac{7}{8} N_\nu \xi_\nu^4 + 1\right)\rho_\gamma \,, \tag{6.8}$$

where, for instance, $\rho_{\nu+h}$ refers to the energy density of the thermal bath including the neutrinos and the hidden sector (which share a common temperature $T_{\nu+h} = \xi_{\nu+h} T_\gamma$). The neutrino–photon temperature ratio after $e^\pm$-annihilation is determined by conservation of comoving entropy and amounts to

$$\text{(A1)} \quad \xi_\nu = \left(\frac{g_\gamma + g_h}{g_\gamma + g_e + g_h}\right)^{\frac{1}{3}} \left(\frac{N_{\text{eff}}^{\text{SM}}}{N_\nu}\right)^{\frac{1}{4}},$$

$$\text{(A2)} \quad \xi_\nu = \left(\frac{g_\gamma}{g_\gamma + g_e + g_h}\right)^{\frac{1}{3}} \left(\frac{N_{\text{eff}}^{\text{SM}}}{N_\nu}\right)^{\frac{1}{4}},$$

$$\text{(B1)} \quad \xi_{\nu+h} = \left(\frac{g_\gamma}{g_\gamma + g_e}\right)^{\frac{1}{3}} \left(\frac{N_{\text{eff}}^{\text{SM}}}{N_\nu}\right)^{\frac{1}{4}},$$

$$\text{(B2)} \quad \xi_\nu = \left(\frac{g_\gamma}{g_\gamma + g_e} \frac{g_\nu + g_h}{g_\nu}\right)^{\frac{1}{3}} \left(\frac{N_{\text{eff}}^{\text{SM}}}{N_\nu}\right)^{\frac{1}{4}}, \tag{6.9}$$

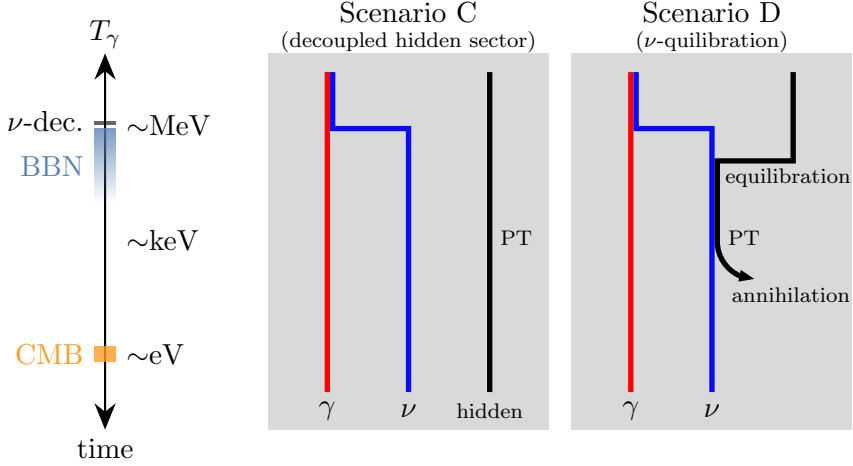

**Figure 6.2:** Thermal evolution of a hidden sector (black path) which (C) is entirely decoupled from the visible sector or which (D) temporarily equilibrates with the neutrinos (blue path). Depending on the number of DOFs in the hidden sector, $g_h$, and on the initial hidden–visible sector temperature ratio, these scenarios can evade the experimental constraints on $N_{\text{eff}}$. Note that the fading of the BBN marker indicates that, while nucleosynthesis is technically ongoing until $T_\gamma \sim 10\,\text{keV}$, the light element abundances are roughly set already earlier [401].

with $N_\nu = 3$, $g_\nu = 2 \times 7/8 N_\nu$, $g_\gamma = 2$, and $g_e = 4 \times 7/8$. The occurrence of the ratio $N_{\text{eff}}^{\text{SM}}/N_\nu$ accounts for the partial reheating of the neutrinos during $e^\pm$-annihilation.[4] Plugging the derived expressions for $\rho_{\text{rad}}$ into Eq. (6.1) yields

$$\text{(A1)} \qquad N_{\text{eff}} = \left(\frac{11}{4}\right)^{\frac{4}{3}} \left( N_{\text{eff}}^{\text{SM}} \left(\frac{4 + 2g_h}{11 + 2g_h}\right)^{\frac{4}{3}} + \frac{2}{7} g_h \right),$$

$$\text{(A2)} \qquad N_{\text{eff}} = N_{\text{eff}}^{\text{SM}} \left(\frac{11}{11 + 2g_h}\right)^{\frac{4}{3}},$$

$$\text{(B1)} \qquad N_{\text{eff}} = N_{\text{eff}}^{\text{SM}} \left(1 + \frac{4}{21} g_h\right),$$

$$\text{(B2)} \qquad N_{\text{eff}} = N_{\text{eff}}^{\text{SM}} \left(1 + \frac{4}{21} g_h\right)^{\frac{4}{3}}.$$

In all scenarios, we obtain $N_{\text{eff}} = N_{\text{eff}}^{\text{SM}}$ for $g_h = 0$, as expected. Furthermore, it turns out that even a single bosonic DOF in the hidden sector ($g_h = 1$) is ruled out by the BBN and CMB bounds given in Eqs. (6.4) to (6.6). For this reason, thermalized hidden sectors below the MeV scale are usually deemed impossible.

---

[4]The partial reheating itself is also affected by additional DOFs, but we neglect this higher-order effect.

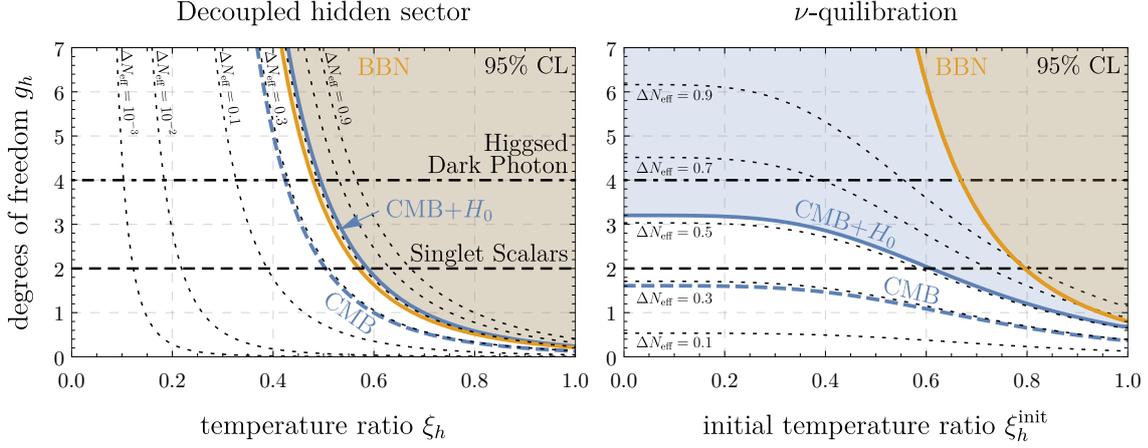

**Figure 6.3:** Upper bounds on the number of effective relativistic DOFs in the hidden sector, $g_h$, as a function of the (initial) hidden–visible sector temperature ratio, $\xi_h$ or $\xi_h^{\rm init}$, under the assumption of different allowed deviations of the effective number of neutrino species from the SM value (black dashed curves). The orange and blue curves refer to the experimental upper bounds on $N_{\rm eff}$ given in Eqs. (6.4) to (6.6). The left panel applies to the case of a hidden sector that is never in thermal equilibrium with SM particles (or has decoupled very early). The right panel is for the $\nu$-quilibration scenario, where the hidden sector (re-)couples with the neutrinos after the latter have decoupled from the photons. The horizontal lines indicate the number of DOFs in our two exemplary toy models presented in Section 6.3.

**Fully decoupled.** Let us now investigate the scenario illustrated in the left panel of Fig. 6.2, where a hidden sector was never in thermal equilibrium with the visible sector (or decoupled very early), and the two sectors have different temperatures, $\xi_h \equiv T_h/T_\gamma \neq 1$. In this case, the relativistic energy density amounts to

$$(\text{C}) \qquad \rho_{\rm rad} = \rho_\nu + \rho_\gamma + \rho_h = \left(\frac{7}{8}N_\nu(\xi_\nu^{\rm SM})^4 + 1 + \frac{g_h}{g_\gamma}\xi_h^4\right)\rho_\gamma\,. \qquad (6.10)$$

Plugged into Eq. (6.1) this gives

$$(\text{C}) \qquad N_{\rm eff} = N_{\rm eff}^{\rm SM} + \frac{4}{7}\left(\frac{11}{4}\right)^{\frac{4}{3}}g_h\xi_h^4\,, \qquad (6.11)$$

which relaxes to $N_{\rm eff}^{\rm SM}$ for $g_h = 0$ or in the limit $\xi_h \to 0$, as expected. The left panel of Fig. 6.3 displays the allowed values for $g_h$ as a function of $\xi_h$ for different upper bounds on $N_{\rm eff}$. The orange and blue regions are excluded at 95% CL by the discussed BBN and CMB+$H_0$ constraints, respectively. We see that hidden sectors with more DOFs (larger $g_h$) must be colder (smaller $\xi_h$) in order to satisfy the constraints on the additional relativistic energy density. For instance, a hidden sector with $g_h = 1$ must have a temperature smaller than $\sim 0.7$ times the photon temperature.

Note that such a scenario only works if the decoupled hidden sector may decay into some kind of "dark radiation" as soon as it becomes non-relativistic. Otherwise, the total matter density would be increased and the Universe overclosed.

**Temporary equilibration with neutrinos ($\nu$-quilibration).** As wee have seen, a hidden sector that becomes massive and reheats the neutrinos after $\nu$-decoupling is ruled out, because $N_{\text{eff}}$ would be increased in a noticeable way. However, there is the possibility that the hidden sector is initially decoupled from the neutrinos, then – while still relativistic – equilibrates with the neutrinos after $\nu$-decoupling, before it finally becomes massive and reheats the neutrinos again [402, 403]. If the hidden sector is initially colder than the neutrinos, $\xi_h^{\text{init}} < 0$, the equilibration cools the neutrinos in comparison to the photons. When the hidden sector then becomes non-relativistic at some later time, it annihilates and heats up the neutrinos again, but – thanks to the cooling effect of the earlier equilibration – the resulting $N_{\text{eff}}$ can still be in agreement with cosmological constraints. We call this scenario "$\nu$-quilibration ".

The scenario is illustrated in the right panel of Fig. 6.2 and consists of two stages:

(D1) hidden sector has equilibrated with neutrinos (after $\nu$-decoupling),

(D2) hidden sector has reheated neutrinos.

To determine $N_{\text{eff}}$, we again start with the relativistic energy density

$$\text{(D1)} \qquad \rho_{\text{rad}} = \rho_{\nu+h} + \rho_\gamma = \left(\left(\frac{7}{8}N_\nu + \frac{g_h}{g_\gamma}\right)(\xi_{\nu+h})^4 + 1\right)\rho_\gamma,$$

$$\text{(D2)} \qquad \rho_{\text{rad}} = \rho_\nu + \rho_\gamma = \left(\frac{7}{8}N_\nu \xi_\nu^4 + 1\right)\rho_\gamma. \tag{6.12}$$

Note that these expressions are identical to those of the excluded scenarios (B1) and (B2) given in Eq. (6.8). However, the neutrino–photon temperature ratio is now different and amounts to

$$\text{(D1)} \qquad \xi_{\nu+h} = \left(\frac{g_\nu + (\xi_h^{\text{init}})^4 g_h}{g_\nu + g_h}\right)^{\frac{1}{4}} \left(\frac{g_\gamma}{g_\gamma + g_e}\right)^{\frac{1}{3}} \left(\frac{N_{\text{eff}}^{\text{SM}}}{N_\nu}\right)^{\frac{1}{4}},$$

$$\text{(D2)} \qquad \xi_\nu = \xi_{\nu+h}\left(\frac{g_\nu + g_h}{g_\nu}\right)^{\frac{1}{3}}, \tag{6.13}$$

which is derived by considering comoving energy (entropy) conservation during equilibration (annihilation). $\xi_h^{\text{init}}$ denotes the hidden–visible sector temperature ratio, before the two sectors equilibrate and is evaluated prior to $e^\pm$-annihilation. The cooling effect that arises for $\xi_h^{\text{init}} < 1$, compared to Eq. (6.9), manifest itself in the first factor of $\xi_{\nu+h}$.

Finally, plugging the energy density into Eq. (6.1) yields

(D1) $$N_{\text{eff}} = N_{\text{eff}}^{\text{SM}}\left(1 + \frac{4}{21}(\xi_h^{\text{init}})^4 g_h\right),$$

(D2) $$N_{\text{eff}} = N_{\text{eff}}^{\text{SM}}\left(1 + \frac{4}{21}(\xi_h^{\text{init}})^4 g_h\right)\left(1 + \frac{4}{21}g_h\right)^{\frac{1}{3}}. \qquad (6.14)$$

If the hidden sector equilibrates between $\nu$-decoupling and the end of BBN[5] and becomes massive between BBN and recombination, i.e. as illustrated in the right panel of Fig. 6.2, then the BBN bounds on $N_{\text{eff}}$ apply to stage (D1) while the CMB bounds constrain stage (D2). The right panel of Fig. 6.3 displays the allowed values for $g_h$ as a function of $\xi_h^{\text{init}}$. The orange and blue regions are excluded at 95% CL by the BBN and CMB+$H_0$ constraints, respectively. As can be gathered from the plot, a model with $g_h = 2$, for instance the Singlet Scalars model we will discuss, is allowed if $\xi_h^{\text{init}} \lesssim 0.6$.

As we will see in the following section, GWs produced in cold hidden sectors are suppressed. Interestingly, the amount of suppression is determined by the temperature ratio *at the time of the PT*, which in our case is given by $\xi_{\nu+h}$ in Eq. (6.13). This quantity turns out to take values not far from one, even if $\xi_h^{\text{init}} \ll 1$. The $N_{\text{eff}}$ constraints, on the other hand, strongly depend on $\xi_h^{\text{init}}$. This makes our $\nu$-quilibration scenario particularly promising: Cosmological constraints can be avoided easily by choosing a very small $\xi_h^{\text{init}}$, while still preserving a sizable GW signal.

### 6.2.2 Gravitational Waves

As pointed out in the last section, hidden sectors that remain relativistic until the SM thermal bath has reached sub-MeV temperatures must be colder and decoupled to not violate constraints from cosmological observations. Probing decoupled sectors is extremely challenging if not impossible using conventional astrophysical or laboratory methods. Fortunately, gravity couples universally to any kind of matter. If the hidden sector undergoes a first-order PT, the collisions of true-vacuum bubbles produce gravitational radiation that might be detectable today as part of the stochastic GW background. The physics of first-order PTs and the accompanying gravitational radiation was reviewed in detail in Chapter 3. If the PT occurs in a hidden sector that is colder than the photons, the resulting GW spectrum differs from the one described by the standard formalism. In the following, we will discuss how the PT parameters that determine the GW spectrum must be modified in this case.

---

[5] While nucleosynthesis is technically ongoing until $T_\gamma \sim 10\,\text{keV}$, the light element abundances are roughly set already somewhat earlier [401]. An equilibration during late stages of BBN would thus loosen the constraint on $N_{\text{eff}}$, making our results conservative in that regard.

**Implications of a cold hidden sector.** The first modification affects the total effective number of relativistic energy and entropy DOFs[6]

$$\begin{aligned} g_\star &= g_\star^{\text{SM}} + g_h \xi_h^4 \,, \\ g_{\star s} &= g_{\star s}^{\text{SM}} + g_h \xi_h^3 \,, \\ g_{\star s}^0 &= g_{\star s}^{\text{SM},0} + g_h (\xi_h^0)^3 \,, \end{aligned} \qquad (6.15)$$

which now include $g_h$ additional DOFs at a temperature $T_h = \xi_h T_\gamma$. The label "0" refers to values today – at the time of possible GW observation – while all other quantities correspond to the time of the PT. This modification alters the amplitude and frequency redshift of the GW spectrum, according to Eqs. (3.23) and (3.25).

The strength parameter $\alpha$ – originally defined in Eq. (3.18) – is also affected. It represents the latent heat of the PT, $\varepsilon$, normalized w.r.t. the total relativistic energy density at the time of the PT. In a scenario with two temperatures, this parameter reads

$$\begin{aligned} \alpha &= \frac{\varepsilon}{\frac{\pi^2}{30} g_\star (T_{\gamma,n})^4} \\ &= \frac{\varepsilon}{\frac{\pi^2}{30} (g_\star^{\text{SM}} \xi_h^{-4} + g_h)(T_{h,n})^4} \,, \end{aligned} \qquad (6.16)$$

where $T_{\gamma,n}$ and $T_{h,n}$ are the temperatures of the visible and hidden sectors at the time of bubble nucleation and $g_\star^{\text{SM}}$ represents the number of effective DOFs in the SM. For $\xi_h \ll 1$ and fixed $T_{h,n}$, we observe the scaling

$$\alpha \propto \xi_h^4 \,, \qquad (6.17)$$

which indicates a *suppression of the GW spectrum* produced in a colder hidden sector. This can be understood intuitively: The energy budget of the PT, which is naturally related to $T_h$, has to compete against the energy in the visible sector, which is set by $T_\gamma > T_h$. We apply the same modification to the formula for the runaway threshold $\alpha_{\text{run}}$, defined in Eq. (3.20). The energy conversion efficiency factors $\kappa_\phi$, $\kappa_{\text{sw}}$, and $\kappa_{\text{turb}}$ are functions of $\alpha$ and $\alpha_{\text{run}}$ and thus depend on $\xi_h$ as well.[7]

The temperature ratio has another minor effect via the temperature dependence of $g_\star^{\text{SM}}$ and $g_{\star s}^{\text{SM}}$. Taking this into account leads to a further suppression of the GW spectrum in case of a colder hidden sector: Smaller $\xi_h$ (assuming fixed $T_h$) imply larger $T_\gamma$ and thus potentially more relativistic SM DOFs that compete with the GW energy budget.

Another important parameter that determines the GW spectrum is $\beta$ – originally defined in Eq. (3.19) – the inverse time scale of the PT, usually expressed in terms of the Hubble rate, $H$. This quantity characterizes the rate of change of the tunneling action that underlies the PT. Since the visible sector is not involved in the hidden PT, $\beta$ is independent of $\xi_h$ (for fixed $T_h$).

---

[6] These formulas are not exact in case of the $\nu$-quilibration scenario, where the neutrino temperature slightly differs from the SM value. However, the error is small compared to the effect of $\xi_h \ll 1$.

[7] We calculate $\kappa_{\text{sw}}$ based on $\alpha$, the latent heat normalized to the *total* energy in the Universe. However, after finalizing our work, we realized that the sound-wave efficiency is probably rather sensitive to the latent heat normalized to the *hidden* energy density. This makes our GW sensitivity projections conservative.

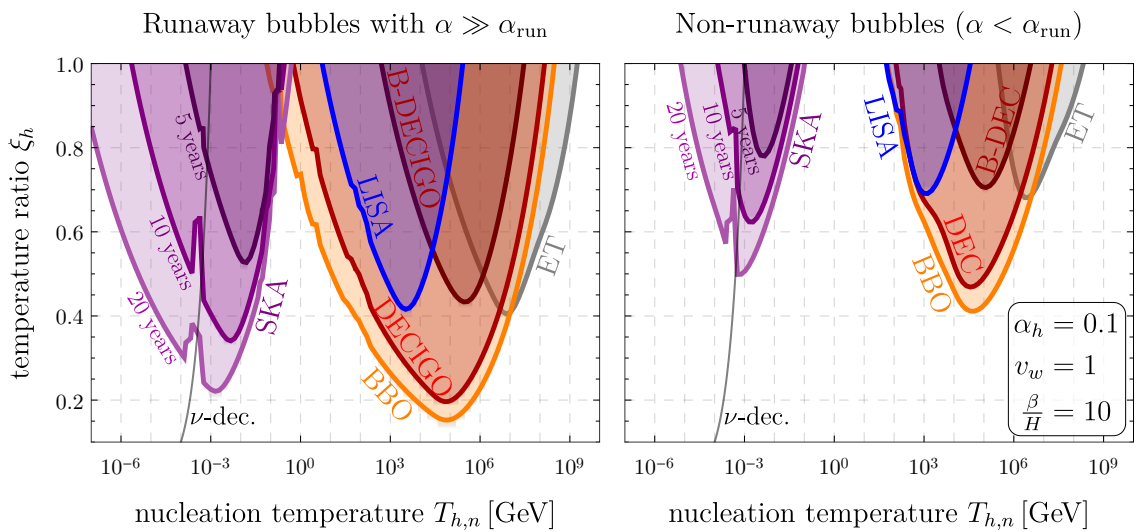

**Figure 6.4:** Anticipated sensitivity of various future GW observatories to a hidden sector PT. In the left panel we assume bubbles that are deep in the runaway regime, while the right panel shows the case of non-runaway bubbles. We show the sensitivities as functions of the hidden sector nucleation temperature, $T_{h,n}$, and the hidden–visible sector temperature ratio, $\xi_h$. For the sake of generality, we assume $g_h \ll g_\star$. Figure 6.5 displays the sensitivity if this assumption is relaxed. The discontinuities (best visible for *SKA*) originate from our step function approximations of $g_\star^{\rm SM}$ and $g_{\star s}^{\rm SM}$ [206].

**Experimental sensitivity.** To illustrate the impact of a colder hidden sector, $\xi_h < 0$, on the anticipated experimental sensitivity to GWs from hidden sector PTs, in Fig. 6.4 we plot the regions where the signal-to-noise ratio (SNR) exceeds the experimental thresholds in the $T_{h,n}$–$\xi_h$-plane. This figure is created in analogy to Fig. 3.4, where we plotted the sensitivity in the $T_n$–$\alpha$- and $T_n$–$\beta$-planes. The exemplary values chosen for $\alpha_h \equiv \alpha(\xi_h = 1)$ and $\beta$ are indicated in the figure. The plot illustrates the $\alpha \propto \xi_h^4$ suppression for $\xi_h \ll 1$. The PTA *SKA* turns out to be sensitive to PTs occuring at $\sim$ MeV temperatures. Note that the discussed cosmological constraints on $N_{\rm eff}$ apply only to PTs occurring after $\nu$-decoupling, i.e. for $T_{\gamma,n} \lesssim 1$ MeV. For completeness, we show a wide temperature range beyond our main region of interest.

Figure 6.5 displays the sensitivity of *SKA* to GWs from a PT in a fully decoupled hidden sector, assuming that the number of relativistic DOFs in the hidden sector, $g_h$, saturates the $N_{\rm eff}$ constraints. As can be inferred from Eq. (6.11), this implies the correspondence

$$\xi_h(g_h) = \left(\frac{4}{11}\right)^{\frac{1}{3}} \left(\frac{7}{4} \frac{N_{\rm eff}^{\max} - N_{\rm eff}^{\rm SM}}{g_h}\right)^{\frac{1}{4}}, \qquad (6.18)$$

which is reflected by the double vertical axis in the figure. In the fully decoupled scenario,

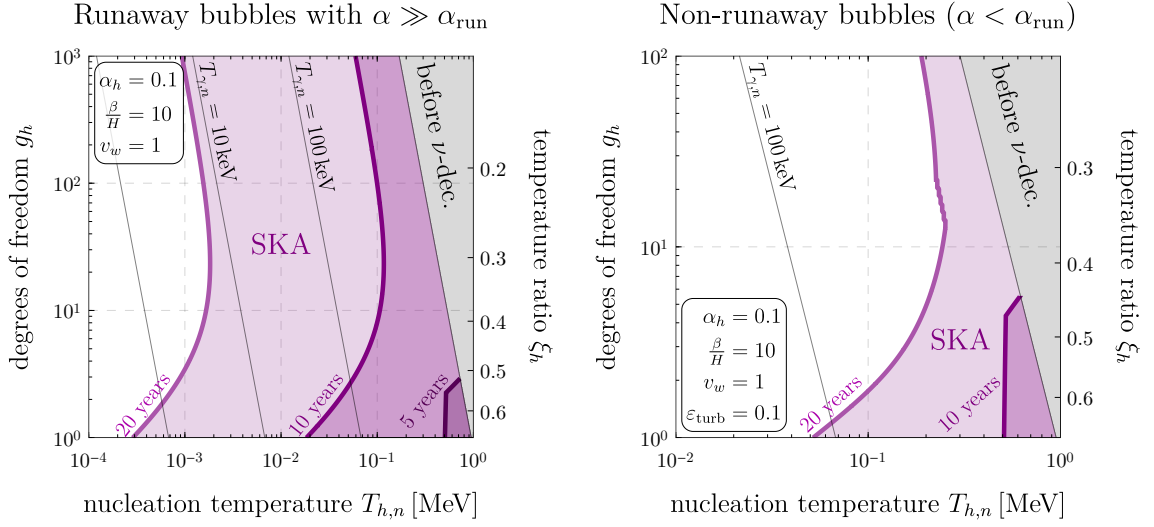

**Figure 6.5:** Sensitivity of *SKA* to PTs in fully decoupled hidden sectors, assuming that the number of relativistic DOFs saturates the $N_\text{eff}$ constraints at a given temperature ratio. The gray shaded region corresponds to PTs occurring before $\nu$-decoupling, where no cosmological constraints apply.

the BBN value of $N_\text{eff}$ yields the tightest constraint with $N_\text{eff}^\text{max} = 3.51$. We see that in the case of runaway bubbles (left panel), *SKA* will be sensitive to an $\mathcal{O}(1)$ number of hidden DOFs already after 5 years of observation time. Discovery prospects are best for PTs that occur early, i.e. right after $\nu$-decoupling. For non-runaway bubbles (right panel), at least 10 years of observation time are required to probe models with an order $\mathcal{O}(1)$ number of DOFs. In either case, hidden sectors with $g_h \gg 1$ will be accessible after 20 years of observation. The shape of the sensitivity regions in Fig. 6.5 can be understood by noting that, according to Eq. (6.18), saturating the $N_\text{eff}$ constraints implies $\xi_h \propto g_h^{-1/4}$. Small $\xi_h$ thus correspond to large $g_h$, which limits the suppression of $\alpha$. As a consequence, the GW spectrum does not become arbitrarily weak and the sensitivity extends towards $\xi_h = 0$. Similar plots for the $\nu$-quilibration scenario are impossible to produce due to the more complex relations that determine $N_\text{eff}$.

Note that the sensitivity of *SKA* in Figs. 6.4 and 6.5 will only be reached if the expected background from supermassive black hole binaries (SMBHBs) can be resolved.

## 6.3 Toy Models

We will now become more specific and investigate two minimal hidden sector toy models. Both of these models can give rise to first-order PTs during the time of sub-MeV photon temperatures and are therefore testable via PTAs. In order to allow for a hidden sector that is colder than the SM sector, as explained in the previous sections, we have to prohibit thermal equilibrium between the two. This is achieved by choosing sufficiently small portal couplings.[8] As a consequence, the PT dynamics are unaffected by the SM and we can treat our toy models as entirely sequestered. In light of the discussed stringent $N_\mathrm{eff}$ bounds, the number of hidden sector DOFs is highly constrained. Therefore, any UV-complete theory with non-trivial dynamics at scales $\lesssim$ MeV should effectively reduce to a model similar to one of our minimal toy models.[9]

### 6.3.1 Singlet Scalars

We start with a model that contains $g_h = 2$ DOFs, consisting of two scalar SM singlets: $\phi$, which will acquire a vacuum expectation value (VEV) and thereby give rise to a PT, and $\tilde{\phi}$, an auxiliary field whose purpose will become clear momentarily. The tree-level potential of the model reads

$$V_\mathrm{tree}(\phi, \tilde{\phi}) = \frac{\mu_\phi^2}{2}\phi^2 + \frac{\kappa}{3}\phi^3 + \frac{\lambda_\phi}{4}\phi^4 + \frac{\mu_{\tilde{\phi}}^2}{2}\tilde{\phi}^2 + \frac{\lambda_{\tilde{\phi}}}{4}\tilde{\phi}^4 + \kappa_{\phi\tilde{\phi}}\phi\tilde{\phi}^2 + \frac{\lambda_{\phi\tilde{\phi}}}{2}\phi^2\tilde{\phi}^2 . \quad (6.19)$$

We impose a $\mathbb{Z}_2$ on $\tilde{\phi}$ to eliminate terms involving odd powers of $\tilde{\phi}$ that are unnecessary for our mechanism, we require $\mu_{\tilde{\phi}}^2, \kappa_{\phi\tilde{\phi}} \geq 0$ to avoid spontaneous breaking of the $\mathbb{Z}_2$ symmetry, and we demand $\lambda_\phi, \lambda_{\tilde{\phi}} > 0$ to ensure stability of the potential, i.e. $V_\mathrm{tree} \to \infty$ for $\phi, \tilde{\phi} \to \infty$. From now on, we suppress any dependence on the field $\tilde{\phi}$, as it never obtains a non-zero VEV.

If the tree-level potential has a minimum at $\langle\phi\rangle^\infty$, the model parameter $\mu_\phi$ can be rewritten based on $V'_\mathrm{tree}(\phi = \langle\phi\rangle^\infty) = 0$:

$$\mu_\phi^2 = -[\kappa + \lambda_\phi \langle\phi\rangle^\infty] \langle\phi\rangle^\infty . \quad (6.20)$$

Note that $\langle\phi\rangle^\infty$ is just an input parameter: Depending on other model parameters, the vacuum can remain stable or metastable at $\phi = 0$ and never undergo a PT. Appendix 6.A.1 lists further details about the toy model, such as the derived field-dependent and thermal masses as well as the one-loop finite counterterms.

**Requirements for a first-order transition.** In the following, we want to take a closer look at the PT dynamics in the Singlet Scalars model. To do so, we consider the effective potential including the tree-level potential, the one-loop zero-temperature Coleman–Weinberg and finite temperature potential, as well as the ring diagram (Daisy) contributions, which we all discussed in Section 3.1.1. A first-order PT can occur if the tree-level potential has a local minimum at $\phi = 0$ and a global one at $\phi = \langle\phi\rangle^\infty$. This situation is given if the model parameters fulfill

$$1 < \bar{\kappa} < \frac{3}{2} \quad \text{with} \quad \bar{\kappa} \equiv -\frac{\kappa}{\lambda_\phi \langle\phi\rangle^\infty} , \quad (6.21)$$

---

[8] Avoiding thermal equilibration until sub-MeV temperatures via the Higgs portal, for instance, requires a portal coupling $\lesssim 10^{-11}$.

[9] There also exists the class of conformal models, where strong transitions can be generated entirely from loop effects [172–191]. We do not consider this special limiting case here.

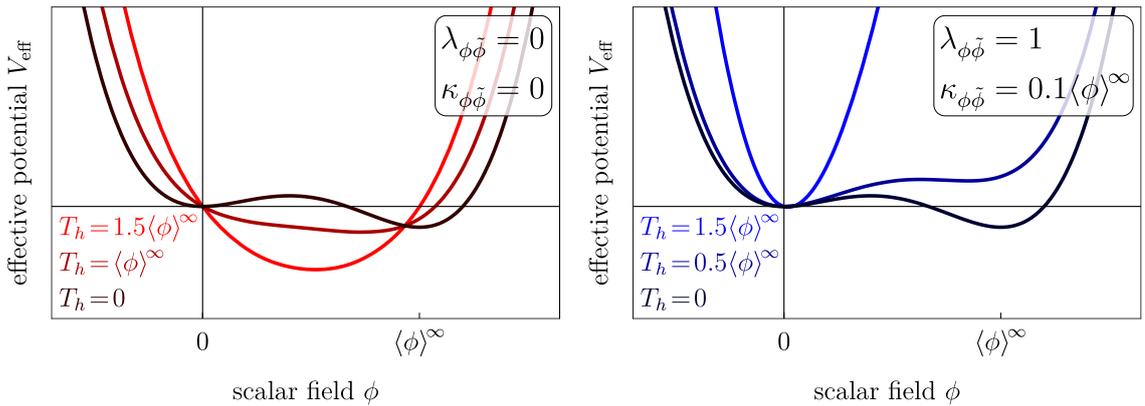

**Figure 6.6:** Behavior of the effective potential in the Singlet Scalars model without (left panel) and with (right panel) the auxiliary scalar field $\tilde{\phi}$. Only in the latter case does the vacuum temporarily occupy a local minimum from which it can transition to the global minimum via a first-order PT.

as can be easily verified. For $\bar{\kappa} < 1$, there is no minimum at $\phi = 0$, while for $\bar{\kappa} > 3/2$, the minimum at $\phi = 0$ is global. In the former case, the PT would be a higher-order transition, while in the latter case, a PT would never occur.

There is another prerequisite for a first-order PT: The vacuum must be trapped in the local minimum first, so that it can tunnel to the global one. To see how this can be achieved, first consider a version of the model without the auxiliary field $\tilde{\phi}$, which is equivalent to setting $\kappa_{\phi\tilde{\phi}} = \lambda_{\phi\tilde{\phi}} = 0$. At high temperatures, the effective potential is dominated by the thermal one-loop potential, which takes the form of a parabola with a minimum that is shifted away from the origin, towards the global minimum of the tree-level potential. Coming from high temperatures, the vacuum therefore never occupies the local minimum at $\phi = 0$ and no tunneling occurs, as illustrated in the left panel of Fig. 6.6.

The situation is different if we include the auxiliary field $\tilde{\phi}$. The thermal potential now contains terms proportional to $\lambda_{\phi\tilde{\phi}}\phi^2$ and $\kappa_{\phi\tilde{\phi}}\phi$, which shift the minimum of the high-temperature parabola towards smaller $\phi$. In this case, as the Universe cools down, the vacuum occupies the local minimum. This behavior is illustrated in the right panel of Fig. 6.6.

**Model parameter space.** To track the evolution of the effective potential and the scalar VEV numerically, we employ the Python package CosmoTransitions [128, 404–406]. This software consecutively calculates the tunneling action – see Eq. (3.12) – while slowly decreasing the hidden sector temperature, $T_h$, and signals when the bubble nucleation condition – Eq. (3.16) – is fulfilled, which determines the nucleation temperature, $T_{h,n}$. Note that in Chapters 4 and 5, we used a generic proxy function for the bubble wall profile. The CosmoTransitions package used for this analysis, on the other hand, determines an actual solution to the bounce equation, which is required to compute the tunneling action. We augmented the software to obtain the PT strength $\alpha$ and inverse time scale $\beta$ as well as derived quantities such as the GW efficiency factors as outputs. Figure 6.7 plots the obtained $\alpha$ and $\beta$ in the $\lambda_{\phi\tilde\phi}$–$\bar\kappa$-plane, assuming $\langle\phi\rangle^\infty = 50\,\mathrm{keV}$ (corresponding to $T_{h,n} \sim 50\,\mathrm{keV}$) together with $\xi_h = 0.66$ at the time of the PT. Note that the value $\xi_h \approx 0.66$ arises naturally in the $\nu$-quilibration scenario, independent of the initial $\xi_h^\mathrm{init}$.

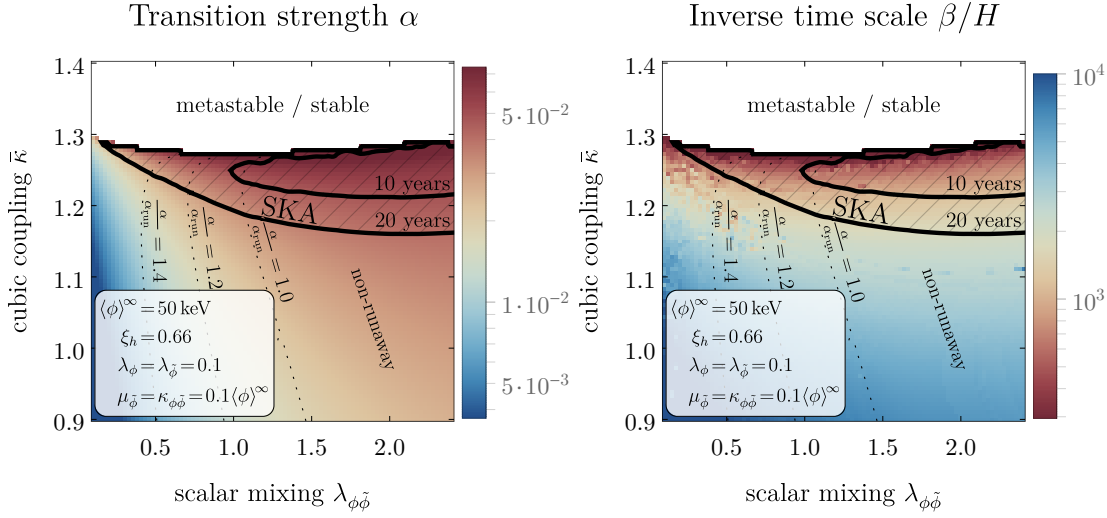

**Figure 6.7:** The strength $\alpha$ (left panel) and inverse time scale $\beta$ (right panel) of a hidden sector PT for the Singlet Scalars model defined in Eq. (6.19), assuming $\langle\phi\rangle^\infty = 50\,\mathrm{keV}$ (corresponding to $T_{h,n} \sim 50\,\mathrm{keV}$). The hatched areas mark the region where the GW signal would be detectable by *SKA* after the indicated periods of observation time. To not violate the cosmological constraints on $N_\mathrm{eff}$, we impose that the hidden sector is colder than the visible one by a factor of $\xi_h = 0.66$. Note that in the $\nu$-quilibration scenario, the temperature ratio at the time of the transition is naturally $\xi_h \approx 0.66$, independent of the initial $\xi_h^\mathrm{init}$.

The parameter $\bar\kappa$ controls the term $\propto \phi^3$ in the tree-level potential and thereby the size of the barrier that separates the true and false vacua. A larger barrier results in a slower PT, which explains the negative correlation between $\bar\kappa$ and $\beta$. Above a threshold at $\bar\kappa \approx 1.3$, the PT never occurs as the initial vacuum state becomes metastable or stable. This resembles our tree-level estimate, Eq. (6.21), which predicted a threshold at $\bar\kappa = 1.5$.

The parameter $\lambda_{\phi\tilde\phi}$ enters the effective potential via the thermal one-loop contribution. At high temperatures, the thermal potential is dominated by a term proportional to $\lambda_{\phi\tilde\phi} T_h^2 \phi^2$. This parabolic contribution is responsible for the stability of the $\langle\phi\rangle = 0$ phase prior to the PT. Larger values of $\lambda_{\phi\tilde\phi}$ thus delay the PT, which in turn corresponds to a further diluted Universe. The relative strength of the PT, $\alpha$, thus correlates positively with $\lambda_{\phi\tilde\phi}$.

The dotted lines in Fig. 6.7 indicate that the Singlet Scalars model sources non-runaway bubbles ($\alpha < \alpha_\mathrm{run}$) in case of small $\lambda_{\phi\tilde\phi}$ and runaway bubbles ($\alpha > \alpha_\mathrm{run}$) for large $\lambda_{\phi\tilde\phi}$. The obtained values for $\alpha$ and $\beta$ can be used to determine the corresponding GW spectra and the expected SNRs for the different GW observatories, following the procedure explained in Section 3.2. The hatched areas in Fig. 6.7 mark the parameter region that can be probed by *SKA* after 10 and 20 years of data taking, respectively.

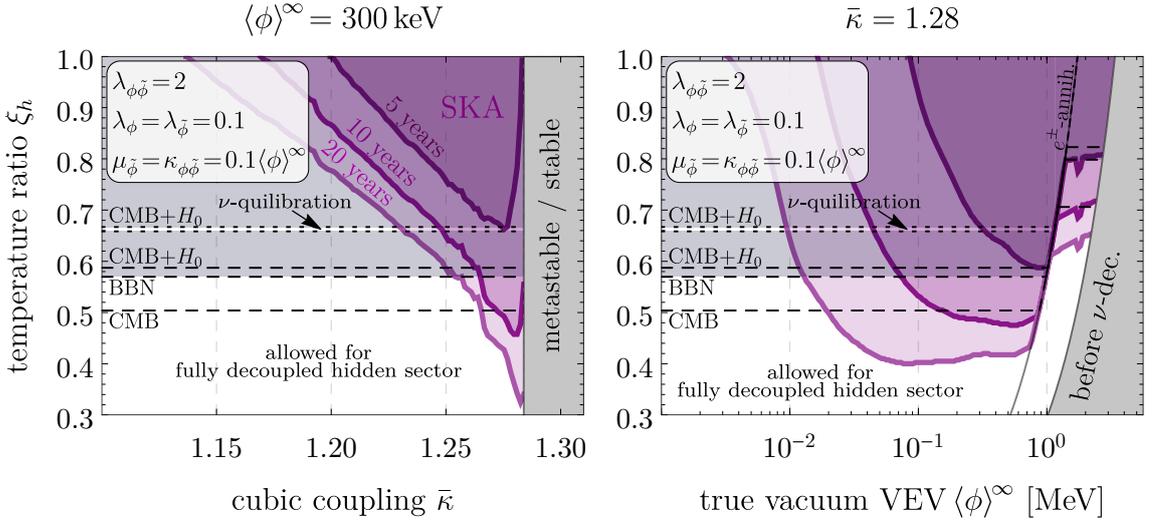

**Figure 6.8:** Dependence of *SKA*'s sensitivity on the hidden–visible sector temperature ratio at the time of the PT, $\xi_h$, for varying $\bar{\kappa}$ (left) and $\langle\phi\rangle^\infty$ (right) in the Singlet Scalars model. Between the dotted lines or below the horizontal dashed lines, the hidden sector evades the discussed $N_{\text{eff}}$ constraints assuming $\nu$-quilibration or the fully decoupled scenario, respectively. Note that the narrowness of the allowed $\nu$-quilibration band is not a sign of fine-tuning, but rather represents the range of values that corresponds to initial ratios $0 < \xi_h^{\text{init}} < 0.61$, where the upper bound is imposed by CMB+$H_0$.

**Impact of a temperature difference.** Figure 6.8 visualizes the impact of a temperature ratio $\xi_h < 1$ on the projected sensitivity of *SKA* after 5, 10, and 20 years of operation, respectively. As expected, a hidden sector colder than the visible sector implies a smaller detectable parameter region. For $\xi_h \lesssim 0.57$, $\lesssim 0.50$, and $\lesssim 0.59$, the hidden sector is cold enough to evade the BBN, CMB, and CMB+$H_0$ constraints discussed in Section 6.2, assuming a hidden sector that remains decoupled throughout the post-BBN evolution of the Universe.[10] Furthermore, the number of DOFs in this model, $g_h = 2$, is small enough to accommodate the $\nu$-quilibration scenario, if one considers the less stringent CMB+$H_0$ instead of the CMB bound. In this scenario, the hidden sector naturally has a temperature ratio of $\xi_h \approx 0.66$ at the time of the PT. More precisely, the narrow accessible band of $\xi_h$ is obtained by varying the initial temperature ratio in the range $0 < \xi_h^{\text{init}} < 0.61$, where the upper limit is dictated by CMB+$H_0$. For the plotted choice of parameters, a PT only occurs if $\bar{\kappa} \lesssim 1.28$.

**Realization.** For the parameter regions displayed in Figs. 6.7 and 6.8, $\tilde{\phi}$ is heavier than $\phi$, while their zero-temperature masses are both of the order $\langle\phi\rangle^\infty$. This is a requirement of the $\nu$-quilibration scenario, which demands that $\phi$ (and not $\tilde{\phi}$) is the stable relic that first cools and then heats the SM neutrinos. One possibility to realize $\nu$-quilibration with the Singlet Scalars is via a small Yukawa-like coupling between $\phi$ and the SM neutrinos, $\nu$, for instance via a heavy right-handed neutrino, $N_R$, together with an interaction term of the form $\phi \overline{N_R} N_R$. Assuming a type-I seesaw then yields the suppressed coupling $m_\nu/m_N \phi \bar{\nu} \nu$. As long as both $\phi$ and all neutrinos are relativistic, the interaction rate between the hidden and visible sectors is proportional to $T_\gamma$, the only relevant energy scale at that time. Therefore, the rate is initially smaller than the Hubble rate, $H \propto T_\gamma^2$, but may become larger at late times, e.g. after BBN. This is required to allow the hidden sector to annihilate and reheat the SM neutrinos, as suggested by the $\nu$-quilibration scenario.

---

[10]The CMB constraints only apply if the hidden sector remains relativistic until recombination.

### 6.3.2 Higgsed Dark Photon

Our second toy model is based on an Abelian hidden sector gauge symmetry $U(1)'$, which gives rise to a dark photon field $A'_\mu$. While GW signatures of $U(1)'$-breaking scenarios have been studied above MeV-scale temperatures [407–409], we instead focus on PTs occurring below 1 MeV. We add a complex scalar field $\phi$ – a "dark Higgs" – which is singlet under the SM gauge groups but charged under $U(1)'$. The model consists of $2+2$ DOFs before and $3+1$ DOFs after the breaking of $U(1)'$, i.e. $g_h = 4$ in total. The relevant terms in the Lagrangian are

$$\mathcal{L} \supset |D_\mu \phi|^2 + |D_\mu H|^2 - \frac{1}{4} F'_{\mu\nu} F'^{\mu\nu} - \frac{\epsilon}{2} F'_{\mu\nu} F^{\mu\nu} - V_{\text{tree}}(\phi, H) \,, \quad (6.22)$$

where $F_{\mu\nu}$ and $F'_{\mu\nu}$ are the field strength tensors of the visible and dark photons, respectively, and $H$ is the SM Higgs field. The kinetic term of $\phi$ contains a covariant derivative,

$$D_\mu \phi = (\partial_\mu + ig' A'_\mu)\phi \,, \quad (6.23)$$

which connects $\phi$ with $A'$ via the gauge coupling $g'$. We consider the tree-level potential

$$V_{\text{tree}}(\phi, H) = -\mu_\phi^2 \phi^\dagger \phi + \frac{\lambda_\phi}{2}(\phi^\dagger \phi)^2 - \mu_H^2 H^\dagger H + \frac{\lambda_H}{2}(H^\dagger H)^2 + \kappa_{\phi H}(\phi^\dagger \phi)(H^\dagger H) \,, \quad (6.24)$$

which is the most generic renormalizable scalar potential under the given symmetries. While the SM quantities $\mu_H$ and $\lambda_H$ are measured quantities, we demand $\mu_\phi^2 > 0$ to allow for $U(1)'$ symmetry breaking and $\lambda_\phi > 0$ to ensure vacuum stability. The position of the non-zero minimum in the $\phi$ direction, $\langle \phi \rangle^\infty$, can be treated as a free parameter by writing

$$\mu_\phi^2 = \frac{\lambda_\phi}{2}(\langle \phi \rangle^\infty)^2 \,. \quad (6.25)$$

See Appendix 6.A.2 for further details on the model

A hidden sector constituted by the Higgsed Dark Photon model has two connections to the visible sector: the kinetic mixing in Eq. (6.22), connecting the $U(1)'$ and $U(1)$ gauge bosons, and the Higgs portal in Eq. (6.24), coupling the dark and SM Higgses. However, to allow for our non-standard thermal scenarios, we impose a decoupled hidden sector by setting $\kappa_{\phi H} = \epsilon = 0$.

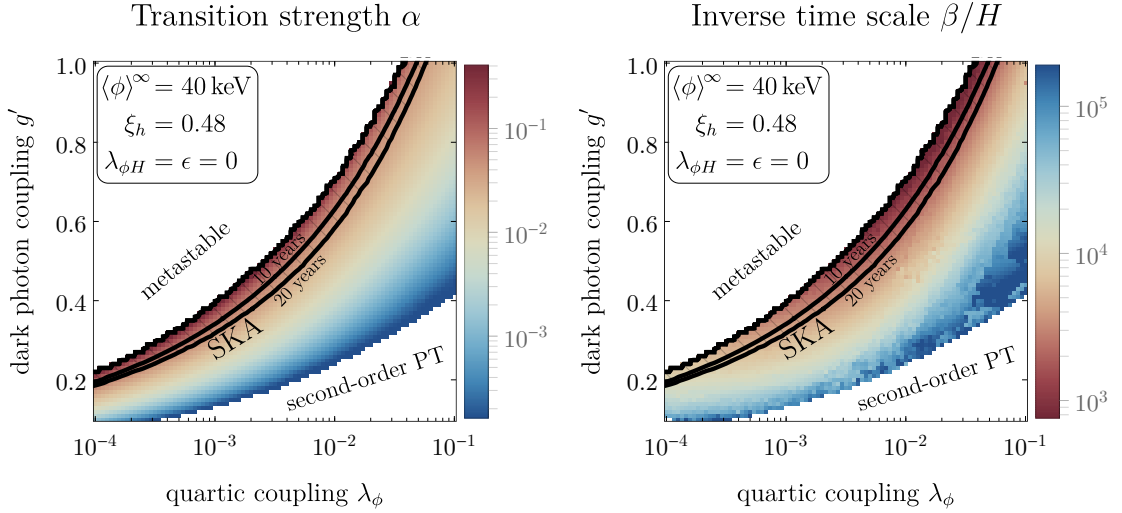

**Figure 6.9:** The strength $\alpha$ (left panel) and inverse time scale $\beta$ (right panel) of a hidden sector PT for the Higgsed Dark Photon model defined in Eqs. (6.22) to (6.24), assuming $\langle\phi\rangle^\infty = 40\,\text{keV}$ (corresponding to $T_{h,n} \sim 40\,\text{keV}$). The hatched areas mark the region where the GW signal will be detectable by *SKA* after the indicated periods of observation time. To not violate the BBN constraint on $N_{\text{eff}}$, we impose that the hidden sector is fully decoupled and colder than the visible sector by a factor of $\xi_h = 0.48$. The noisy behavior in the right plot stems from numerical instabilities in the computation of $\beta$.

**Requirements for a first-order transition.** Similar to the Singlet Scalars model described in the previous section, the Higgsed Dark Photon model allows for a first-order PT. In this model, the PT is responsible for the breaking of the $U(1)'$ gauge symmetry. The transition is of first order if it involves a tunneling through a potential barrier. In the Singlet Scalars model, the barrier was provided by a cubic term ($\propto \phi^3$) in the tree-level potential. The imposed $U(1)'$ symmetry in this case, however, forbids such an operator. A first-order transition is still possible by a barrier that is induced by the interaction between $\phi$ and $A'$, whose strength depends on $g'$ and which is included in our analysis via the thermal one-loop potential, see Section 3.1.1.

**Model parameter space.** The computation of the PT parameters as well as the GW spectrum and its detectability is conducted entirely analogous to the previous toy model. Figure 6.9 shows the obtained PT strength and inverse time scale in the $\lambda_\phi$–$g'$-plane for $\langle\phi\rangle^\infty = 40\,\text{keV}$ (corresponding to $T_{h,n} \sim 40\,\text{keV}$) and with $\xi_h = 0.48$.

Because $g'$ sets the size of the potential barrier, larger values imply more energetic and slower transitions, i.e. larger $\alpha$ and smaller $\beta$. Above a certain threshold for $g'$, which depends on $\lambda_\phi$, the barrier becomes so large that the vacuum remains metastable and the PT never occurs. For small $g'$, on the other hand, the PT becomes weaker and ultimately numerically indistinguishable from a higher-order transition without GW emission.

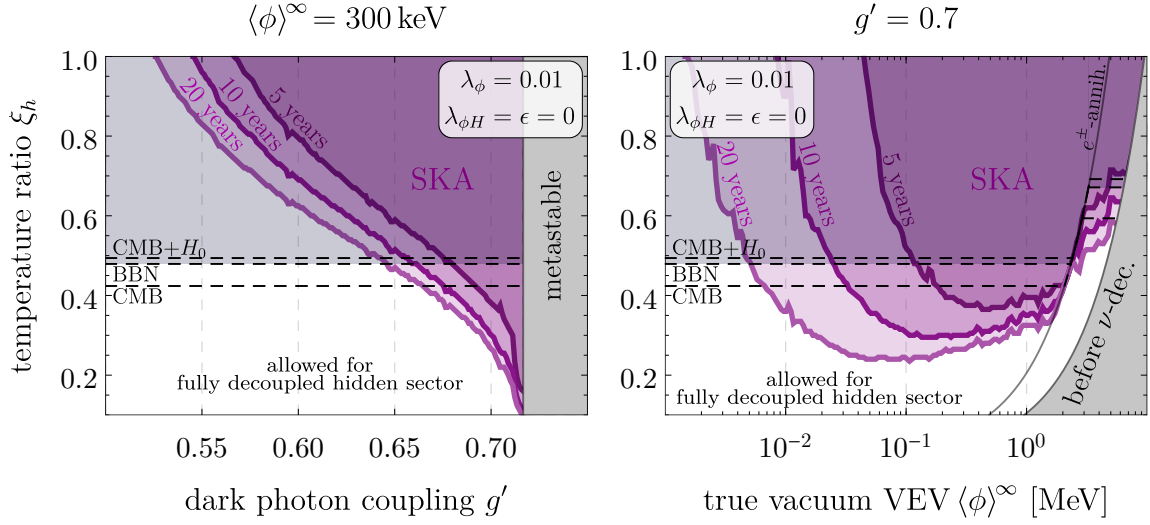

**Figure 6.10:** Dependence of *SKA*'s sensitivity on the hidden–visible sector temperature ratio at the time of the PT, $\xi_h$, for varying $g'$ (left) and $\langle\phi\rangle^\infty$ (right) in the Higgsed Dark Photon model. Below the horizontal dashed lines, the hidden sector evades the discussed $N_{\text{eff}}$ constraints in the fully decoupled scenario.

Increasing $\lambda_\phi$ shifts the region of successful first-order transitions towards higher values of $g'$. This is, because the derived parameter $\mu_\phi^2$ – controlling te depth of the potential in the true vacuum minimum – is proportional to $\lambda_\phi$. A deeper tree-level minimum (larger $\lambda_\phi$) has to be paired with a larger barrier (larger $g'$) in order to leave the dynamics of the PT qualitatively unchanged. This explains the shape of the viable region in Fig. 6.9.[11]

Interactions with the dark photon exert friction on the bubble walls and prevents them from accelerating indefinitely, a general feature of PTs involving vector bosons [124, 203]. The Higgsed Dark Photon model thus sources non-runaway bubbles exclusively. This makes the scalar field contribution in the GW spectrum negligible, so that only the sound-wave and turbulent contributions are relevant.

The hatched areas in Fig. 6.9 again mark the region which can be probed by *SKA* after 10 and 20 years of operation.

**Impact of a temperature difference.** Similar as with the Singlet Scalars, a colder hidden sector leads to smaller parameter ranges that feature detectable GW emission, as can be seen in Fig. 6.10. For $\xi_h \lesssim 0.48$, $\xi_h \lesssim 0.42$, and $\xi_h \lesssim 0.49$, the hidden sector is cold enough to evade the BBN, CMB, and CMB+$H_0$ constraints, respectively, assuming a fully decoupled hidden sector. Considering the model's number of DOFs, $g_h = 4$, the $\nu$-quilibration scenario is excluded for any $\xi_h^{\text{init}}$, as we concluded from the right panel of Fig. 6.3.

---

[11] The observed shape can also be derived analytically. Refs. [164, 165] use an approach of consistent power counting and predict a first-order (higher-order) transition for $\lambda_\phi \sim g'^3$ ($\lambda_\phi \sim g'^2$), which agrees with our findings.

| Singlet Scalars | | | | | |
|---|---|---|---|---|---|
| $\log_{10}\!\big(\frac{\mu_{\tilde\phi}}{\langle\phi\rangle^\infty}\big)$ | $\log_{10}\!\big(\frac{\kappa_{\phi\tilde\phi}}{\langle\phi\rangle^\infty}\big)$ | $\log_{10}(\lambda_\phi)$ | $\log_{10}(\lambda_{\tilde\phi})$ | $\lambda_{\phi\tilde\phi}$ | $\bar\kappa$ |
| $-3 \sim 0$ | $-3 \sim 0$ | $-3 \sim 0$ | $-3 \sim 0$ | $0 \sim 3$ | $0.7 \sim 1.5$ |
| Higgsed Dark Photon | | | | | |
| | | $\log_{10}(\lambda_\phi)$ | $g'$ | | |
| | | $-4 \sim -1$ | $0 \sim 1$ | | |

**Table 6.1:** Model parameter ranges used for the random scans presented in Fig. 6.11.

### 6.3.3 Random Parameter Scans

In the previous sections, we presented numerical results for specific slices of the model parameter spaces to illustrate the behavior of the GW signal and its detectability with respect to the most relevant model parameters. For a more global picture we plot in Fig. 6.11 the results of random parameter scans, overlaid with the sensitivity of different future GW observatories. To produce these plots, we generate 4000 random parameter points per model from the intervals listed in Table 6.1. For each parameter point, we proceed as follows: We set the input parameter $\langle\phi\rangle^\infty$ to 50 keV for the first three plots and to 200 GeV for the last one and determine the corresponding $\alpha_h \equiv \alpha(\xi_h = 1)$ and $\beta$. The numerically obtained nucleation temperature, $T_{h,n}$, is similar but not exactly equal to $\langle\phi\rangle^\infty$. We thus rescale the results for the considered parameter point such that the nucleation temperature matches the desired values (50 keV or 200 GeV), which affects $\alpha_h$ via the temperature dependence of $g_\star^{\rm SM}$. Finally, we compute the suppressed $\alpha$ under the assumption of fixed $T_{h,n}$ but for different temperature ratios $\xi_h \leq 1$, as indicated in the figure. The resulting random points are displayed in Fig. 6.11, together with the sensitive regions of different GW observatories (shaded regions). The temperature ratio affects the shape of these regions by altering the redshift of signal amplitude and peak frequency according to Eq. (6.15).

Figure 6.11 reveals that the PT in the Singlet Scalars model tends to be weaker (smaller $\alpha$) than in the Higgsed Dark Photon model in most of the examined parameter space. In general, the inverse time scale, $\beta$, usually has a lower limit that corresponds to the point where the potential barrier becomes too large for the PT to occur. This sharp bound was already visible in Figs. 6.7 and 6.9 but can also be seen in our random parameter scan for the Higgsed Dark Photon. The effect is less pronounced for the Singlet Scalars, which is due to the larger number of randomly varied model parameters (six vs. two) in this case. At a nucleation temperature $T_{h,n} = 50$ keV (first three plots) *SKA* will be sensitive to significant portions of the considered parameter spaces of both models, while *EPTA* and *NANOGrav* can only probe a tiny fraction of points. As already discussed, the cosmological $N_{\rm eff}$ constraints at $T_{h,n} = 50$ keV require a temperature ratio $\xi_h < 1$. The $\nu$-quilibration scenario is viable in connection with the Singlet Scalars if the less stringent CMB+$H_0$ bound on $N_{\rm eff}$ is applied. In this scenario, a temperature ratio of $\xi_h \approx 0.66$ naturally arises, which we therefore present in the upper right panel of Fig. 6.11. When assuming a fully decoupled and even colder hidden sector with $\xi_h = 0.48$, as shown in the lower left panel, both models are allowed by the CMB and BBN bounds.

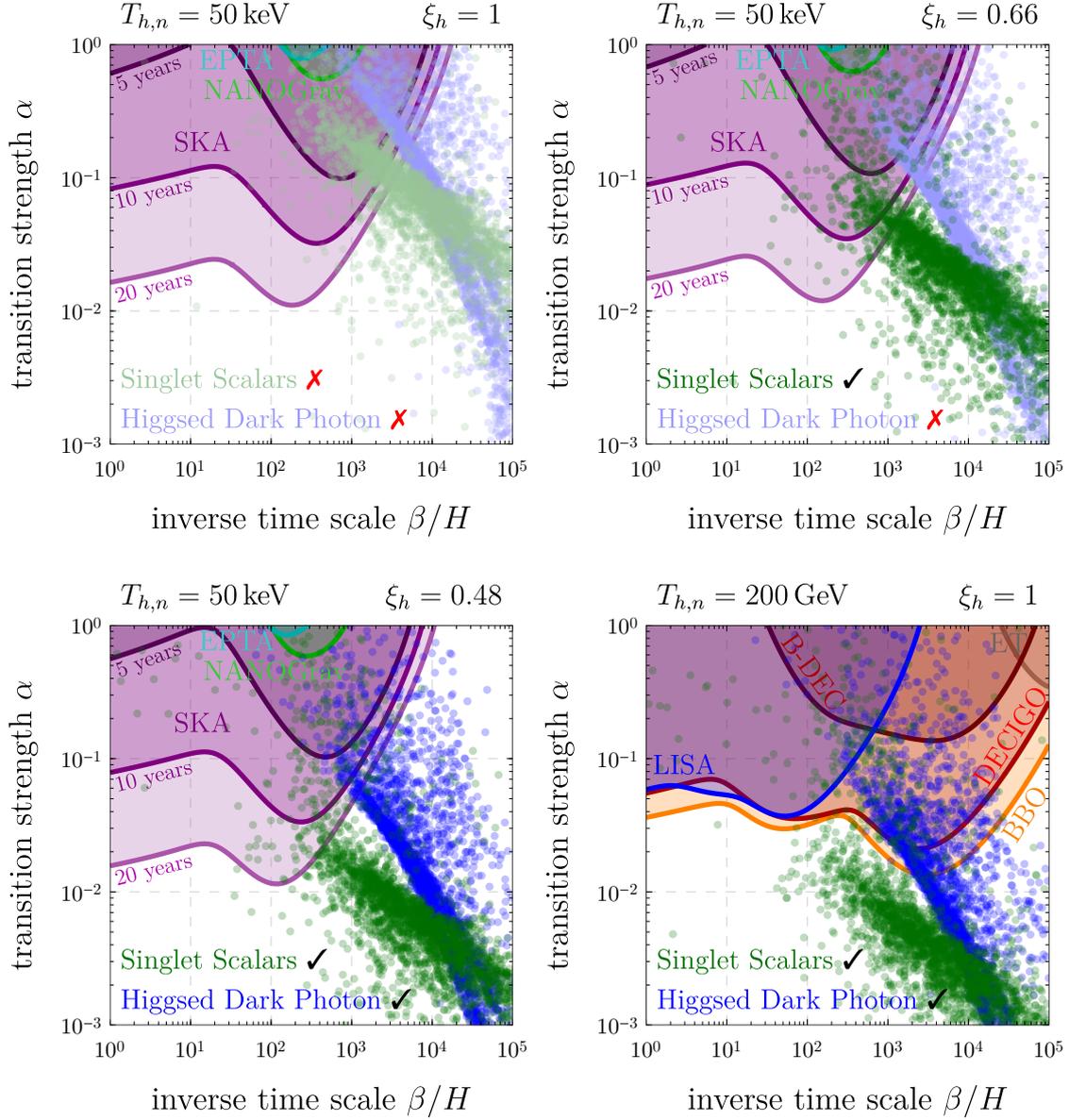

**Figure 6.11:** PT strength $\alpha$ and inverse time scale $\beta$ for the Singlet Scalars (green points) and for the Higgsed Dark Photon (blue points) in a random parameter scan. The scanned parameter intervals are listed in Table 6.1. We compare to the expected sensitivities of various future GW observatories, assuming bubble walls in the non-runaway regime ($\alpha < \alpha_{\rm run}$), which is justified (approximately justified) for the Higgsed Dark Photon (Singlet Scalars). The four panels correspond to different bubble nucleation temperatures, $T_{h,n}$, and to different hidden–visible sector temperature ratios, $\xi_h$, as indicated in the figure. A tick mark (✓) means that the model is cosmologically allowed for the chosen $\xi_h$ in at least one of the two scenarios (fully decoupled or $\nu$-quilibration). A red cross (✗) indicates that the model cannot be reconciled with the $N_{\rm eff}$ constraints at all for the respective value of $\xi_h$.

For comparison, we also include results for a PT at $T_{h,n} = 200\,\text{GeV}$ in the lower right panel, where $\xi_h$ is unconstrained. At this scale, the PT strength $\alpha$ is suppressed by the large number of relativistic SM DOFs. As a result, only *DECIGO* and *BBO* will be partly sensitive to transitions of the Higgsed Dark Photon model, while the more imminent *LISA* and *B-DECIGO* experiments would cover only small portions of the examined parameter regions. The Singlet Scalars model, characterized by even weaker transitions, is mostly undetectable at $T_{h,n} = 200\,\text{GeV}$.

## 6.4 Conclusions

In the present chapter we investigated the possibility of cold and light hidden sectors that feature masses and PT dynamics below $1\,\text{MeV}$. This regime is usually not considered, as relativistic DOFs beyond the SM are tightly constrained by the ratio of light element abundances produced during BBN and by properties of the CMB. Furthermore, light DM is hard to probe via conventional direct detection methods. If the hidden sector undergoes a first-order PT at sub-MeV temperatures, the GW spectrum sourced by bubble collisions would – redshifted until today – exhibit frequencies below $\sim 10^{-7}\,\text{Hz}$ and could be probed by (future) PTAs.

To address the cosmological problems of conventional hidden sectors at sub-MeV scales, we discussed two non-standard thermal scenarios: (i) a fully decoupled hidden sector and (ii) a hidden sector that is initially decoupled from the SM, then equilibrates with the neutrinos, and finally becomes massive and annihilates back into neutrinos ($\nu$-quilibration). Both scenarios can reconcile the existence of sub-MeV hidden sectors with cosmological observations, if the additional DOFs are (initially) *colder* than the visible sector. A fully decoupled hidden sector that is colder makes a smaller contribution to the total energy and thus has a lighter cosmological footprint. In the $\nu$-quilibration scenario, on the other hand, the equilibration with the neutrinos cools them down, which decreases the overall impact on BBN and the CMB. We derived all required relations to quantify these statements, depending on the number of DOFs in the hidden sector.

GW signals from hidden sectors are suppressed if they are cold. This is, because the dimensionless amplitude of the GW spectrum depends on the *relative* PT strength, $\alpha$, which is defined as the latent heat of the hidden PT divided by the total energy density of the Universe. Gravitational radiation from a colder hidden sector has to compete against a larger overall energy content and is thus harder to detect. We illustrated how the sensitivity of various GW observatories is affected by a hidden–visible sector temperature ratio smaller than one.

We applied our theoretical considerations to two minimal toy models, which can be seen as low-energy proxies of almost any UV-complete theory: The Singlet Scalars and the Higgsed Dark Photon. For both models, we argued analytically that a first-order PT can occur, which we then confirmed by numerical simulations of the effective potential (including tree-level, one-loop, and ring diagram contributions). We also determined the nucleation temperature as well as the strength and time scale of the PT for different slices of the parameter spaces and pointed out the regions where PTAs could probe the accompanying gravitational radiation.

Our findings show that both models can be reconciled with the cosmological constraints, if they are colder than the visible sector by $\mathcal{O}(1)$ factors in case of the fully decoupled scenario. The $\nu$-quilibration scenario only works with the Singlet Scalars, which introduce only two additional DOFs and are thus easier to accommodate than than the Higgsed Dark Photon with four DOFs. Despite the suppression of the GW signal, significant parts of the parameter space can be probed by the future PTA *SKA*, assuming that the expected SMBHB background will be resolvable.

# Appendix of Chapter 6

## 6.A Toy Model Details

### 6.A.1 Singlet Scalars

Following the approach discussed in Section 3.1.1 and based on the tree-level potential in Eq. (6.19), we obtain the field-dependent masses

$$
\begin{aligned}
m_\phi^2(\phi) &= \kappa\Big[2\phi - \langle\phi\rangle^\infty\Big] + \lambda_\phi\Big[3\phi^2 - (\langle\phi\rangle^\infty)^2\Big]\,, \\
m_{\tilde\phi}^2(\phi) &= \mu_{\tilde\phi}^2 + 2\kappa_{\phi\tilde\phi}\phi + \lambda_{\phi\tilde\phi}\langle\phi\rangle^2\,,
\end{aligned}
\qquad(6.26)
$$

and the thermal Debye masses

$$
\begin{aligned}
\Pi_\phi(T_h) &= \left[\frac{\lambda_\phi}{4} + \frac{\lambda_{\phi\tilde\phi}}{12}\right]T_h^2\,, \\
\Pi_{\tilde\phi}(T_h) &= \left[\frac{\lambda_{\tilde\phi}}{4} + \frac{\lambda_{\phi\tilde\phi}}{12}\right]T_h^2\,.
\end{aligned}
\qquad(6.27)
$$

The one-loop Coleman–Weinberg potential $V_{\text{CW}}$, see Section 3.1.1, contains counterterms that cancel arising infinities. The finite part of these counterterms ensures that the structure of the potential, i.e. minima and derivatives, remains unchanged w.r.t. the tree-level potential. For the Singlet Scalars model, the finite counterterms read

$$
V_{\text{ct}}(\phi) = \frac{\delta\mu_\phi^2}{2}\phi^2 + \frac{\delta\kappa}{3}\phi^3 + \frac{\delta\lambda_\phi}{4}\phi^4\,, \qquad(6.28)
$$

where the couplings are determined by requiring

$$
V_{\text{CW}}(0) = V_{\text{CW}}(\langle\phi\rangle^\infty)\,, \qquad V'_{\text{CW}}(\langle\phi\rangle^\infty) = 0\,, \qquad V''_{\text{CW}}(\langle\phi\rangle^\infty) = 0\,. \qquad(6.29)
$$

### 6.A.2 Higgsed Dark Photon

The field-dependent masses of the two dark Higgs DOFs and of the dark photon are derived based on the Lagrangian, Eqs. (6.22) to (6.24), and amount to

$$
\begin{aligned}
m_{\phi_m}^2(\phi) &= \frac{\lambda_\phi}{2}\Big[3\phi^2 - (\langle\phi\rangle^\infty)^2\Big]\,, \\
m_{\phi_g}^2(\phi) &= \frac{\lambda_\phi}{2}\Big[\phi^2 - (\langle\phi\rangle^\infty)^2\Big]\,, \\
m_{A'}^2(\phi) &= g'^2\phi^2\,.
\end{aligned}
\qquad(6.30)
$$

We see that in the true vacuum, where the background field occupies the value $\langle\phi\rangle^\infty$, $\phi_m$ is massive while $\phi_g$ remains massless. The latter is a would-be Goldstone boson and takes the form of a longitudinal dark photon polarization mode. The thermal Debye masses of $\phi$ and the longitudinal mode of $A'$ are given by

$$\Pi_\phi(T_h) = \left[\frac{\lambda_\phi}{6} + \frac{g'^2}{4}\right]T_h^2,$$
$$\Pi_{A'}(T_h) = \frac{g'^2}{3}T_h^2. \tag{6.31}$$

The finite counterterms take the form

$$V_{\text{ct}}(\phi) = -\frac{\delta\mu_\phi^2}{2}\phi^2 + \frac{\delta\lambda_\phi}{8}\phi^4, \tag{6.32}$$

where the couplings are chosen such that

$$V'_{\text{CW}}(\langle\phi\rangle^\infty) = 0, \qquad V''_{\text{CW}}(\langle\phi\rangle^\infty) = 0. \tag{6.33}$$

# Part II

# New Physics Searches at the DUNE Near Detector

# 7 The DUNE Experiment

*DUNE* ("Deep Underground Neutrino Experiment") is a long-baseline neutrino experiment at Fermilab in Illinois, which is currently under construction. *DUNE* will make use of an intense neutrino beam and operate two detector sites, one close to the production point and one at a distance of 1300 km, as illustrated in Fig. 7.1. The projects aims to shed light on a variety of unsolved physics puzzles. The components and goals of *DUNE* will be briefly summarized in the following.

## 7.1 Components

**Neutrino production facility.** The neutrino production starts with a beam of protons passing through Fermilab's chain of particle accelerators. The protons with a final energy of 120 GeV, corresponding to a total beam power of 1.2 MW, will be collided with a fixed graphite target at a rate of $1.1 \times 10^{21}$ protons on target (POT) per year [410, 411]. The scattering on target nuclei produces large amounts of charged pions and kaons via hard QCD interactions. The mesons are then, depending on the sign of their charge, focussed or defocussed by two magnetic horns. The focussed mesons subsequently enter a 194 m long decay pipe filled with helium, where most of the pions decay weakly via[1]

$$\pi^+ \to \mu^+ \nu_\mu \qquad \text{or} \qquad \pi^- \to \mu^- \bar\nu_\mu \,.$$

---

[1] Charged pion decays into electrons are possible but helicity suppressed due to the electron's small mass. Decays into tau leptons, on the other hand, are kinematically forbidden since $m_{\pi^\pm} < m_\tau$.

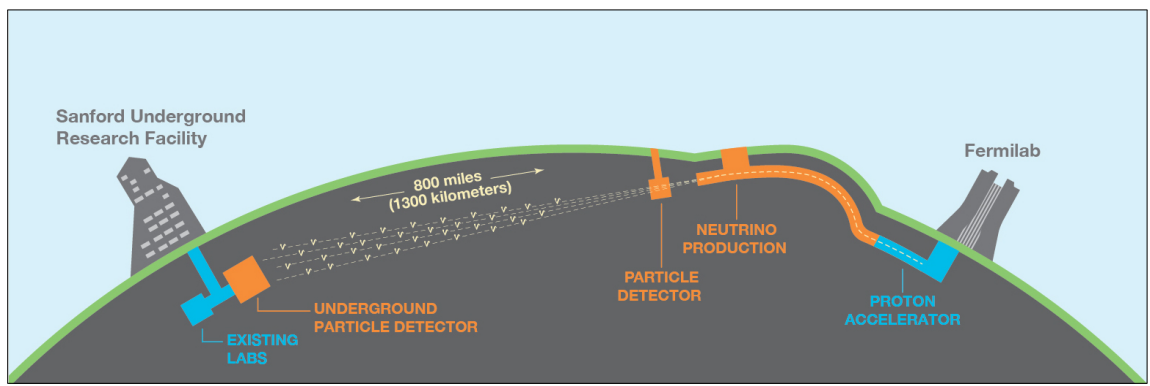

**Figure 7.1:** Illustration of the *DUNE* experiment, which consists of a neutrino production facility as well as near and far underground detectors. Image taken from Ref. [410].

By changing the polarity of the magnetic horns, the facility can switch between neutrino and anti-neutrino production mode. Due to the large kinetic energy of the pions compared to their rest mass ($m_{\pi^\pm} \approx 140\,\text{MeV}$), the neutrinos are highly boosted in the forward direction and thus form a collimated beam.

**Far detector.** The main detector of *DUNE* will be located in South Dakota, 1300 km from the neutrino production point and 1.5 km below the surface. Its main part is a tank of 70 kt cooled liquid argon, of which 40 kt are sensitive and instrumented with time projection chambers (TPCs). The purpose of this far detector is to precisely measure the energy and flavor dependent fluxes of the oscillated neutrino beam [412, 413].

**Near detector.** A three component near detector will be installed at a distance of 574 m from the neutrino production point. It will be comprised of a magnetized beam monitor, a liquid argon TPC and a magnetized gaseous argon TPC [414]. The near detector's purpose is to monitor the unoscillated neutrino flux. This constrains systematic uncertainties in the long-baseline oscillation measurement and furthermore allows measuring the neutrino cross sections. The TPCs will be mounted on a movable platform allowing for a displacement up to about 30 m off-axis, a concept dubbed *DUNE-PRISM* ("*DUNE* Precision Reaction-Independent Spectrum Measurement"). This additional degree of freedom helps to disentangle the uncertainties of the measured neutrino cross sections and fluxes, as obviously only the latter depends on the detector position.

In Chapter 8 we study the possibility of conducting new physics searches employing the *DUNE-PRISM* concept, without interfering with the original purposes of the near detector.

## 7.2 Main Objectives

**Neutrino oscillations.** With its near and far detectors, *DUNE* is capable of examining long-baseline neutrino oscillations with high precision. One of the unsolved questions in this context is the hierarchy of the neutrino mass eigenstates, i.e. the sign of $m_3^2 - m_1^2$. A difference in the oscillation rates $\nu_\mu \to \nu_e$ and $\nu_\mu \to \overline{\nu}_e$ is expected due to the presence of electrons and absence of positrons in the Earth, an effect that is sensitive to the mass hierarchy but also to a possible CP-violating phase ($\delta_{\text{CP}} \neq 0, \pi$) in the neutrino mixing (PMNS matrix). As discussed in Section 2.2.1, the violation of the charge-parity symmetry (CP) is a necessary ingredient for baryogenesis. Thanks to its long baseline, the degeneracy between the two aforementioned effects is large enough to shed light on both the mass hierarchy and $\delta_{\text{CP}}$ [415].[2] On top of that, *DUNE* allows for precision measurements of the PMNS matrix and tests the three-flavor paradigm that was called into question by experimental evidence [417–420].

---

[2]First hints of a CP violation in the neutrino sector were reported by the *T2K* collaboration [49], which however could not be confirmed by *NOνA* [416], the predecessor of *DUNE*.

**Proton lifetime.** Independent from the other components of *DUNE*, the liquid argon TPC in the far detector is capable of measuring or constraining the decay of protons inside the detector. Decays like $p \to K^+ \overline{\nu}$ and $p \to e^+ \pi^0$ could be identified due to their distinctive signatures and would hint at the existence of a (supersymmetric) grand unified theory [30, 421, 422]. *DUNE* will supersede the current lower bound on the proton lifetime, which was set by the water Cherenkov detector Super-Kamiokande [423, 424]

**Supernova neutrinos.** At the end of their lifespans, heavy stars undergo spectacular supernova explosions, such as the famous SN 1987A in the Large Magellanic Cloud [425, 426]. If a burnt-out star's iron core exceeds the critical limit of $\sim 1.4$ solar masses [427], the stabilizing electron degeneracy pressure is overcome by the inward gravitational pull, causing the core to collapse into a neutron star or black hole. A part of the potential energy is released in the form of intense neutrino bursts of all flavors and with neutrino energies of a few MeV. Being sensitive to neutrinos in this energy regime, the *DUNE* far detector is ideally suited to contribute to the understanding of core-collapse supernovae occurring in the vicinitiy of the Milky Way [415].

# 8 New Physics Searches with DUNE-PRISM

*This chapter is based on the publication [5] of the author and his collaborators. In this project, the author was responsible for the simulation of the background events. Furthermore, he worked out the statistical procedure and determined the projected experimental sensitivities, which were crosschecked against independently obtained results by LM. The author produced all figures that appear in present chapter. Note that the publication contains a section about heavy neutral leptons. The author was not directly involved in the corresponding analysis, wherefore this topic will not be addressed.*

## 8.1 Introduction

As already alluded to several times in this work, the standard weakly interacting massive particle (WIMP) provides a tempting dark matter (DM) concept but is getting increasingly constrained experimentally in its vanilla form. In the first part of this thesis, we already went beyond the WIMP paradigm in several ways: We discussed a non-thermal DM mechanism in Chapter 4, the possibility of DM constituted by black holes (BHs) in Chapter 5, and non-standard dark sector scenarios that may exist below the MeV scale in Chapter 6. In the present chapter, we will focus on light DM in the MeV to few-GeV mass range, which is still partly unexplored. Only quite recently, the semiconductor-based tabletop experiment *SENSEI* [428] reported its first direct detection constraints on sub-GeV DM as a proof of concept [386]. Furthermore, specific light DM models are constrained by the non-observation of invisible decays in collider experiments [429–432].

Another approach to probing new physics in the sub-MeV regime is provided by long-baseline neutrino experiments such as *DUNE* (reviewed in the previous chapter) which is currently under construction at Fermilab. This experiment includes an argon-based near detector – 574 m downstream from the neutrino facility – whose main purpose is the measurement of unoscillated neutrino fluxes and cross sections. At the same time and without interfering with its original purpose, the data collected at the near detector can be searched for hints of DM–electron or DM–nucleon scattering, with DM particles that are possibly created by the proton beam that is responsible for the neutrino production [433–436]. An exceptional feature of the *DUNE* near detector will be its mobility, allowing for measurements off the proton beam axis and which will reduce systematic uncertainties in the neutrino flux measurements. However, taking off-axis data is also beneficial for new physics searches. We will focus our analyses on the *DUNE-PRISM* concept, which includes consecutive measurements at seven different on- and off-axis positions (with a maximum offset of 36 m or 62.7 mrad).

DM candidates lighter than $\sim$ GeV typically require additional mediators to avoid an overclosure of the Universe when produced in a thermal context [94, 95]. We will discuss two models with new massive gauge bosons and complex scalar DM candidates: the Dark Photon Model in Section 8.2 and the Leptophobic Model in Section 8.3, which can both be probed by *DUNE-PRISM*, as we will show. For both models, we define the basic properties and interactions, discuss the DM production and detection channels as well as the neutrino induced background, and provide sensitivity projections for *DUNE-PRISM* based on our signal and background simulations. We summarize our findings in Section 8.4.

## 8.2 Dark Photon and Light Dark Matter

DM candidates within the conventional thermal freeze-out paradigm exhibit masses of at least $\mathcal{O}(\text{GeV})$ [94], which makes them ideal subjects for nuclear recoil direct detection searches. Viable DM models in the mass regime MeV $\sim$ GeV, on the other hand, require additional ingredients such as massive mediators [94, 95]. One of the simplest and most generic light DM model is composed of a scalar particle $\phi$ – our DM candidate – and a new gauge symmetry $U(1)'$ with the dark photon $A'$ as force carrier. The corresponding Lagrangian density can be written as

$$\mathcal{L} \supset \mathcal{L}_{A'} + \mathcal{L}_\phi \,, \tag{8.1}$$

with

$$\mathcal{L}_{A'} = -\frac{1}{4} F'_{\mu\nu} F'^{\mu\nu} + \frac{m_{A'}^2}{2} A'^\mu A'_\mu - \frac{\epsilon}{2} F'_{\mu\nu} F^{\mu\nu} \,, \tag{8.2}$$

and

$$\mathcal{L}_\phi = ig' A'^\mu J_\mu^\phi + (\partial_\mu \phi^\dagger)(\partial^\mu \phi) - m_\phi^2 \phi^\dagger \phi \,, \tag{8.3}$$

where $J_\mu^\phi = [(\partial_\mu \phi^\dagger)\phi - \phi^\dagger(\partial_\mu \phi)]$ is the DM current, $g'$ is the $U(1)'$ gauge coupling, and $F_{\mu\nu}$ and $F'_{\mu\nu}$ are the $U(1)$ and $U(1)'$ field strength tensors, respectively. Note that in the Lagrangian, we already expanded the covariant derivative in the kinetic term of $\phi$ and we added an effective mass term for $A'$. This mass could be generated via a Higgs mechanism (by $\phi$ or a different scalar) or via a Stückelberg mechanism [437], which we do not specify. The small kinetic mixing $\epsilon$, connecting the dark and visible photons, parameterizes the interaction strength between the dark and visible sectors. In case of a massless dark photon – corresponding to a long-range force – $\epsilon$ is highly constrained by measurements of positronium decays [438]. We will instead consider a massive dark photon, which is partly constrained experimentally but not yet fully ruled out [439].

Assuming $m_\phi < m_{A'}/2$, which ensures that all $A'$ decay into $\phi\phi^\dagger$, the thermal relic abundance of $\phi$ is determined by its annihilation cross section to fermions of the Standard Model of Particle Physics (SM),

$$\sigma(\phi\phi^\dagger \to \bar{f}f)\, v_{\text{rel}} \sim 8\pi \frac{v_{\text{rel}}^2}{m_\phi^2} Y \,, \tag{8.4}$$

where $v_{\text{rel}}$ denotes the relative velocity of the two annihilating particles. Here, we defined

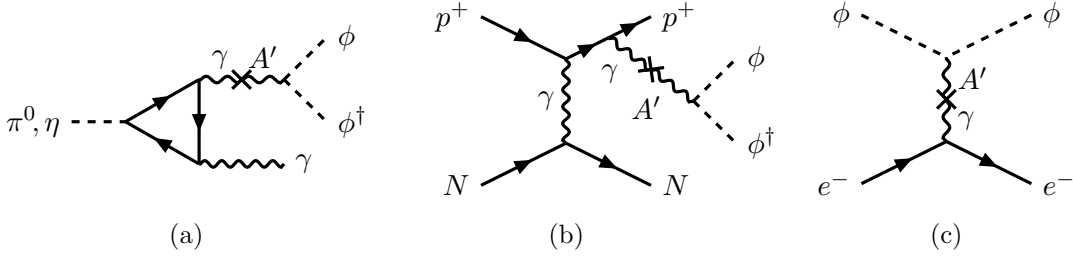

**Figure 8.1:** DM production channels via meson decay (a) and proton bremsstrahlung (b), together with the considered detection channel (c) for the Dark Photon Model. Each mixing between dark and visible photons, indicated by a cross, contributes $\epsilon^2$ to the rate or cross section of the process. $N$ represents a nucleon in the target material. The triangle in (a) sums over all lepton generations and is induced by the chiral anomaly of quantum electrodynamics (QED) [448, 449].

the effective coupling strength [440]

$$Y \equiv \epsilon^2 \alpha_D \left(\frac{m_\phi}{m_{A'}}\right)^4, \tag{8.5}$$

with $\alpha_D \equiv g'^2/(4\pi)$. As part of our results, we will present constraints on $Y$, allowing us to highlight the regions of model parameter space where the observed DM abundance would be obtained thermally.

Note that measurements of the anisotropies in the cosmic microwave background (CMB) set strong constraints on thermal freeze-out DM candidates with masses below $\sim 10 \,\text{GeV}$ [441–444]. However, in case of scalar or Majorana fermion DM, the annihilation rate is $v_{\text{rel}}^2$ suppressed by the spin structure of the process. Thanks to this effect, the DM in the model at hand can be light while still being compatible with the CMB measurements.

### 8.2.1 Dark Matter Production and Detection

The kinetic mixing between dark and visible photons, $A'$ and $\gamma$, implies that dark photons can be produced by any SM process with a final state $\gamma$, if allowed kinematically. This makes the neutrino production facility employed by *DUNE* a potential dark photon source. Given that $m_\phi < m_{A'}/2$, the produced dark photons will decay into pairs of $\phi$. A beam of DM would thus travel alongside the *DUNE* neutrino beam and eventually reach the near detector, where it could scatter on nuclei and electrons and thereby be made visible. Our analysis will focus on the detection via DM–electron scattering (Fig. 8.1c), as this channel competes with relatively small neutrino-induced backgrounds [445–447]. The signal strength scales as $\epsilon^4 \alpha_D$, with a factor of $\epsilon^2$ from $A'$ production and a factor $\epsilon^2 \alpha_D$ arising from DM detection. We will consider the following DM production channels:

1. **Meson decay.** A proton beam dump produces not only charged mesons – crucial for neutrino production – but also sizable amounts of neutral mesons. The latter will decay quickly[1] into DM via the process in Fig. 8.1a, which is active in the regime $m_{A'} < m_\eta \approx 0.5\,\text{GeV}$. Decays of heavier mesons play only a subdominant role and will thus be neglected [450, 451].

We use GEANT4 [452] according to Ref. [453] to model the $\pi^0$ and $\eta$ spectra based on primary interactions as well as the interactions of secondary particles propagating in the *DUNE* target. We compared the resulting spectra to the ones presented in Ref. [454], which were obtained considering primary interactions only using PYTHIA [455]. While we find good agreement at small scattering angles, we observe a growing discrepancy at larger angles, where the spectrum is dominated by low-energy particles, making secondary interactions more important. As we want to harness the advantages of the *DUNE-PRISM* off-axis measurements, it is thus crucial to include secondary interactions in our analysis.

The meson decay DM production channel is dominated by intermediate on-shell dark photons. We therefore make use of the narrow-width approximation[2] where the branching ratio for meson decay into dark photons is given by [445, 457]

$$\frac{\text{BR}(\pi^0, \eta \to \gamma A')}{\text{BR}(\pi^0, \eta \to \gamma\gamma)} \approx 2\,\epsilon^2 \left(1 - \frac{m_{A'}^2}{m_{\pi^0,\eta}^2}\right)^3. \tag{8.6}$$

2. **Proton bremsstrahlung.** Electromagnetic interactions between the beam protons and the target material in the neutrino facility lead to bremsstrahlung, shown in Fig. 8.1b, which dominates DM production in the window $0.5\,\text{GeV} \lesssim m_{A'} \lesssim 1\,\text{GeV}$. The dark photon is preferentially emitted in the forward direction. In this limit, the process can be modeled on the assumption of on-shell exchange bosons [458–460]. For the proton form factor, which enters the bremsstrahlung production rate, we use an effective parameterization that takes $\rho$ and $\eta$ resonances into account [461]. This leads to a resonance peak in the production rate for dark photon masses around 780 MeV [462].

The Drell–Yan process is another dark photon and DM production channel, which is however only relevant for $m_{A'} \gtrsim 1\,\text{GeV}$ [463]. We will also neglect the minor contribution from production processes involving leptonic secondary particles [453].

The DM–electron scattering signal events that enter our analyses are generated using MADDUMP [464]. This software simulates meson decays and bremsstrahlung [465, 466], and takes geometric acceptances as well as finite-detector-size effects into account. For simplicity, we consider a detector of cylindrical shape with radius 3.5 m, oriented along the beam axis.

### 8.2.2 Backgrounds

As discussed, our analysis focuses on $\phi$–$e^-$ scattering as DM detection process, with the corresponding signature being a single, energetic recoil electron. Events with similar sig-nature are produced by neutrinos from the *DUNE* beam via two kinds of interactions: elastic $\nu$–$e^-$ scattering (Figs. 8.2a to 8.2c) and charged current (CC) $\nu_e$–nucleon interactions (Fig. 8.2d). In the latter background process, the striking of the nucleon may induce a hadronic shower. However, we will conservatively assume that this hadronic activity remains unidentified.

---

[1] As opposed to charged mesons – which decay via intermediate off-shell $W^\pm$ – neutral mesons decay electromagnetically almost immediately after being produced.

[2] See Ref. [456] for a discussion of a full treatment, including off-shell decays.

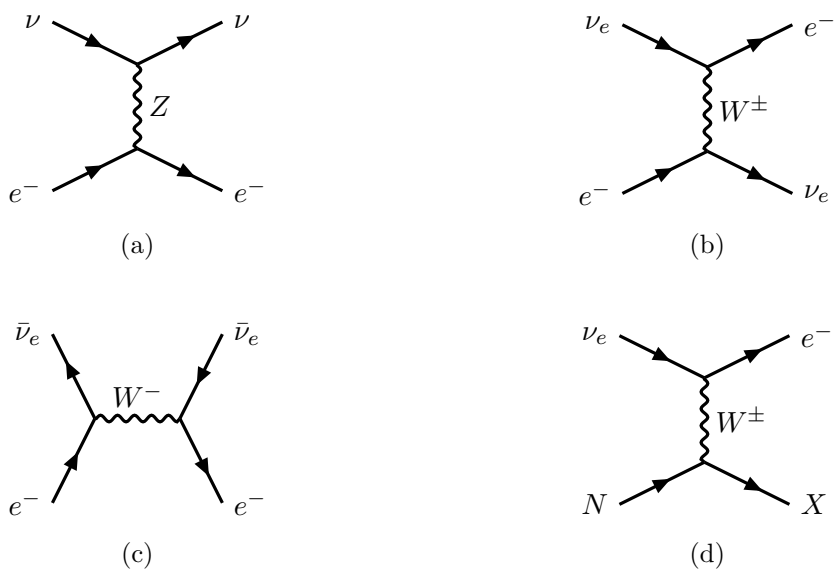

**Figure 8.2:** Main background processes to the DM detection channel for the Dark Photon Model, with the signature of a single, energetic recoil electron. We will conservatively assume that the final hadronic system, $X$, remains unresolved.

We employ the software GENIE [468] to generate background events from the above-mentioned processes, using argon as the simulated target material. GENIE has all relevant scattering cross sections implemented and outputs a list of initial and final state particles together with corresponding kinematic observables for each single interaction. After simulating a sufficient number of events, we reweight them according to the energy-dependent neutrino fluxes for different on- and off-axis positions of the *DUNE* near detector, as provided by Ref. [467]. These fluxes, which have been generated via Monte Carlo techniques, exhibit low statistics for large neutrino energies, especially at the farthest off-axis angles. We therefore fit the high-energy tail of the fluxes linearly in log-space and extrapolate to the regions with insufficient data, as shown in Fig. 8.3. For our analysis, we will assume that *DUNE* is running in *neutrino mode*, which means that the production facility focuses positive mesons and makes use of the production channel $\pi^+ \to \mu^+ \nu_\mu$. Despite the name "neutrino mode", rarer decays such as $K^+ \to \pi^+\pi^+\pi^-$ or $\mu^+ \to e^+ \nu_e \bar{\nu}_\mu$ give rise to $\bar{\nu}_\mu$, $\nu_e$, and $\bar{\nu}_e$ components in the spectrum as well, as can be seen in the neutrino flux predictions plotted in Fig. 8.3.

We notice that the considered DM detection process $\phi$–$e^-$ scattering obeys the kinematic relation [469]
$$E_e \theta_e^2 \lesssim 2 m_e \,, \tag{8.7}$$

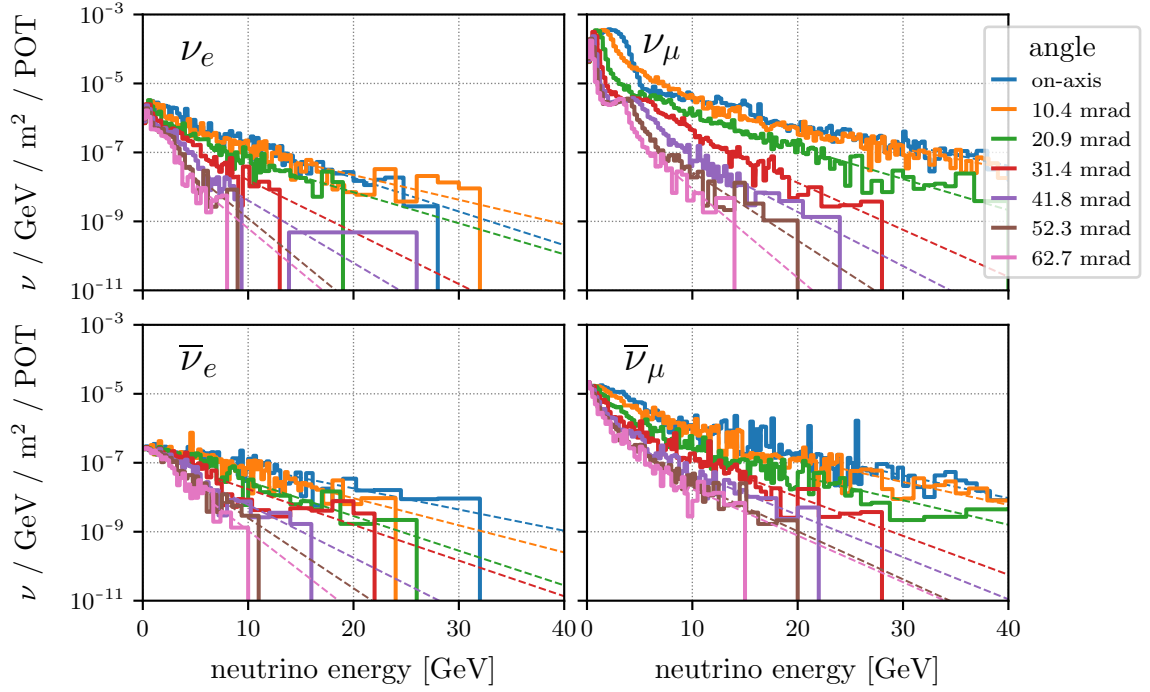

**Figure 8.3:** Neutrino flux predictions at different on- and off-axis angles for *DUNE* running in neutrino mode, as provided by Ref. [467]. Of the different neutrino types, which are shown separately in the four panels, $\nu_\mu$ is the most dominant one. The dashed lines are the fit curves we use for extrapolation.

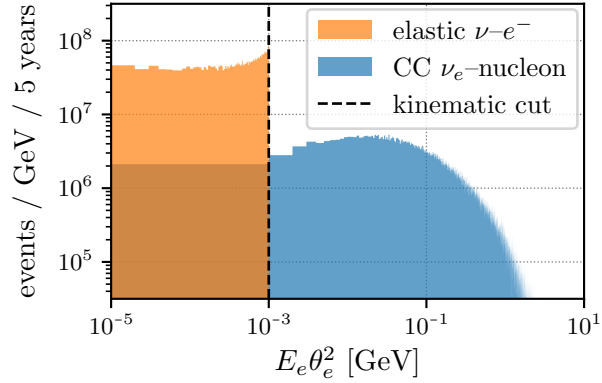

**Figure 8.4:** Expected distribution of backgrounds events as a function of the recoil electron's scattering angle in the Dark Photon Model for 5 years of measurement ($5.5 \times 10^{21}$ POT) in neutrino mode. The DM signal from $\phi$–$e^-$ scattering and the $\nu$–$e^-$ scattering background are both caused by elastic processes and thus satisfy $E_e \theta_e^2 \lesssim 2 m_e \sim 10^{-3}\,\text{GeV}$. Imposing this bound as a kinematic cut is approximately equivalent to disregarding the CC $\nu_e$–nucleon scattering background entirely.

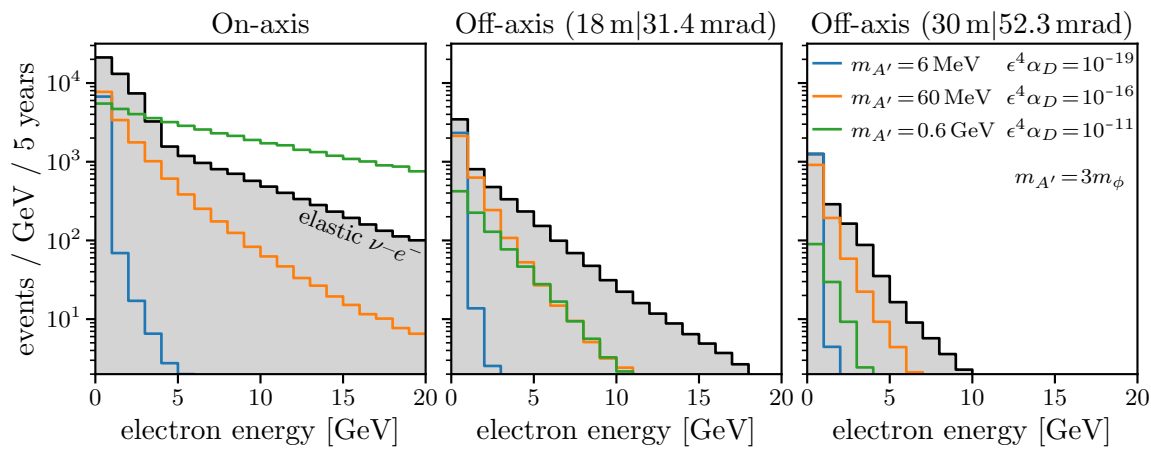

**Figure 8.5:** Expected signal and background spectra for the Dark Photon Model after 5 years of measurement ($5.5 \times 10^{21}$ POT) in neutrino mode. The colored histograms show the expected DM signal from $\phi$–$e^-$ scattering for three exemplary model parameter points. The shaded histogram represents the simulated background from elastic $\nu$–$e^-$ scattering. Comparing the spectra for the on-axis detector position (left) to the ones off-axis (center and right), we see a suppression off-axis in both signal and background, especially at high energies.

where $E_e$ and $\theta_e$ are the recoil electron's energy and scattering angle, respectively. As the expected angular resolution of *DUNE-PRISM* is sufficiently good [415], we make use of this fact and cut the backgrounds accordingly. While the contribution from elastic $\nu$–$e^-$ scattering is untouched by this kinematic cut – it also obeys Eq. (8.7) – most of the $\nu_e$–nucleon events are removed, as can be seen in Fig. 8.4. It is thus well justified to drop the nucleon scattering contribution entirely in order to speed up the calculations.

Figure 8.5 shows the distribution of the recoil electron's energy for signal and background events, comparing three different model parameter points (assuming $m_{A'} = 3m_\phi$) and three different detector positions. We find that signal and background are largest for small electron energies, and that the overall signal strength scales with $\epsilon^4 \alpha_D$, as expected. Furthermore, we observe that for larger dark photon masses, the signal falls off slower with energy but consists of fewer events in total (when comparing for constant $\epsilon^4 \alpha_D$). This is because heavier dark photons are kinematically more difficult to produce with a constant beam energy. When moving the detector off-axis, as shown in the second and third panel of the figure, the event rates of signal and background fall off quicker with energy. This can be explained by the fact that more energetic DM particles or neutrinos tend to emerge from highly boosted production events and thus preferentially show up in measurements that are conducted closer to the beam axis. Note that the total signal-to-background ratio is largest *off-axis*, whereas the spectral shapes of signal and background are most distinct *on-axis*. These competing effects become apparent in the behavior of the exclusion limits that will be presented further below.

### 8.2.3 Statistical Analysis

The goal of our analysis is the derivation of sensitivity limits on $\epsilon^4 \alpha_D$ – the quantity that controls the signal strength – based on the predicted signal and background rates. To do so, we test the signal-plus-background hypothesis against background-only by comparing a likelihood ratio to a $\chi^2$-distribution.

We start with a Poissonian likelihood function,

$$\log \mathcal{L}(\mu, \boldsymbol{X}) \equiv -\sum_{j=1}^{n_{\text{pos}}} \sum_{i=1}^{n_{\text{bins}}} \left\{ B_{ij}(\boldsymbol{X}) + S_{ij}(\mu, \boldsymbol{X}) + B_{ij}(0) \left[ \log\left(\frac{B_{ij}(0)}{B_{ij}(\boldsymbol{X}) + S_{ij}(\mu, \boldsymbol{X})}\right) - 1 \right] \right\}$$
$$- \frac{1}{2} \sum_{c=S,B} \left\{ \left(\frac{X_{c,\text{correl}}^{\text{norm}}}{\sigma_{\text{correl}}}\right)^2 + \left(\frac{X_{c,\text{correl}}^{\text{tilt}}}{\sigma_{\text{correl}}}\right)^2 + \sum_{j=1}^{n_{\text{pos}}} \left[\left(\frac{X_{c,j}^{\text{norm}}}{\sigma_{\text{uncorrel}}}\right)^2 + \left(\frac{X_{c,j}^{\text{tilt}}}{\sigma_{\text{uncorrel}}}\right)^2\right] \right\}, \quad (8.8)$$

which depends on the signal strength, $\mu \equiv \epsilon^4 \alpha_D$, and a set of nuisance parameters, $\boldsymbol{X}$. The first line of Eq. (8.8) sums over the Poissonian contributions from $n_{\text{pos}} = 7$ on-/off-axis positions as well as $n_{\text{bins}} = 80$ energy bins. The signal and background rates used in the likelihood function,

$$S_{ij}(\mu, \boldsymbol{X}) = \left[1 + X_{S,\text{correl}}^{\text{norm}}\right]\left[1 + X_{S,j}^{\text{norm}}\right]\left[1 + X_{S,\text{correl}}^{\text{tilt}}(\tfrac{2i}{n_{\text{bins}}} - 1)\right]\left[1 + X_{S,j}^{\text{tilt}}(\tfrac{2i}{n_{\text{bins}}} - 1)\right] S_{ij}(\mu, 0),$$
$$B_{ij}(\boldsymbol{X}) = \left[1 + X_{B,\text{correl}}^{\text{norm}}\right]\left[1 + X_{B,j}^{\text{norm}}\right]\left[1 + X_{B,\text{correl}}^{\text{tilt}}(\tfrac{2i}{n_{\text{bins}}} - 1)\right]\left[1 + X_{B,j}^{\text{tilt}}(\tfrac{2i}{n_{\text{bins}}} - 1)\right] B_{ij}(0),$$
$$(8.9)$$

are based on the simulated rates, $S_{ij}(\mu, 0)$ and $B_{ij}(0)$, but modified by the nuisance parameters $\boldsymbol{X}$. The latter are introduced to account for various systematic uncertainties, which are modeled as Gaussian and also contribute to the likelihood function, see the second line of Eq. (8.8). We include nuisance parameters for normalization (superscript "norm") and spectral tilt ("tilt") uncertainties in the signal (subscript "$S$") and background ("$B$"). We furthermore distinguish between uncertainties that are uncorrelated (subscript $j = 1, \ldots, n_{\text{pos}}$) as well as correlated ("correl") between all on-/off-axis positions. Similar as in Ref. [454], we assume uncorrelated (correlated) uncertainties of $\sigma_{\text{uncorrel}} = 1\%$ ($\sigma_{\text{correl}} = 10\%$). The fact that $\sigma_{\text{uncorrel}} \ll \sigma_{\text{correl}}$ makes the *DUNE-PRISM* concept potentially superior to fixed-position measurements, as we will see.

With the likelihood function at hand, the statistical significance for the exclusion of a particular $\mu$ can be calculated via [470]

$$Z(\mu) \equiv -2 \log\left(\frac{\mathcal{L}(\mu, \hat{\hat{\boldsymbol{X}}})}{\mathcal{L}(\hat{\mu}, \hat{\boldsymbol{X}})}\right). \quad (8.10)$$

In this expression, $\hat{\hat{\boldsymbol{X}}}$ are the nuisance parameters that maximize the likelihood for fixed $\mu$, whereas $(\hat{\mu}, \hat{\boldsymbol{X}})$ is the combination that maximizes it globally. While obtaining $\hat{\hat{\boldsymbol{X}}}$ is a computationally intensive step, we trivially find $(\hat{\mu}, \hat{\boldsymbol{X}}) = \boldsymbol{0}$ for the likelihood function in use. To finally obtain the sensitivity limits for given $m_{A'}$ and $m_\phi$, we vary $\mu = \epsilon^4 \alpha_D$ until $Z(\mu)$ matches the 90% quantile of a $\chi^2$-distribution for one degree of freedom, $Z \approx \sqrt{2.71}$.

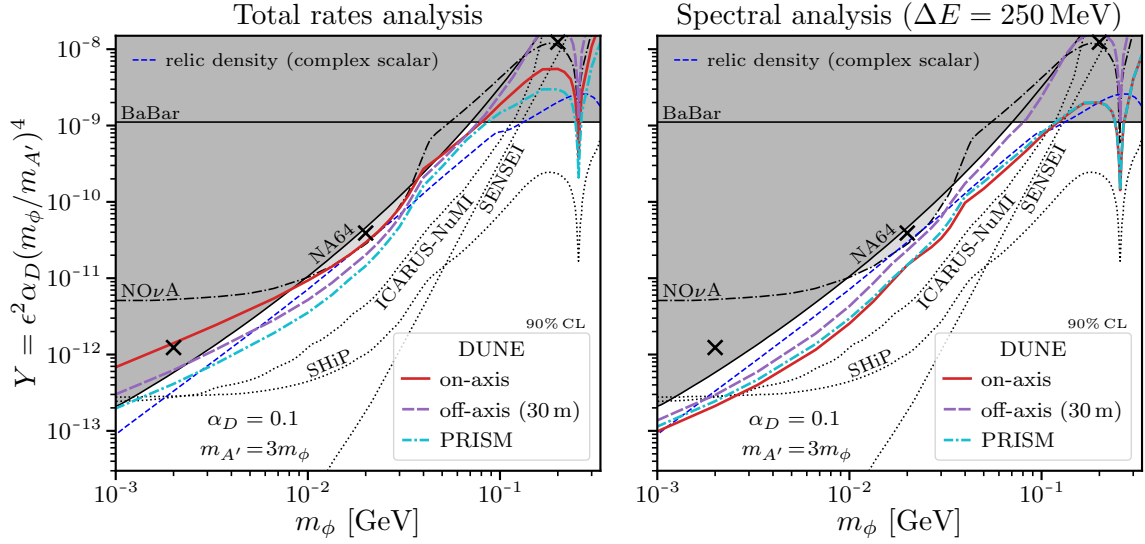

**Figure 8.6:** Expected upper exclusion limits for the Dark Photon Model after 5 years ($5.5 \times 10^{21}$ POT) in neutrino mode, assuming $\alpha_D = 0.1$ and $m_{A'} = 3m_\phi$. The red solid and purple dashed lines indicate the sensitivity of fixed-position measurements on-axis and 30 m (52.3 mrad) off-axis, respectively. The cyan dot-dashed line shows the sensitivity of the *DUNE-PRISM* strategy, where data is taken at seven different positions. The left panel corresponds to an analysis using total event rates, while the right panel includes spectral information. Gray shaded regions and dotted lines indicate exclusions from existing and future experiments, respectively. The observed DM relic density would be generated via thermal freeze-out along the blue dashed line. Black crosses mark the exemplary model parameter points presented in Fig. 8.5.

### 8.2.4 Sensitivity Projection

In the last section we explained how the simulated signal and background rates can be translated into an experimental sensitivity as a function of the signal strength $\epsilon^4 \alpha_D$. We present the projected exclusion limits for 5 years of measurement in terms of the quantity $Y \equiv \epsilon^2 \alpha_D (m_\phi/m_{A'})^4$ in Fig. 8.6. In this figure, we compare three different strategies: all data taken on-axis, all data taken 30 m (52.3 mrad) off-axis, and data taken consecutively at seven different positions (*DUNE-PRISM*).

In the left panel we present a total rate analysis, i.e. with $n_\text{bins} = 1$ in Eq. (8.8). In agreement to Ref. [454], we find that the *DUNE-PRISM* strategy is advantageous compared to fixed-position measurements. This is due to the increased off-axis signal-to-background ratio, but also thanks to the combination of different positions which reduces correlated uncertainties. The situation is however different if we take the full spectral information into account, i.e. with $n_\text{bins} = 80$ corresponding to 250 MeV-wide energy bins from 0 to 20 GeV, as presented in the right panel of Fig. 8.6. In this case, the best sensitivity is achieved by an on-axis-only measurement, closely followed by the *DUNE-PRISM* strategy. The reason for this behavior lies in the spectral shapes (see Fig. 8.5):

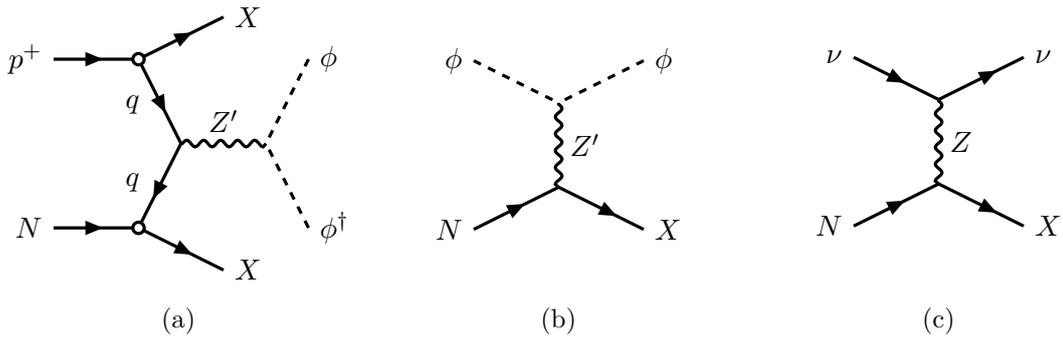

**Figure 8.7:** Drell–Yan-like DM production channel (a), DM detection channel (b), and the considered background NC DIS (c) for the Leptophobic Model. $N$ represents a nucleon in the target material of the neutrino facility and the $X$ sum up the hadronic final states.

On-axis, the signal has a very pronounced high-energy tail, whereas off-axis, most events are captured by the lowest energy bins. The on-axis measurement therefore profits most from the enhanced statistical power of a spectral analysis.

The typical shape of the exclusion curves is due to the fact that for larger $m_\phi$, fewer $\phi$ can be produced, reducing the signal rate and thereby the sensitivity. The characteristic spike at $m_\phi \sim 260\,\text{MeV}$ is caused by the $\rho$ resonance, i.e. the occurrence of maximal mixing between dark photons and $\rho$ vector bosons (which share the same quantum numbers) at $m'_A = 3m_\phi \sim 780\,\text{MeV} \sim m_\rho$.

We compare the *DUNE* sensitivity to existing limits from *BaBar* [429], *NA64* [430], unofficial recasts of *NuMI* off-axis data from *NOνA* [446], the expected sensitivities of *ICARUS-NuMI* off-axis [447], *SHiP* [451], and *SENSEI* [95, 471].[3] We see that *DUNE-PRISM* can probe new regions of parameter space, with a sensitivity that is up to half an order of magnitude below existing constraints for some $m_\phi$. Note that the values of $Y$ and $m_\phi$ that lead to the observed DM relic density in a standard thermal scenario (blue dashed curve in Fig. 8.6) will be almost entirely covered by *DUNE-PRISM* in the surveyed mass range.

## 8.3 Leptophobic Dark Matter

As a second scenario we consider a complex scalar DM candidate $\phi$ and a new gauge boson $Z'$ with gauge coupling $g_Z$. As opposed to the dark photon, which we assumed to be connected to the SM via kinetic mixing, $Z'$ does directly couple to visible matter. However, we consider the scenario where only the quarks are charged under the new gauge symmetry and therefore call $\phi$ "leptophobic". Note that the new local symmetry effectively gauges baryon number and will thus be denoted by $U(1)_B$. A spontaneous breaking of this symmetry results in baryon number violation and can be part of theories on baryogenesis. The Lagrangian of the Leptophobic Model reads

$$\mathcal{L} \supset ig_Z z_\phi Z'^\mu J_\mu^\phi + (\partial_\mu \phi^\dagger)(\partial^\mu \phi) - m_\phi^2 \phi^\dagger \phi + g_Z z_q \sum_q \bar{q} \gamma^\mu q \, Z'_\mu \,, \qquad (8.11)$$

where $J_\mu^\phi = [(\partial_\mu \phi^\dagger)\phi - \phi^\dagger(\partial_\mu \phi)]$ and $z_\phi$, $z_q$ are the $U(1)_B$ charges of $\phi$ and the quarks, respectively. While the lack of leptonic interactions makes this model hard to probe experimentally, it can still be constrained by *DUNE*, as our analysis will show. Note that theories with $U(1)$-charged fermions suffer from chiral anomalies, i.e. a symmetry violation that arises from quantum corrections and that would be unphysical in presence of the gauge symmetry [448, 449]. The anomaly can be cured by adding a specific amount of extra fermions to the model [472–475]. However, we will not include these, as we do not specify a UV completion of the model.

---

[3] We have rescaled the *ICARUS-NuMI* limit to an integrated luminosity of $2.5 \times 10^{21}$ POT, corresponding to 5 years of *NuMI* runtime at the nominal beam power of 700 kW. Furthermore, note that the strong expected *SENSEI* limit applies to scalar DM only, while we expect the *DUNE-PRISM* limits to remain similar for fermion DM.

### 8.3.1 Dark Matter Production and Detection

For the Leptophobic Model, we will consider DM production via the Drell–Yan-like process shown in Fig. 8.7a, which dominates in the mass regime $m_{Z'} \gtrsim 2\,\text{GeV}$. In this production channel, a beam proton hits a nucleus of the graphite target, producing a $Z'$ which immediately decays into DM. Due to $\phi$'s leptophobic nature, it may only be detected via hadronic interactions in the near detector. We will focus on the signature of neutral current (NC) deep inelastic scattering (DIS) on nuclei, which is the most relevant contribution for multi-GeV energies. The signal strength of leptophobic DM scales as $g_Z^6$, with two powers from the production and four powers from the detection via $\phi$–nucleon scattering (Fig. 8.7b).

The DM production can be described by standard methods of perturbative quantum chromodynamics (QCD). For each proton that hits the target, there will be

$$\frac{N_\phi}{N_{\text{POT}}} = \frac{A}{A^{0.71}} \frac{\sigma_{pN \to \phi\phi^\dagger}}{\sigma_{pN,\text{tot}}} \qquad (8.12)$$

outgoing DM particles, where $A = 12$ is the mass number of the target material carbon. We assume that the cross section for $\phi\phi^\dagger$ production scales linearly with $A$, while we use an effective scaling $\propto A^{0.71}$ for the total proton–carbon scattering cross section [476]. For the total proton–nucleon cross section, we adopt the approximation $\sigma_{pN,\text{tot}} \approx 40\,\text{mb}$ [477] and use MADDUMP as done in Ref. [464] to obtain the DM production cross section numerically. In this step, we make use of leading-order parton distribution functions (PDFs) with a factorization scale equal to $m_{Z'}$. As a benchmark point for our analysis, we choose a DM charge and mass of $z_\phi = 3$ and $m_\phi = 750\,\text{MeV}$, respectively, and vary $m_{Z'}$ between $2\,\text{GeV}$ and $7\,\text{GeV}$. With this choice of parameters, almost all $Z'$ decay into DM [478].

The conservation of angular momentum in the DM production process implies the scaling relation $\mathrm{d}\sigma_{pN \to \phi\phi^\dagger}/\mathrm{d}\theta \propto \sin^3\theta$ in the center-of-mass frame, where $\theta$ denotes the DM scattering angle relative to the proton beam axis [478]. As a consequence, we observe an improved geometric acceptance at larger angles.[4] This makes off-axis measurements particularly well-suited for constraining leptophobic DM. For this reason, we also consider a hypothetical detector location at a distance of 60 m (104.2 mrad) from the beam axis. This angle resembles the *ICARUS-NuMI* detector setup and, according to Ref. [433], corresponds to an almost optimal sensitivity. If hints for leptophobic DM would be found in *DUNE-PRISM*, it might be worthwhile to expand the *DUNE* near detector and conduct even farther displaced off-axis measurements. Table 8.1 lists the main parameters of the different *DUNE* and *ICARUS-NuMI* configurations under consideration. We simulate *ICARUS-NuMI* employing the *NuMI* flux from Ref. [479].

| Experiment | Baseline [m] | Mass [t] | Off-axis [mrad] | Exposure [POT] | Background [events] |
|---|---|---|---|---|---|
| *DUNE* on-axis | 574 | 68 | 0 | $5.5 \times 10^{21}$ | $2.6 \times 10^7$ |
| *DUNE* @ 30 m | 574 | 68 | 52.3 | $5.5 \times 10^{21}$ | $2.9 \times 10^5$ |
| *DUNE* @ 60 m | 574 | 68 | 104.2 | $5.5 \times 10^{21}$ | 3950 |
| *ICARUS-NuMI* | 789 | 480 | ~100 | $2.5 \times 10^{21}$ | 1600 |

**Table 8.1:** Main parameters of the different considered experimental setups, assuming 120 GeV proton beam energy and 5 years of exposure. The number of background events is determined after a cut on the visible energy, $E_{\text{vis}} > 3\,\text{GeV}$, and is significantly reduced at far off-axis positions.

### 8.3.2 Backgrounds

As discussed, the signature of leptophobic DM is that of DIS on nuclei, *without* distinctive charged leptons in the final state. The relevant background events are thus likewise of hadronic nature, i.e. $\nu$–nucleon scattering. Similar as for the Dark Photon Model discussed in Section 8.2, we simulate the backgrounds with GENIE and reweight the events according to the expected neutrino fluxes.[5] We will cut all signal and background events with a total visible energy deposit smaller than 3 GeV. This removes backgrounds involving resonant scattering, quasi-elastic scattering, as well as coherent scattering. The only remaining non-negligible background is $\nu$–nucleon NC DIS (Fig. 8.7). Note that CC DIS may also lead to signal-like signatures, in case of a final-state charged lepton that is misidentified as a part of the hadronic system. Due to the similar size of the NC and CC cross sections and a misidentification probability of $\ll 1$, we expect the CC background to be subdominant and hence neglect it.

Figure 8.8 shows the distribution of the total visible energy for signal and background events, comparing three different model parameter points (assuming $z_\phi = 3$, $m_\phi = 750\,\text{MeV}$) and three different detector positions. For the signal, we observe a similar behavior as in the Dark Photon Model: For larger masses of the force mediator $Z'$, there are fewer events in total which are however more spread out towards high energies, as expected for kinematic reasons. Off-axis, both the signal and the background fall off quicker with energy because production processes of energetic particles are boosted in the forward direction. Furthermore, off-axis, the total number of background events is highly suppressed, whereas the signal becomes even stronger due to the increased geometric acceptance (see discussion in Section 8.3.1). The combination of the aforementioned effects grants us a highly improved signal-to-background ratio in off-axis measurements, especially for large $m_{Z'}$.

---

[4] This applies only to the upper half of the considered $m_{Z'}$ range. In the case of lighter mediators, the DM beam is more strongly boosted in the forward direction and the off-axis advantage disappears.

[5] For our analysis at the 60 m (104.2 mrad) off-axis position, the neutrino fluxes provided by Ref. [467] were unusable due to poor statistics. We hence computed the fluxes based on the raw data provided by the same reference.

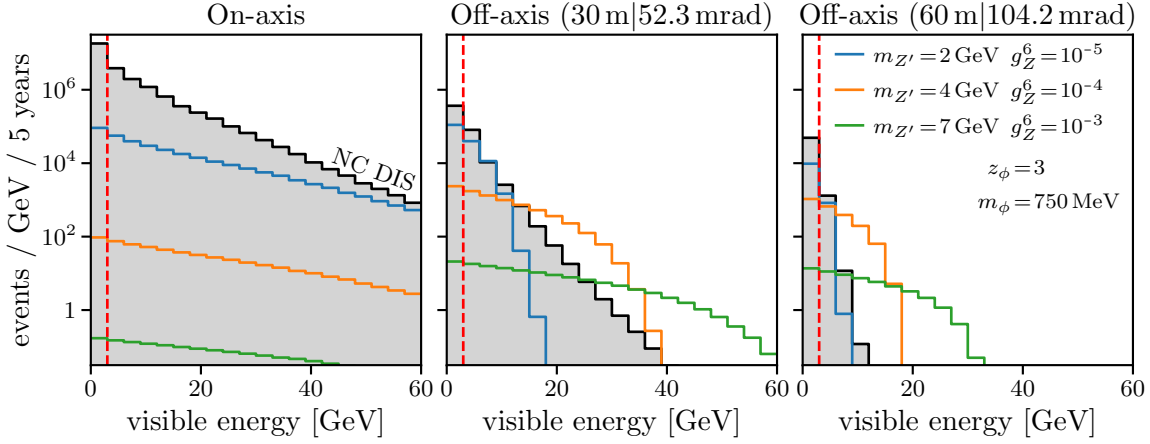

**Figure 8.8:** Expected signal and background spectra for the Leptophobic Model after 5 years of measurement ($5.5 \times 10^{21}$ POT) in neutrino mode. The colored histograms show the expected DM signal from $\phi$–nucleon scattering for three exemplary model parameter points. The shaded histogram represents the simulated background from $\nu$–nucleon NC DIS. Comparing the spectra for the on-axis detector position (left) to the ones off-axis (center and right), we see a suppression off-axis in the background, especially at high energies. The signal, on the other hand, is overall enhanced off-axis due to the increased geometric acceptance. The red dashed line indicates the applied lower kinematic cut.

### 8.3.3 Sensitivity Projection

To estimate *DUNE*'s sensitivity to leptophobic DM, we apply the same statistical procedure as for the Dark Photon Model, with $g_Z^6$ as the signal strength parameter. In the left panel of Fig. 8.9 we present the results of our total rate analysis, i.e. with $n_{\text{bins}} = 1$. As for the Dark Photon Model, the *DUNE-PRISM* setup yields the strongest constraints on leptophobic DM, which can be attributed to the increased off-axis signal-to-background ratio and the combination of different positions which reduces correlated uncertainties. The right panel of the figure shows an analysis including spectral information, i.e. with $n_{\text{bins}} = 57$ corresponding to 1 GeV-wide energy bins from 3 to 60 GeV. The spectral analysis turns out to be highly beneficial for all configurations. Contrary to the situa-tion with the Dark Photon Model, the on-axis measurement is now *not* the best option when performing a spectral analysis: The long high-energy tail of the on-axis signal is still advantageous to the sensitivity, but this effect is outweighed by the increased geometric acceptance off-axis. As a result, the 30 m (52.5 mrad) off-axis measurement and the *DUNE-PRISM* strategy yield the strongest constraints, superseding existing limits from invisible $J/\psi$ and $\Upsilon$ decays at *BaBar* [431, 432] for masses around $m_{Z'} \sim 2$ GeV and $m_{Z'} \sim 4$ GeV.

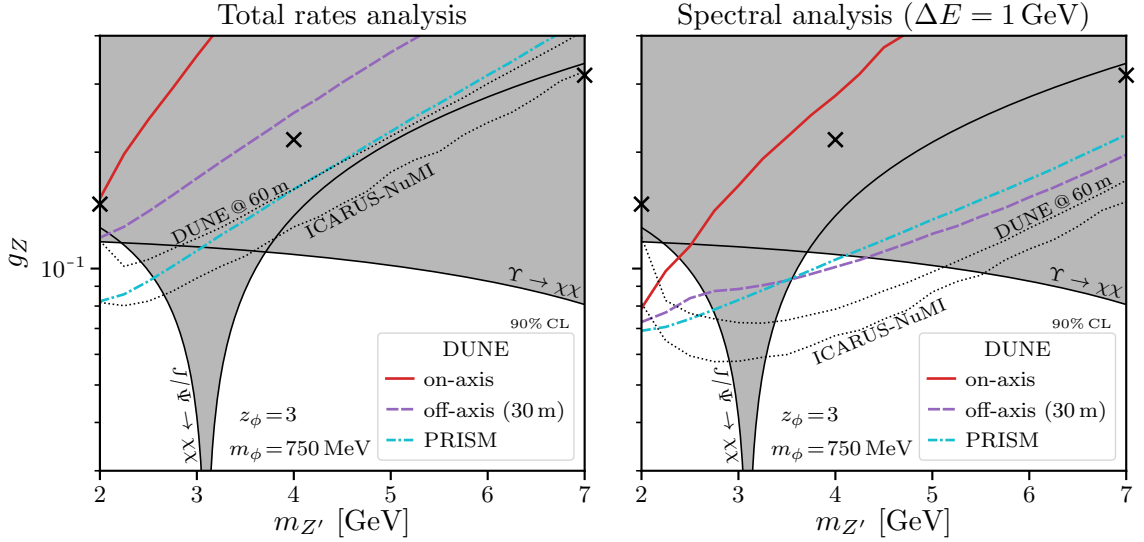

**Figure 8.9:** Expected upper exclusion limits for the Leptophobic Model after 5 years ($5.5 \times 10^{21}$ POT) in neutrino mode, assuming DM with a $U(1)_B$ charge of $z_\phi = 3$ and a mass of $m_\phi = 750$ MeV. The red solid and purple dashed lines indicate the sensitivity for a fixed-position measurements on-axis and 30 m (52.3 mrad) off-axis, respectively. The cyan dot-dashed line shows the sensitivity of the *DUNE-PRISM* strategy, where data is taken at seven different positions. We additionally present dotted exclusion curves that could be obtained at a hypothetical 60 m (104.2 mrad) off-axis position at *DUNE* and for a $\sim 100$ mrad measurement at *ICARUS-NuMI*. The left panel corresponds to an analysis using total event rates, while the right panel includes full spectral information. Gray shaded regions indicate exclusions from existing experiments. Black crosses mark the exemplary model parameter points presented in Fig. 8.8.

Also in Fig. 8.9, we present the sensitivities that could be achieved at a hypothetical 60 m (104.2 mrad) off-axis position at *DUNE* and for a $\sim 100$ mrad measurement at *ICARUS-NuMI* as dotted curves. Both scenarios are superior to any *DUNE* measurements closer to the beam axis, with *ICARUS-NuMI* yielding the strongest constraints. The superior sensitivity of *ICARUS-NuMI* is owed to its large fiducial mass of 480 t (vs. 68 t in case of *DUNE*) and its softer off-axis neutrino fluxes.

## 8.4 Conclusions

In this chapter, we investigated the possibilities for new physics searches at the *DUNE* near detector. Our focus was on the *DUNE-PRISM* concept – measurements conducted at different positions on and off the beam axis – and we found that this setup is beneficial to the constraining power regarding different new physics scenarios.

In our analysis, we determined the expected signal and background rates based on publicly available neutrino flux predictions, applied appropriate kinematic cuts, and derived the expected sensitivity of different *DUNE* near detector configurations. For all projections, we assumed 5 years of measurement ($5.5 \times 10^{21}$ POT) in neutrino mode.

For scalar DM with MeV to GeV masses in context of the Dark Photon Model, we found that the expected signal and background rates drop significantly when moving the detector away from the beam axis. The signal-to-background ratio greatly increases off axis. Therefore, we observe a great sensitivity enhancement for *DUNE-PRISM* in a total rate analysis, exceeding existing constraints on the kinetic mixing parameter. The situation is different when spectral information is taken into account, where the distinctness of the signal and background distributions adds statistical power. This improves the on-axis sensitivity, which in this case is on par with the *DUNE-PRISM* projection.

Similar to the Dark Photon Model, in the Leptophobic Model the neutrino scattering background rates are suppressed off-axis as well. However, the number of signal events for a relatively heavy leptophobic gauge boson *increases* off-axis due to the geometric behavior of the scattering cross section. As a consequence, the *DUNE-PRISM* strategy yields the highest sensitivity in both the total rates analysis and the spectral analysis. In the latter case, existing limits on the leptophobic gauge coupling will be surpassed for gauge boson masses around 2 and 4 GeV. A measurement at a hypothetical 60 m off-axis position would result in an even better sensitivity.

The *DUNE-PRISM* strategy neither interferes with the original purpose of the *DUNE* near detector nor requires additional construction or equipment. We showed that it is well-suited for the exploration of different DM models, with notable sensitivity improvements over existing experiments.

# Epilogue

# 9 Summary and Outlook

The first part of this thesis was dedicated to cosmological first-order PT, their rich phenomenology and novel applications. In Chapter 2, we reviewed the physics of first-order phase transitions and discussed the stochastic GW background that is produced during the collisions of true-vacuum bubbles. We explored the PT parameter space in terms of GW detectability and illustrated that strong and slow transitions are most promising in this regard. PTs occurring in different epochs during the Big Bang produce GWs across a wide frequency range, which will be covered by (future) earth-bound and space-based observatories as well as PTAs. In Chapter 3, we reviewed the effective scalar potential where we discussed the most important contributions and pointed out the ingredient that can induce a first-order transition: a potential barrier, which may be a part of the theory at tree level or induced via quantum corrections in a gauged theory.

In Chapter 4 we proposed a new mechanism that explains the observed BAU and the DM abundance simultaneously. We made use of the Filtered DM scenario, which is based on a special class of PTs with large order parameter. During a dark sector PT, a fermion species gains a mass that is large compared to the temperature. As a consequence, the bulk of the thermal distribution is "filtered out", leaving behind a tiny fraction of DM particles which immediately freezes out and constitutes the abundance observed today. To achieve baryogenesis, we proposed a mechanism similar to conventional EWBG, but which occurs in context of the same dark PT as in the Filtered DM mechanism. A CP-violating DM Yukawa coupling generates a chiral DM asymmetry in front of the moving bubble walls. A portal operator – at EFT dimensions 6 or 8 – transfers the chiral asymmetry into a SM lepton asymmetry, which freezes out in the true vacuum thanks to the large order parameter. Electroweak sphalerons finally convert the leptons into baryons, which constitute the BAU observed today. By our numerical analyses, we found that sizable portions of the model parameter space can simultaneously explain the BAU and the DM abundance in case of the dimension-6 portal, requiring $\mathcal{O}(1)$ DM Yukawa couplings and a DM mass $30 \sim 60$ times larger than the temperature of the Universe. The combined mechanism is viable for any DM mass above $\sim 500\,\text{TeV}$ (exceeding the Griest–Kamionkowski bound) and can be tested by future direct detection experiments for a DM mass up to $\sim 2\,\text{PeV}$.

We employed the same class of PTs (with a large order parameter) to construct and test a PBH production mechanism in Chapter 5. The key difference to the above scenario is the DM Yukawa coupling, which needs to be tiny in this case. As a consequence, the particles that are reflected off the walls accumulate in the remaining and shrinking false-vacuum pockets, where they eventually become so dense that they form BHs by fulfilling the Schwarzschild criterion. A detailed formulation of the problem in terms of a Boltzmann equation allowed us to test our scenario in a sophisticated simulation of the phase space. We found that during the PT the squeezed particles must gain a mass of at least ten times the temperature, that the false-vacuum pockets must initially be as large as a few Hubble volumes, and that a large range of tiny Yukawa couplings is viable. The mechanism is independent of the PT temperature and can correspondingly produce PBHs over a wide range of masses. BHs with around $10^{-13}$ solar masses could possibly constitute the entirety of DM.

In Chapter 6 we investigated light and cold dark sectors which contain sub-MeV DM candidates. Thermal DM in this regime is usually subject to tight constraints imposed by BBN and the CMB. However, if the hidden sector is decoupled from the SM thermal bath and colder by an $\mathcal{O}(1)$ factor, the constraints can be evaded for models with a small number of DOFs. We explicitly constructed two minimal toy models that feature first-order PTs and hence give rise to a stochastic GW background in the course of bubble collisions. The fact that the hidden sector is colder (and hence less energetic) makes the GWs weaker and harder to detect. For both models there are portions of parameter space where the cosmological constraints are satisfied and a GW spectrum is generated that would be detectable by the future PTA *SKA*.

In the second part of this thesis, we discussed the *DUNE* experiment, whose components and scientific goals we reviewed briefly in Chapter 7. In Chapter 8 we computed the anticipated sensitivity of the *DUNE* near detector regarding scalar DM with masses in the MeV to few-GeV range in context of the Dark Photon Model and the Leptophobic Model. We found that the *DUNE-PRISM* concept – which suggests consecutive measurements at different on- and off-axis positions – yields similar or superior sensitivities compared to single-position measurements, depending on the DM mass and the assumed energy resolution. This new physics search program can be conducted independently from the main purpose of the near detector.

<div align="center">* * *</div>

The present thesis has made the attempt to explore new avenues in the fields of cosmology and particle physics and tried to contribute to the enormous efforts undertaken by the science community in order to solve the mysteries of our Universe. We hope that our ideas and findings inspire other scientists and serve as a basis for further theoretical considerations. It will be exciting to see the outcome of future experiments and the implications for our proposals. The quest for the fundamental truths of nature might seem arduous and long-winded, but I am convinced that the effort is worthwhile and will eventually lead to success.

# Danksagung

Zunächst möchte ich mich bei meinem Doktorvater Joachim Kopp für die außergewöhnlich intensive und persönliche Betreuung, sowie für die Möglichkeit zur Promotion in diesem spannenden Bereich der Physik bedanken. Auch allen anderen, die an unseren umfangreichen und fordernden Projekten mitgewirkt haben, möchte ich für die angenehme Kooperation danken. Die endlosen Nachmittage, die ich gemeinsam mit Lukas Mittnacht der Fehlersuche in Formeln und Programmcode gewidmet habe, waren letzten Endes sehr erfolgreich, wie ich finde. Ich danke Lukas Mittnacht, Matthias Heller, Wolfram Ratzinger und Felix Schlapp für das Korrekturlesen dieser Arbeit sowie Eric Madge für nützliche Tipps technischer Art. Die Atmosphäre in der THEP-Arbeitsgruppe habe ich immer als sehr angenehm empfunden und ich möchte mich bei all ihren Mitgliedern für die schöne Zeit bedanken. Zu guter Letzt danke ich meinen Freunden, meiner Familie, meiner Partnerin Alexandra, und meiner wundervollen Tochter Leonie für die liebevolle moralische Unterstützung.

# List of Abbreviations

**ΛCDM**  Lambda-cold-dark-matter Theory

**ALP**  axion-like particle

**BAO**  baryon acoustic oscillations

**BAU**  baryon asymmetry of the Universe

**BBN**  Big Bang nucleosynthesis

**BH**  black hole

**C**  charge symmetry

**CB**  compact binary

**CC**  charged current

**CL**  confidence level

**CMB**  cosmic microwave background

**CP**  charge-parity symmetry

**CPT**  charge-parity-time symmetry

**DIS**  deep inelastic scattering

**DM**  dark matter

**DOF**  degree of freedom

**EFT**  effective field theory

**EWBG**  electroweak baryogenesis

**EWPT**  electroweak phase transition

**GUT**  grand unified theory

**GW**  gravitational wave

**IR**  infrared

**LSS**  large-scale structure

**MACHO** massive astrophysical compact halo object

**MOND** modified Newtonian dynamics

**NC** neutral current

**ODE** ordinary differential equation

**P** parity symmetry

**PBH** primordial black hole

**PDE** partial differential equation

**PDF** parton distribution function

**POT** protons on target

**PT** phase transition

**PTA** pulsar timing array

**QCD** quantum chromodynamics

**QED** quantum electrodynamics

**QFT** quantum field theory

**SM** Standard Model of Particle Physics

**SMBH** supermassive black hole

**SMBHB** supermassive black hole binary

**SNR** signal-to-noise ratio

**TPC** time projection chamber

**UV** ultraviolet

**VEV** vacuum expectation value

**WIMP** weakly interacting massive particle

# List of Experiments and Collaborations

**BaBar** $B$–$\bar{B}$ Experiment

**B-DECIGO** Basic *DECIGO*

**BBO** Big Bang Observer

**DECIGO** Decihertz Interferometer Gravitational-Wave Observatory

**DUNE** Deep Underground Neutrino Experiment

**DUNE-PRISM** *DUNE* Precision Reaction-Independent Spectrum Measurement

**EPTA** European Pulsar Timing Array

**ET** Einstein Telescope

**ICARUS** Imaging Cosmic And Rare Underground Signals

**KAGRA** Kamioka Gravitational-Wave Detector

**LHC** Large Hadron Collider

**LIGO** Laser Interferometer Gravitational-Wave Observatory

**LISA** Laser Interferometer Space Antenna

**LZ** Large Underground Xenon & Zoned Proportional Scintillation in Liquid Noble Gases Experiment

**NA64** North Area 64 Experiment

**NANOGrav** North American Nanohertz Observatory for Gravitational Waves

**NOνA** *NuMI* Off-Axis $\nu_e$ Appearance Experiment

**NuMI** Neutrinos at the Main Injector Experiment

**OGLE** Optical Gravitational Lensing Experiment

**Planck** Planck Satellite

**SENSEI** Sub-Electron-Noise Skipper-CCD Experimental Instrument

**SHiP** Search for Hidden Particles

**SKA** Square Kilometre Array

**T2K** Tokai to Kamioka Experiment

**Virgo** Virgo Gravitational-Wave Detector

**XENONnT** Xenon n-Ton Dark Matter Project

# Bibliography

[1] M.J. Baker, M. Breitbach, J. Kopp, L. Mittnacht, and Y. Soreq, *Filtered Baryogenesis*, *JHEP* **08** (2022) 010 [2112.08987].

[2] M.J. Baker, M. Breitbach, J. Kopp, and L. Mittnacht, *Primordial Black Holes from First-Order Cosmological Phase Transitions*, [2105.07481].

[3] M.J. Baker, M. Breitbach, J. Kopp, and L. Mittnacht, *Detailed Calculation of Primordial Black Hole Formation During First-Order Cosmological Phase Transitions*, [2110.00005].

[4] M. Breitbach, J. Kopp, E. Madge, T. Opferkuch, and P. Schwaller, *Dark, Cold, and Noisy: Constraining Secluded Hidden Sectors with Gravitational Waves*, *JCAP* **07** (2019) 007 [1811.11175].

[5] M. Breitbach, L. Buonocore, C. Frugiuele, J. Kopp, and L. Mittnacht, *Searching for Physics Beyond the Standard Model in an Off-Axis DUNE Near Detector*, *JHEP* **01** (2022) 048 [2102.03383].

[6] Y. Aoki, G. Endrodi, Z. Fodor, S.D. Katz, and K.K. Szabo, *The Order of the quantum chromodynamics transition predicted by the standard model of particle physics*, *Nature* **443** (2006) 675 [hep-lat/0611014].

[7] T. Bhattacharya et al., *QCD Phase Transition with Chiral Quarks and Physical Quark Masses*, *Phys. Rev. Lett.* **113** (2014) 082001 [1402.5175].

[8] M. Claudson, E. Farhi, and R.L. Jaffe, *The Strongly Coupled Standard Model*, *Phys. Rev. D* **34** (1986) 873.

[9] C.S. Wu, E. Ambler, R.W. Hayward, D.D. Hoppes, and R.P. Hudson, *Experimental Test of Parity Conservation in $\beta$ Decay*, *Phys. Rev.* **105** (1957) 1413.

[10] J. Goldstone, A. Salam, and S. Weinberg, *Broken Symmetries*, *Phys. Rev.* **127** (1962) 965.

[11] K. Kajantie, M. Laine, K. Rummukainen, and M.E. Shaposhnikov, *The Electroweak phase transition: A Nonperturbative analysis*, *Nucl. Phys. B* **466** (1996) 189 [hep-lat/9510020].

[12] K. Kajantie, M. Laine, K. Rummukainen, and M.E. Shaposhnikov, *Is there a hot electroweak phase transition at $m_H \gtrsim m_W$?*, *Phys. Rev. Lett.* **77** (1996) 2887 [hep-ph/9605288].

[13] M. Gurtler, E.-M. Ilgenfritz, and A. Schiller, *Where the electroweak phase transition ends*, *Phys. Rev. D* **56** (1997) 3888 [hep-lat/9704013].

[14] Z. Fodor, F. Csikor, J. Heitger, Y. Aoki, and A. Ukawa, *End point of the electroweak phase transition*, in *3rd International Conference on Strong and Electroweak Matter*, pp. 190–195, 12, 1998 [`hep-ph/9901307`].

[15] K. Rummukainen, M. Tsypin, K. Kajantie, M. Laine, and M.E. Shaposhnikov, *The Universality class of the electroweak theory*, *Nucl. Phys. B* **532** (1998) 283 [`hep-lat/9805013`].

[16] F. Csikor, Z. Fodor, and J. Heitger, *Endpoint of the hot electroweak phase transition*, *Phys. Rev. Lett.* **82** (1999) 21 [`hep-ph/9809291`].

[17] N. Cabibbo, *Unitary Symmetry and Leptonic Decays*, *Phys. Rev. Lett.* **10** (1963) 531.

[18] M. Kobayashi, and T. Maskawa, *CP Violation in the Renormalizable Theory of Weak Interaction*, *Prog. Theor. Phys.* **49** (1973) 652.

[19] Super-Kamiokande collaboration, *Evidence for oscillation of atmospheric neutrinos*, *Phys. Rev. Lett.* **81** (1998) 1562 [`hep-ex/9807003`].

[20] B. Pontecorvo, *Inverse beta processes and nonconservation of lepton charge*, *Zh. Eksp. Teor. Fiz.* **34** (1957) 247.

[21] Z. Maki, M. Nakagawa, and S. Sakata, *Remarks on the unified model of elementary particles*, *Prog. Theor. Phys.* **28** (1962) 870.

[22] Particle Data Group collaboration, *Review of Particle Physics*, *PTEP* **2020** (2020) 083C01.

[23] I. Esteban, M.C. Gonzalez-Garcia, M. Maltoni, T. Schwetz, and A. Zhou, *The fate of hints: updated global analysis of three-flavor neutrino oscillations*, *JHEP* **09** (2020) 178 [`2007.14792`].

[24] D. Hanneke, S.F. Hoogerheide, and G. Gabrielse, *Cavity Control of a Single-Electron Quantum Cyclotron: Measuring the Electron Magnetic Moment*, *Phys. Rev. A* **83** (2011) 052122 [`1009.4831`].

[25] J. Woithe, G.J. Wiener, and F.F. Van der Veken, *Let's have a coffee with the Standard Model of particle physics!*, *Phys. Educ.* **52** (2017) 034001.

[26] G. Altarelli, *The Higgs and the Excessive Success of the Standard Model*, *Frascati Phys. Ser.* **58** (2014) 102 [`1407.2122`].

[27] R.J. Adler, B. Casey, and O.C. Jacob, *Vacuum catastrophe: An Elementary exposition of the cosmological constant problem*, *Am. J. Phys.* **63** (1995) 620.

[28] N. Arkani-Hamed, S. Dimopoulos, and G.R. Dvali, *The Hierarchy problem and new dimensions at a millimeter*, *Phys. Lett. B* **429** (1998) 263 [`hep-ph/9803315`].

[29] S. Alekhin, A. Djouadi, and S. Moch, *The top quark and Higgs boson masses and the stability of the electroweak vacuum*, *Phys. Lett. B* **716** (2012) 214 [`1207.0980`].

[30] H. Georgi, and S.L. Glashow, *Unity of All Elementary Particle Forces*, *Phys. Rev. Lett.* **32** (1974) 438.

[31] C. Rovelli, *Quantum gravity*, *Scholarpedia* **3** (2008) 7117.

[32] T. Mannel, *Theory and phenomenology of CP violation*, Nucl. Phys. B Proc. Suppl. **167** (2007) 115.

[33] P.J.E. Peebles, and B. Ratra, *The Cosmological Constant and Dark Energy*, Rev. Mod. Phys. **75** (2003) 559 [`astro-ph/0207347`].

[34] Muon g-2 collaboration, *Measurement of the Positive Muon Anomalous Magnetic Moment to 0.46 ppm*, Phys. Rev. Lett. **126** (2021) 141801 [`2104.03281`].

[35] R.H. Cyburt, B.D. Fields, and K.A. Olive, *Primordial nucleosynthesis in light of WMAP*, Phys. Lett. B **567** (2003) 227 [`astro-ph/0302431`].

[36] Planck collaboration, *Planck 2018 results. VI. Cosmological parameters*, Astron. Astrophys. **641** (2020) A6 [`1807.06209`], [Erratum: Astron.Astrophys. 652, C4 (2021)].

[37] L. Canetti, M. Drewes, and M. Shaposhnikov, *Matter and Antimatter in the Universe*, New J. Phys. **14** (2012) 095012 [`1204.4186`].

[38] A.D. Sakharov, *Violation of CP Invariance, C asymmetry, and baryon asymmetry of the universe*, Pisma Zh. Eksp. Teor. Fiz. **5** (1967) 32.

[39] I. Bah, and F. Bonetti, *Anomaly Inflow, Accidental Symmetry, and Spontaneous Symmetry Breaking*, JHEP **01** (2020) 117 [`1910.07549`].

[40] M. Trodden, *Baryogenesis and leptogenesis*, eConf **C040802** (2004) L018 [`hep-ph/0411301`].

[41] F.R. Klinkhamer, and N.S. Manton, *A Saddle Point Solution in the Weinberg-Salam Theory*, Phys. Rev. D **30** (1984) 2212.

[42] J. de Vries, M. Postma, J. van de Vis, and G. White, *Electroweak Baryogenesis and the Standard Model Effective Field Theory*, JHEP **01** (2018) 089 [`1710.04061`].

[43] KTeV collaboration, *Observation of direct CP violation in $K_{S,L} \to \pi\pi$ decays*, Phys. Rev. Lett. **83** (1999) 22 [`hep-ex/9905060`].

[44] NA48 collaboration, *A New measurement of direct CP violation in two pion decays of the neutral kaon*, Phys. Lett. B **465** (1999) 335 [`hep-ex/9909022`].

[45] BaBar collaboration, *Measurement of CP violating asymmetries in $B^0$ decays to CP eigenstates*, Phys. Rev. Lett. **86** (2001) 2515 [`hep-ex/0102030`].

[46] Belle collaboration, *Observation of large CP violation in the neutral B meson system*, Phys. Rev. Lett. **87** (2001) 091802 [`hep-ex/0107061`].

[47] LHCb collaboration, *First observation of CP violation in the decays of $B_s^0$ mesons*, Phys. Rev. Lett. **110** (2013) 221601 [`1304.6173`].

[48] LHCb collaboration, *Observation of CP Violation in Charm Decays*, Phys. Rev. Lett. **122** (2019) 211803 [`1903.08726`].

[49] T2K collaboration, *Constraint on the matter–antimatter symmetry-violating phase in neutrino oscillations*, Nature **580** (2020) 339 [`1910.03887`], [Erratum: Nature 583, E16 (2020)].

[50] M.B. Gavela, P. Hernandez, J. Orloff, and O. Pene, *Standard model CP violation and baryon asymmetry*, *Mod. Phys. Lett. A* **9** (1994) 795 [hep-ph/9312215].

[51] A. Riotto, and M. Trodden, *Recent progress in baryogenesis*, *Ann. Rev. Nucl. Part. Sci.* **49** (1999) 35 [hep-ph/9901362].

[52] V.A. Kuzmin, V.A. Rubakov, and M.E. Shaposhnikov, *On the Anomalous Electroweak Baryon Number Nonconservation in the Early Universe*, *Phys. Lett. B* **155** (1985) 36.

[53] M.E. Shaposhnikov, *Possible Appearance of the Baryon Asymmetry of the Universe in an Electroweak Theory*, *JETP Lett.* **44** (1986) 465.

[54] M.E. Shaposhnikov, *Baryon Asymmetry of the Universe in Standard Electroweak Theory*, *Nucl. Phys. B* **287** (1987) 757.

[55] W. Kelvin, *Baltimore Lectures on Molecular Dynamics and the Wave Theory of Light*, C.J. Clay and Sons (1904).

[56] S.M. Faber, and R.E. Jackson, *Velocity dispersions and mass to light ratios for elliptical galaxies*, *Astrophys. J.* **204** (1976) 668.

[57] E. Corbelli, and P. Salucci, *The Extended Rotation Curve and the Dark Matter Halo of M33*, *Mon. Not. Roy. Astron. Soc.* **311** (2000) 441 [astro-ph/9909252].

[58] S.W. Allen, A.E. Evrard, and A.B. Mantz, *Cosmological Parameters from Observations of Galaxy Clusters*, *Ann. Rev. Astron. Astrophys.* **49** (2011) 409 [1103.4829].

[59] X.-P. Wu, T. Chiueh, L.-Z. Fang, and Y.-J. Xue, *A comparison of different cluster mass estimates: consistency or discrepancy ?*, *Mon. Not. Roy. Astron. Soc.* **301** (1998) 861 [astro-ph/9808179].

[60] P. Natarajan et al., *Mapping substructure in the HST Frontier Fields cluster lenses and in cosmological simulations*, *Mon. Not. Roy. Astron. Soc.* **468** (2017) 1962 [1702.04348].

[61] A. Refregier, *Weak gravitational lensing by large scale structure*, *Ann. Rev. Astron. Astrophys.* **41** (2003) 645 [astro-ph/0307212].

[62] WMAP collaboration, *Five-Year Wilkinson Microwave Anisotropy Probe (WMAP) Observations: Cosmological Interpretation*, *Astrophys. J. Suppl.* **180** (2009) 330 [0803.0547].

[63] C.J. Copi, D.N. Schramm, and M.S. Turner, *Big bang nucleosynthesis and the baryon density of the universe*, *Science* **267** (1995) 192 [astro-ph/9407006].

[64] F. Chadha-Day, J. Ellis, and D.J.E. Marsh, *Axion dark matter: What is it and why now?*, *Sci. Adv.* **8** (2022) abj3618 [2105.01406].

[65] A.Y. Ignatiev, V.A. Kuzmin, and M.E. Shaposhnikov, *Is the Electric Charge Conserved?*, *Phys. Lett. B* **84** (1979) 315.

[66] S. Davidson, S. Hannestad, and G. Raffelt, *Updated bounds on millicharged particles*, *JHEP* **05** (2000) 003 [hep-ph/0001179].

[67] M. Markevitch, A.H. Gonzalez, D. Clowe, A. Vikhlinin, L. David, W. Forman et al., *Direct constraints on the dark matter self-interaction cross-section from the merging galaxy cluster 1E0657-56*, *Astrophys. J.* **606** (2004) 819 [astro-ph/0309303].

[68] G. Jungman, M. Kamionkowski, and K. Griest, *Supersymmetric dark matter*, *Phys. Rept.* **267** (1996) 195 [hep-ph/9506380].

[69] G. Steigman, B. Dasgupta, and J.F. Beacom, *Precise Relic WIMP Abundance and its Impact on Searches for Dark Matter Annihilation*, *Phys. Rev. D* **86** (2012) 023506 [1204.3622].

[70] N. Craig, *The State of Supersymmetry after Run I of the LHC*, in *Beyond the Standard Model after the first run of the LHC*, 9, 2013 [1309.0528].

[71] P.J. Fox, G. Jung, P. Sorensen, and N. Weiner, *Dark Matter in Light of the LUX Results*, *Phys. Rev. D* **89** (2014) 103526 [1401.0216].

[72] L.J. Hall, K. Jedamzik, J. March-Russell, and S.M. West, *Freeze-In Production of FIMP Dark Matter*, *JHEP* **03** (2010) 080 [0911.1120].

[73] K. Griest, and D. Seckel, *Three exceptions in the calculation of relic abundances*, *Phys. Rev. D* **43** (1991) 3191.

[74] M.J. Baker, and J. Kopp, *Dark Matter Decay between Phase Transitions at the Weak Scale*, *Phys. Rev. Lett.* **119** (2017) 061801 [1608.07578].

[75] M.J. Baker, M. Breitbach, J. Kopp, and L. Mittnacht, *Dynamic Freeze-In: Impact of Thermal Masses and Cosmological Phase Transitions on Dark Matter Production*, *JHEP* **03** (2018) 114 [1712.03962].

[76] M.J. Baker, and L. Mittnacht, *Variations on the Vev Flip-Flop: Instantaneous Freeze-out and Decaying Dark Matter*, *JHEP* **05** (2019) 070 [1811.03101].

[77] M.J. Baker, J. Kopp, and A.J. Long, *Filtered Dark Matter at a First Order Phase Transition*, *Phys. Rev. Lett.* **125** (2020) 151102 [1912.02830].

[78] MACHO collaboration, *The MACHO project: Microlensing results from 5.7 years of LMC observations*, *Astrophys. J.* **542** (2000) 281 [astro-ph/0001272].

[79] A. Dar, *Dark matter and big bang nucleosynthesis*, *Astrophys. J.* **449** (1995) 550 [astro-ph/9504082].

[80] B.D. Fields, K. Freese, and D.S. Graff, *Chemical abundance constraints on white dwarfs as halo dark matter*, *Astrophys. J.* **534** (2000) 265 [astro-ph/9904291].

[81] K. Freese, B. Fields, and D. Graff, *Limits on stellar objects as the dark matter of our halo: nonbaryonic dark matter seems to be required*, Nucl. Phys. B Proc. Suppl. **80** (2000) 0305 [astro-ph/9904401].

[82] B. Carr, K. Kohri, Y. Sendouda, and J. Yokoyama, *Constraints on primordial black holes*, *Rept. Prog. Phys.* **84** (2021) 116902 [2002.12778].

[83] P. Minkowski, $\mu \to e\gamma$ *at a Rate of One Out of* $10^9$ *Muon Decays?*, *Phys. Lett. B* **67** (1977) 421.

[84] R.D. Peccei, and H.R. Quinn, *CP Conservation in the Presence of Instantons*, *Phys. Rev. Lett.* **38** (1977) 1440.

[85] M. Milgrom, *A Modification of the Newtonian dynamics as a possible alternative to the hidden mass hypothesis*, *Astrophys. J.* **270** (1983) 365.

[86] M. Milgrom, *A Modification of the Newtonian dynamics: Implications for galaxies*, *Astrophys. J.* **270** (1983) 371.

[87] M. Milgrom, *A modification of the Newtonian dynamics: implications for galaxy systems*, *Astrophys. J.* **270** (1983) 384.

[88] S.S. McGaugh, *A tale of two paradigms: the mutual incommensurability of ΛCDM and MOND*, *Can. J. Phys.* **93** (2015) 250 [1404.7525].

[89] S. Boran, S. Desai, E.O. Kahya, and R.P. Woodard, *GW170817 Falsifies Dark Matter Emulators*, *Phys. Rev. D* **97** (2018) 041501 [1710.06168].

[90] N. Menci, A. Merle, M. Totzauer, A. Schneider, A. Grazian, M. Castellano et al., *Fundamental physics with the Hubble Frontier Fields: constraining Dark Matter models with the abundance of extremely faint and distant galaxies*, *Astrophys. J.* **836** (2017) 61 [1701.01339].

[91] C. Di Paolo, F. Nesti, and F.L. Villante, *Phase space mass bound for fermionic dark matter from dwarf spheroidal galaxies*, *Mon. Not. Roy. Astron. Soc.* **475** (2018) 5385 [1704.06644].

[92] V. Iršič et al., *New Constraints on the free-streaming of warm dark matter from intermediate and small scale Lyman-α forest data*, *Phys. Rev. D* **96** (2017) 023522 [1702.01764].

[93] C.M. Ho, and R.J. Scherrer, *Limits on MeV Dark Matter from the Effective Number of Neutrinos*, *Phys. Rev. D* **87** (2013) 023505 [1208.4347].

[94] B.W. Lee, and S. Weinberg, *Cosmological Lower Bound on Heavy Neutrino Masses*, *Phys. Rev. Lett.* **39** (1977) 165.

[95] M. Battaglieri et al., *US Cosmic Visions: New Ideas in Dark Matter 2017: Community Report*, in *U.S. Cosmic Visions: New Ideas in Dark Matter*, 7, 2017 [1707.04591].

[96] K. Griest, and M. Kamionkowski, *Unitarity Limits on the Mass and Radius of Dark Matter Particles*, *Phys. Rev. Lett.* **64** (1990) 615.

[97] T.D. Brandt, *Constraints on MACHO Dark Matter from Compact Stellar Systems in Ultra-Faint Dwarf Galaxies*, *Astrophys. J. Lett.* **824** (2016) L31 [1605.03665].

[98] HESS collaboration, *Dark matter gamma-ray line searches toward the Galactic Center halo with H.E.S.S. I*, *PoS* **ICRC2017** (2018) 893 [1708.08358].

[99] SuperCDMS collaboration, *First Dark Matter Constraints from a SuperCDMS Single-Charge Sensitive Detector*, *Phys. Rev. Lett.* **121** (2018) 051301 [1804.10697], [Erratum: Phys.Rev.Lett. 122, 069901 (2019)].

[100] DAMIC collaboration, *Constraints on Light Dark Matter Particles Interacting with Electrons from DAMIC at SNOLAB*, *Phys. Rev. Lett.* **123** (2019) 181802 [1907.12628].

[101] SENSEI collaboration, *SENSEI: Direct-Detection Constraints on Sub-GeV Dark Matter from a Shallow Underground Run Using a Prototype Skipper-CCD*, *Phys. Rev. Lett.* **122** (2019) 161801 [1901.10478].

[102] MAGIC, Fermi-LAT collaboration, *Limits to Dark Matter Annihilation Cross-Section from a Combined Analysis of MAGIC and Fermi-LAT Observations of Dwarf Satellite Galaxies*, *JCAP* **02** (2016) 039 [1601.06590].

[103] Fermi-LAT collaboration, *Updated search for spectral lines from Galactic dark matter interactions with pass 8 data from the Fermi Large Area Telescope*, *Phys. Rev. D* **91** (2015) 122002 [1506.00013].

[104] H.E.S.S. collaboration, *Search for dark matter annihilations towards the inner Galactic halo from 10 years of observations with H.E.S.S*, *Phys. Rev. Lett.* **117** (2016) 111301 [1607.08142].

[105] CMS collaboration, *Search for dark matter produced with an energetic jet or a hadronically decaying W or Z boson at $\sqrt{s} = 13$ TeV*, *JHEP* **07** (2017) 014 [1703.01651].

[106] ATLAS collaboration, *Search for dark matter at $\sqrt{s} = 13$ TeV in final states containing an energetic photon and large missing transverse momentum with the ATLAS detector*, *Eur. Phys. J. C* **77** (2017) 393 [1704.03848].

[107] CMS collaboration, *Search for new physics in the monophoton final state in proton-proton collisions at $\sqrt{s} = 13$ TeV*, *JHEP* **10** (2017) 073 [1706.03794].

[108] ATLAS collaboration, *Search for dark matter and other new phenomena in events with an energetic jet and large missing transverse momentum using the ATLAS detector*, *JHEP* **01** (2018) 126 [1711.03301].

[109] Fermi-LAT collaboration, *Fermi-LAT Observations of High-Energy $\gamma$-Ray Emission Toward the Galactic Center*, *Astrophys. J.* **819** (2016) 44 [1511.02938].

[110] I. Cholis, T. Linden, and D. Hooper, *A Robust Excess in the Cosmic-Ray Antiproton Spectrum: Implications for Annihilating Dark Matter*, *Phys. Rev. D* **99** (2019) 103026 [1903.02549].

[111] R. Bernabei et al., *First model independent results from DAMA/LIBRA-phase2*, *Nucl. Phys. Atom. Energy* **19** (2018) 307 [1805.10486].

[112] T. Marrodán Undagoitia, and L. Rauch, *Dark matter direct-detection experiments*, *J. Phys. G* **43** (2016) 013001 [1509.08767].

[113] T.R. Slatyer, *Indirect Detection of Dark Matter*, in *Theoretical Advanced Study Institute in Elementary Particle Physics: Anticipating the Next Discoveries in Particle Physics*, pp. 297–353, 2018, DOI [1710.05137].

[114] T.M. Hong, *Dark matter searches at the LHC*, in *5th Large Hadron Collider Physics Conference*, 9, 2017 [1709.02304].

[115] LIGO Scientific, Virgo collaboration, *Observation of Gravitational Waves from a Binary Black Hole Merger*, *Phys. Rev. Lett.* **116** (2016) 061102 [1602.03837].

[116] C.J. Hogan, *NUCLEATION OF COSMOLOGICAL PHASE TRANSITIONS*, *Phys. Lett. B* **133** (1983) 172.

[117] E. Witten, *Cosmic Separation of Phases*, *Phys. Rev. D* **30** (1984) 272.

[118] M.S. Turner, and F. Wilczek, *Relic gravitational waves and extended inflation*, *Phys. Rev. Lett.* **65** (1990) 3080.

[119] P. Schwaller, *Gravitational Waves from a Dark Phase Transition*, *Phys. Rev. Lett.* **115** (2015) 181101 [1504.07263].

[120] C. Caprini et al., *Science with the space-based interferometer eLISA. II: Gravitational waves from cosmological phase transitions*, *JCAP* **04** (2016) 001 [1512.06239].

[121] C. Caprini et al., *Detecting gravitational waves from cosmological phase transitions with LISA: an update*, *JCAP* **03** (2020) 024 [1910.13125].

[122] M.B. Hindmarsh, M. Lüben, J. Lumma, and M. Pauly, *Phase transitions in the early universe*, *SciPost Phys. Lect. Notes* **24** (2021) 1 [2008.09136].

[123] D. Bodeker, and G.D. Moore, *Can electroweak bubble walls run away?*, *JCAP* **05** (2009) 009 [0903.4099].

[124] D. Bodeker, and G.D. Moore, *Electroweak Bubble Wall Speed Limit*, *JCAP* **05** (2017) 025 [1703.08215].

[125] J.R. Espinosa, *Tunneling without Bounce*, *Phys. Rev. D* **100** (2019) 105002 [1908.01730].

[126] V. Mukhanov, and A. Sorin, *Instantons with Quantum Core*, [2105.01996].

[127] J.R. Espinosa, and J. Huertas, *Pseudo-bounces vs. new instantons*, *JCAP* **12** (2021) 029 [2106.04541].

[128] C.L. Wainwright, *CosmoTransitions: Computing Cosmological Phase Transition Temperatures and Bubble Profiles with Multiple Fields*, *Comput. Phys. Commun.* **183** (2012) 2006 [1109.4189].

[129] A. Masoumi, K.D. Olum, and J.M. Wachter, *Approximating tunneling rates in multi-dimensional field spaces*, *JCAP* **10** (2017) 022 [1702.00356].

[130] J.R. Espinosa, *A Fresh Look at the Calculation of Tunneling Actions*, *JCAP* **07** (2018) 036 [1805.03680].

[131] J.R. Espinosa, *Fresh look at the calculation of tunneling actions including gravitational effects*, *Phys. Rev. D* **100** (2019) 104007 [1808.00420].

[132] J.R. Espinosa, and T. Konstandin, *A Fresh Look at the Calculation of Tunneling Actions in Multi-Field Potentials*, *JCAP* **01** (2019) 051 [1811.09185].

[133] V. Guada, A. Maiezza, and M. Nemevšek, *Multifield Polygonal Bounces*, *Phys. Rev. D* **99** (2019) 056020 [1803.02227].

[134] V. Guada, M. Nemevšek, and M. Pintar, *FindBounce: Package for multi-field bounce actions*, *Comput. Phys. Commun.* **256** (2020) 107480 [2002.00881].

[135] M. Carena, M. Quiros, and C.E.M. Wagner, *Opening the window for electroweak baryogenesis*, *Phys. Lett. B* **380** (1996) 81 [hep-ph/9603420].

[136] A. Riotto, *The More relaxed supersymmetric electroweak baryogenesis*, *Phys. Rev. D* **58** (1998) 095009 [hep-ph/9803357].

[137] S.J. Huber, T. Konstandin, T. Prokopec, and M.G. Schmidt, *Electroweak Phase Transition and Baryogenesis in the nMSSM*, *Nucl. Phys. B* **757** (2006) 172 [hep-ph/0606298].

[138] J.M. Cline, *Baryogenesis*, in *Les Houches Summer School - Session 86: Particle Physics and Cosmology: The Fabric of Spacetime*, 9, 2006 [hep-ph/0609145].

[139] D.E. Morrissey, and M.J. Ramsey-Musolf, *Electroweak baryogenesis*, *New J. Phys.* **14** (2012) 125003 [1206.2942].

[140] V. Vaskonen, *Electroweak baryogenesis and gravitational waves from a real scalar singlet*, *Phys. Rev. D* **95** (2017) 123515 [1611.02073].

[141] B. Garbrecht, *Why is there more matter than antimatter? Calculational methods for leptogenesis and electroweak baryogenesis*, *Prog. Part. Nucl. Phys.* **110** (2020) 103727 [1812.02651].

[142] J. De Vries, M. Postma, and J. van de Vis, *The role of leptons in electroweak baryogenesis*, *JHEP* **04** (2019) 024 [1811.11104].

[143] J.M. Cline, and K. Kainulainen, *Electroweak baryogenesis at high bubble wall velocities*, *Phys. Rev. D* **101** (2020) 063525 [2001.00568].

[144] E. Fuchs, M. Losada, Y. Nir, and Y. Viernik, *Analytic techniques for solving the transport equations in electroweak baryogenesis*, *JHEP* **07** (2021) 060 [2007.06940].

[145] B. Dutta, and J. Kumar, *Hidden sector baryogenesis*, *Phys. Lett. B* **643** (2006) 284 [hep-th/0608188].

[146] J. Shelton, and K.M. Zurek, *Darkogenesis: A baryon asymmetry from the dark matter sector*, *Phys. Rev. D* **82** (2010) 123512 [1008.1997].

[147] E. Hall, T. Konstandin, R. McGehee, H. Murayama, and G. Servant, *Baryogenesis From a Dark First-Order Phase Transition*, *JHEP* **04** (2020) 042 [1910.08068].

[148] J.M. Cline, K. Kainulainen, and D. Tucker-Smith, *Electroweak baryogenesis from a dark sector*, *Phys. Rev. D* **95** (2017) 115006 [1702.08909].

[149] F.P. Huang, and C.S. Li, *Probing the baryogenesis and dark matter relaxed in phase transition by gravitational waves and colliders*, *Phys. Rev. D* **96** (2017) 095028 [1709.09691].

[150] J. Arakawa, A. Rajaraman, and T.M.P. Tait, *Annihilogenesis*, [2109.13941].

[151] M. Breitbach, *Gravitational Waves from Cosmological Phase Transitions*, master thesis, Johannes Gutenberg University Mainz, 4, 2018, [2204.09661].

[152] S. Weinberg, *The quantum theory of fields. Vol. 2: Modern applications*, Cambridge University Press (8, 2013).

[153] M. Quiros, *Finite temperature field theory and phase transitions*, in *ICTP Summer School in High-Energy Physics and Cosmology*, pp. 187–259, 1, 1999 [hep-ph/9901312].

[154] S.R. Coleman, and E.J. Weinberg, *Radiative Corrections as the Origin of Spontaneous Symmetry Breaking*, *Phys. Rev. D* **7** (1973) 1888.

[155] C. Delaunay, C. Grojean, and J.D. Wells, *Dynamics of Non-renormalizable Electroweak Symmetry Breaking*, *JHEP* **04** (2008) 029 [0711.2511].

[156] G. 't Hooft, and M.J.G. Veltman, *Regularization and Renormalization of Gauge Fields*, *Nucl. Phys. B* **44** (1972) 189.

[157] S. Weinberg, *Gauge and Global Symmetries at High Temperature*, *Phys. Rev. D* **9** (1974) 3357.

[158] A. Nieto, *Evaluating sums over the Matsubara frequencies*, *Comput. Phys. Commun.* **92** (1995) 54 [hep-ph/9311210].

[159] K. Enqvist, A. Riotto, and I. Vilja, *Baryogenesis and the thermalization rate of stop*, *Phys. Lett. B* **438** (1998) 273 [hep-ph/9710373].

[160] D. Comelli, and J.R. Espinosa, *Bosonic thermal masses in supersymmetry*, *Phys. Rev. D* **55** (1997) 6253 [hep-ph/9606438].

[161] D. Curtin, P. Meade, and H. Ramani, *Thermal Resummation and Phase Transitions*, *Eur. Phys. J. C* **78** (2018) 787 [1612.00466].

[162] D. Croon, O. Gould, P. Schicho, T.V.I. Tenkanen, and G. White, *Theoretical uncertainties for cosmological first-order phase transitions*, *JHEP* **04** (2021) 055 [2009.10080].

[163] M.E. Carrington, *The Effective potential at finite temperature in the Standard Model*, *Phys. Rev. D* **45** (1992) 2933.

[164] A. Andreassen, W. Frost, and M.D. Schwartz, *Consistent Use of Effective Potentials*, *Phys. Rev. D* **91** (2015) 016009 [1408.0287].

[165] A. Ekstedt, and J. Löfgren, *A Critical Look at the Electroweak Phase Transition*, *JHEP* **12** (2020) 136 [2006.12614].

[166] C. Ford, I. Jack, and D.R.T. Jones, *The Standard model effective potential at two loops*, *Nucl. Phys. B* **387** (1992) 373 [hep-ph/0111190], [Erratum: Nucl.Phys.B 504, 551–552 (1997)].

[167] S.P. Martin, *Two Loop Effective Potential for a General Renormalizable Theory and Softly Broken Supersymmetry*, *Phys. Rev. D* **65** (2002) 116003 [hep-ph/0111209].

[168] S.P. Martin, *Three-Loop Standard Model Effective Potential at Leading Order in Strong and Top Yukawa Couplings*, *Phys. Rev. D* **89** (2014) 013003 [1310.7553].

[169] S.P. Martin, *Four-Loop Standard Model Effective Potential at Leading Order in QCD*, *Phys. Rev. D* **92** (2015) 054029 [1508.00912].

[170] S.P. Martin, *Effective potential at three loops*, *Phys. Rev. D* **96** (2017) 096005 [1709.02397].

[171] M. Sher, *Electroweak Higgs Potentials and Vacuum Stability*, *Phys. Rept.* **179** (1989) 273.

[172] J.R. Espinosa, T. Konstandin, J.M. No, and M. Quiros, *Some Cosmological Implications of Hidden Sectors*, *Phys. Rev. D* **78** (2008) 123528 [0809.3215].

[173] A. Mégevand, and S. Ramírez, *Bubble nucleation and growth in slow cosmological phase transitions*, *Nucl. Phys. B* **928** (2018) 38 [1710.06279].

[174] T. Konstandin, and G. Servant, *Cosmological Consequences of Nearly Conformal Dynamics at the TeV scale*, *JCAP* **12** (2011) 009 [1104.4791].

[175] F. Sannino, and J. Virkajärvi, *First Order Electroweak Phase Transition from (Non)Conformal Extensions of the Standard Model*, *Phys. Rev. D* **92** (2015) 045015 [1505.05872].

[176] J. Jaeckel, V.V. Khoze, and M. Spannowsky, *Hearing the signal of dark sectors with gravitational wave detectors*, *Phys. Rev. D* **94** (2016) 103519 [1602.03901].

[177] L. Marzola, A. Racioppi, and V. Vaskonen, *Phase transition and gravitational wave phenomenology of scalar conformal extensions of the Standard Model*, *Eur. Phys. J. C* **77** (2017) 484 [1704.01034].

[178] S. Iso, P.D. Serpico, and K. Shimada, *QCD-Electroweak First-Order Phase Transition in a Supercooled Universe*, *Phys. Rev. Lett.* **119** (2017) 141301 [1704.04955].

[179] C. Marzo, L. Marzola, and V. Vaskonen, *Phase transition and vacuum stability in the classically conformal B–L model*, *Eur. Phys. J. C* **79** (2019) 601 [1811.11169].

[180] T. Hambye, A. Strumia, and D. Teresi, *Super-cool Dark Matter*, *JHEP* **08** (2018) 188 [1805.01473].

[181] J. Ellis, M. Lewicki, and J.M. No, *On the Maximal Strength of a First-Order Electroweak Phase Transition and its Gravitational Wave Signal*, *JCAP* **04** (2019) 003 [1809.08242].

[182] S. Bruggisser, B. Von Harling, O. Matsedonskyi, and G. Servant, *Electroweak Phase Transition and Baryogenesis in Composite Higgs Models*, *JHEP* **12** (2018) 099 [1804.07314].

[183] I. Baldes, and C. Garcia-Cely, *Strong gravitational radiation from a simple dark matter model*, *JHEP* **05** (2019) 190 [1809.01198].

[184] V. Brdar, A.J. Helmboldt, and J. Kubo, *Gravitational Waves from First-Order Phase Transitions: LIGO as a Window to Unexplored Seesaw Scales*, *JCAP* **02** (2019) 021 [1810.12306].

[185] T. Prokopec, J. Rezacek, and B. Świeżewska, *Gravitational waves from conformal symmetry breaking*, *JCAP* **02** (2019) 009 [1809.11129].

[186] J. Ellis, M. Lewicki, J.M. No, and V. Vaskonen, *Gravitational wave energy budget in strongly supercooled phase transitions*, *JCAP* **06** (2019) 024 [1903.09642].

[187] M. Aoki, and J. Kubo, *Gravitational waves from chiral phase transition in a conformally extended standard model*, *JCAP* **04** (2020) 001 [1910.05025].

[188] A. Mohamadnejad, *Gravitational waves from scale-invariant vector dark matter model: Probing below the neutrino-floor*, *Eur. Phys. J. C* **80** (2020) 197 [1907.08899].

[189] V. Brdar, A.J. Helmboldt, and M. Lindner, *Strong Supercooling as a Consequence of Renormalization Group Consistency*, *JHEP* **12** (2019) 158 [1910.13460].

[190] L. Delle Rose, G. Panico, M. Redi, and A. Tesi, *Gravitational Waves from Supercool Axions*, *JHEP* **04** (2020) 025 [1912.06139].

[191] J. Ellis, M. Lewicki, and V. Vaskonen, *Updated predictions for gravitational waves produced in a strongly supercooled phase transition*, *JCAP* **11** (2020) 020 [2007.15586].

[192] S.R. Coleman, *The Fate of the False Vacuum. 1. Semiclassical Theory*, *Phys. Rev. D* **15** (1977) 2929 [Erratum: Phys.Rev.D 16, 1248 (1977)].

[193] H. Widyan, and M. Al-Wardat, *Classical Solution for the Bounce Up to Second Order*, *Chin. J. Phys.* **48** (2010) 736 [1206.2734].

[194] A.D. Linde, *Fate of the False Vacuum at Finite Temperature: Theory and Applications*, *Phys. Lett. B* **100** (1981) 37.

[195] A.D. Linde, *Decay of the False Vacuum at Finite Temperature*, *Nucl. Phys. B* **216** (1983) 421 [Erratum: Nucl.Phys.B 223, 544 (1983)].

[196] D. Baumann, *Primordial Cosmology*, *PoS* **TASI2017** (2018) 009 [1807.03098].

[197] J.R. Espinosa, T. Konstandin, J.M. No, and G. Servant, *Energy Budget of Cosmological First-order Phase Transitions*, *JCAP* **06** (2010) 028 [1004.4187].

[198] C. Grojean, and G. Servant, *Gravitational Waves from Phase Transitions at the Electroweak Scale and Beyond*, *Phys. Rev. D* **75** (2007) 043507 [hep-ph/0607107].

[199] S. Weinberg, *Gravitation and Cosmology: Principles and Applications of the General Theory of Relativity*, John Wiley and Sons, New York (1972).

[200] P.J. Steinhardt, *Relativistic Detonation Waves and Bubble Growth in False Vacuum Decay*, *Phys. Rev. D* **25** (1982) 2074.

[201] M. Laine, *Bubble growth as a detonation*, *Phys. Rev. D* **49** (1994) 3847 [hep-ph/9309242].

[202] H. Kurki-Suonio, and M. Laine, *Supersonic deflagrations in cosmological phase transitions*, *Phys. Rev. D* **51** (1995) 5431 [hep-ph/9501216].

[203] S. Höche, J. Kozaczuk, A.J. Long, J. Turner, and Y. Wang, *Towards an all-orders calculation of the electroweak bubble wall velocity*, *JCAP* **03** (2021) 009 [2007.10343].

[204] A. Megevand, and A.D. Sanchez, *Velocity of electroweak bubble walls*, *Nucl. Phys. B* **825** (2010) 151 [`0908.3663`].

[205] D.J. Fixsen, *The Temperature of the Cosmic Microwave Background*, *Astrophys. J.* **707** (2009) 916 [`0911.1955`].

[206] L. Husdal, *On Effective Degrees of Freedom in the Early Universe*, *Galaxies* **4** (2016) 78 [`1609.04979`].

[207] A. Kosowsky, and M.S. Turner, *Gravitational radiation from colliding vacuum bubbles: envelope approximation to many bubble collisions*, *Phys. Rev. D* **47** (1993) 4372 [`astro-ph/9211004`].

[208] S.J. Huber, and T. Konstandin, *Gravitational Wave Production by Collisions: More Bubbles*, *JCAP* **09** (2008) 022 [`0806.1828`].

[209] M. Hindmarsh, S.J. Huber, K. Rummukainen, and D.J. Weir, *Numerical simulations of acoustically generated gravitational waves at a first order phase transition*, *Phys. Rev. D* **92** (2015) 123009 [`1504.03291`].

[210] C. Caprini, R. Durrer, and G. Servant, *The stochastic gravitational wave background from turbulence and magnetic fields generated by a first-order phase transition*, *JCAP* **12** (2009) 024 [`0909.0622`].

[211] M. Hindmarsh, S.J. Huber, K. Rummukainen, and D.J. Weir, *Shape of the acoustic gravitational wave power spectrum from a first order phase transition*, *Phys. Rev. D* **96** (2017) 103520 [`1704.05871`], [Erratum: Phys.Rev.D 101, 089902 (2020)].

[212] D. Cutting, M. Hindmarsh, and D.J. Weir, *Gravitational waves from vacuum first-order phase transitions: from the envelope to the lattice*, *Phys. Rev. D* **97** (2018) 123513 [`1802.05712`].

[213] D. Cutting, M. Hindmarsh, and D.J. Weir, *Vorticity, kinetic energy, and suppressed gravitational wave production in strong first order phase transitions*, *Phys. Rev. Lett.* **125** (2020) 021302 [`1906.00480`].

[214] J. Ellis, M. Lewicki, and J.M. No, *Gravitational waves from first-order cosmological phase transitions: lifetime of the sound wave source*, *JCAP* **07** (2020) 050 [`2003.07360`].

[215] F. Giese, T. Konstandin, and J. van de Vis, *Model-independent energy budget of cosmological first-order phase transitions—A sound argument to go beyond the bag model*, *JCAP* **07** (2020) 057 [`2004.06995`].

[216] K. Kotake, K. Sato, and K. Takahashi, *Explosion mechanism, neutrino burst, and gravitational wave in core-collapse supernovae*, *Rept. Prog. Phys.* **69** (2006) 971 [`astro-ph/0509456`].

[217] LIGO Scientific collaboration, *Coherent searches for periodic gravitational waves from unknown isolated sources and Scorpius X-1: Results from the second LIGO science run*, *Phys. Rev. D* **76** (2007) 082001 [`gr-qc/0605028`].

[218] G. Nelemans, *The Galactic Gravitational wave foreground*, *Class. Quant. Grav.* **26** (2009) 094030 [`0901.1778`].

[219] A. Stroeer, and A. Vecchio, *The LISA verification binaries*, *Class. Quant. Grav.* **23** (2006) S809 [`astro-ph/0605227`].

[220] LIGO Scientific, VIRGO collaboration, *Predictions for the Rates of Compact Binary Coalescences Observable by Ground-based Gravitational-wave Detectors*, *Class. Quant. Grav.* **27** (2010) 173001 [`1003.2480`].

[221] M. Volonteri, F. Haardt, and P. Madau, *The Assembly and merging history of supermassive black holes in hierarchical models of galaxy formation*, *Astrophys. J.* **582** (2003) 559 [`astro-ph/0207276`].

[222] A. Sesana, A. Vecchio, and C.N. Colacino, *The stochastic gravitational-wave background from massive black hole binary systems: implications for observations with Pulsar Timing Arrays*, *Mon. Not. Roy. Astron. Soc.* **390** (2008) 192 [`0804.4476`].

[223] P. Amaro-Seoane et al., *Low-frequency gravitational-wave science with eLISA/NGO*, *Class. Quant. Grav.* **29** (2012) 124016 [`1202.0839`].

[224] P. Amaro-Seoane, *Relativistic dynamics and extreme mass ratio inspirals*, *Living Rev. Rel.* **21** (2018) 4 [`1205.5240`].

[225] C.P.L. Berry, and J.R. Gair, *Observing the Galaxy's massive black hole with gravitational wave bursts*, *Mon. Not. Roy. Astron. Soc.* **429** (2013) 589 [`1210.2778`].

[226] P. Amaro-Seoane, J.R. Gair, M. Freitag, M. Coleman Miller, I. Mandel, C.J. Cutler et al., *Astrophysics, detection and science applications of intermediate- and extreme mass-ratio inspirals*, *Class. Quant. Grav.* **24** (2007) R113 [`astro-ph/0703495`].

[227] T. Damour, and A. Vilenkin, *Gravitational radiation from cosmic (super)strings: Bursts, stochastic background, and observational windows*, *Phys. Rev. D* **71** (2005) 063510 [`hep-th/0410222`].

[228] P. Binetruy, A. Bohe, C. Caprini, and J.-F. Dufaux, *Cosmological Backgrounds of Gravitational Waves and eLISA/NGO: Phase Transitions, Cosmic Strings and Other Sources*, *JCAP* **06** (2012) 027 [`1201.0983`].

[229] LIGO Scientific collaboration, *Advanced LIGO: The next generation of gravitational wave detectors*, *Class. Quant. Grav.* **27** (2010) 084006.

[230] Virgo collaboration, *The Advanced Virgo detector*, *J. Phys. Conf. Ser.* **610** (2015) 012014.

[231] KAGRA collaboration, *Detector configuration of KAGRA: The Japanese cryogenic gravitational-wave detector*, *Class. Quant. Grav.* **29** (2012) 124007 [`1111.7185`].

[232] B. Sathyaprakash et al., *Scientific Objectives of Einstein Telescope*, *Class. Quant. Grav.* **29** (2012) 124013 [`1206.0331`], [Erratum: Class.Quant.Grav. 30, 079501 (2013)].

[233] LISA collaboration, *Laser Interferometer Space Antenna*, [`1702.00786`].

[234] J. Crowder, and N.J. Cornish, *Beyond LISA: Exploring future gravitational wave missions*, *Phys. Rev. D* **72** (2005) 083005 [`gr-qc/0506015`].

[235] S. Isoyama, H. Nakano, and T. Nakamura, *Multiband Gravitational-Wave Astronomy: Observing binary inspirals with a decihertz detector, B-DECIGO*, *PTEP* **2018** (2018) 073E01 [1802.06977].

[236] N. Seto, S. Kawamura, and T. Nakamura, *Possibility of direct measurement of the acceleration of the universe using 0.1-Hz band laser interferometer gravitational wave antenna in space*, *Phys. Rev. Lett.* **87** (2001) 221103 [astro-ph/0108011].

[237] S. Sato et al., *The status of DECIGO*, *J. Phys. Conf. Ser.* **840** (2017) 012010.

[238] R.D. Ferdman et al., *The European Pulsar Timing Array: current efforts and a LEAP toward the future*, *Class. Quant. Grav.* **27** (2010) 084014 [1003.3405].

[239] L. Lentati et al., *European Pulsar Timing Array Limits On An Isotropic Stochastic Gravitational-Wave Background*, *Mon. Not. Roy. Astron. Soc.* **453** (2015) 2576 [1504.03692].

[240] F. Jenet et al., *The North American Nanohertz Observatory for Gravitational Waves*, [0909.1058].

[241] G. Janssen et al., *Gravitational wave astronomy with the SKA*, *PoS* **AASKA14** (2015) 037 [1501.00127].

[242] SKA collaboration, *Cosmology with Phase 1 of the Square Kilometre Array: Red Book 2018: Technical specifications and performance forecasts*, *Publ. Astron. Soc. Austral.* **37** (2020) e007 [1811.02743].

[243] NANOGrav collaboration, *The NANOGrav 12.5 yr Data Set: Search for an Isotropic Stochastic Gravitational-wave Background*, *Astrophys. J. Lett.* **905** (2020) L34 [2009.04496].

[244] A. Sesana et al., *Unveiling the gravitational universe at µ-Hz frequencies*, *Exper. Astron.* **51** (2021) 1333 [1908.11391].

[245] M.A. Fedderke, P.W. Graham, and S. Rajendran, *Asteroids for µHz gravitational-wave detection*, [2112.11431].

[246] NANOGRAV collaboration, *The NANOGrav 11-year Data Set: Pulsar-timing Constraints On The Stochastic Gravitational-wave Background*, *Astrophys. J.* **859** (2018) 47 [1801.02617].

[247] K. Yagi, and N. Seto, *Detector configuration of DECIGO/BBO and identification of cosmological neutron-star binaries*, *Phys. Rev. D* **83** (2011) 044011 [1101.3940], [Erratum: Phys.Rev.D 95, 109901 (2017)].

[248] E. Thrane, and J.D. Romano, *Sensitivity curves for searches for gravitational-wave backgrounds*, *Phys. Rev. D* **88** (2013) 124032 [1310.5300].

[249] D. Chway, T.H. Jung, and C.S. Shin, *Dark matter filtering-out effect during a first-order phase transition*, *Phys. Rev. D* **101** (2020) 095019 [1912.04238].

[250] J. Choi, and R.R. Volkas, *Real Higgs singlet and the electroweak phase transition in the Standard Model*, *Phys. Lett. B* **317** (1993) 385 [hep-ph/9308234].

[251] J.R. Espinosa, T. Konstandin, and F. Riva, *Strong Electroweak Phase Transitions in the Standard Model with a Singlet*, *Nucl. Phys. B* **854** (2012) 592 [1107.5441].

[252] J.M. Cline, and K. Kainulainen, *Electroweak baryogenesis and dark matter from a singlet Higgs*, *JCAP* **01** (2013) 012 [`1210.4196`].

[253] A. Azatov, M. Vanvlasselaer, and W. Yin, *Baryogenesis via relativistic bubble walls*, *JHEP* **10** (2021) 043 [`2106.14913`].

[254] I. Baldes, S. Blasi, A. Mariotti, A. Sevrin, and K. Turbang, *Baryogenesis via relativistic bubble expansion*, *Phys. Rev. D* **104** (2021) 115029 [`2106.15602`].

[255] S. Nussinov, *TECHNOCOSMOLOGY: COULD A TECHNIBARYON EXCESS PROVIDE A 'NATURAL' MISSING MASS CANDIDATE?*, *Phys. Lett. B* **165** (1985) 55.

[256] S.M. Barr, R.S. Chivukula, and E. Farhi, *Electroweak Fermion Number Violation and the Production of Stable Particles in the Early Universe*, *Phys. Lett. B* **241** (1990) 387.

[257] S.M. Barr, *Baryogenesis, sphalerons and the cogeneration of dark matter*, *Phys. Rev. D* **44** (1991) 3062.

[258] D.E. Kaplan, M.A. Luty, and K.M. Zurek, *Asymmetric Dark Matter*, *Phys. Rev. D* **79** (2009) 115016 [`0901.4117`].

[259] H. An, S.-L. Chen, R.N. Mohapatra, and Y. Zhang, *Leptogenesis as a Common Origin for Matter and Dark Matter*, *JHEP* **03** (2010) 124 [`0911.4463`].

[260] N. Haba, and S. Matsumoto, *Baryogenesis from Dark Sector*, *Prog. Theor. Phys.* **125** (2011) 1311 [`1008.2487`].

[261] H. Davoudiasl, D.E. Morrissey, K. Sigurdson, and S. Tulin, *Hylogenesis: A Unified Origin for Baryonic Visible Matter and Antibaryonic Dark Matter*, *Phys. Rev. Lett.* **105** (2010) 211304 [`1008.2399`].

[262] M.R. Buckley, and L. Randall, *Xogenesis*, *JHEP* **09** (2011) 009 [`1009.0270`].

[263] P.-H. Gu, M. Lindner, U. Sarkar, and X. Zhang, *WIMP Dark Matter and Baryogenesis*, *Phys. Rev. D* **83** (2011) 055008 [`1009.2690`].

[264] M. Blennow, B. Dasgupta, E. Fernandez-Martinez, and N. Rius, *Aidnogenesis via Leptogenesis and Dark Sphalerons*, *JHEP* **03** (2011) 014 [`1009.3159`].

[265] R. Allahverdi, B. Dutta, and K. Sinha, *Cladogenesis: Baryon-Dark Matter Coincidence from Branchings in Moduli Decay*, *Phys. Rev. D* **83** (2011) 083502 [`1011.1286`].

[266] B. Dutta, and J. Kumar, *Asymmetric Dark Matter from Hidden Sector Baryogenesis*, *Phys. Lett. B* **699** (2011) 364 [`1012.1341`].

[267] A. Falkowski, J.T. Ruderman, and T. Volansky, *Asymmetric Dark Matter from Leptogenesis*, *JHEP* **05** (2011) 106 [`1101.4936`].

[268] M.L. Graesser, I.M. Shoemaker, and L. Vecchi, *Asymmetric WIMP dark matter*, *JHEP* **10** (2011) 110 [`1103.2771`].

[269] M.R. Buckley, *Asymmetric Dark Matter and Effective Operators*, *Phys. Rev. D* **84** (2011) 043510 [`1104.1429`].

[270] N.F. Bell, K. Petraki, I.M. Shoemaker, and R.R. Volkas, *Pangenesis in a Baryon-Symmetric Universe: Dark and Visible Matter via the Affleck-Dine Mechanism*, *Phys. Rev. D* **84** (2011) 123505 [1105.3730].

[271] C. Cheung, and K.M. Zurek, *Affleck-Dine Cogenesis*, *Phys. Rev. D* **84** (2011) 035007 [1105.4612].

[272] J. March-Russell, and M. McCullough, *Asymmetric Dark Matter via Spontaneous Co-Genesis*, *JCAP* **03** (2012) 019 [1106.4319].

[273] Y. Cui, L. Randall, and B. Shuve, *Emergent Dark Matter, Baryon, and Lepton Numbers*, *JHEP* **08** (2011) 073 [1106.4834].

[274] Y. Cui, L. Randall, and B. Shuve, *A WIMPy Baryogenesis Miracle*, *JHEP* **04** (2012) 075 [1112.2704].

[275] H. Davoudiasl, and R.N. Mohapatra, *On Relating the Genesis of Cosmic Baryons and Dark Matter*, *New J. Phys.* **14** (2012) 095011 [1203.1247].

[276] J. Unwin, *Exodus: Hidden origin of dark matter and baryons*, *JHEP* **06** (2013) 090 [1212.1425].

[277] Y. Cui, and R. Sundrum, *Baryogenesis for weakly interacting massive particles*, *Phys. Rev. D* **87** (2013) 116013 [1212.2973].

[278] S.M. Barr, and H.-Y. Chen, *Cogeneration of Dark Matter and Baryons by Non-Standard-Model Sphalerons in Unified Models*, *JHEP* **10** (2013) 129 [1309.0020].

[279] G. Servant, and S. Tulin, *Baryogenesis and Dark Matter through a Higgs Asymmetry*, *Phys. Rev. Lett.* **111** (2013) 151601 [1304.3464].

[280] W.-Z. Feng, A. Mazumdar, and P. Nath, *Baryogenesis from dark matter*, *Phys. Rev. D* **88** (2013) 036014 [1302.0012].

[281] E. Hall, T. Konstandin, R. McGehee, and H. Murayama, *Asymmetric Matters from a Dark First-Order Phase Transition*, [1911.12342].

[282] P. John, *Bubble wall profiles with more than one scalar field: A Numerical approach*, *Phys. Lett. B* **452** (1999) 221 [hep-ph/9810499].

[283] P. Gondolo, and G. Gelmini, *Cosmic abundances of stable particles: Improved analysis*, *Nucl. Phys. B* **360** (1991) 145.

[284] R. Apreda, M. Maggiore, A. Nicolis, and A. Riotto, *Gravitational waves from electroweak phase transitions*, *Nucl. Phys. B* **631** (2002) 342 [gr-qc/0107033].

[285] V. Silveira, and A. Zee, *SCALAR PHANTOMS*, *Phys. Lett. B* **161** (1985) 136.

[286] C.P. Burgess, M. Pospelov, and T. ter Veldhuis, *The Minimal model of nonbaryonic dark matter: A Singlet scalar*, *Nucl. Phys. B* **619** (2001) 709 [hep-ph/0011335].

[287] B. Patt, and F. Wilczek, *Higgs-field portal into hidden sectors*, [hep-ph/0605188].

[288] CMS collaboration, *Search for invisible decays of a Higgs boson produced through vector boson fusion in proton-proton collisions at $\sqrt{s} = 13$ TeV*, *Phys. Lett. B* **793** (2019) 520 [1809.05937].

[289] J. Liu, X.-P. Wang, and F. Yu, *A Tale of Two Portals: Testing Light, Hidden New Physics at Future $e^+e^-$ Colliders*, *JHEP* **06** (2017) 077 [`1704.00730`].

[290] K. Kumar, R. Vega-Morales, and F. Yu, *Effects from New Colored States and the Higgs Portal on Gluon Fusion and Higgs Decays*, *Phys. Rev. D* **86** (2012) 113002 [`1205.4244`], [Erratum: Phys.Rev.D 87, 119903 (2013)].

[291] M. Joyce, T. Prokopec, and N. Turok, *Nonlocal electroweak baryogenesis. Part 1: Thin wall regime*, *Phys. Rev. D* **53** (1996) 2930 [`hep-ph/9410281`].

[292] V. Cirigliano, M.J. Ramsey-Musolf, S. Tulin, and C. Lee, *Yukawa and tri-scalar processes in electroweak baryogenesis*, *Phys. Rev. D* **73** (2006) 115009 [`hep-ph/0603058`].

[293] S. Weinberg, *Baryon and Lepton Nonconserving Processes*, *Phys. Rev. Lett.* **43** (1979) 1566.

[294] G. Mangano, G. Miele, S. Pastor, O. Pisanti, and S. Sarikas, *Updated BBN bounds on the cosmological lepton asymmetry for non-zero $\theta_{13}$*, *Phys. Lett. B* **708** (2012) 1 [`1110.4335`].

[295] G. Barenboim, W.H. Kinney, and W.-I. Park, *Resurrection of large lepton number asymmetries from neutrino flavor oscillations*, *Phys. Rev. D* **95** (2017) 043506 [`1609.01584`].

[296] G. Barenboim, and W.-I. Park, *A full picture of large lepton number asymmetries of the Universe*, *JCAP* **04** (2017) 048 [`1703.08258`].

[297] C. Lee, V. Cirigliano, and M.J. Ramsey-Musolf, *Resonant relaxation in electroweak baryogenesis*, *Phys. Rev. D* **71** (2005) 075010 [`hep-ph/0412354`].

[298] R.N. Mohapatra, and J.W.F. Valle, *Solar Neutrino Oscillations From Superstrings*, *Phys. Lett. B* **177** (1986) 47.

[299] S. Centelles Chuliá, R. Srivastava, and A. Vicente, *The inverse seesaw family: Dirac and Majorana*, *JHEP* **03** (2021) 248 [`2011.06609`].

[300] C. Bardos, E. Bernard, F. Golse, and R. Sentis, *The Diffusion Approximation for the Linear Boltzmann Equation with Vanishing Scattering Coefficient*, *Communications in Mathematical Sciences* **13** (2015) pp. 641 30 pages.

[301] E. Fuchs, M. Losada, Y. Nir, and Y. Viernik, *CP violation from $\tau$, $t$ and $b$ dimension-6 Yukawa couplings - interplay of baryogenesis, EDM and Higgs physics*, *JHEP* **05** (2020) 056 [`2003.00099`].

[302] F. Sattin, *Fick's law and fokker–planck equation in inhomogeneous environments*, *Physics Letters A* **372** (2008) 3941–3945.

[303] J.M. Cline, and K. Kainulainen, *A New source for electroweak baryogenesis in the MSSM*, *Phys. Rev. Lett.* **85** (2000) 5519 [`hep-ph/0002272`].

[304] K. Kainulainen, *CP-violating transport theory for electroweak baryogenesis with thermal corrections*, *JCAP* **11** (2021) 042 [`2108.08336`].

[305] J.M. Cline, M. Joyce, and K. Kainulainen, *Supersymmetric electroweak baryogenesis in the WKB approximation*, Phys. Lett. B **417** (1998) 79 [hep-ph/9708393], [Erratum: Phys.Lett.B 448, 321–321 (1999)].

[306] J.M. Cline, M. Joyce, and K. Kainulainen, *Supersymmetric electroweak baryogenesis*, JHEP **07** (2000) 018 [hep-ph/0006119].

[307] J.M. Cline, M. Joyce, and K. Kainulainen, *Erratum for 'Supersymmetric electroweak baryogenesis'*, [hep-ph/0110031].

[308] K. Kainulainen, T. Prokopec, M.G. Schmidt, and S. Weinstock, *First principle derivation of semiclassical force for electroweak baryogenesis*, JHEP **06** (2001) 031 [hep-ph/0105295].

[309] K. Kainulainen, T. Prokopec, M.G. Schmidt, and S. Weinstock, *Semiclassical force for electroweak baryogenesis: Three-dimensional derivation*, Phys. Rev. D **66** (2002) 043502 [hep-ph/0202177].

[310] K. Kainulainen, and O. Koskivaara, *Non-equilibrium dynamics of a scalar field with quantum backreaction*, JHEP **12** (2021) 190 [2105.09598].

[311] L. Fromme, and S.J. Huber, *Top transport in electroweak baryogenesis*, JHEP **03** (2007) 049 [hep-ph/0604159].

[312] J.M. Cline, *Is electroweak baryogenesis dead?*, [1704.08911].

[313] P. Virtanen, R. Gommers, T.E. Oliphant, M. Haberland, T. Reddy, D. Cournapeau et al., *SciPy 1.0: Fundamental Algorithms for Scientific Computing in Python*, Nature Methods **17** (2020) 261.

[314] A. Gutlein et al., *Solar and atmospheric neutrinos: Background sources for the direct dark matter search*, Astropart. Phys. **34** (2010) 90 [1003.5530].

[315] J. Kopp, *New Signals in Dark Matter Detectors*, J. Phys. Conf. Ser. **485** (2014) 012032 [1210.2703].

[316] J. Billard, L. Strigari, and E. Figueroa-Feliciano, *Implication of neutrino backgrounds on the reach of next generation dark matter direct detection experiments*, Phys. Rev. D **89** (2014) 023524 [1307.5458].

[317] S. Matsumoto, Y.-L.S. Tsai, and P.-Y. Tseng, *Light Fermionic WIMP Dark Matter with Light Scalar Mediator*, JHEP **07** (2019) 050 [1811.03292].

[318] M. Escudero, A. Berlin, D. Hooper, and M.-X. Lin, *Toward (Finally!) Ruling Out Z and Higgs Mediated Dark Matter Models*, JCAP **12** (2016) 029 [1609.09079].

[319] A. Djouadi, O. Lebedev, Y. Mambrini, and J. Quevillon, *Implications of LHC searches for Higgs–portal dark matter*, Phys. Lett. B **709** (2012) 65 [1112.3299].

[320] G.D. Moore, and M. Tassler, *The Sphaleron Rate in SU(N) Gauge Theory*, JHEP **02** (2011) 105 [1011.1167].

[321] A.J. Long, A. Tesi, and L.-T. Wang, *Baryogenesis at a Lepton-Number-Breaking Phase Transition*, JHEP **10** (2017) 095 [1703.04902].

[322] M. Joyce, T. Prokopec, and N. Turok, *Efficient electroweak baryogenesis from lepton transport*, *Phys. Lett. B* **338** (1994) 269 [`hep-ph/9401352`].

[323] Y.B..N. Zel'dovich, I. D., *The Hypothesis of Cores Retarded during Expansion and the Hot Cosmological Model*, *Soviet Astron. AJ (Engl. Transl. )*, **10** (1967) 602.

[324] B. Carr, F. Kuhnel, and M. Sandstad, *Primordial Black Holes as Dark Matter*, *Phys. Rev. D* **94** (2016) 083504 [`1607.06077`].

[325] A.M. Green, and B.J. Kavanagh, *Primordial Black Holes as a dark matter candidate*, *J. Phys. G* **48** (2021) 043001 [`2007.10722`].

[326] B. Carr, and F. Kuhnel, *Primordial Black Holes as Dark Matter: Recent Developments*, *Ann. Rev. Nucl. Part. Sci.* **70** (2020) 355 [`2006.02838`].

[327] P. Villanueva-Domingo, O. Mena, and S. Palomares-Ruiz, *A brief review on primordial black holes as dark matter*, *Front. Astron. Space Sci.* **8** (2021) 87 [`2103.12087`].

[328] A. Katz, J. Kopp, S. Sibiryakov, and W. Xue, *Femtolensing by Dark Matter Revisited*, *JCAP* **12** (2018) 005 [`1807.11495`].

[329] A.M. Green, *Supersymmetry and primordial black hole abundance constraints*, *Phys. Rev. D* **60** (1999) 063516 [`astro-ph/9903484`].

[330] M.Y. Khlopov, A. Barrau, and J. Grain, *Gravitino production by primordial black hole evaporation and constraints on the inhomogeneity of the early universe*, *Class. Quant. Grav.* **23** (2006) 1875 [`astro-ph/0406621`].

[331] T. Fujita, M. Kawasaki, K. Harigaya, and R. Matsuda, *Baryon asymmetry, dark matter, and density perturbation from primordial black holes*, *Phys. Rev. D* **89** (2014) 103501 [`1401.1909`].

[332] R. Allahverdi, J. Dent, and J. Osinski, *Nonthermal production of dark matter from primordial black holes*, *Phys. Rev. D* **97** (2018) 055013 [`1711.10511`].

[333] O. Lennon, J. March-Russell, R. Petrossian-Byrne, and H. Tillim, *Black Hole Genesis of Dark Matter*, *JCAP* **04** (2018) 009 [`1712.07664`].

[334] L. Morrison, S. Profumo, and Y. Yu, *Melanopogenesis: Dark Matter of (almost) any Mass and Baryonic Matter from the Evaporation of Primordial Black Holes weighing a Ton (or less)*, *JCAP* **05** (2019) 005 [`1812.10606`].

[335] D. Hooper, G. Krnjaic, and S.D. McDermott, *Dark Radiation and Superheavy Dark Matter from Black Hole Domination*, *JHEP* **08** (2019) 001 [`1905.01301`].

[336] I. Masina, *Dark matter and dark radiation from evaporating primordial black holes*, *Eur. Phys. J. Plus* **135** (2020) 552 [`2004.04740`].

[337] I. Baldes, Q. Decant, D.C. Hooper, and L. Lopez-Honorez, *Non-Cold Dark Matter from Primordial Black Hole Evaporation*, *JCAP* **08** (2020) 045 [`2004.14773`].

[338] P. Gondolo, P. Sandick, and B. Shams Es Haghi, *Effects of primordial black holes on dark matter models*, *Phys. Rev. D* **102** (2020) 095018 [`2009.02424`].

[339] N. Bernal, and O. Zapata, *Self-interacting Dark Matter from Primordial Black Holes*, *JCAP* **03** (2021) 007 [`2010.09725`].

[340] N. Bernal, and O. Zapata, *Dark Matter in the Time of Primordial Black Holes*, *JCAP* **03** (2021) 015 [`2011.12306`].

[341] A. Chaudhuri, and A. Dolgov, *PBH Evaporation, Baryon Asymmetry, and Dark Matter*, *J. Exp. Theor. Phys.* **133** (2021) 552 [`2001.11219`].

[342] D. Stojkovic, and K. Freese, *A Black hole solution to the cosmological monopole problem*, *Phys. Lett. B* **606** (2005) 251 [`hep-ph/0403248`].

[343] D. Stojkovic, K. Freese, and G.D. Starkman, *Holes in the walls: Primordial black holes as a solution to the cosmological domain wall problem*, *Phys. Rev. D* **72** (2005) 045012 [`hep-ph/0505026`].

[344] R. Bean, and J. Magueijo, *Could supermassive black holes be quintessential primordial black holes?*, *Phys. Rev. D* **66** (2002) 063505 [`astro-ph/0204486`].

[345] F. Hoyle, and J.V. Narlikar, *On the Formation of Elliptical Galaxies*, *Proceedings of the Royal Society of London Series A* **290** (1966) 177.

[346] J. Ryan, Michael P., *Is the Existence of a Galaxy Evidence for a Black Hole at its Center?*, *Astrophys. J. Lett.* **177** (1972) L79.

[347] B.J. Carr, and M.J. Rees, *Can pregalactic objects generate galaxies?*, *MNRAS* **206** (1984) 801.

[348] N. Afshordi, P. McDonald, and D.N. Spergel, *Primordial black holes as dark matter: The Power spectrum and evaporation of early structures*, *Astrophys. J. Lett.* **594** (2003) L71 [`astro-ph/0302035`].

[349] LIGO Scientific, Virgo collaboration, *Observation of Gravitational Waves from a Binary Black Hole Merger*, *Phys. Rev. Lett.* **116** (2016) 061102 [`1602.03837`].

[350] B.J. Carr, *The Primordial black hole mass spectrum*, *Astrophys. J.* **201** (1975) 1.

[351] P. Ivanov, P. Naselsky, and I. Novikov, *Inflation and primordial black holes as dark matter*, *Phys. Rev. D* **50** (1994) 7173.

[352] J. Garcia-Bellido, A.D. Linde, and D. Wands, *Density perturbations and black hole formation in hybrid inflation*, *Phys. Rev. D* **54** (1996) 6040 [`astro-ph/9605094`].

[353] J. Silk, and M.S. Turner, *Double Inflation*, *Phys. Rev. D* **35** (1987) 419.

[354] M. Kawasaki, N. Sugiyama, and T. Yanagida, *Primordial black hole formation in a double inflation model in supergravity*, *Phys. Rev. D* **57** (1998) 6050 [`hep-ph/9710259`].

[355] J. Yokoyama, *Formation of MACHO primordial black holes in inflationary cosmology*, *Astron. Astrophys.* **318** (1997) 673 [`astro-ph/9509027`].

[356] S. Pi, Y.-l. Zhang, Q.-G. Huang, and M. Sasaki, *Scalaron from $R^2$-gravity as a heavy field*, *JCAP* **05** (2018) 042 [`1712.09896`].

[357] S.W. Hawking, *Black Holes From Cosmic Strings*, *Phys. Lett. B* **231** (1989) 237.

[358] A. Polnarev, and R. Zembowicz, *Formation of Primordial Black Holes by Cosmic Strings*, *Phys. Rev. D* **43** (1991) 1106.

[359] J.H. MacGibbon, R.H. Brandenberger, and U.F. Wichoski, *Limits on black hole formation from cosmic string loops*, *Phys. Rev. D* **57** (1998) 2158 [astro-ph/9707146].

[360] S.G. Rubin, M.Y. Khlopov, and A.S. Sakharov, *Primordial black holes from nonequilibrium second order phase transition*, *Grav. Cosmol.* **6** (2000) 51 [hep-ph/0005271].

[361] S.G. Rubin, A.S. Sakharov, and M.Y. Khlopov, *The Formation of primary galactic nuclei during phase transitions in the early universe*, *J. Exp. Theor. Phys.* **91** (2001) 921 [hep-ph/0106187].

[362] A. Ashoorioon, A. Rostami, and J.T. Firouzjaee, *Examining the end of inflation with primordial black holes mass distribution and gravitational waves*, *Phys. Rev. D* **103** (2021) 123512 [2012.02817].

[363] R. Brandenberger, B. Cyr, and H. Jiao, *Intermediate mass black hole seeds from cosmic string loops*, *Phys. Rev. D* **104** (2021) 123501 [2103.14057].

[364] E. Cotner, and A. Kusenko, *Primordial black holes from supersymmetry in the early universe*, *Phys. Rev. Lett.* **119** (2017) 031103 [1612.02529].

[365] E. Cotner, A. Kusenko, M. Sasaki, and V. Takhistov, *Analytic Description of Primordial Black Hole Formation from Scalar Field Fragmentation*, *JCAP* **10** (2019) 077 [1907.10613].

[366] M. Crawford, and D.N. Schramm, *Spontaneous Generation of Density Perturbations in the Early Universe*, *Nature* **298** (1982) 538.

[367] H. Kodama, M. Sasaki, and K. Sato, *Abundance of Primordial Holes Produced by Cosmological First Order Phase Transition*, *Prog. Theor. Phys.* **68** (1982) 1979.

[368] I.G. Moss, *Black hole formation from colliding bubbles*, [gr-qc/9405045].

[369] B. Freivogel, G.T. Horowitz, and S. Shenker, *Colliding with a crunching bubble*, *JHEP* **05** (2007) 090 [hep-th/0703146].

[370] S.W. Hawking, I.G. Moss, and J.M. Stewart, *Bubble Collisions in the Very Early Universe*, *Phys. Rev. D* **26** (1982) 2681.

[371] M.C. Johnson, H.V. Peiris, and L. Lehner, *Determining the outcome of cosmic bubble collisions in full General Relativity*, *Phys. Rev. D* **85** (2012) 083516 [1112.4487].

[372] A. Kusenko, M. Sasaki, S. Sugiyama, M. Takada, V. Takhistov, and E. Vitagliano, *Exploring Primordial Black Holes from the Multiverse with Optical Telescopes*, *Phys. Rev. Lett.* **125** (2020) 181304 [2001.09160].

[373] J.-P. Hong, S. Jung, and K.-P. Xie, *Fermi-ball dark matter from a first-order phase transition*, *Phys. Rev. D* **102** (2020) 075028 [2008.04430].

[374] C. Gross, G. Landini, A. Strumia, and D. Teresi, *Dark Matter as dark dwarfs and other macroscopic objects: multiverse relics?*, *JHEP* **09** (2021) 033 [2105.02840].

[375] P. Asadi, E.D. Kramer, E. Kuflik, G.W. Ridgway, T.R. Slatyer, and J. Smirnov, *Thermal squeezeout of dark matter*, *Phys. Rev. D* **104** (2021) 095013 [`2103.09827`].

[376] R. Kippenhahn, and A. Weigert, *Stellar Structure and Evolution*, Springer-Verlag Berlin (1994).

[377] J.M. Gerard, and S. Pireaux, *The Observable light deflection angle*, [`gr-qc/9907034`].

[378] G.D. Moore, and T. Prokopec, *How fast can the wall move? A Study of the electroweak phase transition dynamics*, *Phys. Rev. D* **52** (1995) 7182 [`hep-ph/9506475`].

[379] G.D. Moore, and T. Prokopec, *Bubble wall velocity in a first order electroweak phase transition*, *Phys. Rev. Lett.* **75** (1995) 777 [`hep-ph/9503296`].

[380] J. Dormand, and P. Prince, *A family of embedded runge-kutta formulae*, *Journal of Computational and Applied Mathematics* **6** (1980) 19.

[381] F. Capela, M. Pshirkov, and P. Tinyakov, *Constraints on Primordial Black Holes as Dark Matter Candidates from Star Formation*, *Phys. Rev. D* **87** (2013) 023507 [`1209.6021`].

[382] F. Capela, M. Pshirkov, and P. Tinyakov, *Constraints on primordial black holes as dark matter candidates from capture by neutron stars*, *Phys. Rev. D* **87** (2013) 123524 [`1301.4984`].

[383] P.W. Graham, S. Rajendran, and J. Varela, *Dark Matter Triggers of Supernovae*, *Phys. Rev. D* **92** (2015) 063007 [`1505.04444`].

[384] P. Mróz, A. Udalski, J. Skowron, R. Poleski, S. Kozłowski, M.K. Szymański et al., *No large population of unbound or wide-orbit jupiter-mass planets*, *Nature* **548** (2017) 183–186.

[385] H. Niikura, M. Takada, S. Yokoyama, T. Sumi, and S. Masaki, *Constraints on Earth-mass primordial black holes from OGLE 5-year microlensing events*, *Phys. Rev. D* **99** (2019) 083503 [`1901.07120`].

[386] SENSEI collaboration, *SENSEI: Direct-Detection Results on sub-GeV Dark Matter from a New Skipper-CCD*, *Phys. Rev. Lett.* **125** (2020) 171802 [`2004.11378`].

[387] T. Lin, *Dark matter models and direct detection*, *PoS* **333** (2019) 009 [`1904.07915`].

[388] K. Schutz, and K.M. Zurek, *Detectability of Light Dark Matter with Superfluid Helium*, *Phys. Rev. Lett.* **117** (2016) 121302 [`1604.08206`].

[389] S. Knapen, T. Lin, M. Pyle, and K.M. Zurek, *Detection of Light Dark Matter With Optical Phonons in Polar Materials*, *Phys. Lett. B* **785** (2018) 386 [`1712.06598`].

[390] R.A. Alpher, H. Bethe, and G. Gamow, *The origin of chemical elements*, *Phys. Rev.* **73** (1948) 803.

[391] R.V. Wagoner, W.A. Fowler, and F. Hoyle, *On the Synthesis of elements at very high temperatures*, *Astrophys. J.* **148** (1967) 3.

[392] V.F. Shvartsman, *Density of relict particles with zero rest mass in the universe*, *Pisma Zh. Eksp. Teor. Fiz.* **9** (1969) 315.

[393] R.J. Scherrer, and M.S. Turner, *Primordial Nucleosynthesis with Decaying Particles. 1. Entropy Producing Decays. 2. Inert Decays*, *Astrophys. J.* **331** (1988) 19.

[394] B.D. Fields, P. Molaro, and S. Sarkar, *Big-Bang Nucleosynthesis*, *Chin. Phys. C* **38** (2014) 339 [1412.1408].

[395] Particle Data Group collaboration, *Review of Particle Physics*, *Phys. Rev. D* **98** (2018) 030001.

[396] G. Mangano, G. Miele, S. Pastor, T. Pinto, O. Pisanti, and P.D. Serpico, *Relic neutrino decoupling including flavor oscillations*, *Nucl. Phys. B* **729** (2005) 221 [hep-ph/0506164].

[397] G. Steigman, D.N. Schramm, and J.E. Gunn, *Cosmological Limits to the Number of Massive Leptons*, *Phys. Lett. B* **66** (1977) 202.

[398] Particle Data Group collaboration, *Review of Particle Physics*, *Chin. Phys. C* **40** (2016) 100001.

[399] M. Hufnagel, K. Schmidt-Hoberg, and S. Wild, *BBN constraints on MeV-scale dark sectors. Part I. Sterile decays*, *JCAP* **02** (2018) 044 [1712.03972].

[400] A.G. Riess et al., *New Parallaxes of Galactic Cepheids from Spatially Scanning the Hubble Space Telescope: Implications for the Hubble Constant*, *Astrophys. J.* **855** (2018) 136 [1801.01120].

[401] E.W. Kolb, and M.S. Turner, *The Early Universe*, vol. 69, Westview Press (1990), 10.1201/9780429492860.

[402] A. Berlin, and N. Blinov, *Thermal Dark Matter Below an MeV*, *Phys. Rev. Lett.* **120** (2018) 021801 [1706.07046].

[403] A. Berlin, and N. Blinov, *Thermal neutrino portal to sub-MeV dark matter*, *Phys. Rev. D* **99** (2019) 095030 [1807.04282].

[404] J. Kozaczuk, S. Profumo, L.S. Haskins, and C.L. Wainwright, *Cosmological Phase Transitions and their Properties in the NMSSM*, *JHEP* **01** (2015) 144 [1407.4134].

[405] N. Blinov, J. Kozaczuk, D.E. Morrissey, and C. Tamarit, *Electroweak Baryogenesis from Exotic Electroweak Symmetry Breaking*, *Phys. Rev. D* **92** (2015) 035012 [1504.05195].

[406] J. Kozaczuk, *Bubble Expansion and the Viability of Singlet-Driven Electroweak Baryogenesis*, *JHEP* **10** (2015) 135 [1506.04741].

[407] K. Hashino, M. Kakizaki, S. Kanemura, P. Ko, and T. Matsui, *Gravitational waves from first order electroweak phase transition in models with the U(1)$_X$ gauge symmetry*, *JHEP* **06** (2018) 088 [1802.02947].

[408] A. Addazi, and A. Marciano, *Gravitational waves from dark first order phase transitions and dark photons*, *Chin. Phys. C* **42** (2018) 023107 [1703.03248].

[409] E. Madge, and P. Schwaller, *Leptophilic dark matter from gauged lepton number: Phenomenology and gravitational wave signatures*, *JHEP* **02** (2019) 048 [1809.09110].

[410] DUNE collaboration, *Long-Baseline Neutrino Facility (LBNF) and Deep Underground Neutrino Experiment (DUNE): Conceptual Design Report, Volume 1: The LBNF and DUNE Projects*, [1601.05471].

[411] DUNE collaboration, *Long-Baseline Neutrino Facility (LBNF) and Deep Underground Neutrino Experiment (DUNE): Conceptual Design Report, Volume 3: Long-Baseline Neutrino Facility for DUNE June 24, 2015*, [1601.05823].

[412] DUNE collaboration, *Long-Baseline Neutrino Facility (LBNF) and Deep Underground Neutrino Experiment (DUNE): Conceptual Design Report, Volume 4 The DUNE Detectors at LBNF*, [1601.02984].

[413] DUNE collaboration, *Deep Underground Neutrino Experiment (DUNE), Far Detector Technical Design Report, Volume I Introduction to DUNE*, *JINST* **15** (2020) T08008 [2002.02967].

[414] DUNE collaboration, *Deep Underground Neutrino Experiment (DUNE) Near Detector Conceptual Design Report*, *Instruments* **5** (2021) 31 [2103.13910].

[415] DUNE collaboration, *Long-Baseline Neutrino Facility (LBNF) and Deep Underground Neutrino Experiment (DUNE): Conceptual Design Report, Volume 2: The Physics Program for DUNE at LBNF*, [1512.06148].

[416] NOvA collaboration, *An Improved Measurement of Neutrino Oscillation Parameters by the NOvA Experiment*, [2108.08219].

[417] G. Mention, M. Fechner, T. Lasserre, T.A. Mueller, D. Lhuillier, M. Cribier et al., *The Reactor Antineutrino Anomaly*, *Phys. Rev. D* **83** (2011) 073006 [1101.2755].

[418] LSND collaboration, *Evidence for neutrino oscillations from the observation of $\bar{\nu}_e$ appearance in a $\bar{\nu}_\mu$ beam*, *Phys. Rev. D* **64** (2001) 112007 [hep-ex/0104049].

[419] C. Giunti, and M. Laveder, *Statistical Significance of the Gallium Anomaly*, *Phys. Rev. C* **83** (2011) 065504 [1006.3244].

[420] J.N. Abdurashitov et al., *Measurement of the response of a Ga solar neutrino experiment to neutrinos from an Ar-37 source*, *Phys. Rev. C* **73** (2006) 045805 [nucl-ex/0512041].

[421] J.C. Pati, and A. Salam, *Is Baryon Number Conserved?*, *Phys. Rev. Lett.* **31** (1973) 661.

[422] S. Dimopoulos, S. Raby, and F. Wilczek, *Proton Decay in Supersymmetric Models*, *Phys. Lett. B* **112** (1982) 133.

[423] Super-Kamiokande collaboration, *Search for Proton Decay via p —> e+ pi0 and p —> mu+ pi0 in a Large Water Cherenkov Detector*, *Phys. Rev. Lett.* **102** (2009) 141801 [0903.0676].

[424] B. Bajc, J. Hisano, T. Kuwahara, and Y. Omura, *Threshold corrections to dimension-six proton decay operators in non-minimal SUSY SU (5) GUTs*, *Nucl. Phys. B* **910** (2016) 1 [1603.03568].

[425] R.M. Bionta et al., *Observation of a Neutrino Burst in Coincidence with Supernova SN 1987a in the Large Magellanic Cloud*, *Phys. Rev. Lett.* **58** (1987) 1494.

[426] Kamiokande-II collaboration, *Observation of a Neutrino Burst from the Supernova SN 1987a*, *Phys. Rev. Lett.* **58** (1987) 1490.

[427] S. Chandrasekhar, *On stars, their evolution and their stability*, *Rev. Mod. Phys.* **56** (1984) 137.

[428] SENSEI collaboration, *Single-electron and single-photon sensitivity with a silicon Skipper CCD*, *Phys. Rev. Lett.* **119** (2017) 131802 [1706.00028].

[429] BaBar collaboration, *Search for Invisible Decays of a Dark Photon Produced in $e^+e^-$ Collisions at BaBar*, *Phys. Rev. Lett.* **119** (2017) 131804 [1702.03327].

[430] D. Banerjee et al., *Dark matter search in missing energy events with NA64*, *Phys. Rev. Lett.* **123** (2019) 121801 [1906.00176].

[431] M.L. Graesser, I.M. Shoemaker, and L. Vecchi, *A Dark Force for Baryons*, [1107.2666].

[432] BaBar collaboration, *A Search for Invisible Decays of the Upsilon(1S)*, *Phys. Rev. Lett.* **103** (2009) 251801 [0908.2840].

[433] P. Coloma, B.A. Dobrescu, C. Frugiuele, and R. Harnik, *Dark matter beams at LBNF*, *JHEP* **04** (2016) 047 [1512.03852].

[434] P. Ballett, T. Boschi, and S. Pascoli, *Heavy Neutral Leptons from low-scale seesaws at the DUNE Near Detector*, *JHEP* **03** (2020) 111 [1905.00284].

[435] I. Krasnov, *DUNE prospects in the search for sterile neutrinos*, *Phys. Rev. D* **100** (2019) 075023 [1902.06099].

[436] P. Coloma, E. Fernández-Martínez, M. González-López, J. Hernández-García, and Z. Pavlovic, *GeV-scale neutrinos: interactions with mesons and DUNE sensitivity*, *Eur. Phys. J. C* **81** (2021) 78 [2007.03701].

[437] E.C.G. Stueckelberg, *Interaction energy in electrodynamics and in the field theory of nuclear forces*, *Helv. Phys. Acta* **11** (1938) 225.

[438] S.L. Glashow, *Positronium Versus the Mirror Universe*, *Phys. Lett. B* **167** (1986) 35.

[439] R. Essig et al., *Working Group Report: New Light Weakly Coupled Particles*, in *Community Summer Study 2013: Snowmass on the Mississippi*, 10, 2013 [1311.0029].

[440] E. Izaguirre, G. Krnjaic, P. Schuster, and N. Toro, *Analyzing the Discovery Potential for Light Dark Matter*, *Phys. Rev. Lett.* **115** (2015) 251301 [1505.00011].

[441] T. Lin, H.-B. Yu, and K.M. Zurek, *On Symmetric and Asymmetric Light Dark Matter*, *Phys. Rev. D* **85** (2012) 063503 [1111.0293].

[442] PLANCK collaboration, *Planck 2015 results. XIII. Cosmological parameters*, *Astron. Astrophys.* **594** (2016) A13 [1502.01589].

[443] T.R. Slatyer, *Indirect dark matter signatures in the cosmic dark ages. I. Generalizing the bound on s-wave dark matter annihilation from Planck results*, *Phys. Rev. D* **93** (2016) 023527 [1506.03811].

[444] T.R. Slatyer, *Indirect Dark Matter Signatures in the Cosmic Dark Ages II. Ionization, Heating and Photon Production from Arbitrary Energy Injections*, *Phys. Rev. D* **93** (2016) 023521 [1506.03812].

[445] P. deNiverville, M. Pospelov, and A. Ritz, *Observing a light dark matter beam with neutrino experiments*, *Phys. Rev. D* **84** (2011) 075020 [1107.4580].

[446] P. deNiverville, and C. Frugiuele, *Hunting sub-GeV dark matter with the NOνA near detector*, *Phys. Rev. D* **99** (2019) 051701 [1807.06501].

[447] L. Buonocore, C. Frugiuele, and P. deNiverville, *Hunt for sub-GeV dark matter at neutrino facilities: A survey of past and present experiments*, *Phys. Rev. D* **102** (2020) 035006 [1912.09346].

[448] S.L. Adler, *Axial vector vertex in spinor electrodynamics*, *Phys. Rev.* **177** (1969) 2426.

[449] J.S. Bell, and R. Jackiw, *A PCAC puzzle: $\pi^0 \to \gamma\gamma$ in the $\sigma$ model*, *Nuovo Cim. A* **60** (1969) 47.

[450] A. Berlin, S. Gori, P. Schuster, and N. Toro, *Dark Sectors at the Fermilab SeaQuest Experiment*, *Phys. Rev. D* **98** (2018) 035011 [1804.00661].

[451] SHiP collaboration, *Sensitivity of the SHiP experiment to light dark matter*, *JHEP* **04** (2021) 199 [2010.11057].

[452] GEANT4 collaboration, *GEANT4–a simulation toolkit*, *Nucl. Instrum. Meth. A* **506** (2003) 250.

[453] A. Celentano, L. Darmé, L. Marsicano, and E. Nardi, *New production channels for light dark matter in hadronic showers*, *Phys. Rev. D* **102** (2020) 075026 [2006.09419].

[454] V. De Romeri, K.J. Kelly, and P.A.N. Machado, *DUNE-PRISM Sensitivity to Light Dark Matter*, *Phys. Rev. D* **100** (2019) 095010 [1903.10505].

[455] T. Sjöstrand, S. Ask, J.R. Christiansen, R. Corke, N. Desai, P. Ilten et al., *An introduction to PYTHIA 8.2*, *Comput. Phys. Commun.* **191** (2015) 159 [1410.3012].

[456] Y. Kahn, G. Krnjaic, J. Thaler, and M. Toups, *DAEδALUS and dark matter detection*, *Phys. Rev. D* **91** (2015) 055006 [1411.1055].

[457] S. Gardner, R.J. Holt, and A.S. Tadepalli, *New Prospects in Fixed Target Searches for Dark Forces with the SeaQuest Experiment at Fermilab*, *Phys. Rev. D* **93** (2016) 115015 [1509.00050].

[458] E. Fermi, *On the Theory of the impact between atoms and electrically charged particles*, *Z. Phys.* **29** (1924) 315.

[459] E.J. Williams, *Nature of the high-energy particles of penetrating radiation and status of ionization and radiation formulae*, *Phys. Rev.* **45** (1934) 729.

[460] C.F. von Weizsacker, *Radiation emitted in collisions of very fast electrons*, *Z. Phys.* **88** (1934) 612.

[461] A. Faessler, M.I. Krivoruchenko, and B.V. Martemyanov, *Once more on electromagnetic form factors of nucleons in extended vector meson dominance model*, *Phys. Rev. C* **82** (2010) 038201 [0910.5589].

[462] D.E. Morrissey, and A.P. Spray, *New Limits on Light Hidden Sectors from Fixed-Target Experiments*, *JHEP* **06** (2014) 083 [1402.4817].

[463] SHiP collaboration, *Sensitivity of the SHiP experiment to dark photons decaying to a pair of charged particles*, *Eur. Phys. J. C* **81** (2021) 451 [2011.05115].

[464] L. Buonocore, C. Frugiuele, F. Maltoni, O. Mattelaer, and F. Tramontano, *Event generation for beam dump experiments*, *JHEP* **05** (2019) 028 [1812.06771].

[465] J. Blümlein, and J. Brunner, *New Exclusion Limits on Dark Gauge Forces from Proton Bremsstrahlung in Beam-Dump Data*, *Phys. Lett. B* **731** (2014) 320 [1311.3870].

[466] P. deNiverville, C.-Y. Chen, M. Pospelov, and A. Ritz, *Light dark matter in neutrino beams: production modelling and scattering signatures at MiniBooNE, T2K and SHiP*, *Phys. Rev. D* **95** (2017) 035006 [1609.01770].

[467] DUNE collaboration, *Simulated DUNE Neutrino Fluxes*, ROOT Ntuple files available at https://home.fnal.gov/~ljf26/DUNEFluxes/. We use the v3r5p4 release which has also been used for the DUNE TDR.

[468] C. Andreopoulos et al., *The GENIE Neutrino Monte Carlo Generator*, *Nucl. Instrum. Meth. A* **614** (2010) 87 [0905.2517].

[469] O. Tomalak, and R.J. Hill, *Theory of elastic neutrino-electron scattering*, *Phys. Rev. D* **101** (2020) 033006 [1907.03379].

[470] G. Cowan, K. Cranmer, E. Gross, and O. Vitells, *Asymptotic formulae for likelihood-based tests of new physics*, *Eur. Phys. J. C* **71** (2011) 1554 [1007.1727], [Erratum: Eur.Phys.J.C 73, 2501 (2013)].

[471] G. Lanfranchi, M. Pospelov, and P. Schuster, *The Search for Feebly Interacting Particles*, *Ann. Rev. Nucl. Part. Sci.* **71** (2021) 279 [2011.02157].

[472] P. Fileviez Perez, and M.B. Wise, *Breaking Local Baryon and Lepton Number at the TeV Scale*, *JHEP* **08** (2011) 068 [1106.0343].

[473] B.A. Dobrescu, and C. Frugiuele, *Hidden GeV-scale interactions of quarks*, *Phys. Rev. Lett.* **113** (2014) 061801 [1404.3947].

[474] J.A. Dror, R. Lasenby, and M. Pospelov, *Dark forces coupled to nonconserved currents*, *Phys. Rev. D* **96** (2017) 075036 [1707.01503].

[475] L. Michaels, and F. Yu, *Probing new U(1) gauge symmetries via exotic $Z \to Z'\gamma$ decays*, *JHEP* **03** (2021) 120 [2010.00021].

[476] NA61/SHINE collaboration, *Measurements of Cross Sections and Charged Pion Spectra in Proton-Carbon Interactions at 31 GeV/c*, *Phys. Rev. C* **84** (2011) 034604 [1102.0983].

[477] Particle Data Group collaboration, *Review of Particle Physics*, *Phys. Rev. D* **98** (2018) 030001.

[478] B.A. Dobrescu, and C. Frugiuele, *GeV-Scale Dark Matter: Production at the Main Injector*, *JHEP* **02** (2015) 019 [1410.1566].

[479] MiniBooNE, MINOS collaboration, *First Measurement of $\nu_\mu$ and $\nu_e$ Events in an Off-Axis Horn-Focused Neutrino Beam*, *Phys. Rev. Lett.* **102** (2009) 211801 [0809.2447].

*Curriculum vitae removed from online version.*

*Curriculum vitae removed from online version.*